\documentclass[fleqn]{unmeethesis}

\usepackage[usenames,dvipsnames]{xcolor}
\usepackage{graphicx}
\usepackage[numbered,autolinebreaks,useliterate]{mcode}
\usepackage{listings}
\usepackage{amsmath}

\begin{document}

\frontmatter



\title{Development and Characterization of a $^{171}$Yb$^+$ Miniature Ion Trap Frequency Standard}

\author{Heather L. Partner}

\degreesubject{Ph.D., Physics}

\degree{Doctor of Philosophy \\ Physics\\  }

\documenttype{Dissertation}

\previousdegrees{B.S., Physics, Indiana University of Pennsylvania, 2003 \\
                 B.S., Mathematics, Indiana University of Pennsylvania, 2003 \\
                 MASt, Mathematics, University of Cambridge, 2012}

\date{August, \thisyear}

\maketitle

\makecopyright

\begin{dedication}
To my parents Bruce and Melody, and especially to Carlos.\\
 \vspace{0.4in}
\begin{centering}
``Si tu no est\'as aqu\'i a mi lado, de nada vale ya, lo que s\'e y mi doctorado." \\
\end{centering}
 -Tony Dize
 \end{dedication}

\begin{acknowledgments}
   \vspace{1.1in}
    \pagenumbering{roman}
	\setcounter{page}{4}
    
I would first like to thank my advisor, Peter Schwindt, for all his guidance, advice and patience while working with me throughout my time at Sandia. Sandia has been a fantastic environment in which to work, and in particular, it was a pleasure to work with Peter, to experience and learn from his enthusiasm, professionalism, and serious perseverance and creativity for making stuff work.  I would also like to thank Yuan-Yu Jau, who as the postdoc on this project has shared his amazing experimental skills and vast general knowledge with me, as well being a very nice person to work (and travel) with.  The rest of the group at Sandia has also been supportive and enlightening, including Dave Moehring, Dan Stick, Grant Biedermann, Matt Blain, Mike Descour, Todd Barrick, Jon Sterk, Paul Parazzolli, George Burns, Francisco Benito, and many others.  Working at Sandia can be quite different from working at the University and these people made it a welcoming, stimulating, and productive environment.  

At UNM, my opportunity to work at Sandia came through my supporters at CQUIC, especially my formal academic advisor Carl Caves, and Ivan Deutsch.    In particular, Ivan, though he was never officially my advisor, has been a willing and valuable source of advice through many difficult situations, and I am truly grateful for his guidance.  Our department chair, Bernd Basselleck, has also been extremely supportive in times of trouble and otherwise.  Without them I may never have finished.  I have also had many enjoyable interactions (both professional and personal) with the members of CQUIC going back to the beginning of my seven year tenure, including Vaibhav Madhok, Leigh Norris, Joshua Combes, Alexandre Tacla, Jonas Anderson, Chris Cesare, Seth Merkel, Iris Reichenbach, Krittika Goyal, Collin Trail, Bryan Eastin, and Steve Flammia.  In addition, I am very grateful to many people who I met and worked with during my more formative years as a physicist, including Dana Berkeland, Malcolm Boshier, Jayne Giniewicz, and Larry Freeman.

I will not forget those several enjoyable years in a research group with Rob Cook, Ben Baragiola, Brad Chase, Brigette Black, Tom Jones, Thomas Loyd, and Paul Martin, where we had more fun than any other group in the department despite the circumstances.  Rob in particular has shared much of my grad school story and has always been friendly, helpful, and supportive. Besides them I thank Ziya Kalay for his friendship, as well as Steve Tremblay, Fonda Day, Katie Richardson, Laura Zschaechner, Jessica Metcalfe, Daniel Riofr\'io, many people mentioned already, and all the people I've shared these years with, for their friendship and help.

I thank my parents, Bruce \& Melody, for their bottomless support during my academic adventures,  as well and Kristi, Shawn, and especially Charity, who though far away, have always been supportive.  
Finally, I thank Carlos Riofr\'io for sharing his love, wisdom, and patience with me through all these years.
\\
\\
\\

\emph{Sandia National Laboratories is a multi-program laboratory managed and operated by Sandia Corporation, a wholly owned subsidiary of Lockheed Martin Corporation, for the U.S. Department of Energy's National Nuclear Security Administration under contract DE-AC04-94AL85000. }

\emph{ The views expressed are those of the author and do not reflect the official policy or position of the Department of Defense or the U.S. Government.  Approved for Public Release, Distribution Unlimited.}

  \end{acknowledgments}

\maketitleabstract 

\begin{abstract}
   
This dissertation reports on the development of a low-power, high-stability miniature atomic frequency standard based on $^{171}$Yb$^+$ ions.  The ions are buffer-gas cooled and held in a linear quadrupole trap that is integrated into a sealed, getter-pumped vacuum package, and interrogated on the 12.6 GHz hyperfine transition.  We hope to achieve a long-term fractional frequency stability of 10$^{-14}$ with this miniature clock while consuming only 50 mW of power and occupying a volume of 5 cm$^3$, as part of a project funded to rapidly develop an advanced miniaturized frequency standard that has exceptional long-term stability.

I discuss our progress through several years of development on this project. We began by building a relatively conventional tabletop clock system to act as a ``test bed" for future components and for testing new techniques in a controlled environment. We moved on to develop and test several designs of miniature ion-trap vacuum packages, while also developing techniques for various aspects of the clock operation, including ion loading, laser and magnetic field stabilization, and a low power ion trap drive. The ion traps were modeled using boundary element software to assist with the design and parameter optimization of new trap geometries. We expect a novel trap geometry made using a material that is new to ion traps to lead to an exceptionally small ion trap vacuum package in the next phase of the project. 

To achieve the long-term stability required, we have also considered the sensitivity of the clock frequency to magnetic fields. A study of the motion of the individual ions in a room-temperature cloud in the trap was performed. The purpose of this simulation was to understand the effect of both spatially varying and constant magnetic fields on the clock resonance and therefore the operation of the clock.  These effects were studied experimentally and theoretically for several traps. 

In summary, this dissertation is a contribution to the design, development, and testing of a $^{171}$Yb$^+$ ion cloud frequency standard and related techniques, including analyses of trap geometries and parameters, modeling of the ion motion, and the practical operation of the clock.

\end{abstract}

\tableofcontents
\listoffigures
\listoftables


\mainmatter

\chapter{Introduction}
\label{intro}


   
Atomic frequency standards are of fundamental importance on many levels: on the political scale (national standards, Universal Coordinated Time (UTC), leap seconds \cite{Guinot2011,Seidelmann2011}), in research \cite{Diddams19112004}, and for industry and the individual (GPS, telecommunications \cite{Lew2011,Levine2011}).  Besides these  practical applications, frequency standards can be used to probe fundamental physical constants \cite{PhysRevLett.90.150801,0034-4885-70-9-R01}.  A broad array of frequency standards is available commercially, and an even broader array exists only in research settings.  The historical development of atomic references since the pioneering work of Rabi and Ramsey in interrogating atomic frequencies in regions of oscillating fields \cite{0026-1394-42-3-S01,Ramsey1956} has been to make increasingly stable and accurate frequency standards that all operate on the same basic principle, with the same goal: obtaining a signal that indicates an interval of time that is as perfect in its duration as the fundamental separation of the energy levels of a atom.

Historically, atomic clocks have been very large and very stable \cite{1408302}
, or, to accomodate applications that do not have a need for the extreme stability or have limited space, they have been smaller and (usually) less stable \cite{367393}. 
Along with the general trend to make technologies smaller and smaller that started with the transistor and the silicon wafer, there has been a push in recent years to make atomic clocks that are quite small.  The Defense Advanced Research Projects Agency (DARPA) project referred to as CSAC (Chip-Scale Atomic Clock), which ran during the 2000s and  was very successful, aimed to develop a frequency standard on a chip, taking up 1 cm$^3$ and using 30 mW of power, and having 10$^{-11}$ stability at one hour (e.g. \cite{5168247}).  CSACs are now commercially available.    The more recent Integrated Micro Primary Atomic Clock Technology (IMPACT) project (also a DARPA-funded project) has a greater focus on long-term stability.  The volume and power requirements are slightly relaxed compared to the CSAC project, but the goal is to create a standard that has nearly the same stability as a cesium beam clock. CSAC clocks are on the micro scale but have significantly compromised long-term stability; IMPACT clocks will not compromise on long-term stability, but will still deliver small size and low power.  If successful, the result of the IMPACT project will fill a technology gap in the parameter space of stability versus size.
 This dissertation is the result of work done over a period of several years on the development of a miniature ion-trap frequency standard for IMPACT.  
%


Our approach to this problem is to use trapped ions as an atomic source.  Our collaborators from the Jet Propulsion Laboratory (JPL) have decades of experience in developing these types of clocks, and in particular have developed an ion clock using $^{199}$Hg ions, with the eventual goal of use in space \cite{4383369}, that is even more stable than a cesium beam standard.  This JPL space-ion-clock has been miniaturized significantly from the laboratory scale with a physics package that is approximately 1 liter in size, and uses a lamp to do state selection and detection \cite{4623080,5168141}.  Because of the promise in the miniaturization demonstrated with this clock, it acts as a model for our own design for IMPACT. We believe that by trading off some of the stability of the JPL-style ion clock for a reduction in volume, we can make a clock that meets the requirements of the IMPACT project.   Because of power and size requirements, the lamp must be replaced by lasers, and lasers at the mercury wavelength (194 nm) are not readily available.  This is one of the reasons we chose Yb as the ion to use while otherwise following JPL's example.  The ion trap as a system is attractive for many reasons: it is robust, it is capable of storing ions for long times (days or weeks), the ions are well isolated and stable yet the vacuum requirements are more relaxed than for neutral atom trapping, and the ions do not require laser cooing in order to be trapped.  

\begin{figure}
   \centering
  \includegraphics[scale=0.7]{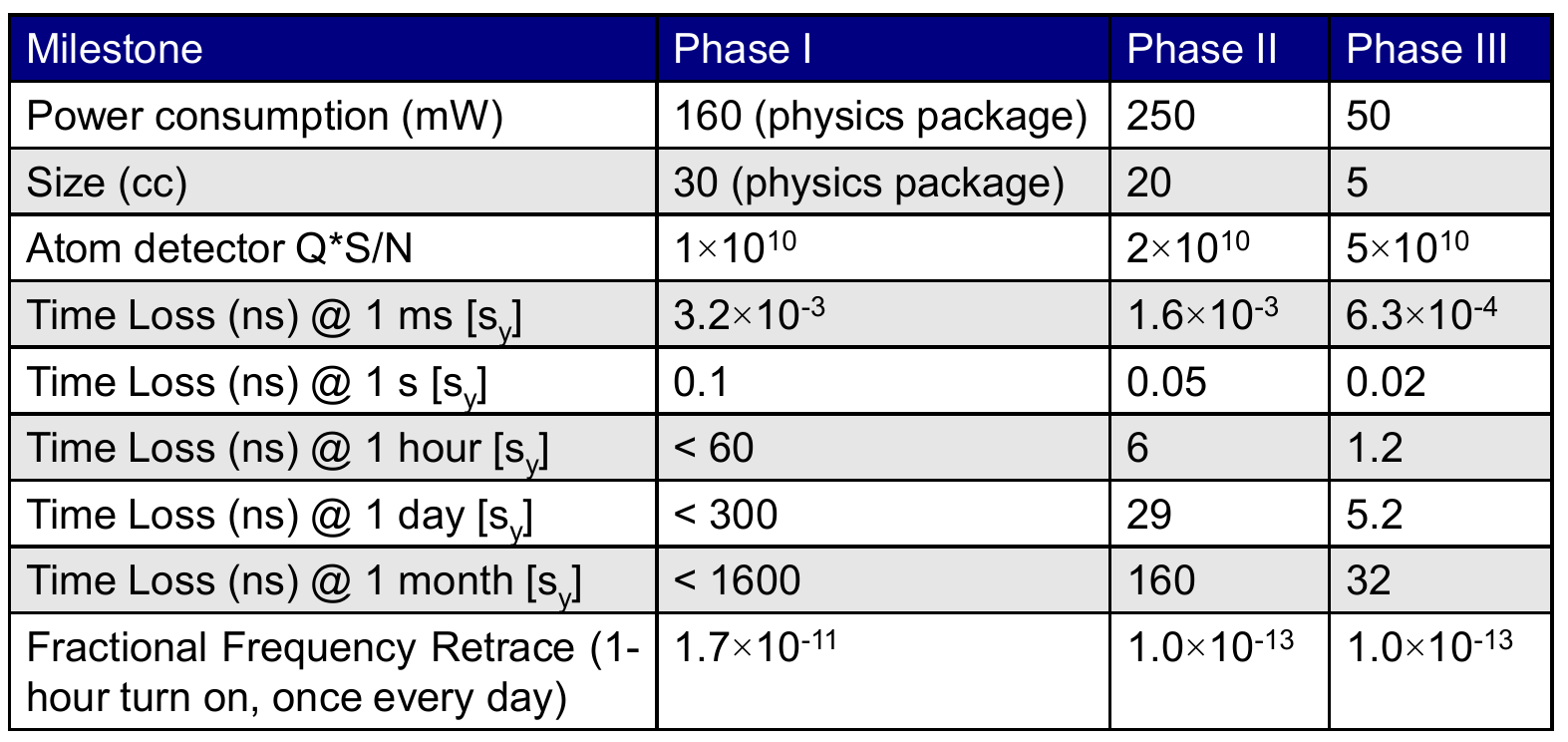}
   \caption{Table of the IMPACT performance goals for the three phases.  Each phase lasts 12-18 months.}
  \label{fig:DARPAgoals}
\end{figure}

The IMPACT project occurs in three Phases; as this dissertation is being written, we are approaching the end of Phase II.  The goals for each phase of the project are summarized in the table in Fig. \ref{fig:DARPAgoals}.  In the first phase of the project, with the collaboration of JPL, we were able to produce a 10 cm$^3$ ion trap vacuum package with no active pumps required, where ions in the trap have trapping lifetimes of more than one month.  To my knowledge, this is an unprecedented accomplishment for miniaturization of a complete ion-trap vacuum system.  Although the clock performance in this package approaches the Phase II requirements, the system for the entire functioning clock takes up a significant volume, since we were only in the elementary stages of integration at the end of Phase I.  In the second phase, however, we have produced a similar package that is even smaller (3 cm$^3$) and is already exhibiting similar behavior to the Phase I package.  The system will be more integrated by the end of Phase II, including a detection system (PMT + filters + lenses), ion loading system (oven + ionization process), low-power RF trap driving system, and miniaturized lasers and a single control electronics board for operating the clock.   Further, a study that we performed on the ion motion in the trap and its effect on the clock stability in the presence of magnetic field gradients will be useful as we focus on long-term stability going forward.   

This dissertation concerns a project that is very technological in nature, and as part of the technological development process, there is a lot of trial and error.  I have tried in the experimental chapters to express this trial and error process, which is the only way, along with simulation and prediction, that we have to understand how to best make our clock smaller without compromising its quality. 

\subsubsection{Organizational Overview}

In Chapters \ref{ch:trapping} and \ref{ch:clocks} I will discuss the background information about the two primary system concepts needed for this project: trapped ions and atomic clocks.  In Chapter \ref{ch:testBed}, I discuss all of the experimental aspects of the ion trap and the clock, in the context of our initial tabletop ``test bed" system, which we built before focusing on miniaturization.  
 There are many aspects to the miniaturization, and since many of these I am only peripherally involved in, I will not discuss all of them in detail.  However, Chapter \ref{ch:packages} includes a detailed description of the small packages that we designed, tested, and iterated to discover the best techniques for building and operating a miniature system, which was my primary experimental focus during my tenure at Sandia.  Our work on the current clock system should lead to a publication in the near future, when we have completed its assembly and testing phases, since it is already showing promising performance. In Chapter \ref{ch:biasfield}, I describe our study of the magnetic field effects on miniaturized ion-trap frequency standards.  This chapter is the subject of a paper in preparation for submission to a peer-reviewed journal.  Chapter \ref{ch:conclusions} briefly concludes.
%
%
%

\chapter{Trapped Ions}
\label{ch:trapping}

Before we can create a trapped ion frequency standard, we need a well understood, reliable system for trapping ions.  In our case in particular, we want to trap a large number of ions so that we maximize the signal to noise ratio of our clock signal.  This is very different from, for example, ion trapping for quantum computation \cite{PhysRevLett.74.4091,Eschner:03,blatt_entangled_2008}, or a single-ion frequency standard \cite{PhysRevLett.104.070802}, in which the focus is to trap one to a few tens of ions in a very controlled way, where the ions are made very cold, almost eliminating all motion. Unlike neutral atom traps, ion trapping predates laser cooling since electromagnetic traps can be made deep enough to trap ions cooled only by a buffer gas \cite{Kellerbauer2001276}, which is the method we use for our clock.  Although we have observed laser cooling in our traps, we do not use it\footnote{It is clear that laser cooling adds another layer of complexity to miniaturization, and since we can operate a clock perfectly well with  room-temperature ions, we choose only to buffer-gas cool.  Not using cold ions also relaxes our vacuum pressure requirements.}.  Ion trapping is a well defined subject, but we must examine the field with an eye to our focus, which is creating a robust, miniature trap that can contain many ions while maintaining good optical and microwave access.

\section{Ion Trapping fundamentals} 
The classical fundamentals of ion traps are covered in detail throughout the literature, and I will only give a broad overview.  Good reviews of the subject can be found in \cite{JMS:JMS512,RevModPhys.62.531,0034-4885-60-8-001,Dehmelt196853}, on which much of this description is based.  The interesting thing about ion traps is that although they were originally conceived for perfect geometries such as a spherically symmetric harmonic trap created with hyperbolic electrodes, in reality one can drastically change the geometries of the electrodes and still create a deep, stable trapping region.  This freedom comes at the cost of any combination of the stability, depth, harmonicity, and shape of the trap, each of which may have varying degrees of importance for a particular application.  This flexibility in trap geometries has led ion trapping to be a very useful and scalable technology for a multitude of applications.  A prime example of the opportunities for scalability are shown in the silicon-based, microfabricated chip traps being used by many groups \cite{stick_ion_2006,DBLP:journals/qic/ChiaveriniBBJLLOW05,PhysRevLett.96.120801} to move toward scalable quantum computation.  It is this flexibility that will ultimately provide us the freedom to successfully shrink our trap to a drastically small size while preserving a high signal to noise ratio by trapping a large cloud of ions. 

\subsection{Trap history and types} 
Wolfgang Paul received half of the Nobel prize along with Hans Demelt in 1989 for work on electromagnetic trapping \cite{RevModPhys.62.531,Dehmelt196853}.  Although Earnshaw's theorem tells us that we cannot trap particles in 3 dimensions using static fields, it has long since been understood that with radio-frequency (RF) oscillating fields, one can create a time-averaged pseudo-potential that can trap charges in two dimensions, which if one dimension is rotated about an axis, becomes a 3 dimensional quadrupole trap. 
 The original idea for this type of trap used purely hyperbolic electrodes, which create a spherically symmetric trapping potential when the appropriate dimensions for the axes are chosen.    A drawback of the spherically symmetric potential was that there is only a single point in which there is a node of the potential.  Trapping ions outside of that point means that the ions face an imperfect potential minimum, leading to increased micromotion\footnote{Micromotion is a motion of the ion in small amplitude at the frequency of the RF driving field.  It is fast on the timescale of the ion's harmonic motion in the pseudopotential well.}. The spherical trap  design was often modified in order to provide better laser and detector access to the trap, of which there is little in a true hyperbolic trap, as it was found that there was a relatively insignificant effect on the trapping qualities of the trap by doing so \cite{RevModPhys.62.531}. In later years the linear trap was widely adopted, in which the RF potential in two dimensions is created using linear rods, while the third, free dimension can be ``plugged" with an additional, static potential. This allowed for a nodal line along which to trap, so that many ions in a row could see a minimum potential at the same time.  Laser and detector access become even easier in this case.   

Further variations included adding more poles to the linear trap.  A conventional linear Paul trap has four rods; however, increasing the number of alternating RF rods to 6, 8, or even 16 or more poles creates a potential that is more uniform in the center, but shallower overall, than a four-rod trap with the same RF voltage applied.  The result for a trapped-ion-based atomic clock is a decreased sensitivity to the $2^{\rm{nd}}$ order Doppler shift \cite{JPLmultipole2000}.  It turns out that a broadly varying arsenal of designs can mimic the quadrupole trap by creating a stable trapping region, such as the ring-and-fork trap, the needle trap, planar surface (chip) traps, examples of which are shown in \cite{PhysRevA.74.063421}.  The shape of the electrodes can be relatively arbitrary, as long as they use RF to create a pseudo-potential trapping region in two dimensions, with ``endcap" electrodes (whose geometry matters even less) that provide confinement in the 3$^{\rm{rd}}$ dimension.  As yet another example of this flexibility, we will see a new trap geometry in this thesis that was developed as a convenience to a particular miniaturization technology.

A large part of the work for this thesis was in designing, testing, and characterizing ion traps for a clock during the miniaturization process.  For the IMPACT project, we have primarily used four-rod linear traps in Phases I and II.  One example (the test bed trap) is shown in Figs. \ref{fig:SWtrap} and \ref{fig:testBedPhoto}; the design changes little between versions except for the dimensions of the rods and the shape of the endcaps.  This very basic geometry has a lot of flexibility.  Although the test bed trap is made with a small, plain 4-rod trap, the original intention of the project was to later switch to a higher-pole (8-pole) geometry, to reap the benefits of a higher-pole trap for frequency standards (namely, reduced second order Doppler shift as mentioned above).  Eventually we decided to continue using the four-rod design, since it is clearly a simpler choice for miniaturization, and we were quite satisfied with the performance of our four-rod traps.  Variations that we did perform on this design included changing the endcap shapes and straying from the conventional  circular electrode cross-section.  These designs will be detailed in the discussion of the small packages in Chapter \ref{ch:packages}.  Our only main departure from the four-rod quadrupole trap was for our hopeful Phase III technology, in which we modify the geometry significantly, but still ultimately create a pseudo-quadrupole potential, albeit rather asymmetric.  This design speaks to the flexibility of ion traps and its development will be discussed in Chapter \ref{ch:packages} as well.   As a last note, one might wonder, for the sake of miniaturization, why we do not consider a ``chip trap" as we look toward miniaturization.  At this point, these microfabricated surface traps are not stable, large, or deep enough to sustain the kinds of ions numbers we need nor the trap lifetimes we desire.  

\subsection{The pseudopotential} 

For completeness I will give a brief discussion of the pseudopotential.  For the case of an idealized trap with hyperbolic ring electrode and hyperbolic endcaps, the equations of motion of the ion in the potential can be derived by considering a sinusoidal electric potential $V_0 \cos(\Omega t)$ and a DC electric potential $ U_0$  on the ring electrode, where the endcaps are grounded. These applied potentials produce an electric potential equal to 
\begin{equation}
\Phi(x,y,z,t) = - \frac{m \Omega}{16} \left( a - 2q \cos(\Omega t) \right) (r^2-2z^2)
\end{equation}
where $r$ and $z$ are the cylindrical coordinates of the hyperbolic trap ($r^2 = x^2+y^2$), and $a$ and $q$ are parameters defined by $a= - \frac{8 q U_0}{m r_0^2 \Omega^2 }$ and $q = \frac{ 4 q V_0}{m r_0^2 \Omega^2}$,  where $r_0$ and $z_0$ are the principal distances of the trap (the ring radius and the distance between endcaps) \cite{springerlink:10.1007/BF00329103,doi:10.1080/09500349214550221}.  The equations of motion due to this potential are equivalent to the Mathieu equation, whose solution describes the motion of the ion in the trap as a superposition of a ``secular" motion at some frequency $\omega$ due to the harmonic pseudopotential and a ``micromotion" which is a fast oscillation at the frequency $\Omega$ of the RF voltage.  This solution is stable on certain domains of $a$ and $q$, corresponding to stable trapping. The equations of motion and the parameters $a$ and $q$ can be calculated for different trap geometries  as a way of understanding the complete motion, but this can be a difficult task for arbitrary geometries.

We are more interested in the ``secular" portion of the motion, which is caused by the thermal motion of the ions in the time-averaged potential known as the pseudopotential.  On the time- and length-scales we are interested in, the micromotion is small and therefore unimportant.  For this reason we ignore the micromotion in our calculations, in favor of the average motion.  As an alternative to the Mathieu solution, we can consider the force on the ion and draw conclusions about the ``ponderomotive force," which is an effect of the time-averaged RF electric fields that governs the secular motion.  This discussion follows the description in \cite{0034-4885-60-8-001}.  

Consider in one dimension an ion in the presence of an oscillating electric field that oscillates with frequency $\Omega/2 \pi$ and amplitude $E$.  If the field is homogeneous in space, the ion will experience a force
\begin{equation}
F = m \ddot{x} = q E_0 \cos( \Omega t)
\end{equation}
leading to an oscillatory motion
\begin{equation} \label{eq:rapidsoln}
x(t) = - \frac{q E_0}{m \Omega^2}\cos( \Omega t) \ .
\end{equation}
If the field is not homogeneous in space, i.e. the electric field is given by $E(x)$, then the ion will not only experience an oscillatory motion but also an additional average force.  At high oscillating field frequencies, the oscillatory motion is much faster and much smaller than the motion due to the average force.  Then we can expand the electric field $E(x) = E(x_0) + \frac{d E(x_0)}{dx} \delta x$ where $x = x_0 + \delta x$ is the sum of the average ion location ($x_0$) and the small perturbation  from its location $\delta x$.
In this approximation \cite{springerlink:10.1007/BF00329103} we can write the force
\begin{equation}
F = q E(x_0) \cos(\Omega t) - \frac{q E(x_0)}{ \Omega^2} \frac{d E(x_0)}{dx} \delta x \cos(\Omega t) \ .
\end{equation}

We can replace $\delta x$ in this equation with the rapidly varying solution (Eq. \ref{eq:rapidsoln}) for a uniform field, where the ``uniform" field is taken to be the field at $x_0$
\begin{equation}
\delta x(t) = - \frac{q E_0(x_0)}{m \Omega^2}\cos( \Omega t) \ 
\end{equation}
which leads to
\begin{equation}
F = q E(x_0) \cos(\Omega t) - \frac{q^2 E(x_0)}{m \Omega^2} \frac{d E(x_0)}{dx} \cos^2(\Omega t) \ .
\end{equation}
This force, averaged over one cycle of the field oscillating at $\Omega$, gives the ponderomotive force
\begin{equation} \label{pond1}
F_{pond} = -\frac{q^2 E(x_0)}{2m \Omega^2} \frac{d E(x_0)}{dx} \ . 
\end{equation}
We can  let the ponderomotive force be given by the ponderomotive potential (or ``pseudopotential") $\Psi(x_0)$ where
\begin{equation}  \label{pond2}
F_{pond} = - q \nabla \Psi_{pond}(x_0) \ .
\end{equation}
Then we find $\Psi_{pond}(x_0)$ by extending Eq. (\ref{pond1}) to three dimensions and equating it with Eq. (\ref{pond2}) to get
\begin{equation}
\Psi(x,y,z) = \frac{ q E^2(x_0,y_0,z_0) }{4 m \Omega^2} =  \frac{ q }{4 m \Omega^2} |\nabla V_0(x_0,y_0,z_0)|^2
\label{eq:pseudopotential}
\end{equation}
where $x_0$, $y_0$, and $z_0$ indicate the average position over one cycle and since $E(x,y,z) = - \nabla V_0(x,y,z)$.  
This is a more practical expression for the potentials we are interested in. Instead of attempting to modify the analytical Mathieu solution to approximate the solution for an arbitrary geometry, we numerically simulate the electric field due to specific potentials applied to a particular electrode geometry.  Then we use Eq.(\ref{eq:pseudopotential}) to calculate the pseudopotential for that geometry.



\section{Trap models} 

A numerical model of the traps can be produced by solving for the fields and potentials surrounding a particular electrode configuration and thus finding the total pseudopotential that an ion sees on average.  In this way we can explore different trap geometries and predict whether they will produce a satisfactory potential minimum, optimize trap parameters, and eventually model the motion of the ions within.  

I used Charged Particle Optics (CPO), a software package developed for use in designing electron-optical systems, to model the trap potentials for many geometries (http://www.electronoptics.com/).  Although a bit clumsy, this software has a history of being used by members of our group and other ion trap groups for effective calculations of trap parameters and properties (for example see \cite{StickThesis2007}). CPO uses the boundary-element method to compute fields and potentials, which is touted by its authors as a faster \emph{and} more accurate method than others such as the finite-element method \cite{Read1976}.  Normally, I draw the electrode configuration in CPO and add unit voltage to the RF electrodes, and compute the electric field, which is used to calculate the trapping pseudopotential in the transverse dimension according to Eq. (\ref{eq:pseudopotential}).  The potential from unit voltage on the endcaps which confines the ions in the third dimension is calculated separately. We calculate these stationary fields and potentials in CPO and use the output to combine the RF and endcap potentials in Mathematica.  This means that we only care about the average secular motion in the trap and not the micromotion as discussed above.

The procedure for modeling potentials with CPO is as follows.
\begin{enumerate}
\item{Draw the trap geometry in CPO.  Place all four RF electrodes and the endcaps so that they are all present during the calculations.  The electrodes are all grounded except for a unit voltage placed on two diagonally opposing RF electrodes, and the electric field in a 3-D grid is saved for all space in and around the trap. This electric field grid is used to calculate the pseudopotential for the time-averaged case with the quadrupole field switching at a certain frequency $\omega_0$.  Since we use unit voltage, this result can be scaled to different values of RF voltage, meaning that the electric field grid need only be calculated once for each trap geometry, unless grids of varying resolutions are needed. }
\item{Set the RF electrodes to ground and calculate the potential due to unit voltage on the endcaps for the same position grid.}
\item{Calculate the scaled pseudopotential and endcap potential and combine them using Mathematica.  }
\end{enumerate}

CPO outputs just the electric field vector values as well as the potential values for each corresponding case.   The output of the program in Mathematica is a grid of the actual time-averaged potential seen by the ion in 3-D space.  This grid can then be used to find the size, depth, and stability of the trap, as well as the trajectory of an ion in that potential.  

\begin{figure}
   \centering
  \includegraphics[scale=0.6]{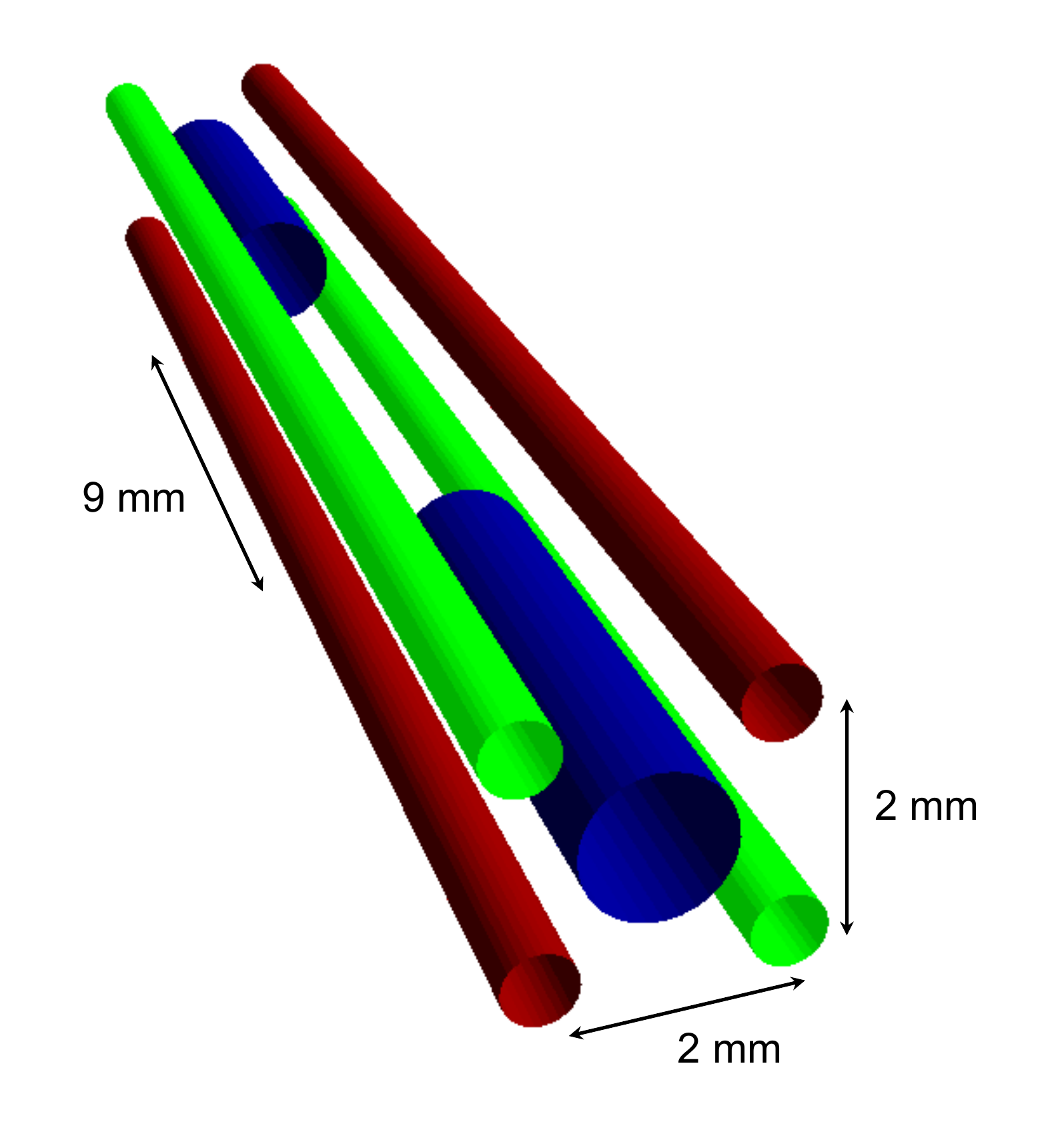}
   \caption{Modeled geometry for the test bed trap model. The red electrodes carry the RF, green are RF grounded, and blue have a DC endcap potential. }
  \label{fig:OldTrapCPO}
\end{figure}

An example of the modeling process is illustrated for the test bed trap, as a reference for the models for different geometries throughout this work.  The electrode geometry as drawn in CPO is shown in Fig. \ref{fig:OldTrapCPO}.  The red electrodes carry RF, the green are RF ground, and the endcap voltage is represented by blue.  This color scheme remains the same for all the geometries in later chapters.  The electrodes are represented by simple shapes in CPO that are a good approximation to reality.  The electrodes must be divided into segments that have a specific voltage applied to facilitate calculation of the fields and potentials.  These segments, as well as the grid for which we calculate field values, must be chosen on the correct scale to get meaningful results.

\begin{figure}
   \centering
  \includegraphics[scale=0.45]{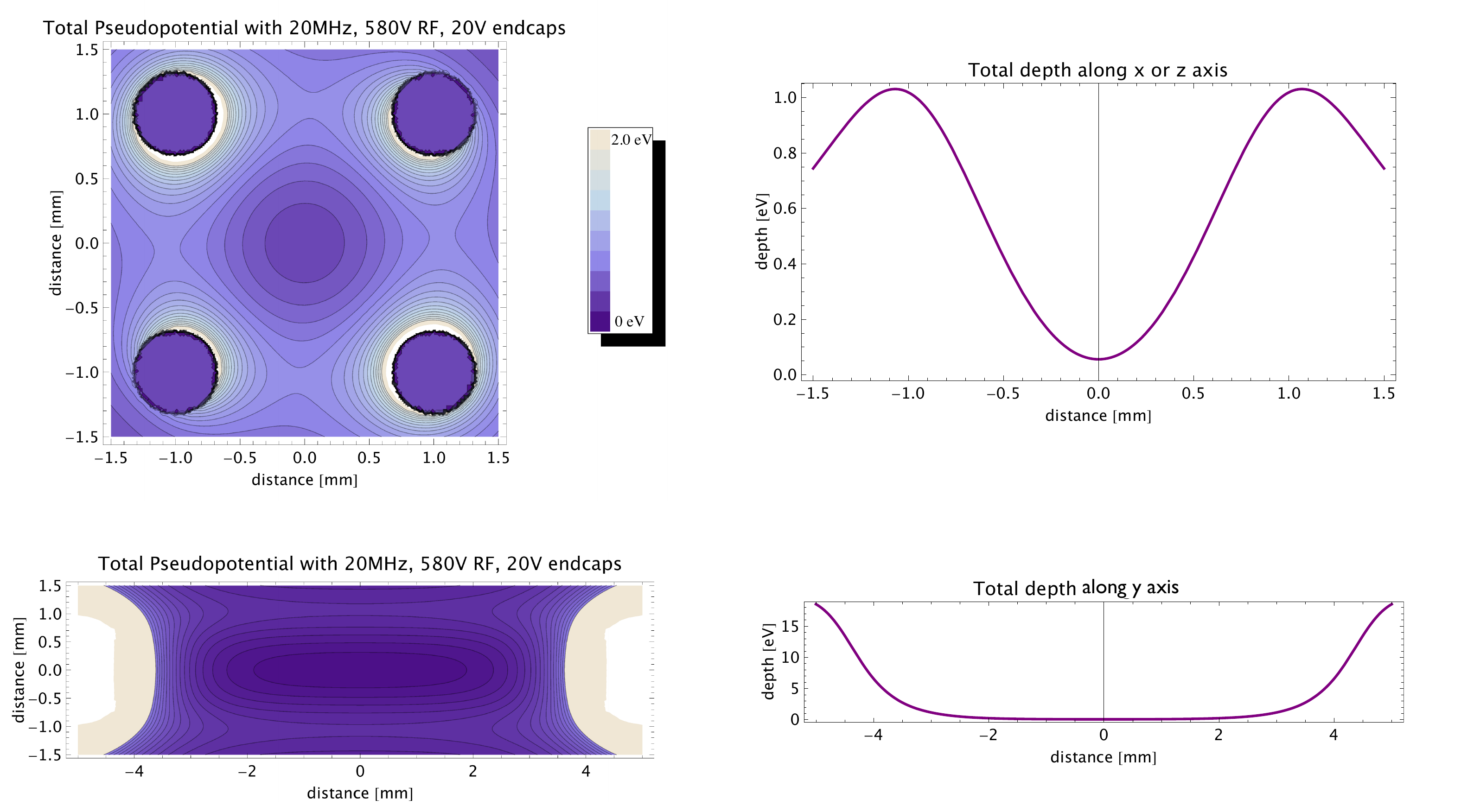}
   \caption{Contour plots of the model of the transverse and longitudinal planes of the pesudopotential as calculated using CPO and Mathematica.  Left: The potential in the transverse and longitudinal planes.  Right: Trap depth in one dimension along the $x$ or $z$ axis (these are the shallowest directions and are equivalent by symmetry), and along the $y$ (long) axis.} 
    \label{fig:OldTrapModel}
 \end{figure}

After processing the CPO data for the fields and potentials in Mathematica, the main result, an illustration of the total pseudopotential seen by the ions (in an empty trap, i.e. not including space charge) is shown in Fig. \ref{fig:OldTrapModel}.  The pseudopotential is calculated on a 3-D grid, but we visualize it by looking at the transverse and longitudinal cross-sections.  The one-dimensional potential is plotted along the ``shallow" axis which gives the depth of the trap.  The depth of the trap is calculated by subtracting the deepest part of the trap (usually the center, and usually not at zero potential) from the lowest part of the maximum surrounding the center of the trap (the first ``leak point" the ions see).

\begin{figure}
   \centering
  \includegraphics[scale=0.42]{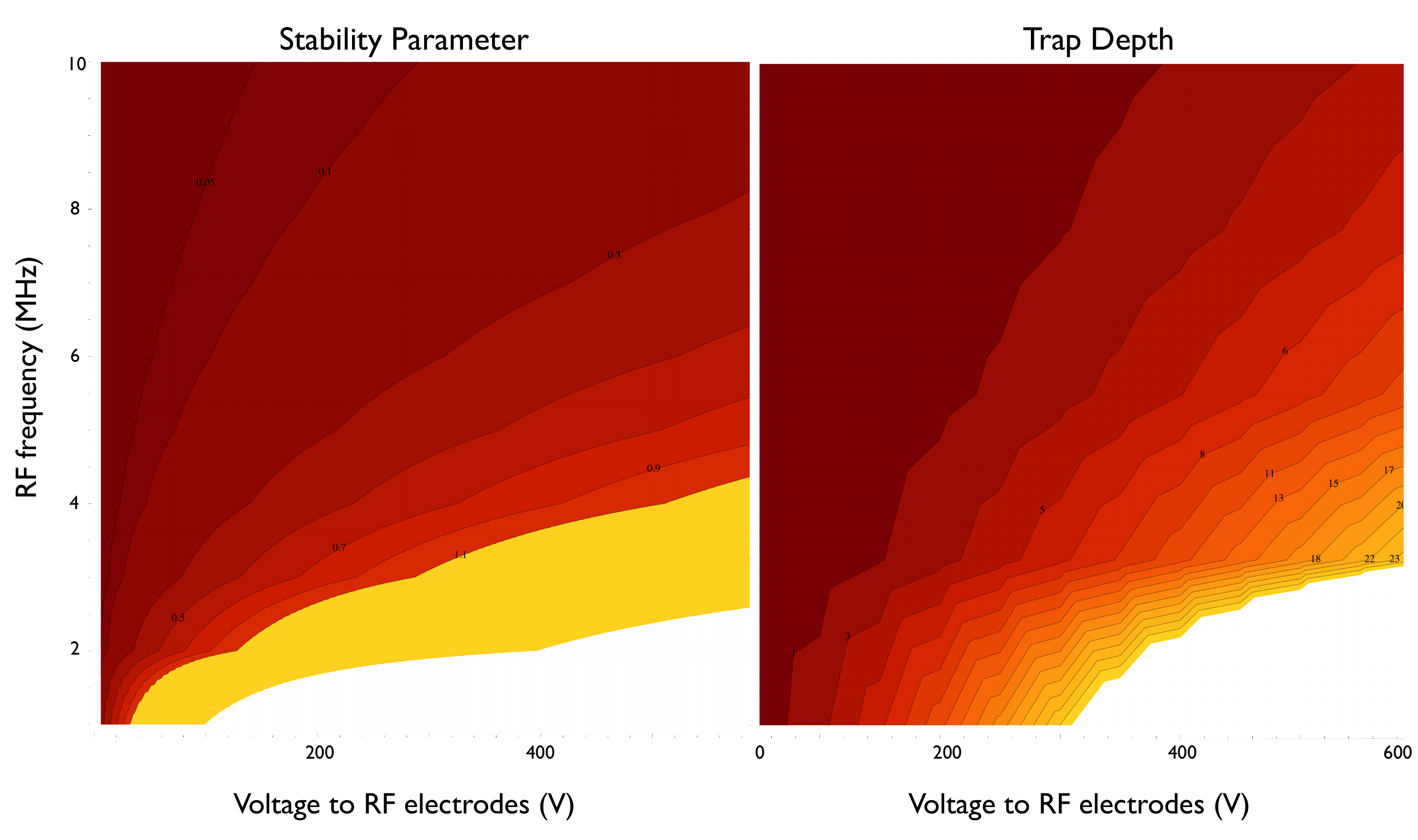} 
   \caption{Plots of the trap parameter space (frequency vs. $V_{\rm{RF}}$) showing the value of the stability parameter (left) and the trap depth (right).}
   \label{fig:StabilityandDepth}
\end{figure}

\begin{figure}
   \centering
  \includegraphics[scale=0.5]{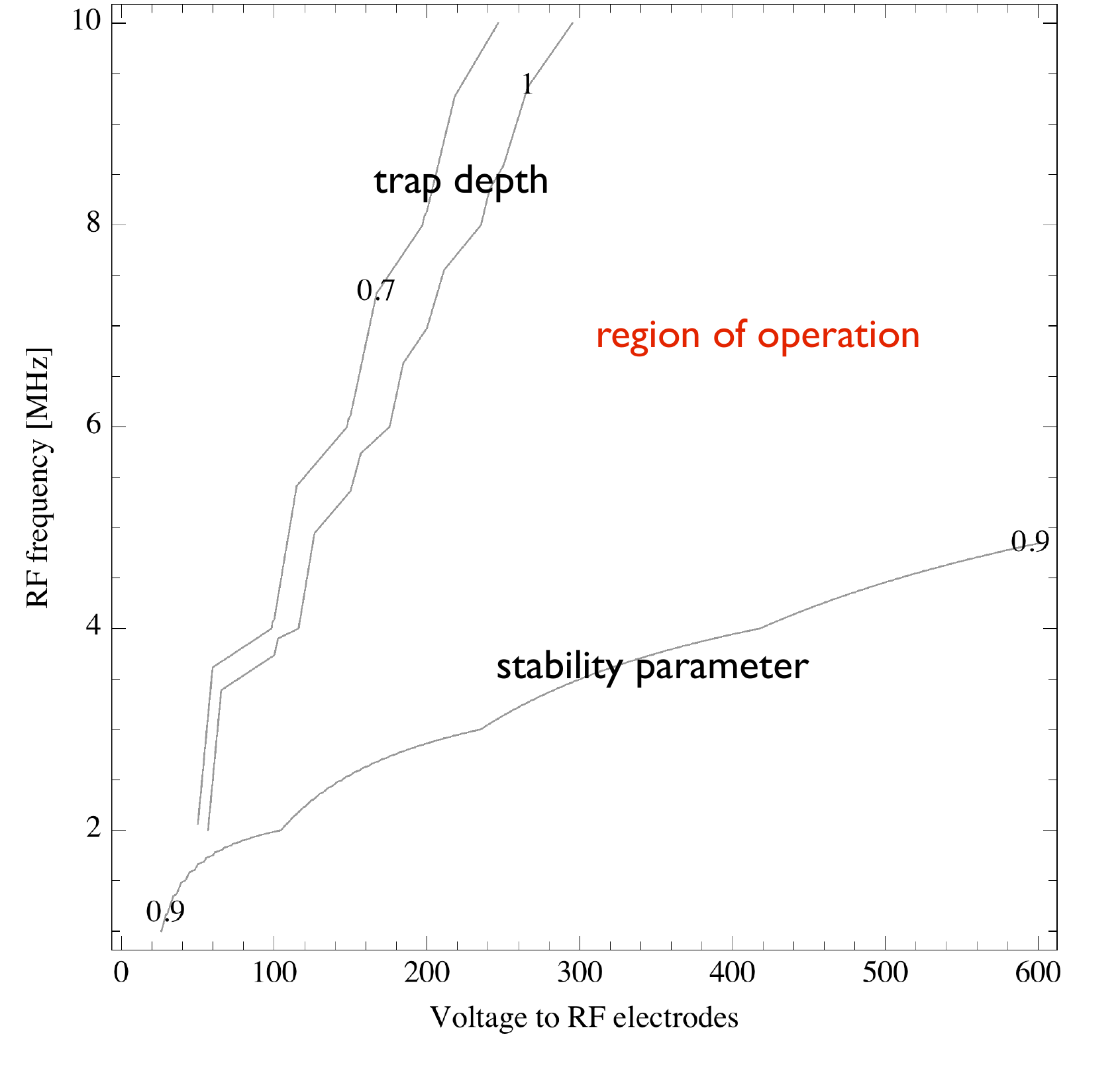} 
   \caption{Important values from the parameter space plots  in Fig. \ref{fig:StabilityandDepth} assembled on one graph as described in the text.  We can use a plot like this to select effective trapping parameters for a new trap.}
   \label{fig:Parameters}
\end{figure}

We can evaluate the stability of our traps as characterized by the stability parameter $q$
\begin{equation}
q = -\frac{4 q_e V_{\rm{RF}}}{m \Omega^2 r_0^2}
\end{equation}
which for a linear quadrupole trap should theoretically be $\leq$ 0.9 for stable trapping \cite{JMS:JMS512}.  In this case $q_e$ is the charge of the electron and $r_0$ is the equivalent radius at which the ion hits an electrode, that is, the direct distance from the trap center to the electrode at its closest point.  For multipole linear traps, this stability parameter is generalized to 
\begin{equation}
\eta = k(k-1) \frac{2 q V_{\rm{RF}}}{m \Omega^2 r_0^2}\ r^{k-2}
\end{equation} 
for a trap with $2k$ poles \cite{JPLmultipole2000}.
For more exotic trap geometries, we calculate the potential for instability directly at each point using the gradient of the potential (see Sec. \ref{sec:LTCC}).  
We can calculate the stability parameter for a range of relevant trap parameters to understand the region of stable trapping (in the frequency and RF voltage parameter space) for our numerically calculated pseudopotential.  An example of a plot demonstrating this type of calculation is shown in Fig. \ref{fig:StabilityandDepth}.  By noting the contour that corresponds to the stability value of 0.9, we have a prediction of the approximate range of frequencies and RF voltages we can use to drive the trap.  This is very useful information for new traps that we have designed when it is time to test them in the lab.  In a similar fashion, we have calculated the depth of the trapping potential for the same trap parameter space.  We know that our test bed trap functions very well with a depth of 1 eV, so we can use this plot to strive for a similar depth in other traps.  These plots are particularly useful when we are having trouble trapping in a new package, to help us rule out poor trap driving parameters as the source of the problem.  Fig. \ref{fig:Parameters} shows the relevant contours from each plot on one graphic; it is easy to see that we need to be above the $\eta = 0.9$ line for stable trapping, and below the 1 eV depth contour line to have sufficient depth\footnote{These plots are plotted on a relatively coarse grid, but they still provide an adequate guideline.}.

\begin{figure}
   \centering
  \includegraphics[scale=0.75]{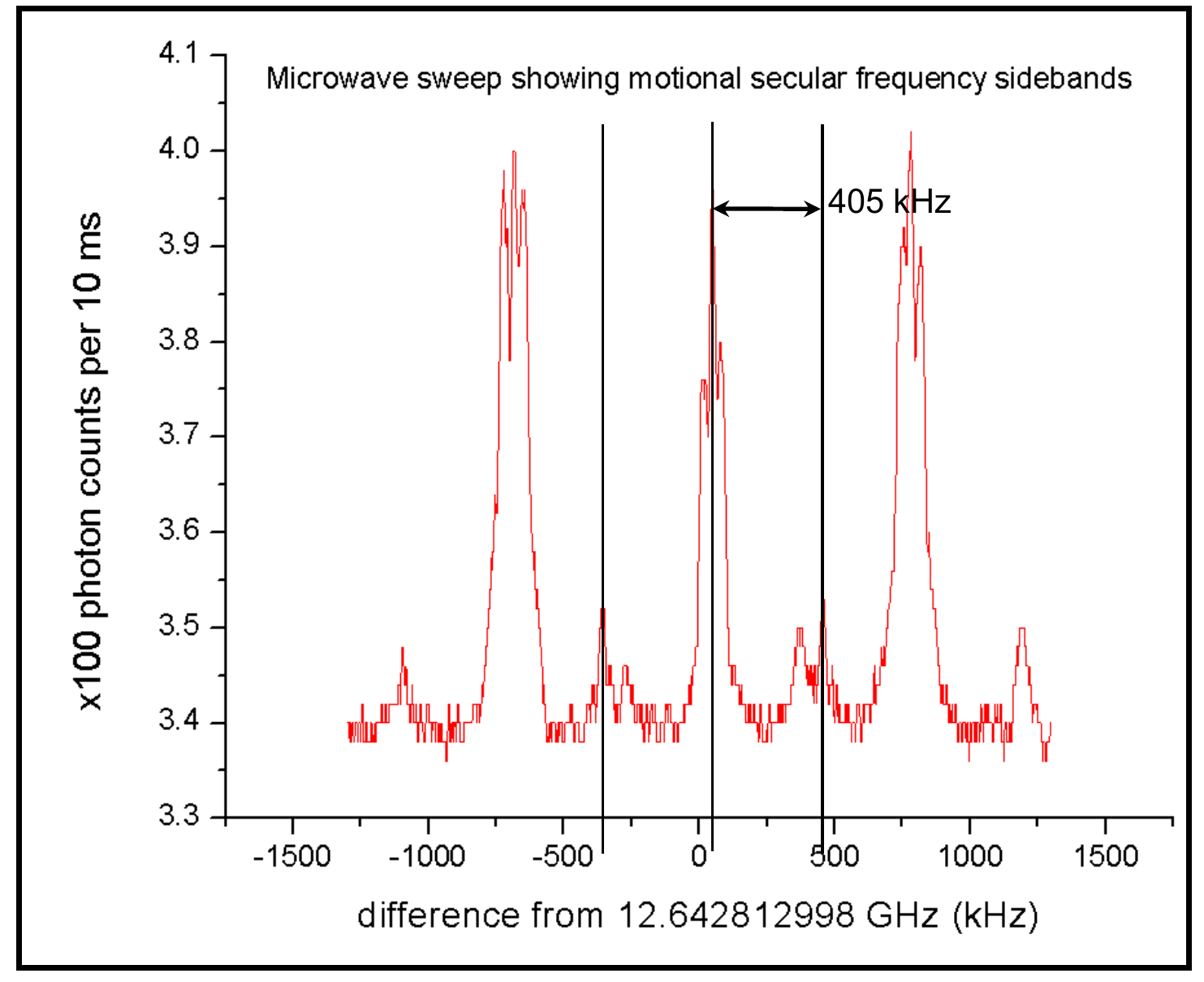} 
   \caption{A wide microwave sweep across resonance showing the three magnetically sensitive levels of $|F=1\rangle$ and the motional sidebands that indicate the secular frequency of the trap.}
   \label{fig:SecFreqExp}
\end{figure}

A last parameter we like to calculate for each trap is the characteristic secular frequency.   We do this in Mathematica once the full potential is calculated.  At the center part of the trap, we compare the shape of the potential well with an equivalent harmonic potential.  When they are matched, the frequency of the perfect harmonic well is taken to be the approximate secular frequency for an empty trap. This can be done by hand or by fitting a harmonic well to the complete pseudopotential well. The secular frequency will in reality be lower for a large cloud of ions based on the fact that the effect of space charge will make the potential shallower and change its shape.  We can experimentally measure the secular frequency in some cases.   The motion of the ions in the trap can modulate the microwave field phase so that sidebands are produced in the microwave spectrum.  By sweeping the frequency of the microwaver radiation near the clock resonance we can see these motional sidebands, whose distance from the resonance indicates the secular frequency of the trap \cite{doi:10.1080/09500349214550321}.  The ability to detect this effect varies from trap to trap, since the small packages limit our control over the microwave propagation direction and polarization..  Fig. \ref{fig:SecFreqExp} shows an example of a frequency sweep showing  secular frequency sidebands in the LTCC trap (described in Sec. \ref{sec:LTCC}).  With this kind of measurement we can confirm the validity of our simulations.

\section{Ion cloud characterization} 
There are a number of other parameters we might like to know about our traps.  One example is how many ions are trapped at any given time.  We have only been able to estimate this using order of magnitude calculations based on the amount of fluorescence, which has a limited accuracy because the laser is likely not interacting with all of the ions\footnote{An estimate of this type in the test bed trap, which includes solid angle of detection, detector efficiency, filter efficiency, and photon scattering rates, predicted the presence of 50,000 ions (in the laser beam).}.  We would like to know the density of the cloud, but a detailed description cannot be obtained from observation and must rely on a model calculation (which we have done, see Chapter \ref{ch:biasfield}).  The cloud size can be estimated based on the trap size and calculated potential, but except for some early attempts with laser cooled ions, we have found it difficult to visibly image the ions on the camera because of excessive background at the primary wavelength, or low signal at the secondary wavelength (the detection wavelengths are discussed in Chapter \ref{ch:testBed}).  In addition to looking for motional sidebands, a broad sweep across resonance with the clock transition and the magnetically sensitive Zeeman states can tell us the magnitude of the (average) magnetic field at the location of the ions, and sometimes further information about gradients.  We have not characterized these aspects of each of our systems thoroughly while focusing on the operational development of our ion clock systems.

We have approximated the temperature of our ions using the doppler broadening of the clock resonance.  This is discussed in Sec. \ref{sec:temp}.  We also estimate the lifetime of the ions on a regular basis.  Generally we estimate this roughly by measuring the signal size at two points and assuming an exponential decay.  Then, the characteristic decay time $\tau$ of the ion signal, and thus the number of ions, is
\begin{equation}
\tau = -\,  \frac {t}{\ln \left(   \frac{ A_{t_1}   }{   A_{t_0}  } \right)   }   
\end{equation}
where $A_t$ is the measurement of the fluorescence amplitude at time $t_1$ or $t_0$, and $t$ is the time between the measurements ($t = t_1 - t_0$).  This approximation gives us a good idea of how well our traps are working.    Occasionally, we take continuous data of the amplitude over some time period to which we can fit an exponential decay and get a better estimate than is possible using just two points.  


\chapter{Atomic clocks}
\label{ch:clocks}


All atom-based frequency standards have at their heart the same concept: extracting the precise frequency separation of the discrete energy levels of an atom, which is an intrinsic and reproducible quantity.  We can extract the energy level separation in the form of a frequency by interacting with the atom through electromagnetic radiation (absorption or emission). An atomic frequency standard is a device that recreates the atomic resonance frequency in a practical form, and delivers the frequency with minimal environmental influences.  
In many clocks, this is done by operating a \emph{local oscillator} (LO) that oscillates or resonates at a frequency near that of the ions (or some signal mixed from that frequency), which is locked to the ions through some physical interaction mechanism (in our case, microwave radiation).  Clocks that follow this model are known as \emph{passive} frequency standards, and most commercially available frequency standards are of this type.  In \emph{active} frequency standards, the atoms perform the primary oscillation themselves, emitting the signal directly through radiation from the atom.  An example of an active standard is the hydrogen maser (although these can also be operated in a passive mode).  In this introductory chapter we will focus on passive microwave standards, i.e. standards that are similar to our Yb ion clock.

\section{Basic functionality} 
In any passive microwave frequency standard, there are three parts to the interrogation of the atomic transition: state preparation, microwave interrogation, and state detection.  When we perform these steps in each duty cycle, we gain information about how to steer the local oscillator to the correct frequency at each step.  

Typically, microwave clocks operate by locking to the frequency of the hyperfine splitting of two ground state energy levels for some species of atom or ion.  For example, in $^{171}$Yb$^+$, the clock transition is between the $|F=0,m_F=0\rangle$ and $|F=1,m_F=0\rangle$ $^2 S_{1/2}$ ground states, which are separated by a microwave frequency of approximately 12.642 GHz.  The atoms are prepared into the lower ground state, then a microwave pulse (designed to be a $\pi$-pulse) is applied to transfer the population from the lower ground state level to the upper one.  If the local oscillator that provides the frequency to the microwave synthesizer is exactly on resonance with the atoms, the entire population will be transferred, and we will see a maximum signal upon state detection.  In our case, since we readout the state using fluorescence from an optical pumping transition, this means that maximum fluorescence will be seen by a detector during the state detection stage.  

If the local oscillator that determines the microwave frequency is not exactly on resonance, then the control electronics for the clock must use the fluorescence information to steer the oscillator toward resonance.  In practice this means that an error signal must be generated for the feedback loop. As is the case for all error signals, we desire a zero-crossing at the center point and a sign change that can be used to readout the necessary direction for the frequency correction. This can be done by shifting the frequency back and forth on either side of the resonance during successive duty cycles, and driving the difference between the two measurements toward zero.  This difference gives the error signal we desire by indicating when the frequency of the LO has drifted above or below resonance and in which direction.

\section{Types of atomic clocks}

The main types of atomic clocks in commercial use today are cesium beam standards, rubidium vapor standards, and hydrogen masers. Neutral atom (fountain) clocks \cite{1408302}, trapped-ion clocks  \cite{PrestageRecentDevelopments}, and various types of optical clocks  are also common in a laboratory setting.

Since cesium-133 is used to define the second based on the separation of its hyperfine ground state transition (at 0 Kelvin), the cesium beam clock is a very important frequency standard for general use. It also provides a very clear example of the state-preparation/interrogation/state-detection process that is used by a standard of this type.  The methods used for manipulating cesium, preparing and detecting the state, and interrogating with microwaves in a beam standard are very conventional and  relatively straightforward.   The cesium standard along with the basis of most microwave frequency standards is reviewed in many articles across the literature, which I will briefly summarize \cite{NIST2001,0034-4885-60-8-001,levine:2567}.  

\begin{figure}
   \centering
  \includegraphics[scale=0.4]{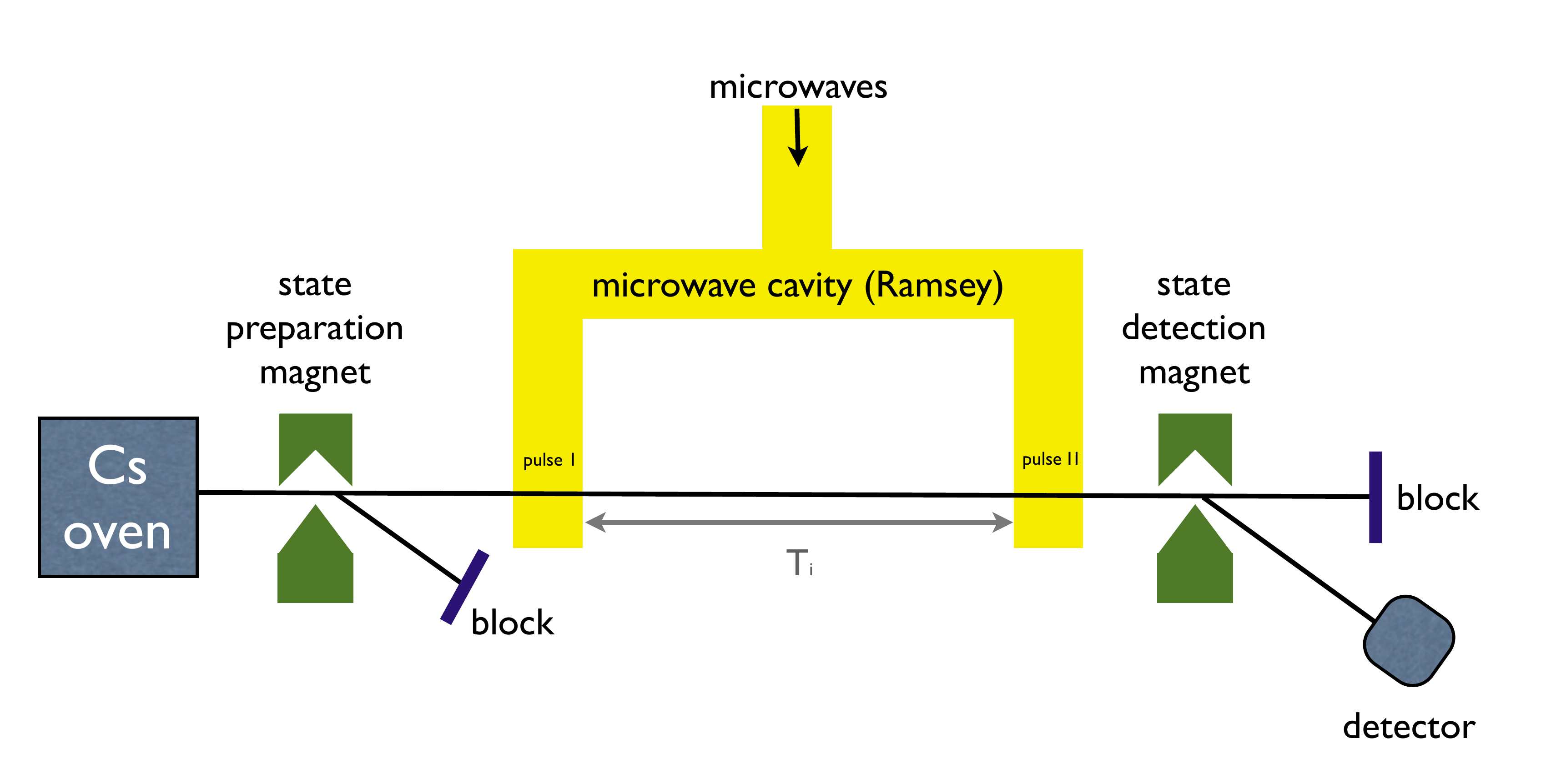} 
   \caption{Schematic of a cesium beam standard as described in the text.}
   \label{fig:CsClock}
\end{figure}

A diagram showing the operation of a cesium beam standard is shown in figure \ref{fig:CsClock}. 
A cesium beam is produced by heating an oven and allowing the Cs atoms to flow out through an opening to create the atomic beam.  The beam passes through a magnet that uses the Stern-Gerlach effect to do state preparation by separating the atoms into two beams depending on their hyperfine state, one of which is blocked from continuing toward the microwave cavity.  In this way, we ``select" the $|F=3\rangle$ state by throwing away the beam of atoms in the $|F=4\rangle$ state. 

Once the preferred atoms have been selected, they proceed to pass through a microwave cavity (while they are simultaneously subjected to a magnetc field that separates the Zeeman levels, known traditionally as the ``C-field").  This could be one long interaction region that would do what is known as \emph{Rabi interrogation} in which the atoms interact just once with a microwave field, or it could be two in-phase interaction regions separated by a distance that allows the interaction of the cesium with the microwaves to be coherent across the two regions.  This is known as \emph{Ramsey interrogation}.  State selection is done by using another Stern-Gerlach magnet to ``select" the $|F=4, m_F=0\rangle$ state by throwing away the beams containing all the atoms not in that particular clock state.  (This means that only about $1/16$ of the atoms coming from the oven are ultimately used to influence the control of the clock.)  The surviving atoms are measured.   Just as in our clock, the signal from these surviving ions is used to tune the microwave frequency until the signal is maximized.

The cesium clock can be improved by using lasers to optically pump for state selection and detection; this improves signal to noise ratio based on the fact that many more of the atoms will successfully be included in the measurement without the need to throw away all but one probabilistic state.  Increasing the interaction time also improves the inherent limiting linewidth of the clock.  This can be done by making a longer Rabi interrogation cavity or a longer space between the two Ramsey interaction regions.  Either method has clear limits based on the fact that a physical beam of atoms is responsible for completing the process, but the Rabi method is limited further by the necessity for a completely uniform C-field in the interaction region.  One way of dealing with is to turn to a fountain configuration, where the atomic beam is in the vertical direction, such that it is slowed by gravity until it falls back toward its origin.  In this scheme, a microwave cavity on the beam path is passed through twice, and the atoms are moving slower than in the conventional device, increasing the interaction time even further.  This scheme has the added benefit that the self-reversing direction of the atoms can eliminate some typical systematic shifts \cite{levine:2567}.  The ``fountain" configuration just mentioned is the basis for the NIST F1 clock which is the current US standard for atomic time \cite{NIST2001}.  For a conventional cesium beam clock, the fractional frequency stability can be in the range of $2\times10^{-14}$ for integration times of one day. The fountain clock at NIST has a fractional frequency stability on the order of $1\times10^{-15}$ or less at one day.

It is easy to transform the concept of spatial interactions of a moving beam of atoms discussed here to temporal interactions with (relatively) stationary atoms.  As an example, in a rubidium standard, instead of an atomic beam, the Rb is localized in a vapor cell.  State preparation is typically done by optical pumping with a lamp, then the cell is interrogated with microwaves, and the state is detected by a convenient scheme involving absorption of the probing lamp light.  The transmitted light is decreased when the microwaves are on resonance, and this phenomenon is used to lock the frequency signal.  Rubidium standards can be very small and fairly accurate, but their long term stability is limited by buffer gas pressure in the cell, light shifts due to the pumping lamp, and the influence of magnetic fields.  Plus, the absorption that is used to lock the clock is a very small effect (because it is measured on the continuous background of lamp light) that cannot provide the stability needed for a long-term frequency reference.   Nice comparisons of several traditional frequency standards (including quartz crystal oscillators) are given in \cite{abcdef} and \cite{84969}.

Similarly to the rubidium standard, an ion clock is interrogated temporally instead of spatially.  The ions are relatively stationary in the trap, and their states are prepared using lasers.  The ions are interrogated with microwaves and the state detection occurs when the fluorescence is read out from an optical cycling transition that we excite.    Ion clocks have the advantage that the ions can be stored for very long periods (we will see a trap in this work which stores ions for more than a month at a time).  If systematic frequency shifts are kept to a minimum, this means that interrogation times can be much longer than in beam or fountain standards, although in a practical sense, our interrogation period is limited by the short-term stability of the LO.  A thorough discussion of the advantages and disadvantages of using ions for an atomic reference is given in \cite{0026-1394-22-3-014}.  There are many current examples of excellent ion trap based clocks, including those developed at JPL \cite{5422506,4623080,PhysRevLett.104.070802}.

Last, research into optical clocks is widespread, and has proceeded at a much faster rate than, for example, the development and improvement of cesium references.  If an optical transition can be used as a clock state, the Q value of the clock resonance benefits from two sides; the clock frequency is increased, and the linewidth can be extremely narrow.  First, for measuring time, a higher frequency will have more cycles to be counted in any given interval than a lower frequency, intrinsically enabling a more precise measurement.  For this reason, a higher frequency for the clock transition is always better, making optical frequencies an optimal resource, far better than microwave transitions (providing 5 orders of magnitude higher in frequency).  Second, one can choose to use a forbidden transition  as the clock transition, making the linewidth extremely narrow. The reason these transitions were not probed in the past is because there was no available frequency counter that could count such high frequencies with sufficient accuracy to take advantage of the optical reference.  This problem has been efficiently solved with optical frequency combs, which when locked to the clock transition frequency can provide the information at a manageable frequency \cite{combs2007,nistcombs}.  Optical frequency standards can be made using a single trapped ion or cold atoms, and have reached fractional frequency stabilities of $10^{-18}$ \cite{PhysRevLett.104.070802}.  This type of clock will likely provide the reference for a new definition of the second in the not too distant future \cite{redefinetheSecond}.

\subsubsection{Ion traps for clock  miniaturization}

Clearly, not all frequency standards are subject to miniaturization.  When considering candidates for an IMPACT-level clock, ion traps stand out as a front-runner.  Conventional standards technologies such as cesium beam clocks cannot be miniaturized because the beams require a considerable interaction region which, when reduced in size, jeopardizes the quality of the atomic resonance linewidth.  Rubidium standards can already be made very small, but they cannot support long-term stability due to the long-term drifts and calibration requirements associated with vapor cells \cite{camparo:33}.  Similar problems exist for the CSAC (chip-scale) clocks, which although low-power and very small, have poor long-term performance \cite{4319292}.   Standards as advanced as fountain clocks and hydrogen masers would have their performance fundamentally degraded to meet the size and power goals of IMPACT, if it were possible to do so.    Optical standards are still being developed and have the difficulty of requiring a frequency comb for readout, although there is one group attempting an IMPACT clock using an optical transition.  

Ion traps present a friendlier system for miniaturization. The ions are very isolated from their environment and the line-Q of the clock resonance does not in principle change as we reduce the trap size.  The technology of the traps is relatively simple and is very robust to miniaturization.  The deep wells that can be made with electric fields allow us to trap a large number of ions at once while needing only to cool them to approximately room temperature using a buffer gas.

To emphasize the advantages of an ion standard for the IMPACT project, it is helpful to compare it with another candidate system: trapped neutral atoms, which are usually confined using a magneto-optical trap (MOT).   In contrast to a MOT, the vacuum requirements for ions are much more relaxed.  The laser requirements are also less strict since there is no need for laser cooling.  A drawback of miniaturizing an ion trap is coping with the issue of the electrical feedthroughs to the trap that are necessary to drive it.  A neutral atom standard needs to have nothing inside of a vacuum except for the atomic vapor (with the caveat that the vapor must be in a very \emph{good} vacuum, with a sufficient amount of vapor available inside).  For making a clock, ions are convenient because when they have a long lifetime, loading of the trap is performed very infrequently, in contrast with a MOT that is reloaded before every measurement. 
Also, ions can be interrogated for an arbitrary length of time, while a neutral standard will have its interrogation time limited by the period of free-fall expansion during which each measurement is made.  For these reasons, we believe that an ion trap is the best choice for miniaturization and has a lot of promise for long-term stability, compared with other approaches.



 \subsubsection{The local oscillator} 
Some attention should be given to the nature of the so-called local oscillator (LO) that is driven to be on resonance with the atomic transition.  By far the most common oscillator in use today is the quartz crystal oscillator, which comes in many forms, from the inexpensive quartz found in wristwatches to highly temperature-stabilized oscillators.  The LO is generally an oscillator that has very good stability at short times but would typically drift and/or age in the long-term.  The LO is responsible for maintaining a stable frequency between feedback loop updates, that is, during the time that the ions are being interrogated.  Thus, the maximum interrogation time is related to the stability of the LO.  We use a small-sized temperature-stabilized quartz oscillator for our clock. 
We had hoped to incorporate a MEMS (micro-electro-mechanical system) resonator or oscillator into the system, but its development is incomplete.  The availability of small, relatively stable quartz makes it difficult to compete given the current state of micro-resonator technologies.  
\subsubsection{Rabi and Ramsey interrogation}
\begin{figure}
   \centering
  \includegraphics[scale=0.55]{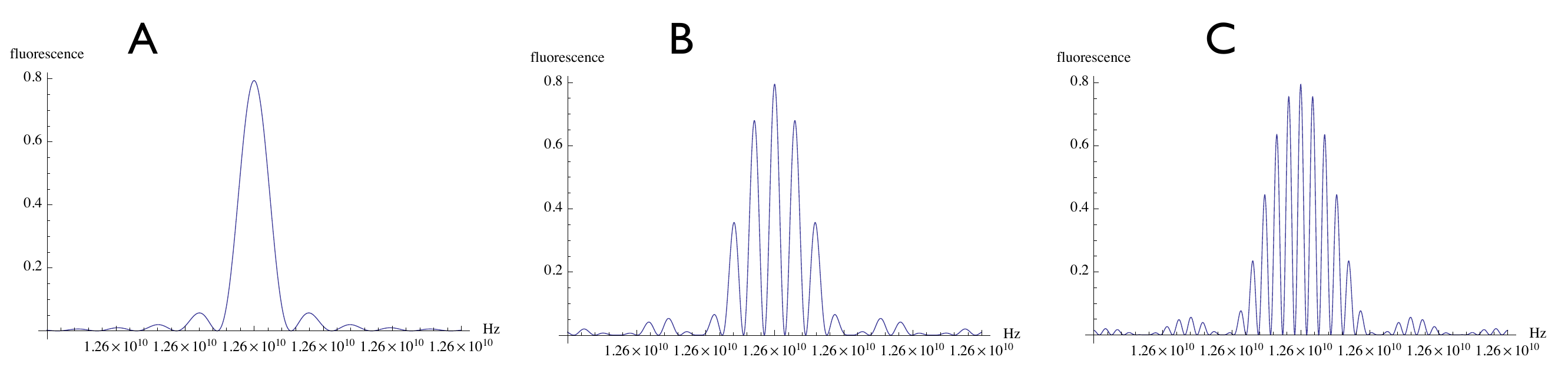}
   \caption{Examples of Rabi and Ramsey fringes.  A: Rabi fringe for microwave interaction pulse of 1100 ms. B: Ramsey fringes for two 550 ms pulses separated by 1.5 seconds.  C: Ramsey fringes for two 550 ms pulses separated by 3 seconds.  Increasing the delay time increases the number of fringes while the envelope stays the same, decreasing the linewidth of the locking peak.  Therefore it is advantageous to make the delay time as long as possible. }
    \label{fig:RabiRamsey}
   \end{figure}

As mentioned above, Rabi interrogation and Ramsey interrogation are two common methods for obtaining the clock signal for locking a frequency standard to an atomic source.  In the 1940s, I. I. Rabi pioneered the techniques that are used now in frequency standards, but Norman Ramsey improved greatly on the method of Rabi interrogation.  In Rabi interrogation, the atoms are subjected to a single pulse of microwaves (in time, or they pass through a single microwave cavity in a beam standard).  The resonance peak that is seen as the frequency is scanned across the resonance (``scanned" in the sense that during a series of pulses the microwave frequency is gradually increased) looks like the Fourier transform of the pulse (see Fig. \ref{fig:RabiRamsey}A).  An experimental example from the test bed clock is shown later in Fig. \ref{fig:OldTrapRabi}. For a given microwave power the center of the peak indicates a $\pi$-pulse is performed during the interrogation. The linewidth of the resonance peak and therefore the line Q of the clock that uses this peak for locking is limited by the pulse length: longer interrogation times lead to narrower linewidth.  

Ramsey's ``method of separated oscillatory fields" \cite{ramsey:25} improves upon this technique by interrogating with two pulses with a specific phase relationship, with a wait time $T_i$ in between the pulses.  In this way, the first pulse acts as a $\pi/2$ pulse (on resonance), the mixed state can evolve during $T_i$, and the final $\pi/2$ pulse will complete the transition.  The Fourier transform of this pulse sequence, which is the resulting curve in a microwave frequency scan, is a series of narrow fringes with the envelope of a Rabi fringe.  This method can provide a much narrower linewidth for locking \cite{MolecularBeams} and makes the standard more forgiving toward inhomogeneities of the C-field \cite{NIST2001, 0034-4885-60-8-001}.  Examples of Rabi and Ramsey interrogation for the same total microwave interaction time are illustrated in Fig. \ref{fig:RabiRamsey}. In our ion trap standards, we use Rabi interrogation because of its simplicity, the ease of locking to the single peak (which may lead to better long-term behavior of the lock), and because in principle, with trapped ions we are free to interrogate for arbitrarily long times.  

\section{Figures of Merit} \label{sec:FOM}
Before going any further it will be helpful to discuss several ways by which people describe the quality of an atomic clock.  There are two main indicators of clock performance that we are concerned with in this work.  The first is Allan Deviation, which is a typical measure of the stability of the frequency being delivered by the device.  The next is time loss, which is related to variations of the phase and therefore the accuracy of the clock.  
 A good introduction to these characteristic quantities and the noise that contributes to their values is found in \cite{levine:2567}.  Conveniently, in the lab we measure frequency differences, Allan deviation, and time loss using an instrument (the Symmetricom 5125A phase noise test set) that takes as input the two signals (reference and measure) and automatically calculates these values over time.

Allan deviation is a commonly quoted way of characterizing the stability of a frequency standard.  It defines the fractional frequency stability (and is therefore unitless) for a particular measurement averaging time.  Thus, a plot of the Allan deviation against the averaging time reveals the stability behavior of the standard on different timescales.   Also known as the two-sample variance, the Allan variance of a discrete set of $M$ frequency measurements, each one given by a frequency $f_1$ which is  averaged over some period $\tau$, is given by \cite{0034-4885-60-8-001}
\begin{equation}
\sigma_y(\tau) \approx \left(  \frac{1}{2(M-1)}\sum^{M-1}_{i=1} (y_{i+1}-y_i)^2  \right)^{1/2}
\end{equation}
where $y(t) = \frac{f_{1i}(t)-f_0}{f_0}$. This is an approximation of the ideal form \cite{1539968} which is true as $M\rightarrow\infty$\ ,
\begin{equation}
\sigma_y(\tau)  = \sqrt{\langle \frac{1}{2} \left(  y(t+\tau) - y(t) \right)^2 \rangle  }  
\end{equation}
where $y(t) = \frac{f_1(t) - f_0}{f_0}$ and the angle brackets average an infinite number of measurements.  In these equations, $f_0$ is the nominal frequency of the device under test, which we take to be much more stable than the measurement $f_1$ and use as a basis for comparison.  That is, each point $y$ is the difference between two frequencies: that of the system under test and a control frequency (the nominal frequency, in this case taken to be $f_0$).  
We must take the difference of these differences to get meaningful information about the variation, and we must intrinsically consider $\tau$, the time between difference measurements, so that we can understand the fractional frequency variance for intervals of different lengths.  We generally plot this variance as a function of this ``averaging time" $\tau$.  Generally, the slope of the Allan variance at different averaging times can reveal information about the dominant noise processes affecting the frequency stability.

An ion frequency standard that is limited only by quantum projection noise has a limiting Allan variance given by \cite{84970,0034-4885-60-8-001}
\begin{equation}
\sigma_y(\tau)  = \frac{1}{2 T_i \omega_0} \frac{1}{\rm{SNR}}\sqrt{\frac{T_c}{\tau}}
\end{equation}
where $T_i$ is the interrogation time, $T_c$ is the cycle time, $\omega_0$ is the clock frequency,  (e.g. $2\pi\times12.642$ GHz in Yb), and SNR represents the signal-to-noise ratio.  Thus the Allan variance will scale as some factor of $1/\tau^{1/2}$.  In our clock, there will be other noise sources so we will not be limited by shot noise.  Additional noise will increase the factor in front of $1/\tau^{1/2}$ but it should still integrate down at the same rate.  The IMPACT project goal in these terms is $2\times 10^{-11} \tau^{-1/2}$.

While Allan deviation is perhaps the most common way to describe a frequency standard's performance, time loss can also be used to describe variations in the frequency signal. Time loss is a measure of the difference between the number of cycles that have occurred in one frequency standard in an interval $\tau$ and the number of cycles occuring in a reference frequency standard during the same interval. 
This number is a function of the accumulated phase information in a sine wave over the measurement period $\tau$; therefore, we can calculate time loss using information about the phase drift over time (note that time loss, like Allan deviation, is always given as a function of $\tau$).  Because time loss indicates the accumulated phase difference, it tells us the final difference in the number of cycles of the device under test versus the reference for period $\tau$.  If we take the reference to be absolute, this difference determines the accuracy, or drift, of the clock away from the ideal frequency over time. The official IMPACT project goals are set in terms of time loss rather than frequency stability.  Our phase noise test set is also able to measure the phase variation.

%

\section{Important shifts} 
I have discussed \emph{stability} in terms of the Allan variance.  The \emph{accuracy} of a frequency standard is the difference of its output frequency from the \emph{true and exact} clock frequency when no external perturbations are present. Any atomic reference will have perturbations due to the environment.  Generally, the offset of the frequency from the \emph{true} frequency is calculated based on specific shifts known to be relevant and physically uncorrected, and the uncertainty posed by this value gives a measure of the accuracy.  A \emph{primary} frequency standard is one whose accuracy is said to be understood sufficiently that its uncertainty (accuracy) can be completely estimated quantitatively.  That is, it is a standard that we ``trust" up to its understood accuracy, without having to compare it to another standard.  An example is the F1 fountain clock at NIST.

The three main shifts of interest to us in our ion-trap standard are the second-order Zeeman shift of the clock-state energy levels, the second-order Doppler effect, and the pressure shift from the buffer gas in the package.  First, while the $|F=(0,1),m_F=0\rangle$ levels are insensitive to first order to magnetic fields, to second order the Zeeman effect causes a shift in  the fractional frequency  as 
\begin{equation}
\frac{\delta\nu}{\nu} = 4.9 \times 10^{-8}\  [\rm{G}^{-2}]\  B \delta B
\end{equation}
where $B$ is the bias field and $\delta B$ is the uncertainty in the field.  Next, the second order Doppler shift is caused by the thermal motion (and micromotion) of the ions in the trap causing a perceived shift in the clock frequency.  The average ion kinetic energy depends on the temperature and the number of ions in the trap, which influences the cloud density and size.  Thus, the second order Doppler shift requires a detailed understanding of the ion motion in a particular trap or relies on an approximate model.  
Last, the buffer gas shift is not well understood theoretically but can be estimated from reported measurements for the appropriate ion and buffer gas species. 
A discussion of the shifts relevant to ion trap standards can be found in \cite{1573975}.  Except for the effect of the second order Zeeman shift, at the time of writing this thesis, we have not reached a point where the second order Doppler and buffer gas shifts are relevant to our work.  We expect that even with the influence of these shifts, for the parameters relevant to our clock, we will be able to meet the stability goals for the project.




\chapter{The Experimental System: Test Bed Clock}
\label{ch:testBed}

Our foray into miniature ion clock design began with the construction of a ``test bed" system, which included a miniature ion trap in a conventional vacuum system, that was used throughout the package evaluation process to improve and compare our understanding of the miniature packages, and as a test system as we worked through miniaturization of several components.  This test bed also served as a proof of principle experiment for our small ion trap, since the ion trap was the same size as we originally intended for later traps (including the final design), despite its being held in a larger vacuum space.  After this step, we had in hand a working clock and a good understanding of the overall Yb ion trapping system, as well as a ``control" system with which we could evaluate new operational ideas.  

Because we must always be preparing for miniaturization, in this project we must continually be evaluating a variety of options or methods for each component of the clock system,  to put ourselves in a position where we can choose the best miniaturization-friendly technique.  In this chapter I will discuss in detail the basic elements of the test bed system we made initially, and the concepts of many of these additional components, some of which are only used later in the small packages.  My intent is to lay out all the experimental concepts here, in the context of the test bed system, while simultaneously preparing for the discussion of the miniature systems later.

\section{The Ytterbium Ion} \label{sec:Yb}

Named for a village on the Swedish island of Resar\"{o} where it was discovered, ytterbium is a rare earth metal of atomic number 70 and atomic weight 173.054 that is silver in color and very malleable.  It oxidizes relatively slowly in air and has a melting point of 824$^\circ$C.  It has seven stable isotopes: $^{168}$Yb, $^{170}$Yb, $^{171}$Yb, $^{172}$Yb, $^{173}$Yb, $^{174}$Yb, and $^{176}$Yb.  The isotopes and their properties are listed in Table \ref{tab:YbIsotopes} and resonant wavelengths for the isotopes we use are shown in Table \ref{tab:YbWavelengths}.  We of course use $^{171}$Yb$^+$ for our clock, but from a naturally-abundant sample, we also frequently use the even isotopes $^{174}$Yb and $^{172}$Yb when optimizing our system, since their structure does not require microwave pumping in order to detect ions.
\begin{table} 
  \begin{center}
  \begin{tabular}{ccccc}
    \hline
    Yb isotope & Abundance & Mass & Spin & Mag. Moment \\
\hline
    168 & 0.13\% & 167.933894 & 0 & 0  \\ 
    170 & 3.05\% & 169.934759 & 0 & 0  \\ 
    171 & 14.3\% & 170.936323 & 1/2 & +0.4919 \\ 
    172 & 21.9\% & 171.936378 & 0 & 0  \\ 
    173 & 16.12\% & 172.938208 & 5/2 & -0.6776 \\
    174 & 31.8\% & 173.938859 & 0  & 0  \\ 
    176 & 12.7\% & 175.942564 & 0 & 0   \\
\hline
  \end{tabular}
  \end{center}
   \caption{Isotopes of ytterbium and their properties \cite{nistData}.}
      \label{tab:YbIsotopes}
\end{table}
\begin{table} \label{tab:YbWavelengths}
  \begin{center}
  \begin{tabular}{cccc}
    \hline
  \multicolumn{4}{r}{Resonant Wavelengths} \\
  \cline{2-4}
   Yb Isotope & 369 nm & 935 nm & 399 nm \\
\hline
    171 & 739.0516 & 935.1875 & 797.8213  \\ 
    172 & 739.0489 & 935.1872 & 797.8218 \\ 
    174 & 739.0500 & 935.1791 & 797.8224 \\
    \hline
  \end{tabular}
  \end{center}
\caption{Resonant wavelengths are listed for the isotopes we use most often, as read from our wavemeter.  The 369 nm and 399 nm lasers are measured from the fundamental beam before doubling and listed as they are read on the wavemeter.  The wavelengths can vary in the last two digits for different packages: those listed here are for the test bed package and are our most commonly used values, but should not be taken as absolutes.  The resonant wavelengths are determined based on the optimal ion signal strength.}
 \label{tab:YbWavelengths}
\end{table}

Ytterbium has been used in the past for high-resolution spectroscopy experiments using the long-lived $D$ and $F$ states \cite{PhysRevA.49.3351,PhysRevA.56.2699}.  Many groups have adopted Yb recently for use in trapped ions systems for studies of quantum information processing \cite{PhysRevA.73.041407,PhysRevLett.100.150404}.  Yb has also previously been used in a Yb trapped ion frequency reference in the Fisk group in Australia \cite{585119}.  Last, Yb has a promising future as an optical frequency reference, using the highly narrow lines \cite{PhysRevLett.94.230801,5544331,1408282}. 

Despite the fact that Yb has a complicated energy level structure, depending on system requirements, the main pumping processes can be dealt with primarily using 3 reasonably achievable laser wavelengths.  In our case, just 2 lasers are sufficient. Also, ytterbium has a favorable subatomic structure for our use.  Since Yb has a nuclear spin of $I =1/2$, there is only one Zeeman level in the lower ground state ($|F=0\rangle$), making the $|F=0,m_F=0\rangle$ and $|F=1,m_F=0\rangle$ hyperfine ground states ideal for use as a clock state with a first-order magnetically-insensitive separation of 12.642812 GHz. Because of the spin-1/2 nucleus, a cycling transition exists between the $^2S_{1/2}$ ($|F=1\rangle$) and $^2P_{1/2}$ ($|F'=0\rangle$). We rely on this cycling transition for a strong signal when we want to detect the population in the $|F=0,m_F=1\rangle$ state.  Preparing the ions back into the $|F=0,m_F=0\rangle$ state to begin the next cycle is straightforward since an ion will eventually be pumped back to the dark state during this process through decay via the $^2D_{3/2}$ state. However, because we also repump out of this state to retain the ions in the cycling transition, an ion scatters an average of 2000 photons before returning to the $|F=0, m_F = 0\rangle$ state.  This makes state preparation and readout very straightforward (with the addition of pumping lasers) since they both occur during the same pulse of 369 nm light.  


\begin{figure}
   \centering
  \includegraphics[scale=0.8]{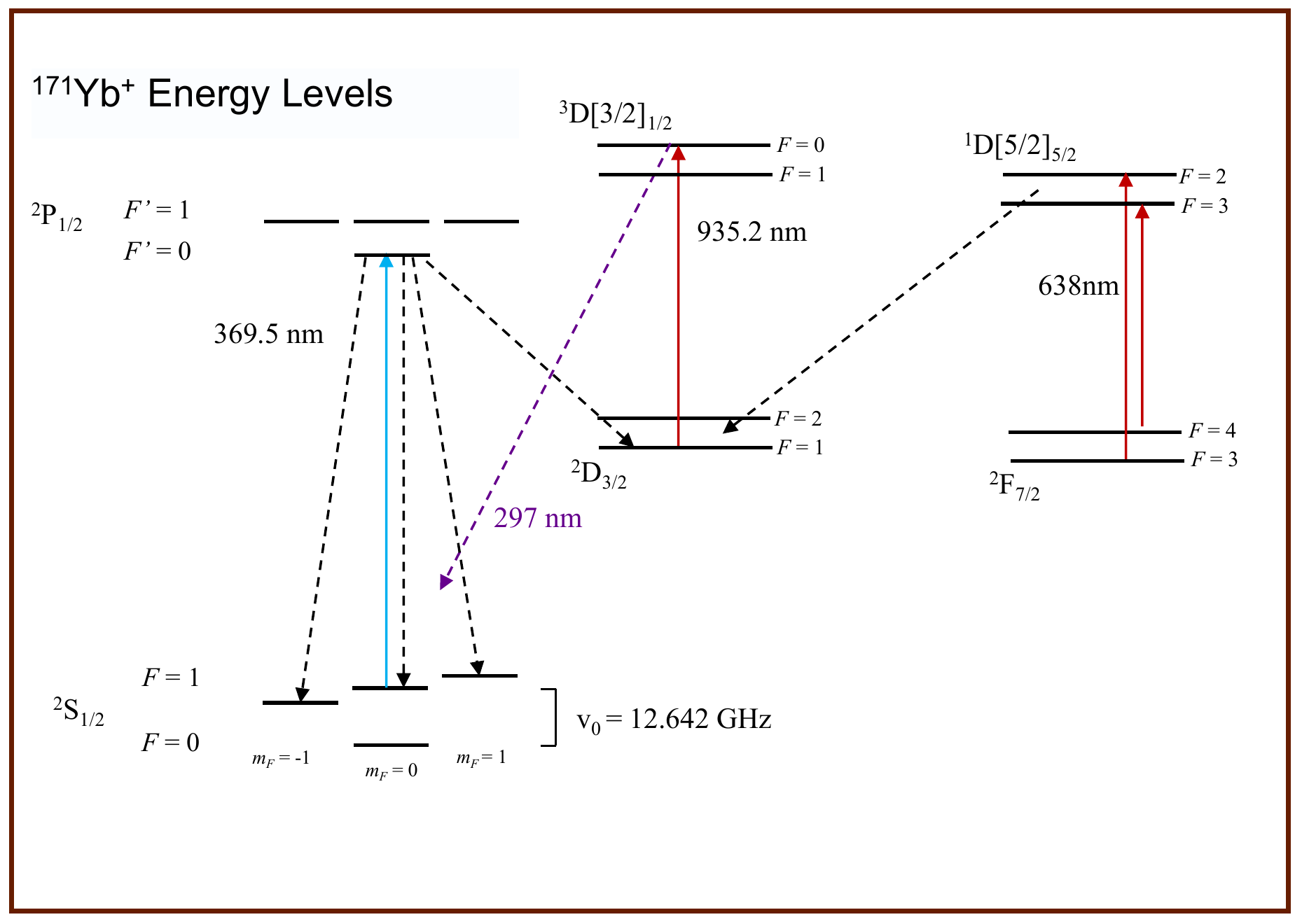}
   \caption{Partial level diagram relevant to our work.  Solid lines represent possible lasers in the experiment.  At least the 369 nm cycling transition and the 935 nm repump must be driven to detect and prepare in the ground state manifold.  The ions enter the $^2 F_{7/2}$ state every few hours, which has a lifetime of 5.4 years, through collisional mechanisms.  Thus depending on conditions, the 638 nm repump may also be required. }
    \label{fig:YbLevels}
   \end{figure}

A diagram of the levels of $^{171}$Yb$^+$ relevant to our clock system appears in Fig. \ref{fig:YbLevels}.  The $|F=0,m_F=0\rangle$ and $|F=1,m_F=0\rangle$ levels of the $^2S_{1/2}$ ground states are separated by 12.642812 
 GHz which provides the magnetically insensitive hyperfine microwave clock transition.  The $^2S_{1/2}$ to $^2P_{1/2}$ cycling transition is used simultaneously for state detection and preparation. This optical transition is typically used for laser cooling, but laser cooling is not necessary in our system.  In addition, a decay to the $^2D_{3/2}$ line occurs with a probability 0.5\% \cite{PhysRevA.61.022507}.
 This metastable state is cleared by a repumping laser at 935.2 nm which returns population to the $^2S_{1/2}$ ground state manifold through the $^3D[3/2]_{1/2}$ levels.  Last, the ions can also fall into the low-lying $^2F_{7/2}$ state which has a lifetime of a few years.  This forbidden transition can occur due to collisions with foreign gases in the trap.  A 638 nm laser is sometimes necessary to pump out of this state if loss to the $F$-state becomes a problem.  Another possibility is, for example, to repump using an 850 nm laser through the $^3[5/2]_{3/2}$ transition. 
  Alternatively, this issue can be dealt with by using a quenching gas which can remove ions from the $F$-state collisionally instead of through a resonant interaction \cite{JPLquenching}. 
 Because of the dependence of this state on collisions, the necessity of the 638 laser depends on the experimental situation.  While we have observed some loss into the $F$-state, we do not require a 638 repump for our miniature clock system to function satisfactorily.

Some other characteristics of ytterbium that are not strictly necessary turn out to be convenient for our work.  One is the fact that although Yb reacts with air, it does not oxidize too quickly, and when it does, an oxidized layer forms on the outside and does not oxidize the whole bulk substance. This is useful for us because it means that with sufficient heat in vacuum the Yb sample can be made usable again after contacting air, by heating and breaking through the oxide layer via sublimation. 
 In practice we find that even ytterbium samples that have been in air for weeks are still useable. (This can be an issue for us since we often change which traps are connected to our system and the ovens are sometimes exposed for long periods, coupled with the fact that replacing the Yb in oven in a package can be a very tricky process.) This slow oxidation is not the case for many other species, for example calcium (we will see a problem with the oxidation of calcium in Sec. \ref{sec:JPLphase1}).  Also, the fact that ytterbium has a very low work function makes it an easy target for ejecting photoelectrons for ionization (see Sec. \ref{sec:neutralAndIonize}). 

There are disadvantages to using ytterbium, of course.  The fact that we must heat it to around 400$^\circ$C in order to achieve adequate vapor pressure to capture ions 
complicates a miniaturized system.  We need to heat part of a tiny clock to this temperature, thereby heating the whole system considerably (however seldom we actually have to do so).  As a contrasting example, JPL's 1-liter mercury clock requires no heating because the vapor pressure of mercury at room temperature is sufficient to trap without an oven or heating element \cite{4623080}
(this scheme has drawbacks as well: a high pressure of background Hg may disturb the system more than in a system like ours where we can ``turn off" the presence of the vapor).  


\section{Vacuum system}

The first system to trap ions for this project already existed when I first arrived at Sandia.  It was assembled by my thesis advisor Peter Schwindt and George Burns, the technologist on the project.  This large, ``conventional" vacuum manifold was originally built for the test bed, and later modified many times to add and test small packages.

\begin{figure}
   \centering
  \includegraphics[scale=0.7]{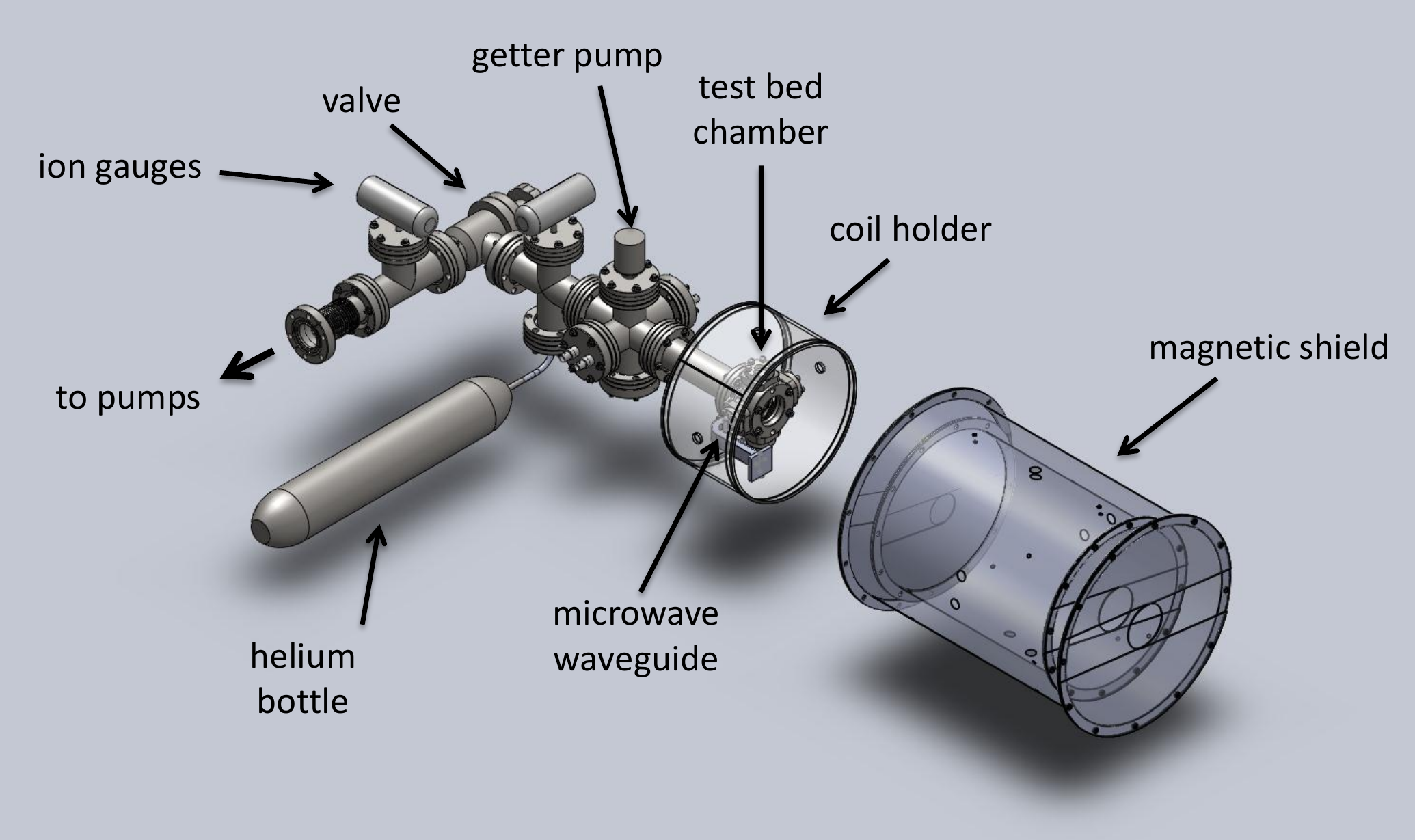}
   \caption{Solidworks image of the original chamber design, showing the test bed chamber, helium bottle, and the coil wrapping and shield assemblies.  (This model courtesy of George Burns) }
   \label{fig:OriginalChamber}
 \end{figure}

\begin{figure}
   \centering
  \includegraphics[scale=0.8]{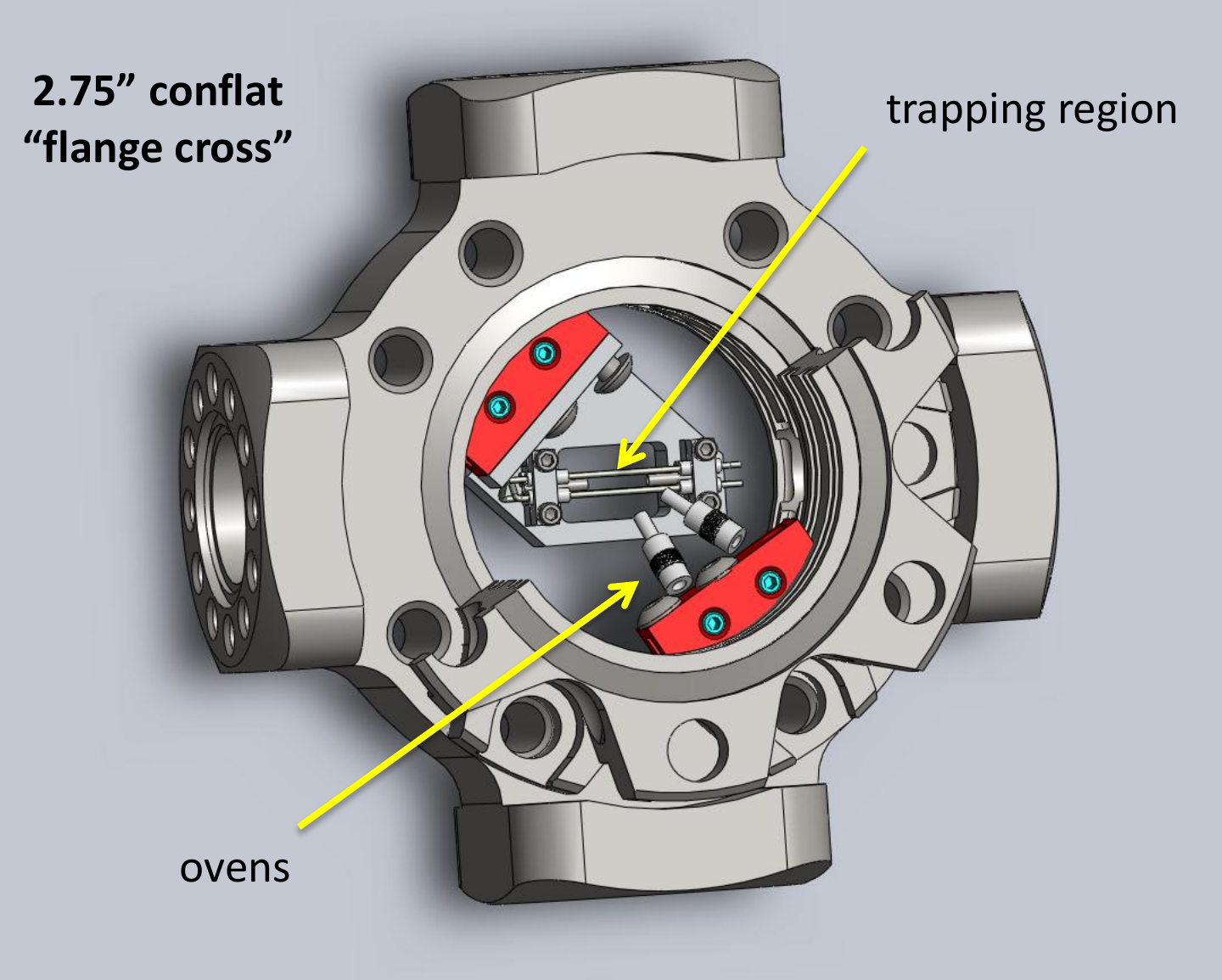}
   \caption{Closeup of the Solidworks version of the test bed trap held in the Kimball close coupler by groove grabbers.}
    \label{fig:OriginalTrap}
 \end{figure}

\begin{figure} 
   \centering
  \includegraphics[scale=0.6]{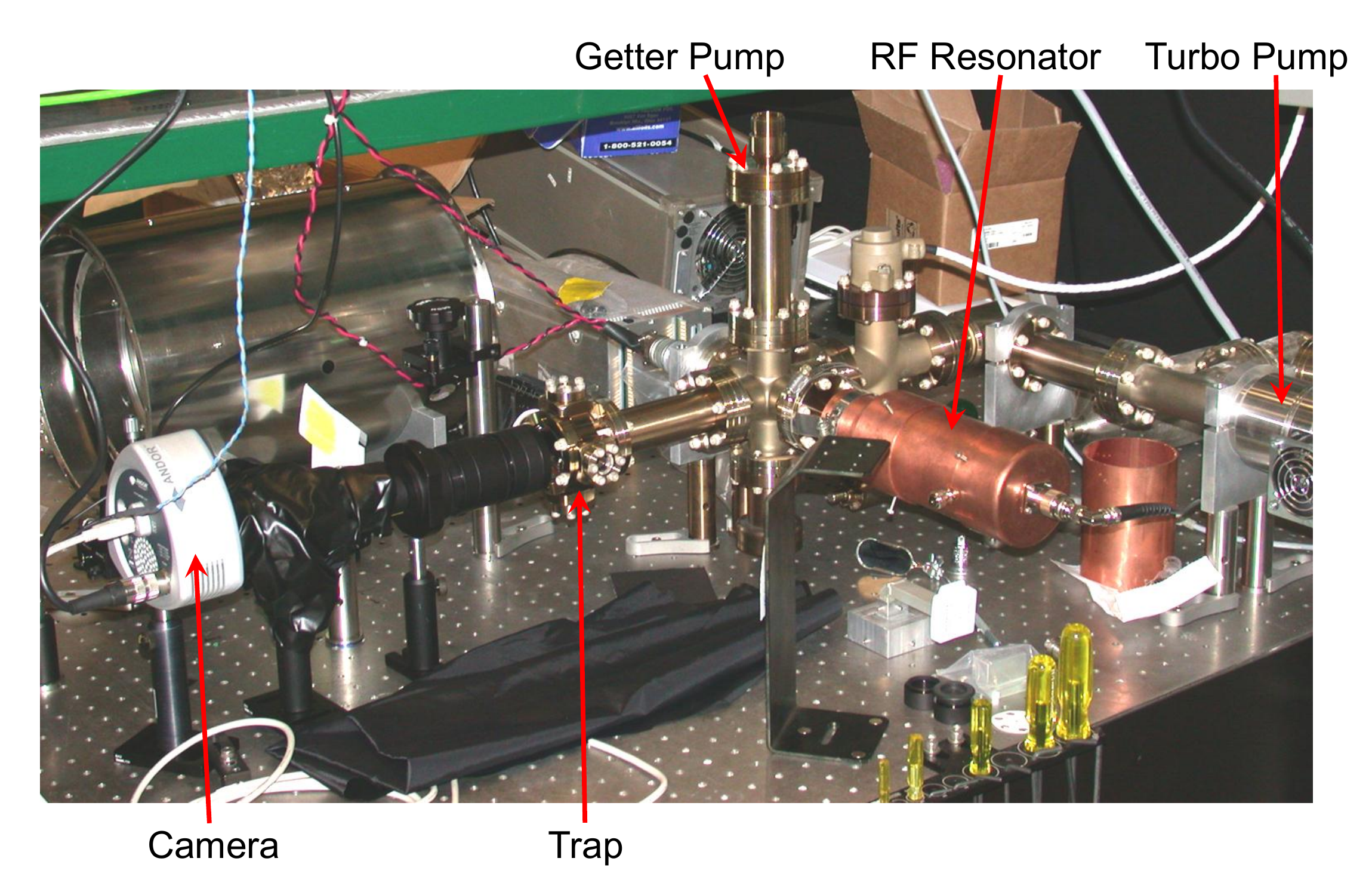}
   \caption{Photo of the original test bed chamber}
   \label{fig:OldChamberPhoto}
\end{figure}

The original version of the vacuum chamber is depicted in Fig. \ref{fig:OriginalChamber}.  Its physics chamber is shown for clarity in Fig. \ref{fig:OriginalTrap} which consists of a small ion trap in a relatively small volume chamber, which I will describe in detail in Sec. \ref{sec:minitrap}.  A helium bottle connected to the system through a calibrated capillary leak element and a valve provides the ability to flood the system with buffer gas, a necessary component of our planned system.  The chamber pressure is controlled by two turbo pumps: a Varian Turbo-V 81M turbo pump backed by a Pfeiffer Hi-Cube Eco pumping station which includes a roughing pump capable of pumping from atmosphere.  This dual turbo system is necessary to be able to continuously flow helium through the chamber and maintain low partial pressure of other gases.  We never need an ion pump, since we are always either actively turbo pumping our traps, or operating them in a sealed state where pumping is performed only by getters (see Sec. \ref{s:getters}).  

Also built into the original system is a valve that can close off the branch of the chamber containing the ion trap (and getter) from the rest of the vacuum manifold containing the pumps and gauges, in order to test the sealed system.  Additionally, two ion gauges (Varian) used for monitoring pressure in the chamber and a Residual Gas Analyzer (RGA) system (Stanford Research Systems) for monitoring partial pressures of individual gases were attached to the system for diagnostic purposes.   A photo of the chamber is shown in Fig. \ref{fig:OldChamberPhoto}.

\subsubsection{Swapout and bakeout procedures}
Since we frequently change small ion trap packages that attach to the vacuum, we needed a reasonable system for swapping out packages for testing and characterization.  We kept the test bed chamber functioning with the standard small ion trap during the first two phases of the project (during which we tested iterations of many different small package designs), which could be sealed by a vacuum valve from the rest of the system, so that we could always go back to it for testing and comparison.  We added an additional prototype package testing area to the chamber, that could utilize the same pumping station, RGA and ion gauge diagnostics, and He source as the test bed chamber, with valves that could seal off new packages as well.  Anytime we connected a package to the system, we would attach it to a valve not only so we could test the stability of vacuum inside the sealed package, but with the hope that we could remove the valve from the main vacuum manifold and still use the sealed chamber.  The ultimate goal was to pinch off each package at the copper pinch-off tube attached to the chamber.  We did pinch off one package in Phase I, but before performing such a permanent procedure we of course  prefer the reversible use of valves to test the validity of the sealed systems.

\begin{figure} 
   \centering
  \includegraphics[scale=0.5]{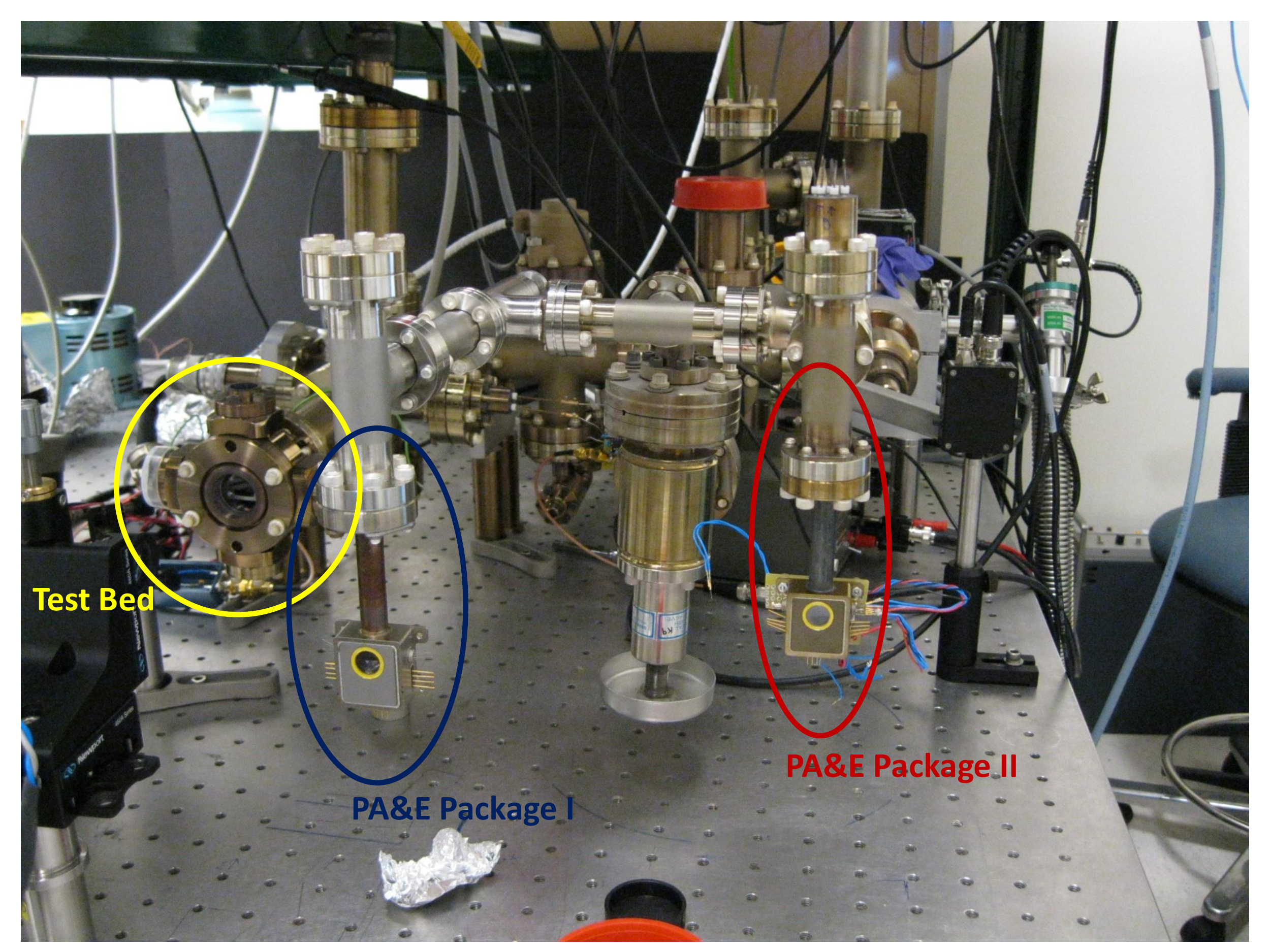}
   \caption{Our chamber after it was setup over time to accomodate numerous packages at once.}
   \label{fig:TripleChamber}
 \end{figure}

We eventually wanted to test more than one package simultaneously, so we split the package testing area with a tee and and were able to attach two packages, with their own individual valves, in a manner that could also accommodate our imaging and laser systems.  An example of one iteration of the vacuum manifold with two packages attached, as well as having the test bed chamber intact, is shown in Fig. \ref{fig:TripleChamber}.  

The bakeout procedures we used for the test bed vacuum chamber were quite severe.  In consideration of the fact that we want to have long trapping lifetimes in a sealed, getter-pumped volume, we made it our aim to construct each chamber or package to be bakeable at 400$^\circ$C.  An experienced vacuum system user will recognize this as a very high bake temperature.  More common baking temperatures, even for achieving UHV ($<10^{11}$ Torr) for ultracold atomic systems for example, are in the range of 200 to 300$^\circ$C under ordinary conditions.   We achieved our goal rather easily in the larger chamber, since many high-temperature-bakeable parts are available. In particular we had to be sure to avoid any organic or lower-temperature-tolerance parts inside the vacuum to make up the trap and the associated electrical connections.  Still, since 400$^\circ$C is much hotter than most ordinary UHV bakes require, although this temperature was within the specifications for the parts used\footnote{Many UHV vacuum parts are bakeable to 450$^\circ$C, except for some windows and electrical feedthroughs which may be limited to around 250$^\circ$C, especially e.g. windows with anti-reflection coatings.  In addition, care must be taken to avoid gradients or rapid changes in temperature while heating and cooling the chamber.}, our bakes were not without failures of some parts.  Particularly, feedthroughs and other connecting parts seem to have an increased failure rate when pushed toward the extreme of their acceptable temperature range.  In addition, the repeated temperature cycling and physical stress\footnote{Each small package test required torqueing and untorqueing valves, changing basic vacuum hardware, exposing part of the system to air, doing a localized bake, etc.} that was necessary due to the swap-in swap-out trial and error procedures that we used when testing the miniature packages also caused some wear and tear on our vacuum system and thus the occasional leak. 

\begin{figure}
   \centering
  \includegraphics[scale=0.6]{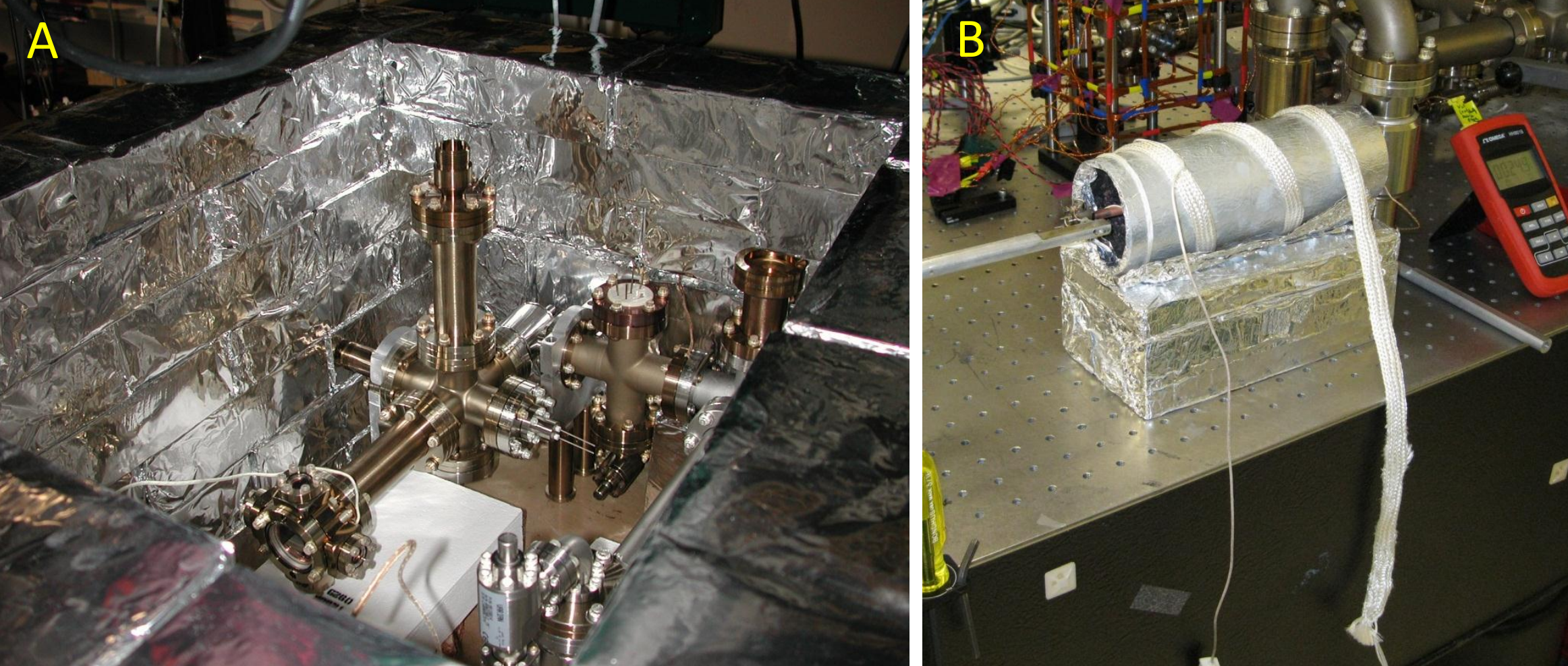}
   \caption{A: Building an oven around the original chamber with foil-covered bricks. A brick-style heater can be seen below the chamber in white. B: Setting up a localized bake on a small package.}
    \label{fig:bakePhotos}
 \end{figure}

Our bakes were done in situ on the optical table using local heaters (heater tapes or heating brick with internal filaments), with oven bricks and foil for insulation.  In particular, the initial bake of the entire system (which was done twice: once before I arrived at Sandia when the first trap was assembled, and a second time after adding more components and the first small package) required building an an entire oven from insulating bricks to enclose the whole chamber.  Even with very powerful brick heaters, it was difficult to reach 400$^\circ$C, but we did so after adding many layers of foil and bricks to provide insulation from the cooler outside air.  A photo of the chamber and part of this oven building process is shown on the left in Fig. \ref{fig:bakePhotos}A.  

Reaching this temperature in the smaller packages was easier, especially since after the initial two bakes we found it was only necessary to bake locally, that is to bake just the newly-placed package itself and the part of the attached vacuum piping leading back toward the pump that was exposed to air during replacement.  In these cases the heating was done with heater tape and small ovens could be made of foil and metal tubes that just fit over the part in question, as seen in Fig. \ref{fig:bakePhotos}B.    In a smaller oven and with plenty of spare bricks to use as multi-layer insulation from the outside, we could reach 400$^\circ$C reliably.

A much greater challenge was developing small packages that could withstand such a hot bakeout.  We were not always successful in this regard, due to manufacturing constraints and procedures required to assemble packages (including windows, feedthroughs, electrical connections etc.) at the scale required.  The all-metal packages designed by JPL were made from all parts bakeable up to 450$^\circ$C, but the PA\&E package and later LTCC/HTCC packages required compromises to the preferred bakeout temperature due to assembly difficulties.  This is discussed more in Chapter \ref{ch:packages} where the packages are described individually.

\subsubsection{Getters}
\label{s:getters}
A getter is a vacuum element that contains a reactive material that can be used to maintain vacuum without active pumps, if the vacuum is sufficiently prepared. Gas molecules inside the vacuum system adsorb or combine chemically with the getter material, removing them from the vacuum space and effectively ``pumping" the vacuum volume.  (Getter materials have been used for many decades to maintain vacuum in the vacuum tubes used for TVs and radios.)  This tool is very important for us in miniaturization, since it allows us to maintain an ion-trapping chamber with no active pumps that can be permanently sealed. 
The large-chamber getter pump is a ``macroscopic" Capaci-Torr getter, which takes up the space of a full 2.75" vacuum nipple.  The getter elements are thermally activated with electrical heating elements raising the temperature of the getter to 450$^\circ$C for 45 minutes. The surface of the getter material is saturated with oxides and carbides after exposure to air or higher pressures (such as during a bake). This layer is removed by heating, which causes the molecules on the surface to either diffuse into the bulk material of the getter or desorb from the surface, leaving a "clean" surface for further adsorption.  This activation step is performed while we are actively pumping on the system, so that any gas molecules ejected from the getter surfaces can be removed through an external pathway.  This activation step is only necessary after bakeout or exposure to air.

\begin{figure}
   \centering
  \includegraphics[scale=0.6]{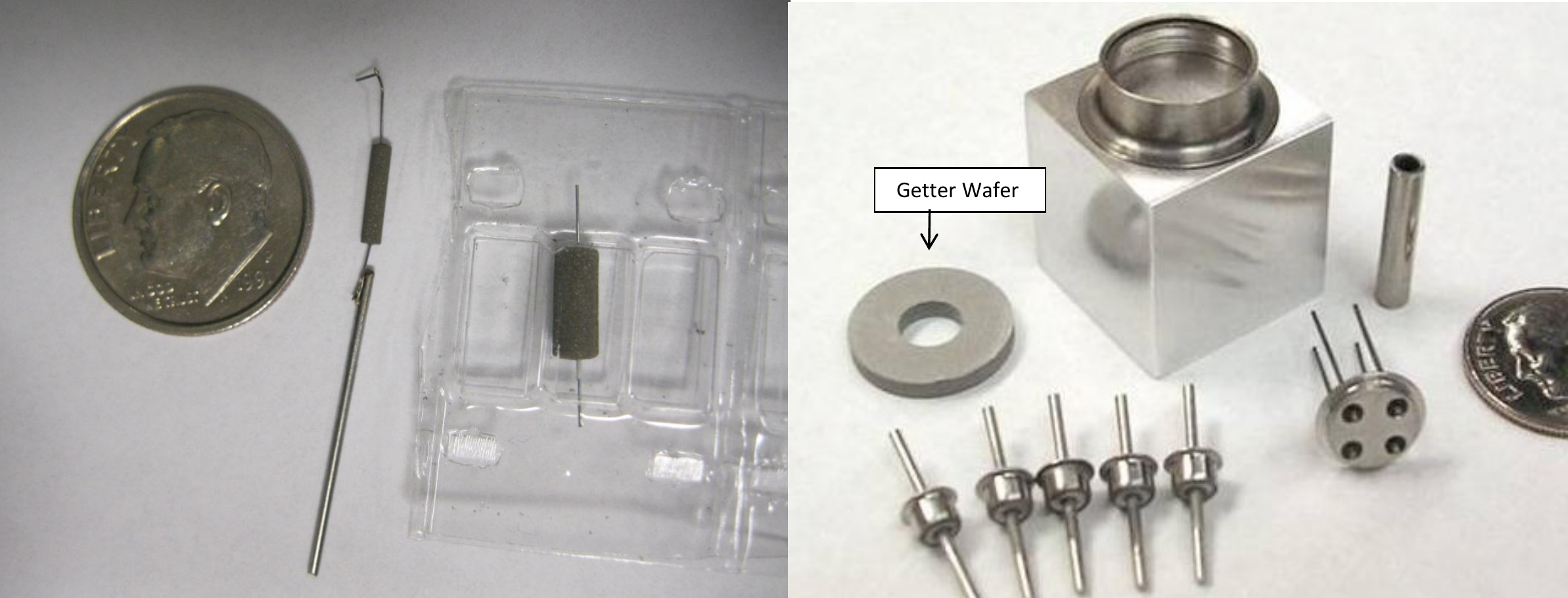}
   \caption{A: Cylindrical getters from SAES Getters that we use in the small packages.  They can be activated by running current through the center heating wire.  B: Donut-shaped getter used in the JPL packages, shown with the other package parts.}
   \label{fig:GetterPics}
 \end{figure}

In the miniature packages, we use a different type of getter, a small volume of reactive material which, by the same principle, can be heated up in order to ``reactivate" its pumping capability after saturation.  It is important for us to activate and/or reactivate the getter either using an electrical heater or through a baking process before temporarily or permanently sealing off any package.  Otherwise, outgassing from surfaces and gases that may permeate the package walls over time will eventually cause a gradual increase in pressure inside the vacuum volume.  The miniature getters used in several of our packages are cylindrical nonevaporable getters made from  St172 (a getter alloy material) from SAES Getters, which we purchased in two sizes (1.5 mm and 3 mm diameters, each 7 mm in length).  Some of the getters we use are pictured in Fig. \ref{fig:GetterPics}A.  These particular getters have a heating wire through the center that can be used for activation after baking, since the saturation from higher pressures during baking can happen especially fast to these small getter elements.  These St172 getters can also be activated passively by baking, which eliminates the need to be concerned about additional feedthroughs for the getter heater.   Baking even at lower temperatures (\textgreater250$^\circ$C) for several days should be sufficient to activate the getter material, and may not simultaneously saturate the material if the chamber is already clean enough.  In fact, in later packages we have used donut shaped getters made from a similar getter material (SAES St175) that do not have a heating wire, therefore they can only be activated during a bake (see Fig. \ref{fig:GetterPics}B).  Even for our 1 cm$^3$ ceramic package we have been able to incorporate off-the-shelf getters, but an alternative for small volumes would be to sputter getter material directly onto some inside surface of the package, where it can perform the same function.  This option is technologically viable now (in fact, this is how the old vacuum tube getters were applied) but we did not find it to be a necessary step through the end of Phase II.




\section{Laser system} \label{sec:lasers}
For our final system, we require only two lasers to run the atomic clock: the 369 nm to excite the optical cycling transition for detection and state preparation, and the 935 nm laser to repump out of the metastable $^2D_{3/2}$ state.  In our lab setup, we also need a 399 nm laser for photoionization, and we also have a 638 nm laser for potentially pumping out of the $^2F_{7/2}$ state, which we have not used with our system at all.  Since we have found no necessity for the 638 nm laser (and we don't intend to photoionize in the small packages), our experiment only requires miniaturization at two wavelengths.  Both of these detection lasers as well as the photoionization light are sent into each of the packages collinearly and from the same side, and a lens focuses them approximately in the center of the trap.  The focal length of the lens varies with different traps and the individual approaches for the lasers into the packages is discussed in Chapter \ref{ch:packages}. 

For the miniature system, two small lasers are being developed at Sandia: a 935 nm VCSEL, which was ready and implemented in Phase I with our portable clock demonstration, and a 369 nm VECSEL, frequency-doubled in a periodically-poled KTP (potassium titanyl phosphate) waveguide, which will be used to demonstrate our Phase II clock. 
The photoionization laser at 399 nm that we use in the lab is replaced in the smaller packages by use of a mercury lamp, free-running laser diode, or LED, using methods other than the 2-photon ionization to load the trap.  Techniques for ionization, including the use of these elements, are described in Sec. \ref{sec:neutralAndIonize}.

\begin{figure} 
   \centering
  \includegraphics[scale=0.6]{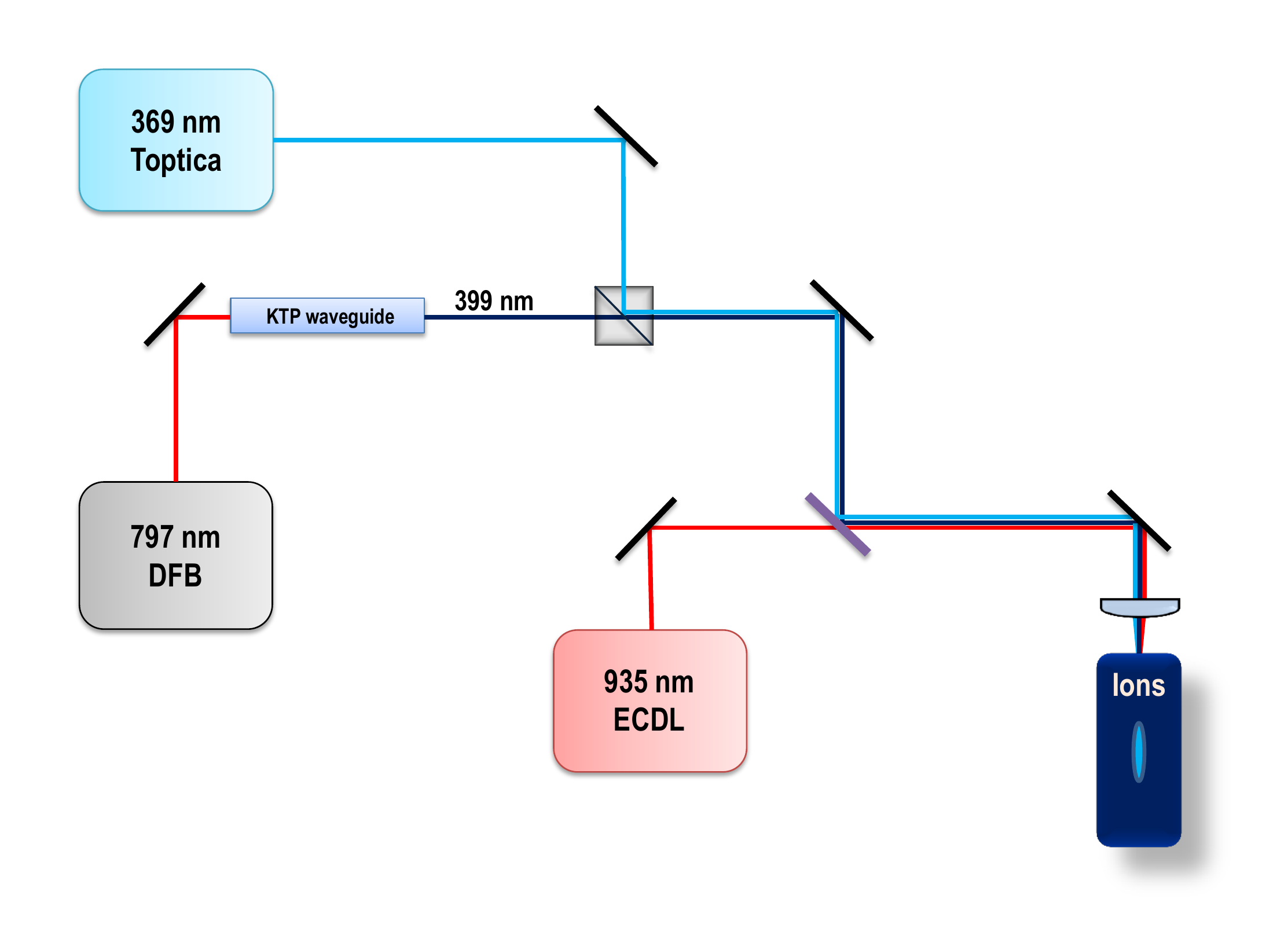}
   \caption{Schematic of our laser sources and their path to the trap}
   \label{fig:LaserSystem}
 \end{figure}

In the lab, the 369 nm laser is a Toptica SHG-10, which is made up of a diode laser at 739 nm and tapered amplifier with a second-harmonic-generation cavity which doubles the frequency to create 369 nm light.  This system is specified to put out 60 mW of 369 nm light, but in reality, due to what we believe is a degrading doubling crystal we usually have less than 10 mW.  Fortunately, we do not require a lot of power to interrogate the ions, and in fact, after optimization we often want to test operation with a very low power, since a miniaturized light source  at 369 nm with a much lower power will ultimately be used to run the clock.  At the trap, the 369 nm light power is between 5 $\mu$W and 1 mW.

The lab 935 nm laser that we use as a repump is a homebuilt external cavity diode laser (ECDL) that puts out around 20 mW of infrared light.   I designed and built this laser based in large part based on a design by Scott Papp \cite{PappThesis2001} (it has also been used to make several other ECDLs here at Sandia). 
At the trap, the power at 935 nm is usually around 10 mW, but in reality we only need a small amount of rempump light ($<$1 mW), 
especially when operating with low 369 nm power.   We decided to make a homebuilt ECDL because attempts to cool a 937 nm DFB (distributed feedback) laser sufficiently to use at 935 nm led to an unstable and mode-hopping source that could not reach and maintain the desired wavelength for an extended period of time. 

Last, the photoionization laser we use at 399 nm is a DFB laser at 795 nm, heated to reach 797 nm, and frequency doubled using a periodically-poled KTP (potassium titanyl phosphate) waveguide from the company AdvR, Inc.  The infrared light is coupled into the crystal and optimized for 399 nm transmission through the crystal.  Typical efficiencies for the frequency doubling are on the order of 0.1\%\footnote{We input around 30 mW at 797 nm and see less than 50 microWatts at 399 nm in recent waveguide crystals.}.  We have received and implemented various crystals, which have had a tendency to fail (probably by becoming unpoled) after operating for a period of days to months.  The more recent crystals have a significantly lower efficiency than the earlier ones.  When I first arrived at Sandia, we could typically get a few hundred microWatts out of this system right after the doubling crystal for approximately 32 mW of input power, with little adjustment necessary. During Phase II, with some adjustment to alignment or temperature every few days, we usually get between 40 and 60 microWatts of power after the crystal for nearly the same amount of input power (28 mW).  Since we have optics that span several wavelengths and 399 nm is our least important wavelength, there are losses on the way to the trap of more than 50\%. 
This means that at the trap we sometimes use $<$20 microWatts of power to load.  Fortunately, this has been enough power to load in the later stages of the project, coupled with the fact that we also can load using other techniques.  The 399 nm laser has always been the most efficient way to optimize the system, however, since it is the only technique that reliably allows continuous simultaneous loading and probing\footnote{This situation changed somewhat in Phase II when were able to load using a 405 nm laser diode, which would not saturate our detector while it was on, meaning we could probe while loading.}.

The scheme for sending the lasers to the trap is shown in Fig. \ref{fig:LaserSystem}.  The 369 nm and 399 nm lasers are combined on a polarizing beamsplitter cube coated for UV.  These blue beams are combined with the IR 935 beam using a dichroic filter which passes the 935 nm light and reflects in the UV.  Then the combined beams are maneuvered by several broadband (Thorlabs silver P01) mirrors to the various traps.  There are of course (acceptable) losses due to the fact that we are not using wavelength-specific optics after combination.  The combined beams are also focused using an uncoated lens, and the different wavelengths will have slightly different focal points inside the trap.  While these details may have some effect on trap performance, we find that in practice, if our three beams are relatively collinear from their initial combination point all the way to the focusing lens, then the lasers perform their respective functions efficiently.  In the end, final optimization of individual beams in order to obtain maximum ion signal is given highest priority.

All the lasers are free running in wavelength, adjusted periodically by hand or by a simple lock to the software of the wavemeter while we work with the system.  When we run the clock, we lock the laser wavelengths directly to the ions, by optimizing the fluorescence signal using each laser separately, in tandem with the clock measurement.  Naturally, the purpose of this procedure is a practical solution for our ultimate miniature system.  More information about this locking process is explained in Sec. \ref{sec:laserlock}.   The modeshapes of the lasers are also not optimal, but due to the nature of our ultimate experiment we do not require extreme uniformity in the beam.  Although our signal could probably be improved in the laboratory setup by giving some attention to these areas, in our final design we will have minimal control over these factors and it is operationally useful to know that the overall conditions on the light sources can be relaxed.

\section{Detection systems} \label{sec:detection} 
The natural choice for the ytterbium-based system is to detect on the cycling transition at 369 nm, because that is the most abundant source of photons.  A signal at 369 nm should therefore provide the highest fidelity peak for locking the clock to the atomic resonance.  However, detecting at the same wavelength as the laser being used to excite the transition causes a serious problem in smaller systems: stray scattered light within a small, (shiny) metal chamber causes a very large background, which in turn makes it harder to resolve the signal for our clock resonance lock, and increases our sensitivity to fluctuations in power and other noise from the laser.  Besides engineering constraints that physically restrict stray light using the geometry of the package, one way to deal with this problem is by ensuring that an imaging system images the ion cloud very well, thereby avoiding a large part of the misplaced background light.  Without a good imaging system in place, although a signal may still be resolved, the background counts can be orders of magnitude larger than the signal size.  However, this imaging also requires additional space, so that the emitted photons from the ion cloud can be collimated and collected in a well defined image.  The distance required to image the ions would make it impossible for us to reach the size goals of the project with conventional optics. 

 Because of these issues, detecting at 369 nm in a miniaturized system is far from optimal.  Instead, in the small packages we often choose to look at another wavelength altogether, which has an almost completely negligible background: 297 nm.  The source of 297 nm is shown in Fig. \ref{fig:YbLevels}.  Ions that have fallen out of the cycling transition into the $^2D_{3/2}$ state, decay back to the ground state, after being repumped by the 935 nm laser into the $^3D[3/2]_{1/2}$ state,  by emitting a 297 nm photon. Since the ions decay to the $^2 D_{3/2}$ state with a 0.5\% probability (not to mention additional optical and detecting losses), this signal is hundreds of times smaller than the 369 nm signal in principle; however, its complete lack of background still makes it a reasonable choice for detecting the clock resonance.  The lack of background also means that we can ultimately eliminate the need for an imaging system altogether, choosing instead to put the detector (PMT) as near to the detection window as possible, since then the only concern is to collect as much light as possible.

\begin{figure}[t]
   \centering
  \includegraphics[scale=0.7]{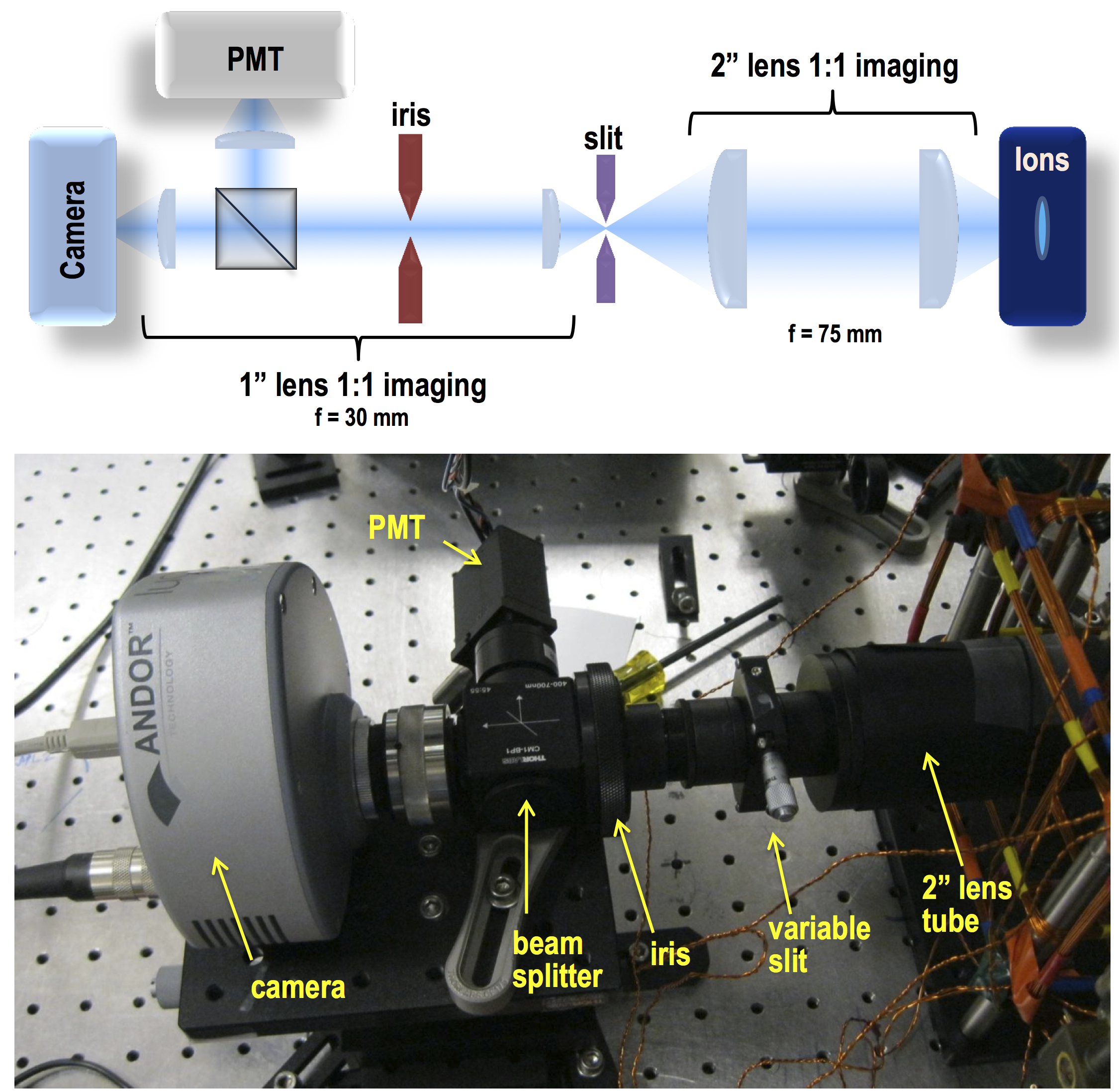}
   \caption{Schematic and photo of our imaging system used for the test bed and for initially evaluating the miniature packages.}
   \label{fig:ImagingSystem}
\end{figure}

For our macroscopic system, the imaging system is shown in Fig. \ref{fig:ImagingSystem}.  We use an Andor Luca R camera to locate the trap and look for neutral fluorescence using the 399 nm laser, and a Hamamatsu photomultiplier tube (PMT) to photon count on selected wavelengths for the clock signal\footnote{We have used the Hamamatsu H7155 and later the H10862-210 PMTs, since the second one has around twice the detection efficiency of the first at 297 nm.}.
 The imaging system consists of two 1:1 imaging configurations, with a slit oriented to the cigar-shaped cloud for spatial filtering of background light at the focus between the two, an iris so that we can reduce the overall light that reaches the detectors, and a 50-50 beamsplitter that allows us to use the camera and the PMT simultaneously.  Normally, the camera has no filtering and can see all wavelengths in its natural range, whereas the PMT has filters in front of it to select the 369 nm or 297 nm light as appropriate.  All these elements are inside Thorlabs optical tubing to minimize effects of room light.

When measuring 369 nm, we must close the slit as much as possible while optimizing the signal in order to reduce the background. In addition, we add baffles that reduce the clear aperture near to the front end of the imaging system to reduce background that we believe to come from light bouncing inside the optical tube and causing even more stray light to reach the detector.  To measure 369 nm we use a 369 nm filter from Semrock in front of the PMT.

When measuring 297 nm, we open the slit and iris as much as possible, and remove the baffles used at 369 nm.  We can usually obtain a larger signal by putting the PMT with filters as close as possible to the detection window of the vacuum package, but this is not practical for optimization procedures when we want to use the camera as well.  However, once the system is characterized, a PMT immediately in front of the window is sufficient for operation.  Also, when we initially made the switch to 297 nm we replaced the conventional BK7 lenses we used to detect at 369 nm with UV-grade fused silica lenses, 
to minimize losses at 297 nm while still being able to detect at both wavelengths.  Other challenges of detecting at the deep UV wavelength of 297 nm include finding a detector whose specifications reach that wavelength and ensuring that optical feedthroughs to the vacuum are compatible (since many glasses are very reflective in the UV).  We use sapphire windows in all of our packages, and sapphire has around 80\% transmission (including reflections) 
  at 297 nm (for 1 mm thickness). 
Last, finding filters that block wavelengths away from 297 nm without reducing the signal significantly was difficult.  We use a filter, Semrock FF01-292/27-25, which is specified to have  \textgreater70\% transmission at 297 nm, but we quickly learned that our powerful 935 nm beam could pass through this filter.  A simple solution was to combine the Semrock filter with a simple color filter from Thorlabs (FGUV11), that also has \textgreater70\% transmission at 297 nm but blocks 935 nm.  In addition, we later found it advantageous to put two of the Semrock filters together with the color filter in order to adequately reduce leakage from other wavelengths.

The typical procedure for setting up the detection of a new package is to use the camera to locate the trap electrodes through the window and find the estimated center of the trap.  Then the camera is used to look for neutral fluorescence to ensure we have Yb available to the trap.  An image of neutral fluorescence can be found in Sec.  \ref{sec:neutralAndIonize} (on ionization).  After focusing on the neutral fluorescence, generally speaking the imaging system must be moved from this position closer to the ion trap in order to see the trapped ion signal.  This is due to the wavelength difference between 297 nm (or 369 nm) and the 399 nm neutral light affecting the focal length of the singlet imaging lenses. After these steps, we move our attention to the PMT and look for trapped ion signals on the photon counting signal.  We do this by sweeping the 369 nm laser continuously across resonance, and adjusting the alignment until we can see periodic peaks on the photon counting signal, meaning that enough ions are trapped to get a visible signal. 
Although we sometimes perform these steps while looking at 369 nm (this is especially reasonable in the test bed trap), for small packages we almost always must look for 297 nm signal first, and then switch to 369 nm, if desired, after optimizing.  It is too difficult to resolve the ion signal on the background at 369 nm when many other trapping parameters are being manipulated at the same time during the search for the initial signal. 

Overall, detecting at 297 nm is the easiest path to finding trapped ion signal.  We do find that this signal can be very small and noisy, so for some systems we still choose to look at 369 nm.  After optimizing with 297 nm the 369 nm signal can be quite good in terms of SNR despite a rather large background. However, in reality detecting at 297 nm is the only viable option for detecting in a small package without considerable additional engineering to minimize stray light background, and we have already shown that we can obtain sufficient signal at 297 nm to run a clock.

\section{The miniature ion trap} \label{sec:minitrap}
\begin{figure} 
   \centering
  \includegraphics[scale=0.6]{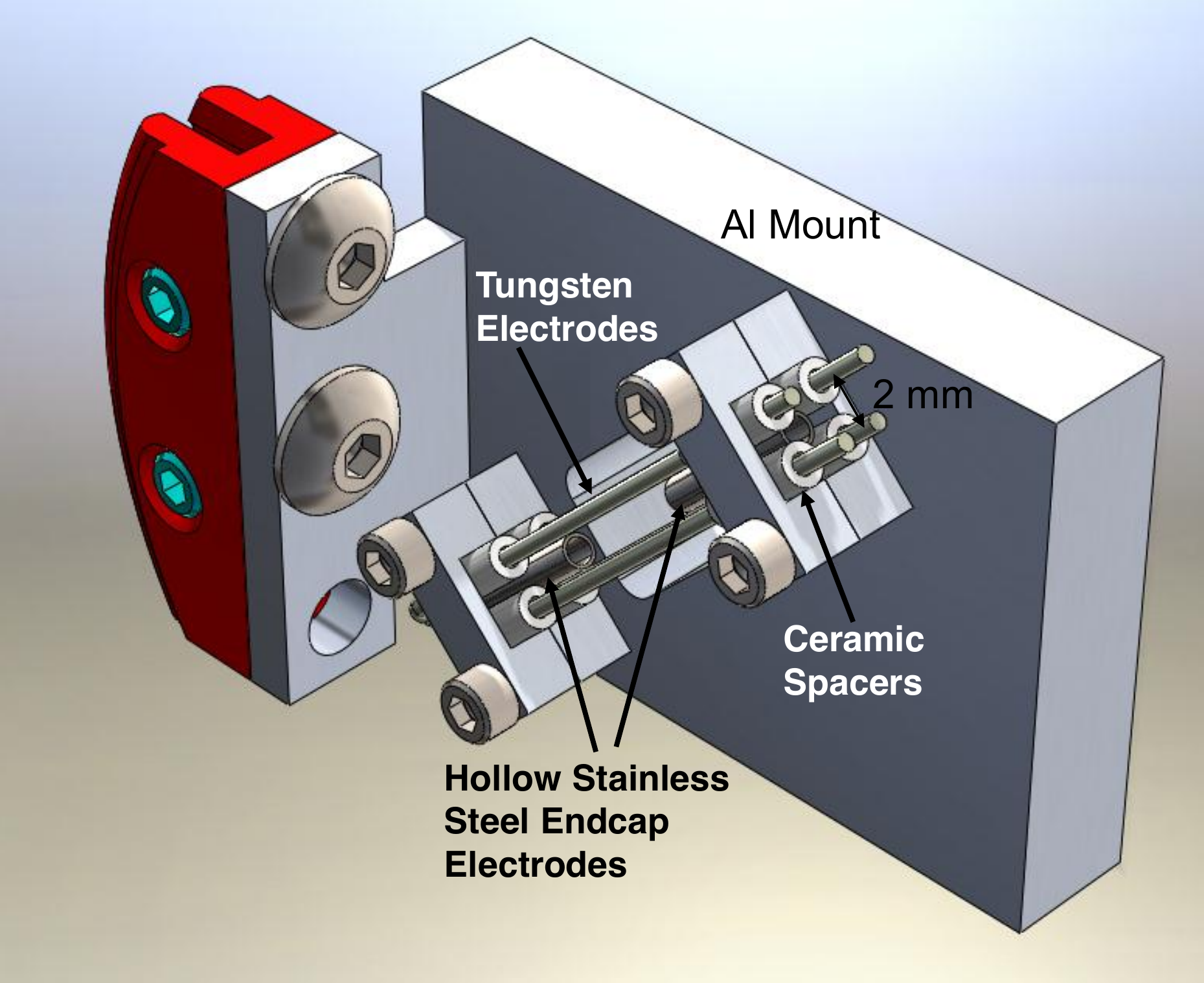}
   \caption{Close-up view of how we mount the test bed trap.}
    \label{fig:SWtrap}
 \end{figure}

\begin{figure} 
   \centering
  \includegraphics[scale=0.6]{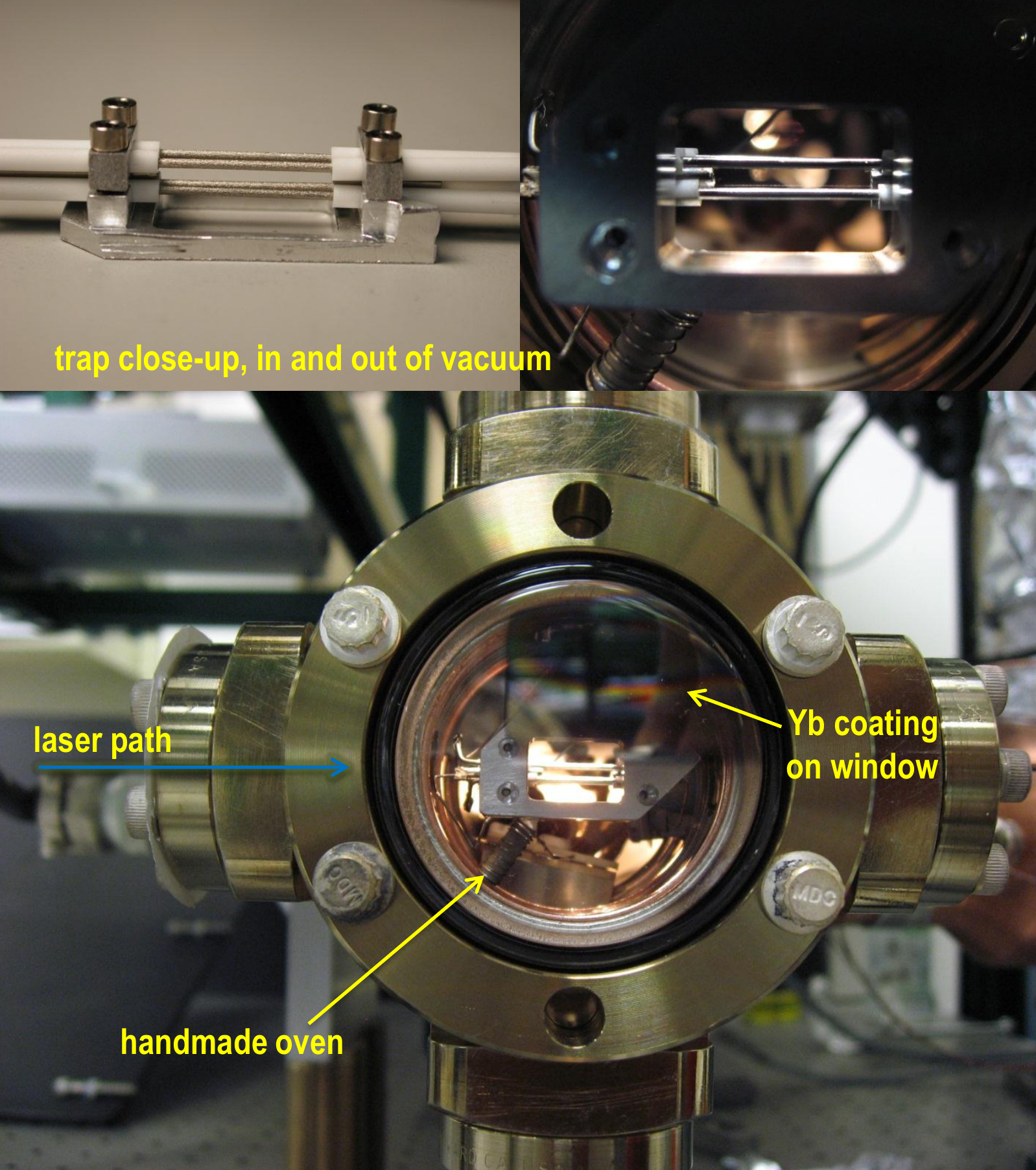}
   \caption{Photos of the test bed and our original small ion trap.}
   \label{fig:testBedPhoto}
\end{figure}

The original ion trap in the test bed system was our jumping-off point for learning about many aspects of a clock system.  From the start, the test bed ``small ion trap" was designed to be approximately the same size as would be a trap in the final 5 cm$^3$ clock prototype.  The aim for the physics package's share of volume in the final prototype is 1 cm$^3$.  The trap is a typical 4-rod linear quadrupole Paul trap, (details are found in Chapter \ref{ch:trapping}) with tubular endcaps on the interior of the four rods.  A closeup of the trap structure is shown schematically in Fig. \ref{fig:SWtrap} and photos of the trap in the chamber appear in Fig. \ref{fig:testBedPhoto}.  In the original configuration, which we used for about two years, the trap rods are 0.64 mm in diameter and made of tungsten, the endcap tubes are stainless steel and 1.28 mm in diameter, and there is about 9 mm longitudinally in the trap between the endcaps. The spacing of the quadrupole rods in the transverse direction is 2 mm. The rods-endcap assembly is held together by clamps that are part of an aluminum baseplate that connects to a Kimball Physics groove grabber.  Some efforts were taken in buying commercial parts that would make a small volume in the region of the trap, although it is still orders of magnitude larger than the goal for future chambers.  The trap on the groove grabber is mounted in a minimal four-way cross (the Kimball Physics 2.75"CF flange cross) which is constructed as a thick 2.75" conflat flange that has 1.33" (miniflange) attachments in four directions.  This is attached to a Kimball Physics close coupler that connects it to a nipple on the vacuum manifold.  Attached to the other (front) side of the cross flange is the detection window, and three out of four of the miniflanges are fitted with windows for laser and microwave access.  It is important to note also that the electrical connections from the rods to the electrical feedthroughs (which are far away in the vacuum manifold) consist of stainless steel foils and wires spot-welded to the electrodes, clamped with stainless steel screws to additional groove grabbers for stress relief, with wires running into the vacuum.  Although the endcaps, electrical connections, and screws are 304 (``nonmagnetic" ) stainless steel, we eventually realized that stainless steel is still too magnetic to be this close to the trap.  This is why we later replaced these parts with titanium before performing the magnetic field measurements discussed in Chapter \ref{ch:biasfield}.

\subsection{Trap model} 

Since this was our first trap, I did extensive modeling in CPO not only to learn the trap depth and working parameters, but also to make decisions about future trap geometries and to examine its stability properties and secular frequencies.  Most of the details about these calculations have already been discussed in Chapter \ref{ch:trapping}.  The example calculations discussed there are for the test bed trap at our initial, high-power parameters. For these parameters, the modeled geometry, a contour plot of the trapping potential in the $x$-$z$ (transverse) plane, and the potential depth in one dimension (along either of the ``shallower" axes) are shown in Figs. \ref{fig:OldTrapCPO} and \ref{fig:OldTrapModel}.  The total potential plotted takes into effect the slight anti-trapping effect of the endcaps.  The depth of the trap for these parameters is about 1 eV as seen in Fig. \ref{fig:OldTrapModel}.  Despite the fact that the plots shown are in 2D, they are taken as a slice of a full simulation that includes the 3D geometry of the electrodes.  We explored a range of trap parameters with this trap, since we began driving it with power limited only by our amplifier and later tried to reduce the power necessary to run the system.  Changing the RF voltage changes the steepness of the harmonic-like trap walls.

\begin{figure}[h]
   \centering
  \includegraphics[scale=0.45]{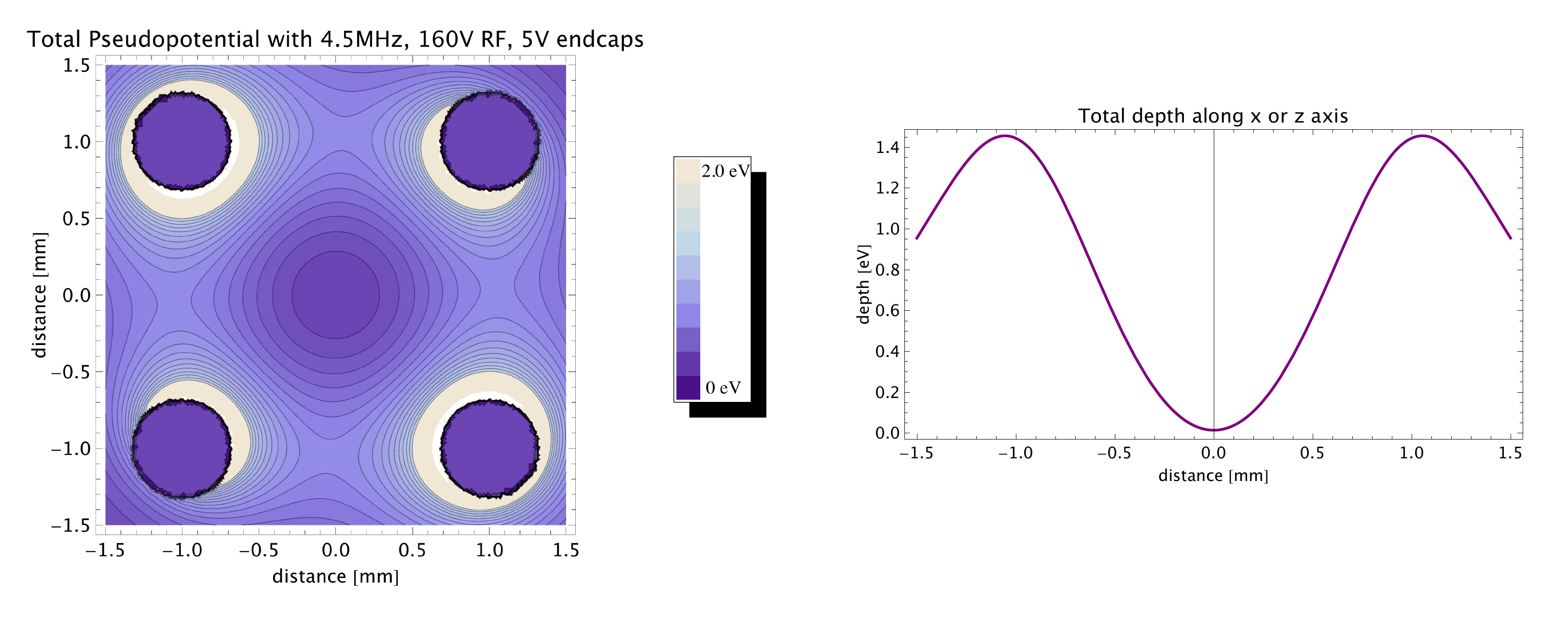}
   \caption{Transverse pseudopotential of the test bed trap for the lower-power parameters.}
  \label{fig:OldTrapModel4.5MHz}
\end{figure}

For the example modeled in Chapter \ref{ch:trapping}, the parameters are: trap driving frequency $\omega_{\rm{trap}} = 20$ MHz,  RF voltage (amplitude) V$_{\rm{RF}} = 580$ V, and endcap voltage V$_{\rm{EC}} = 20$ V (typical trapping parameters for when we used an amplifier as a power source and an RF resonator), giving a predicted depth of about 1 eV at its shallowest point.  At these voltages, the predicted transverse secular frequency is 210 kHz.  After we worked to reduce the trap frequency and thereby the necessary RF voltage, we were able to maintain comparable trap characteristics, for example, at $\omega_{\rm{trap}} = 4.5$ MHz, V$_{\rm{RF}} = 160$ V, and V$_{\rm{EC}} = 5$ V, the depth is increased to 1.4 eV and the transverse secular frequency to 255 kHz\footnote{There is a linear relationship between the frequencies and voltages required maintain a trap of the same depth.  For the second set of parameters, a lower voltage and frequency corresponds to a deeper trap because the frequency and voltage have changed with different ratios.}. Plots of the trapping potential at these parameters, which require a lower total power, are shown in Fig. \ref{fig:OldTrapModel4.5MHz}.

\subsection{Buffer gas cooling} 
\label{sec:temp}

\begin{figure}
   \centering
  \includegraphics[scale=0.5]{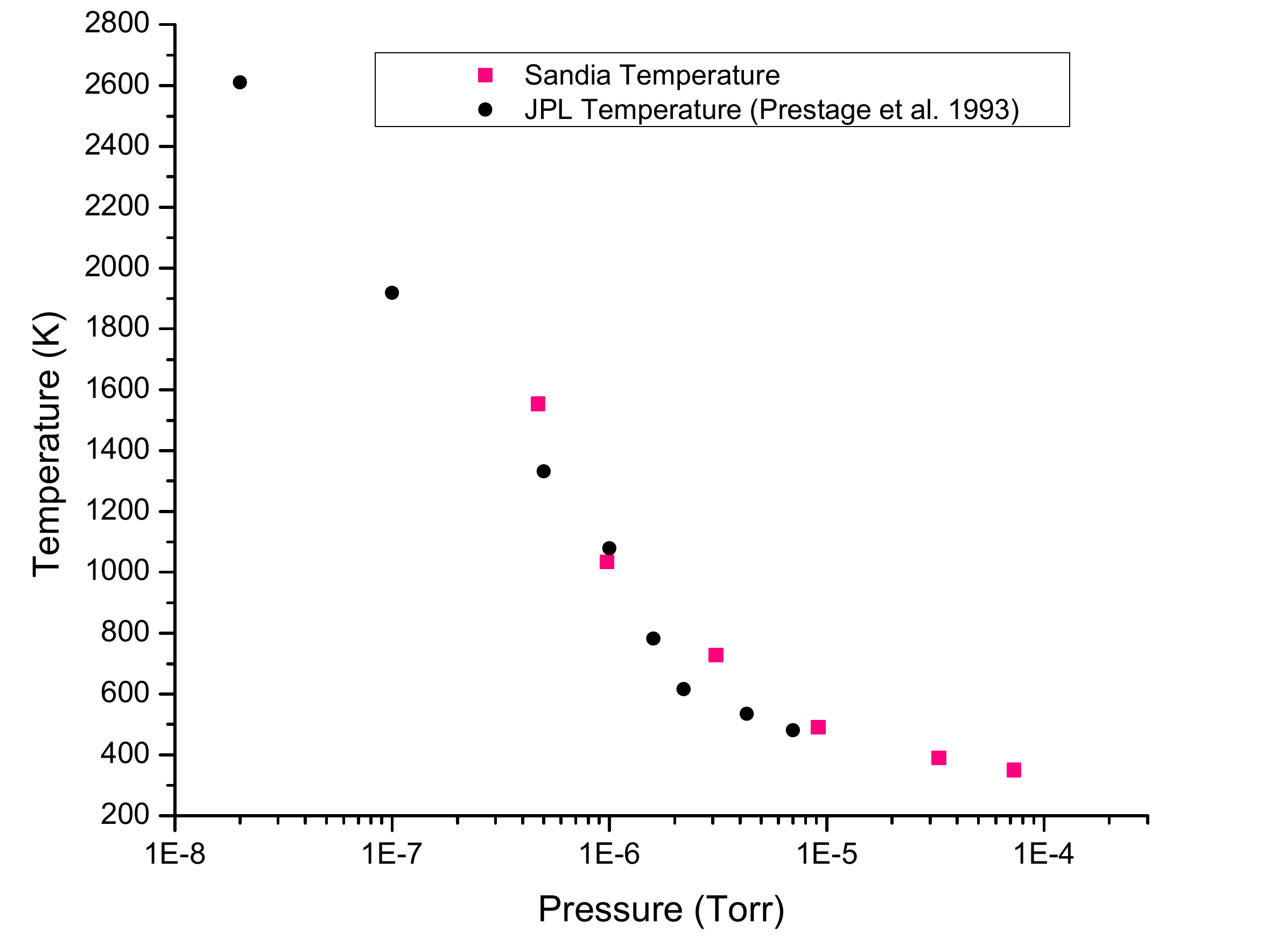}
   \caption{Ion temperature vs. pressure in the test bed chamber.  Since all of the miniature packages are subjected to the same vacuum pressure and He backfill conditions before sealing, we expect these temperatures to be consistent across traps.  Our typical operating pressure is $3\times10^{-6}$ Torr, which corresponds to approximately 730$^{\circ}$C  based on this measurement. }  
    \label{fig:IonTemp}
\end{figure}

For a simple and robust system, and because ion traps can easily be made very deep, we use a buffer gas for cooling in order to trap ions.  Laser cooling could be done but is not necessary.  Our ion traps are on the order of a few eV in depth, where we recall that 1 eV is about 40 times room temperature.  With a helium buffer gas of a few microTorr we can cool the ions to approximately room temperature or a few times room temperature.  A plot of ion temperature vs. pressure measured in the more recent revised (titanium) version of the test bed trap is shown in Fig. \ref{fig:IonTemp}.  The pressure used in this measurement is the overall pressure in the chamber, as measured by the ion gauge\footnote{The ion gauge is calibrated for nitrogen; for helium there is a factor of 0.18.}.   The partial pressure of gases besides helium can be assumed to be less than $1\times10^{-9}$ Torr, based on measurements from the RGA.  The temperature is measured by sweeping across the resonance of the 369 nm transition at different background pressures.  The temperature can then be found by fitting this peak to the doppler-broadened Gaussian profile given by
\begin{equation}
P_{f}(f)d f = \sqrt{\frac{mc^2}
{2\pi k_B Tf_0^2}}\exp\left(-\frac{mc^2(f-f_0)^2}{2kTf_0^2}\right)df\ ,
\end{equation}
which gives the distribution of frequencies seen by ions of temperature $T$.  This profile has a width 
\begin{equation}
\Delta f_{\rm{FWHM}} = \sqrt{\frac{8 k_B T\, \rm{ln}(2)}{m c^2}}f_0 \ ,
\end{equation}
and by assuming that temperature is the primary broadening mechanism, this width can be used to calculate temperature.  I have compared our measurement to a similar one in Fig. \ref{fig:IonTemp} by plotting results found  in a 1993 paper by Prestage et al. in which they measure the temperature versus pressure using another technique (microwave sidebands) and get comparable values \cite{367391}.

We typically use a pressure between 3$\times10^{-6}$ and 4$\times10^{-6}$ Torr for all of our trapping activities.  This is also the pressure we use when we backfill miniature trap packages before sealing.  In these traps we usually pump out, bake, activate getters, and then backfill with helium before closing the valve (or pinching-off) to seal the chamber.  The corresponding typical temperature is approximately  730$^{\circ}$C.

\subsection{Neutral Yb and ionization} \label{sec:neutralAndIonize}


The ytterbium used in our traps comes in two varieties: naturally abundant Yb which contains all isotopes (see Table \ref{tab:YbIsotopes}), which we obtained in the form of a powder and later in a thick foil, and isotopically enriched $^{171}$Yb, which was in the form of a thinner foil.    The natural abundance is much easier to begin trapping with in a new unknown trapping system, because microwave pumping is not required in order to see the ion signal.  This greatly simplifies the parameter set with which we need to experiment while initially searching for signal.  Several even isotopes (most notably the highly abundant 174 and 172 isotopes) have a much larger signal than 171, making the initial signal much clearer and easier to optimize.  Because of this we have always put naturally abundant Yb into our systems.  When possible, we have also included the isotopically enriched $^{171}$Yb in a separate oven in order to do our microwave clock interrogation.  Since the natural abundance of the 171 isotope is quite low (14.3\%), the clock signal is smaller when using the natural abundance source as opposed to the isotopically enriched source.  Although we have only seen about a factor of two or three improvement in signal using the $^{171}$Yb source, it still provides an advantage in terms of the SNR.

The melting point of Yb is 824$^\circ$C, and the Yb must be heated to around 400$^\circ$C to release sufficient vapor to see fluorescence on the camera, which is the only sure sign that we can load the trap.  We do not measure this exact temperature.  This heating is performed in a homemade oven.  The most notable oven design, the oven for the test bed trap (see Fig. \ref{fig:OriginalTrap} or \ref{fig:testBedPhoto}) which was also used in several other traps, is made from a 0.124" outer diameter alumina tube.  The tube is plugged on one end using a ceramic paste known as Sauereisen Ceramic Cement (this is a powder mixed with water that cures in air to a ceramic substance), and after inserting powder or pieces of bulk Yb, the second end is plugged with a smaller tube (0.62") that fits just inside the first one, and is sealed around the edges with the ceramic paste.  This tube is wrapped tightly with a tungsten wire that is used as a heating element.  The exposed nature of the tungsten element requires careful placement of the oven inside the vacuum.  We used this design not only in the test bed but also as a backup oven in many of the traps we tested.  There are also other designs of homemade and microfabricated ovens that will be described for each trap in Chapter \ref{ch:packages}.  Last, the collimation of the neutral Yb beam depends on the design of the oven and needs to be taken into consideration for reasons that will become clearer later on.

\begin{figure} 
   \centering
  \includegraphics[scale=0.6]{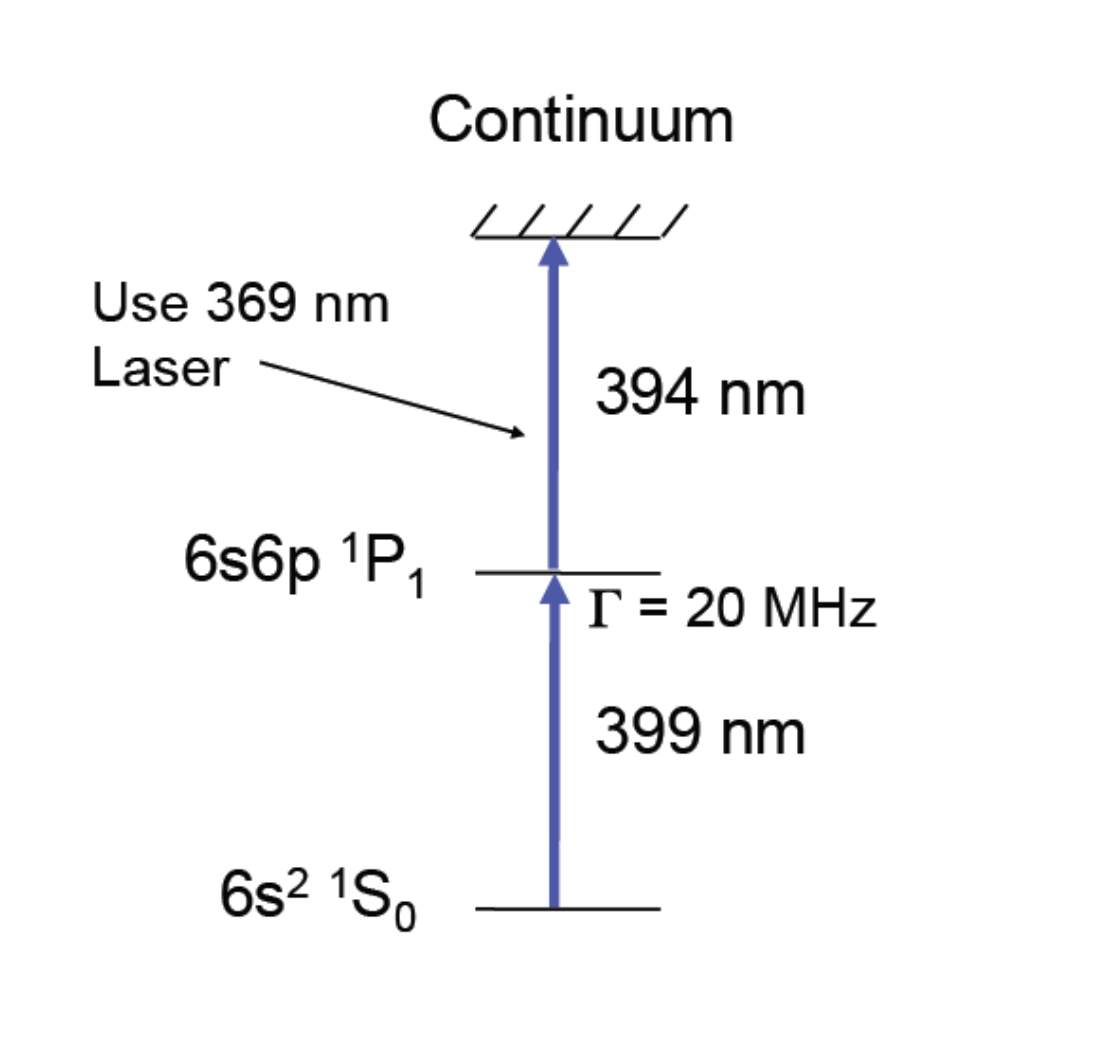}
   \caption{Relevent level diagram for neutral ytterbium showing the 2-photon process for photoionization.}
   \label{fig:NeutralLevels} 
\end{figure}

The Yb atoms in the neutral beam from the oven must be ionized in order to be trapped.  This is performed in several ways, again depending on the trap being used.  Among the methods we have used are photoionization and electron impact ionization using photoelectrons produced using lasers, mercury lamps, and light emitting diodes (LEDs).  A relevant level diagram for neutral Yb is shown in Fig. \ref{fig:NeutralLevels}.  Photoionization can be performed by a two-photon transition, where we use a 399 nm laser for the first transition from the $^1\rm{S}_0$ to $^1\rm{P}_1$, and the second transition to eject the electron can be performed by any wavelength below 394. In our system this role is played by the 369 nm detection laser.  Photoionization is very convenient for trap exploration and testing, since it provides continuous loading as long as the oven is heated, and requires no in-vacuum parts.  However, this method is not convenient when we consider miniaturization because it requires an additional frequency-stable laser.  A method using a broadband source that we can provide using an LED or free-running laser diode is a much more practical method for our later packages.  

The conventional way to ionize with no concern for size or power is to place a filament in the vacuum that ejects an electron beam which bombards the neutral atom beam coming out of the oven in the vicinity of the trap.  Besides the fact that this requires fitting another element in the vacuum and additional electrical feedthroughs, this is not a practical option for small packages because of the possibility of charging up dielectric surfaces around the trap with excess charges from the electron beam.  An alternative that we discovered and made use of is a similar electron-bombardment technique where the electrons are obtained from the trap/package surfaces through the photoelectric effect.  We recall from modern physics that the relation between the work function of a metal and the energy required to eject an electron from the metal is given by
\begin{equation}
\phi = h f_0
\end{equation}
where $h$ is Planck's constant and $f_0$ is the minimum (threshold) frequency of the photon required to produce photoelectric emission. Since Yb has a work function of approximately 2.50 eV \cite{PhysRevB.46.7157}, 
this means that photons with a wavelength shorter than about 495 nm can eject electrons from a Yb surface.  Since the surfaces around our trap are often coated by Yb (after use of the oven for some time), this means that we can use an LED or uncollimated laser to eject electrons from these surfaces, which then bombard the neutral atom beam if the coated surface is sufficiently close to the trap (assuming the beam is within the trapping region).  We utilized this effect initially by accident, and in later traps, by design.  Using this method lets us produce electrons directly, but noninvasively using light. Of course, with this method it might still be possible to charge up surfaces near the trap, but this problem has not yet appeared.  In our smallest package to date, one way we deal with this possibility is by allowing almost no dielectric surfaces to be visible (near) to the trap.   The specifics of how we applied this photoelectric-ionization method with various traps is described later in Chapter \ref{ch:packages}.

\subsection{RF trap drive electronics} 
The test bed system in its first form used an RF resonator constructed from copper tube and helical coils that acts as a transformer to couple the RF signal from a function generator into the trap.  Our RF resonator can be seen in the photo of the vacuum chamber in Fig. \ref{fig:OldChamberPhoto}.  These devices are common in the ion trapping field and a description of their use can be found in \cite{springerlink:10.1007/s00340-011-4837-0} and \cite{4065638}.  The idea is to put an RF oscillation on one end of the resonator and then physically adjust the components of the resonator to find optimal coupling into the ion trap electrical connections. This way the trap is impedence matched to the RF source to make an efficient use of power.  We used this initially, but it clearly does not meet our need for size and power (these typically use several Watts of output from the driver), so we quickly set out to develop a trap drive that would be small and low-power, in the form of an oscillating circuit.  We tested several of these circuits in the test bed and other traps.

After abandoning the helical resonator, whenever we want to use an arbitrary power of RF, we use an RF amplifier to drive a simple LC circuit. 
We can change the frequency of the oscillator by adding to the trap capacitance with a capacitor in parallel.  This method is convenient for starting out with new traps until good parameters are found.


With a great effort by Yuan-Yu Jau, the postdoc on the IMPACT project, we developed and tested a low-power RF driving circuit that takes up about 1 square inch (limited by the size of a high-Q inductor) that uses low-power high-speed CMOS inverters to amplify the RF voltage to the needed value.  We showed that we can operate our trap at lower frequencies and voltages than most ion traps would use, which lets us consume less power. For example, using 2 V to drive the circuit around 2.6 MHz, we can obtain RF voltages above 200 V while consuming less than 50 mW.  

It should be noted that later, for driving the JPL traps, which have no electrically isolated endcaps, we must run our trap drives with a bias.  Placing a uniform negative DC bias on the four RF electrodes makes the grounded package appear at a higher potential than the minimum created by the fields from trap electrodes, so that endcap formations or package walls on either end of the trap can still act as endcaps without any electrical connection or isolation.  Being able to add this bias was another important element to consider when developing the trap driver for those packages.

\subsection{F-state \& quenching gases}
As I mentioned in Sec. \ref{sec:Yb}, the Yb ion can, in the case of collisions, be dropped into a low-lying F-state with an enormously long  lifetime.  This means that a loss into this state is equivalent to losing an ion in the trap if we do not optically pump out of the level.  We do see effects from the F-state under certain experimental conditions (depending on pressure, laser powers, foreign gases (leaks) and loading method), but in the worst case we see that the signal is reduced by a certain percentage (which can be up to and greater than 50\%), but then remains constant.  Thus, we sometimes have a reduced overall signal due to the F-state, but not a vanishing signal. 
If we find the F-state to be an irreconcilable problem in a future package, we do have the possibility to cope with it through the use of a ``quenching gas", which is a gas (for example, nitrogen) that has the tendency to exchange ions between the F and the D states through collisions. We do see an increased effect from the F-state over time in our pinched-off package due to the vacuum becoming cleaner and cleaner (pumped by the getter) which likely means that there is less quenching from foreign gases.  Helium, which is abundant in our system, does not have good quenching characteristics.

\subsection{Ion trap performance} 
The test bed trap, when sealed off from the pumps, had lifetimes on the order of 100 to several hundred hours (the lifetime generally improves over time in the closed volumes as the getter continues to pump) for trap frequencies on the order of a 3-5 MHz.  The original driving parameters at around 20 MHz and RF voltages of more than 1000 V peak-to-peak had shorter lifetimes, indicating that we were actually over-driving the trap by working in a less stable region than when we used lower frequencies and voltages.  A rough estimate of the number of ions in the trap based on the quantity of fluorescence is $10^4$ to $10^5$ ions.  This is therefore an estimate of the number of ions seen by our probe laser, which may not include the entire cloud.  

\begin{figure}
   \centering
  \includegraphics[scale=0.7]{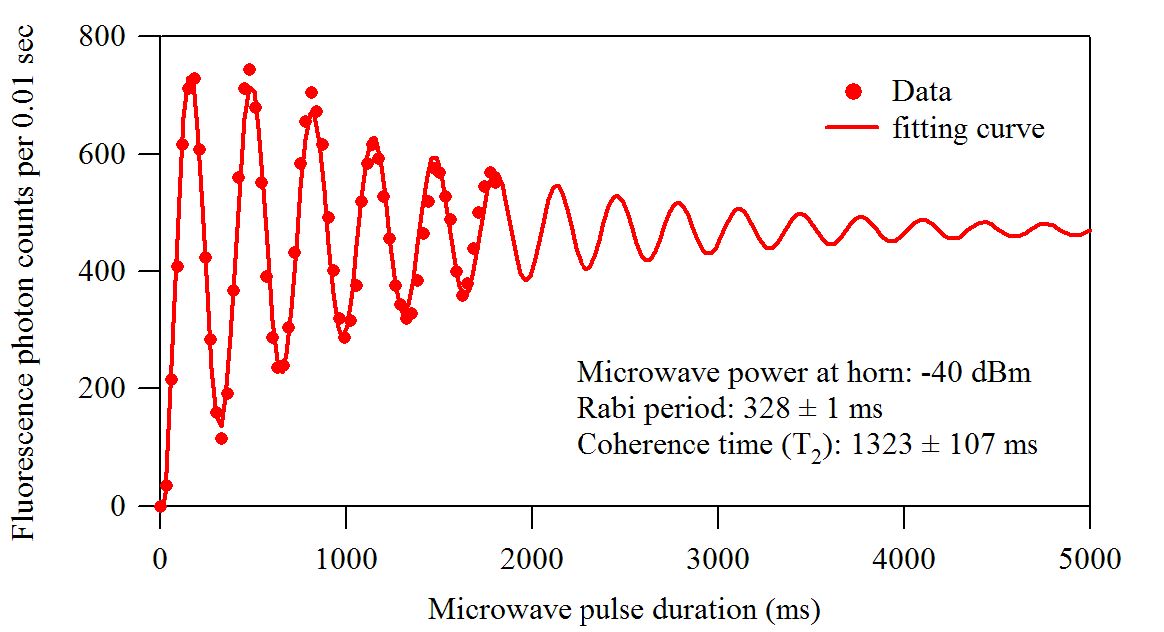}
   \caption{Measurement of decaying Rabi flopping for determination of the $T_2$ decay time. }
   \label{fig:RabiT2}
\end{figure}

Detection in this trap is much more optimal than any of the smaller packages, since reflection of stray light is much less severe, and the large entrance and exit windows do not cause much additional scatter. Also, the tubular endcaps that the laser enters the trap through turned out to perform convenient spatial filtering that we would miss in later traps.   For these reasons, signal to noise ratio, signal size, and background minimization in this trap are phenomenal. We measured coherence times of $\geq$1.3 s by looking at Rabi flopping on the hyperfine transition.  The measurement is shown in Fig. \ref{fig:RabiT2}. 

\section{The $^{171}$Yb$^+$ ion clock}
Once one has done the work of trapping $^{171}$Yb$^+$ ions, making an atomic clock from them is a relatively straightforward task of regulating pulses and a proper electronic feedback loop.  There is also the task of doing everything with the ions that you can to optimize the quality (see figures of merit, Sec. \ref{sec:FOM}) of the clock.  These ideas will be put in the context of our test bed tabletop clock, but are applicable to all of the traps.

\subsection{Making the clock tick: microwave interrogation} 
\begin{figure}
   \centering
  \includegraphics[scale=0.8]{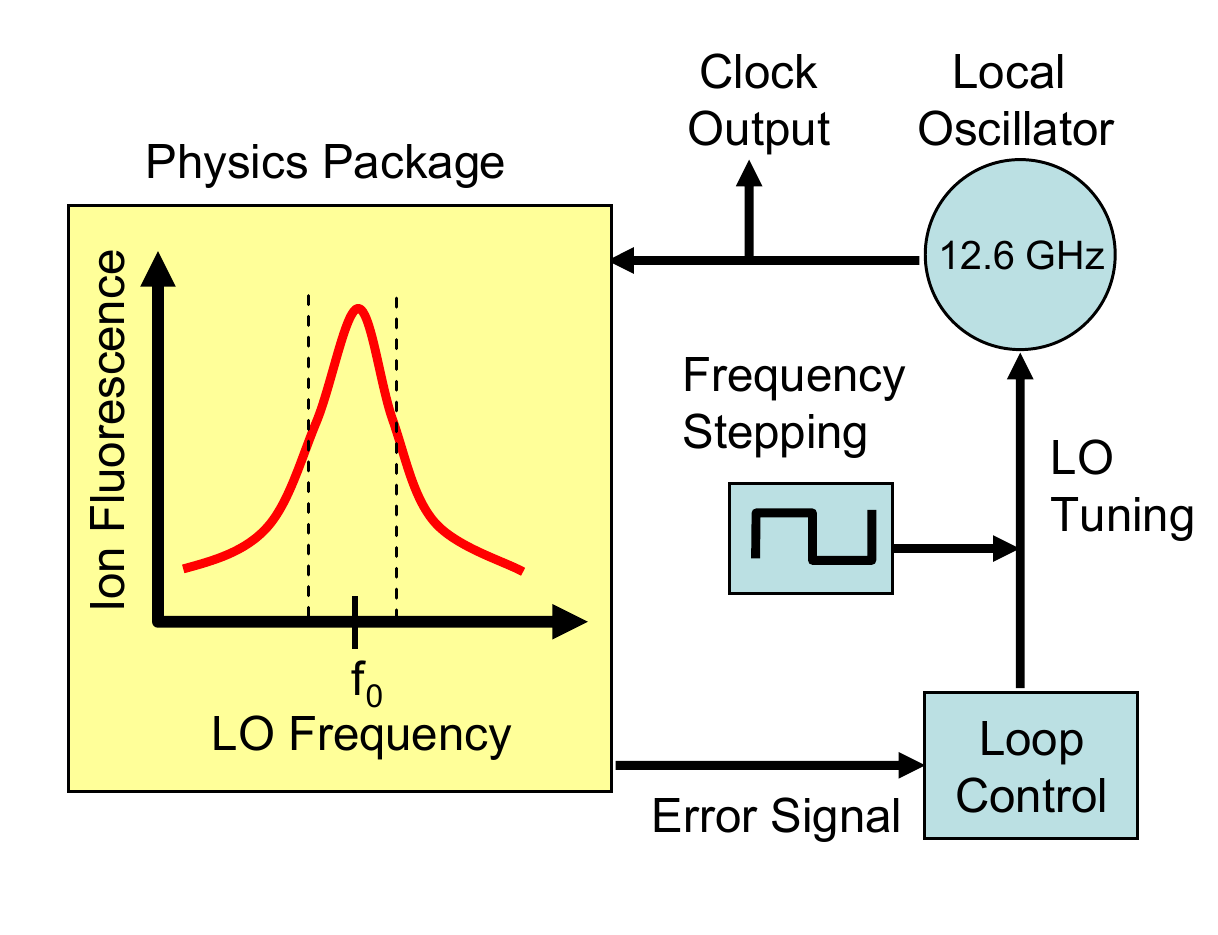}
   \caption{Schematic of the frequency-locking loop for the clock. (Courtesy of Peter Schwindt)}
 \label{fig:ClockLoopSchematic} 
\end{figure}

The anatomy of an atomic clock consists of an atomic source of frequency information (the ions), a local oscillator (LO) that oscillates at its own frequency nearby to the atomic frequency, and a feedback loop that locks the LO to the atomic frequency.  In a microwave clock, which uses the hyperfine transition of an atom as the source of frequency information, uses microwaves as the means of passing information from the LO to the atoms: the LO sets the microwave frequency, and the microwaves illuminate the atoms, inducing behavior that depends on whether or not the microwaves are exactly resonant.  An electronic circuit processes the fluorescence information which is recorded on a photon counter and adjusts the LO accordingly.  Then the cycle begins again.  A schematic of this loop is shown in Fig. \ref{fig:ClockLoopSchematic}.

As discussed in Chapter \ref{ch:clocks}, there are two primary means of microwave interrogation: Rabi and Ramsey.  Rabi interrogation involves one $\pi$-pulse of microwaves while Ramsey involves two $\pi/2$-pulses with a wait time in between.  We use Rabi interrogation for our clock because it is simpler, requires less microwave power, and may provide better long-term behavior. 

\begin{figure}
   \centering
  \includegraphics[scale=0.55]{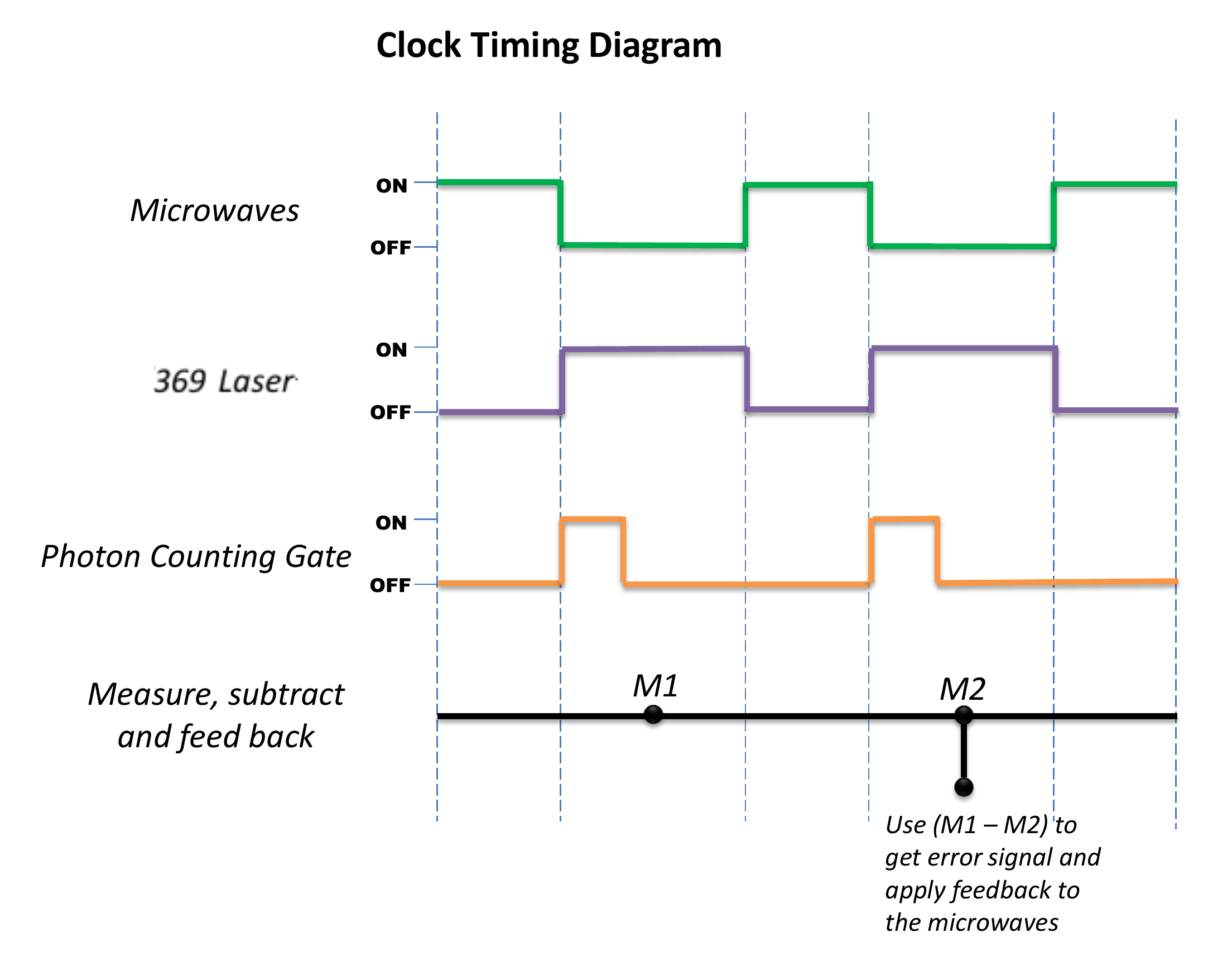}
   \caption{Timing diagram demonstrating a few clock measurement cycles of our system as discussed in the text.}
 \label{fig:TimingDiagram} 
\end{figure}

The basic timing diagram that we use is shown in Fig. \ref{fig:TimingDiagram}.  First, we note the main clock functions: the microwaves and probing laser turning on and off.  We prepare the state in the lower ground state ($|F = 0, m_F = 0\rangle$) and then perform the microwave pulse.  We allow the microwave to evolve the system ``in the dark"; this means that the lasers are blocked during this step.  This is important because when the lasers and the microwaves are on at the same time, broadening occurs that limits the linewidth of the clock resonance due to the fact that light is pumping population out of the upper state as it is still being transferred in.  Similarly, when the lasers are on to read out the population in the upper level ($|F = 1, m_F = 0\rangle$), the microwaves are off so that they don't continue to induce hyperfine transitions while we are reading out the population.  This light pulse must be long enough to prepare the state for the next sequence, which in our $^{171}$Yb$^+$ system is limited to a minimum of about 300 ms because we rely on off-resonant transitions to return all of our population to the ``dark" $|F = 0, m_F = 0\rangle$ state.  We can see if we are pumping adequately by looking at the rate at which the signal decays after a microwave pulse.  The lasers are turned on an off by using a triggered Uniblitz mechanical optical shutter for the Phase I packages.

The photon counter is gated and measures for a gate time of 100 ms during the first part of the 300 ms repumping time.  The microwave source frequency is stepped back and forth about the current center frequency for alternating measurement cycles, so that when two successive photon count measurements (one on each side of the peak: see Sec. \ref{sec:phaseSwitch}) are subtracted, the feedback loop can use the subtracted value as an error signal to stabilize and correct the center microwave frequency and push this difference to zero.  This basic clock loop is the same for all incarnations of the IMPACT clock, with changes only being made to the microwave interrogation and optical interrogation pulse lengths, thus increasing or decreasing the overall duty cycle. 

For the test bed and the first small packages we use an Agilent E8257N microwave synthesizer to generate and control the microwave frequency lock.  The synthesizer accepts a 10 MHz reference input that we use with an oscilloquartz LO, which we steer using the feedback loop, and in this way we change the reference frequency and therefore the output frequency of the synthesizer.   The synthesizer's many functions allow us to control the shutter,  microwave frequency and pulsing, and trigger the photon counter all from one source.  We can also easily sweep the frequency and change the pulse character of the microwave to perform diagnostics on the trap and the clock signal, with the help of a Stanford Research Systems DS345 function generator.  These large laboratory apparatus are replaced in Phase II by an integrated electronics board (developed by our collaborators at Symmetricom) that performs the loop discussed here, and can also control the lasers, shutter, and oven.

\subsection{Microwave system}
A fundamental element of the Yb clock is the necessity of a microwave field at the clock transition frequency, since it provides the direct  communication between the local oscillator and the atoms themselves.  So we must take into account a microwave source, coupling techniques and interrogation techniques in our system design.


\subsubsection{Coupling techniques} 

One of the particular problems we face as we make a miniaturized microwave clock is getting the microwave field itself inside of the package.  Microwave radiation at 12.6 GHz, which is the frequency of our clock transition, has a wavelength of 2.4 cm in vacuum, which is considerably larger than even one side of our final device.  How do we get the microwaves inside of the package to talk to the ions, and how do we control the field orientation and/or polarization?  

There are several ways to get the microwaves inside the package.  The original plan for the small device was to couple microwaves into the package using a rectangular metal waveguide, but we quickly learned that we could get a much better signal by simply using a microwave horn pointed (somewhat haphazardly) at the trap. For the first several versions of the clock we used this technique. Using this method to add the microwave field means that we do not have much control over the field orientation.  Occasionally we can make adjustments based on the ion signal directly, but since the wavelength is large compared to the chamber and the chamber walls are mostly metal, we know that it is extremely difficult for us to control the propagation, phase, or polarization of the microwave field inside the chamber and across the ions.  Fortunately, as long as we have a component of the microwave  polarization aligned with the bias field (C-field) of the ions, we can operate successfully.  When we change the horn's orientation we are simply optimizing the size of this component.  

However, as the packages get smaller, the windows available for coupling microwaves get smaller.  This means that the horn technique requires higher and higher power to be effective.  Eventually we tried a different technique that we intend to use in the final design: coupling the microwaves directly to the trap electrodes.  We can connect the microwave source to one of the RF (ground) electrodes through a 1 pF capacitor so that the microwave signal is transferred to the electrodes without the RF being sunk to ground.  Then the electrode trap rod itself acts as an antenna to deliver the microwaves directly to the ions in the trap. Since we are already not too concerned about microwave orientation, this method should provide sufficient delivery.  In fact, since the microwave source to the ions in this scheme is so close to the ion cloud, in initial testing the power required to saturate the ion signal was almost two orders of magnitude 
 lower than using the horn.

\subsection{Locking to the clock signal: phase-switching vs. frequency hopping} \label{sec:phaseSwitch} 

\begin{figure}
   \centering
  \includegraphics[scale=0.4]{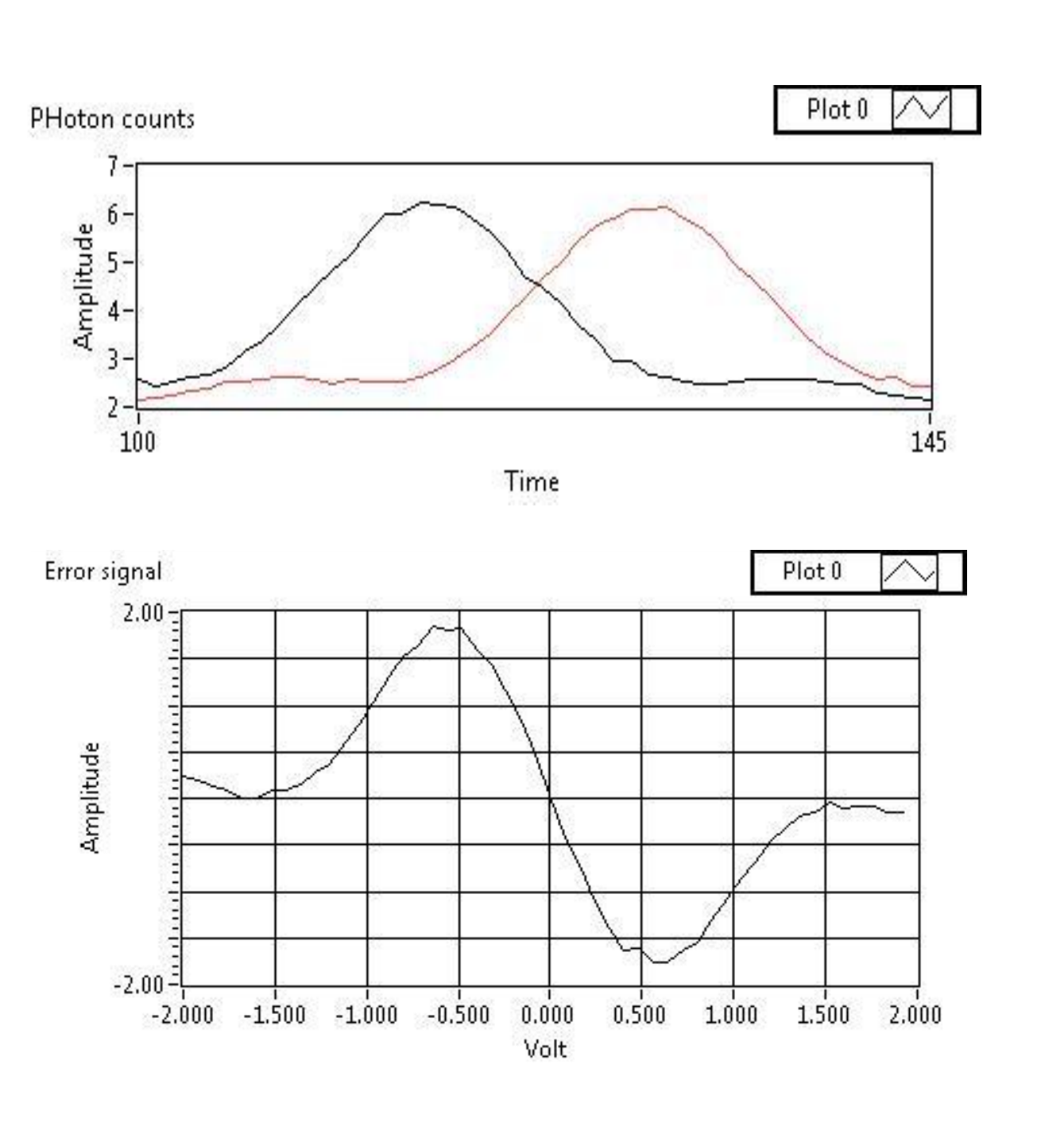}
   \caption{Using frequency hopping to generate an error signal.}
 \label{fig:ErrorSignal} 
\end{figure}

An important practical issue for the clock loop electronics is the method of locking the local oscillator signal to the clock resonance peak signal.  As with any feedback loop, this requires the creation of an error signal by some electronic manipulation.  The standard method to implement this is to hop between the two sides of the LO  resonance peak during measurement, in an alternating pattern.  The resulting values are subtracted, and if the LO center frequency is exactly on resonance with the atomic transition, the difference between the measurements on the two sides will be zero.  If the difference is not zero, the sign and magnitude determine the feedback correction. 
This scheme gives rise to the usual form of an error signal about zero, which can be seen by subtracting a peak that is wholly shifted to the left from one that is shifted to the right (Fig. \ref{fig:ErrorSignal}).

\begin{figure}
   \centering
  \includegraphics[scale=0.7]{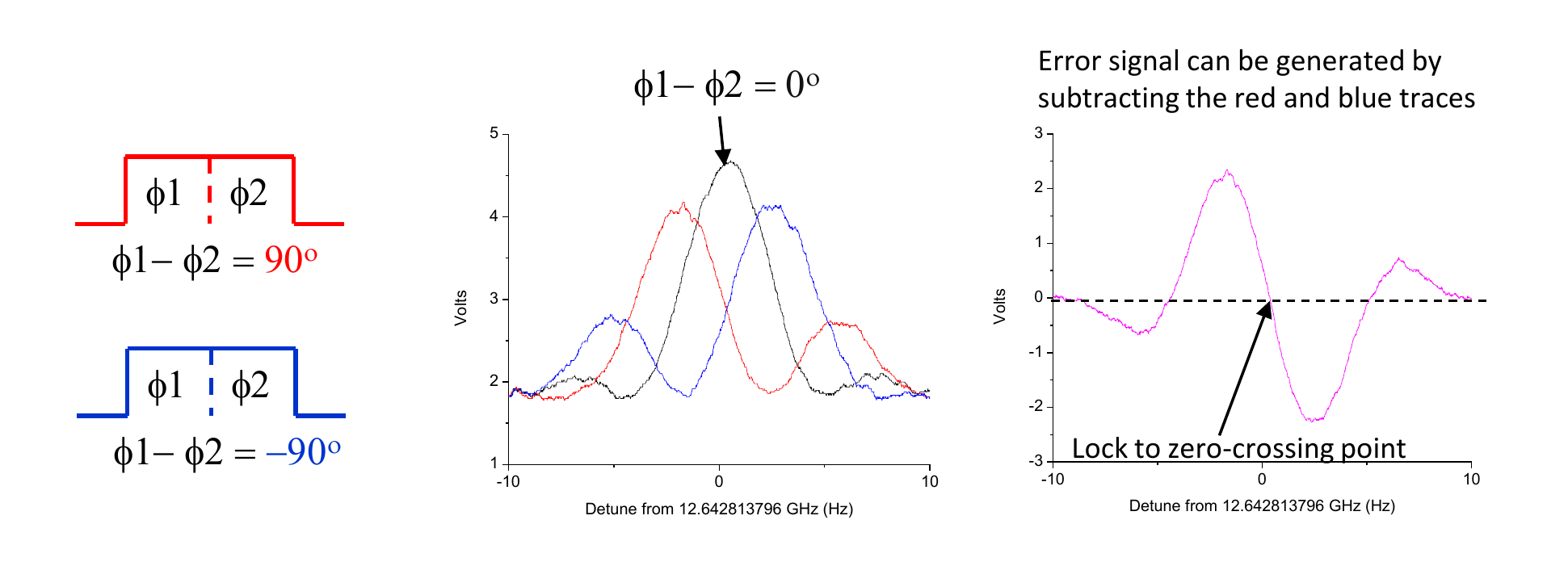}
   \caption{Using phase-hopping for generating an error signal. Left: Phase-switching pulses.  Center: Rabi interrogation signal for different phase differences.  Right: Error signal obtained by subtracting +90$^{\circ}$ and -90$^{\circ}$.}
 \label{fig:PhaseSwitching} 
\end{figure}

For a peak in frequency space, as shown in the cartoon Fig. \ref{fig:ClockLoopSchematic} and in many of the experimental plots of the clock resonance in this work, this process requires hopping the frequency of the LO back and forth to either side of the peak at a rate corresponding to the duty cycle of the clock loop.    Normally, we create this frequency hopping by microwave synthesis, adding and subtracting a frequency on alternating duty cycles to the microwaves which illuminate the ions.  However, as part of our overall miniaturization initiative, we planned to develop a MEMS oscillator for use as our LO.  In exploring the possibilities to use this LO for the frequency lock, we found that we would not need to perform microwave synthesis on our local oscillator, since the intent was to create an LO that already oscillates at 12.6 GHz.  This meant that to hop the frequency, we would have to adjust the frequency of the whole device (not just the output to the ions), which would cause the frequency output from device to be hopping as well.  A hopping output frequency is clearly undesirable.    
This led us to investigate the little used technique of hopping the phase instead of the frequency in order to create an error signal without influencing the output frequency.  
We made some measurements in the lab to confirm the validity of this approach.

\begin{figure}
   \centering
  \includegraphics[scale=0.6]{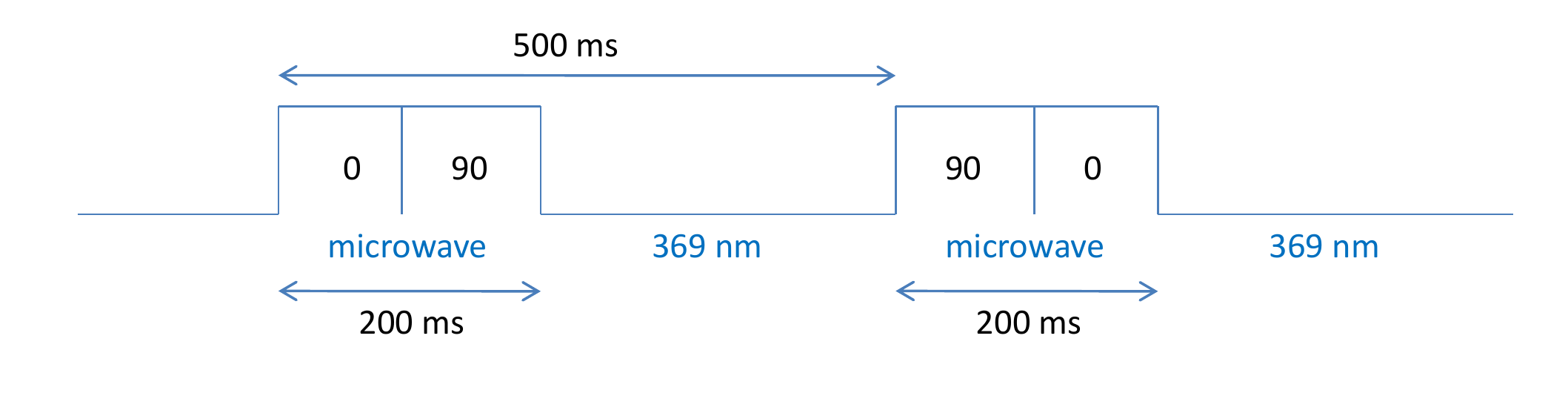}
   \caption{Timing diagram demonstrating the phase-switching scheme for generating an error signal.}
 \label{fig:PhaseSwitchingTiming} 
\end{figure}

Fig. \ref{fig:PhaseSwitchingTiming} shows a timing diagram for the phase-switching scheme.  This process is equivalent to a Ramsey interrogation, with no wait time between pulses, with a fixed phase relation between the pulses.  Instead of changing the frequency in two successive pulses, we reverse the phase relationship between two successive pulses.  The resonance signal provided by this scheme is an asymmetric Rabi-style peak.  The spectrum for each direction of phase-switch is shown in Fig. \ref{fig:PhaseSwitching}, along with the error signal obtained by subtracting them.  These curves were measured in the lab using a simple phase switch on our microwave source.  

Because we did not complete the MEMS oscillator, we did not implement this technique, but it could still be useful for an LO that is physically  more suited to phase switching than frequency hopping.

\subsection{Stabilizing lasers and magnetic field using the fluorescence ion signal} \label{sec:laserlock}
As I stated in Sec. \ref{sec:lasers}, we want to use the ion signal to directly lock the frequencies of the lasers to their optimal frequencies for detection.  In proposing this scheme we assume that the drift of the lasers is on a slow timescale compared with the duty cycle of the microwave clock frequency lock.     We can use alternating pairs of duty cycles to stabilize a laser, by stepping the laser up or down by some small amount $\Delta$ and observing on average whether the signal (using the gated photon measurement as a ruler) has increased or decreased.  This will affect the maximum amplitude of the signal, but will have no effect on the peak's location in frequency.  Therefore it does not disturb the clock lock. 

\begin{figure}
   \centering
  \includegraphics[scale=0.6]{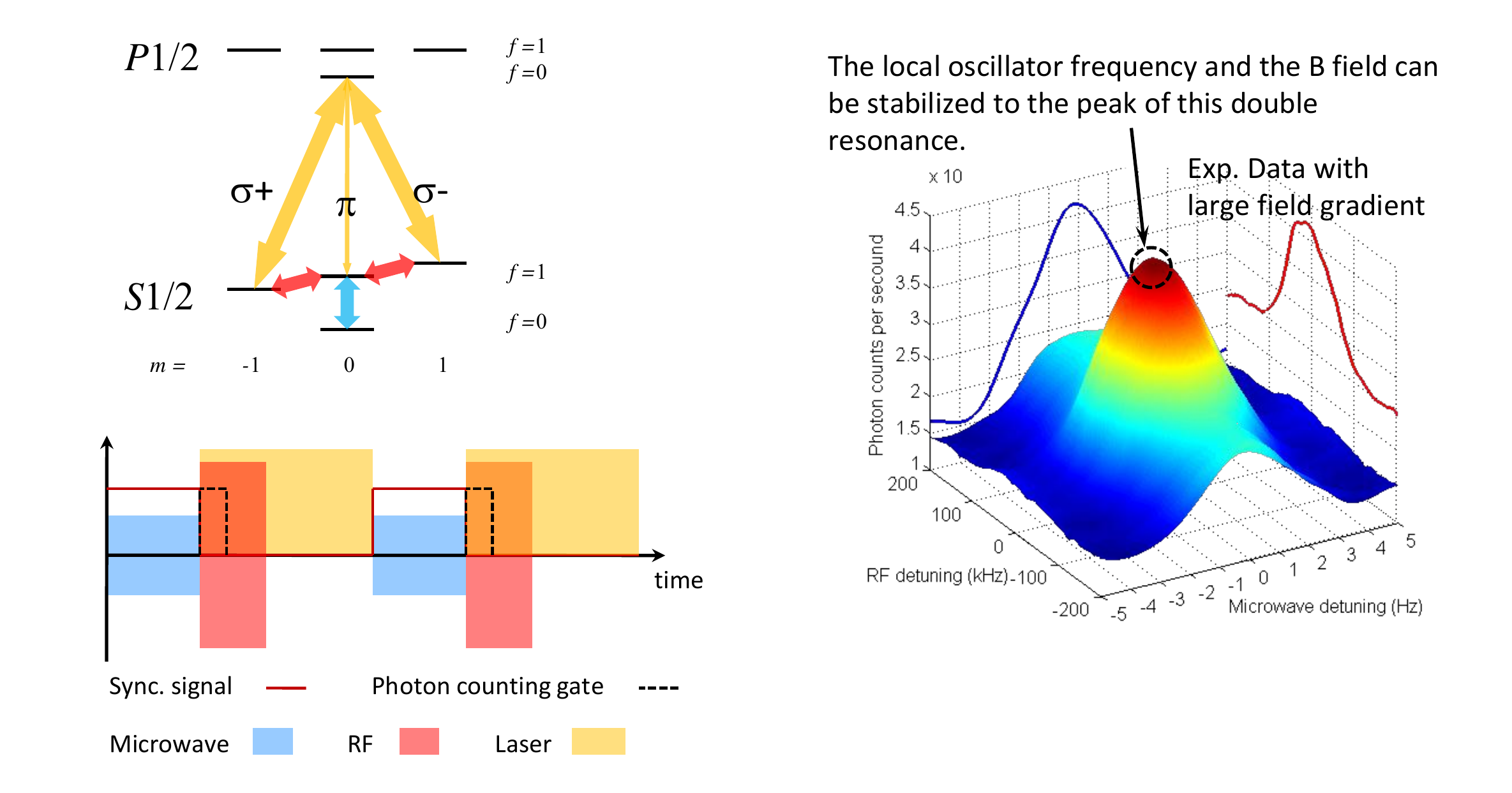}
   \caption{Left: Timing diagram for simultaneous RF and microwave stabilization.  Right: Dual resonance used for locking both (experimental data). (Figure courtesy of Yuan-Yu Jau)}
 \label{fig:RFtiming} 
\end{figure}

A similar procedure could be used to stabilize the magnetic field using the ion signal.  We can use the time in which the microwaves are off and we are photon counting to excite an additional transition probability between the Zeeman levels using an RF antenna inside of the vacuum.  This can be accomplished by placing a loop inside, or even by creating a loop using two connected, diagonally opposing RF-ground trap rods.  
In this way we can probe the bias magnetic field (which determines the resonant RF value), and create a pathway to feed back to the current loops that generate the bias field, simultaneously with our microwave clock probing. A timing scheme for interrogating with both microwaves and RF fields is shown in Fig. \ref{fig:RFtiming}. The RF is activated for a period at least as long as the photon counting gate so that we can probe its effect with the 369 nm laser.  We note that the RF does not interfere with the period for microwave interrogation, which must be left undisturbed for the sake of clock stability.

It is possible to perform both the field stabilization and laser stabilization simultaneously, in addition to the microwave frequency lock, if the signal size and SNR allow.  Note that we need two duty cycles to make an error signal point for the microwave (clock) lock.   Now, to stabilize two lasers, one needs two pairs of duty cycles for each laser in order to get an error value, meaning a total of eight duty cycles before each laser can be adjusted once, equivalent to about every four seconds for a 500 ms duty cycle.  This explains why we need lasers that drift slowly on this timescale. In addition, if we want to add field stabilization (also at a slow timescale), then between cycles that consider the laser optimization, we perform the RF excitation to check the bias field.  Again assuming 2 pairs of duty cycles to measure a change in the field, that means that at least 12 duty cycles must pass to adjust each of the lasers and the magnetic field after probing each twice.  In practice, we will wait even longer, so that we can obtain averages before making feedback adjustments.  
We have implemented both of these locks in the lab and confirmed their operational validity; however, only the laser stabilization has been implemented while running the clock.

\subsection{Magnetic field effects}

A very important effect that we want to consider in our analysis of the clock performance is what happens to when ions are moving in both a DC magnetic field and a magnetic field gradient.  A DC field that defines the quantization axis and provides the Zeeman splitting of the $|F=1\rangle$ states is of course necessary for the functioning of the clock.  However, this DC field must be uniform in space and stable in time, to an extent that magnetic field fluctuations across the ion cloud or in time are below the threshold that would make them a dominant player in the fractional frequency stability of the device.  We will examine this point in detail in Chapter \ref{ch:biasfield} about magnetic field effects on the clock.  Another phenomenon we will observe and explore deeply in Chapter  \ref{ch:biasfield} is the fact that gradients existing in the trap, combined with the natural harmonic motion of the ions about the RF pseudopotential minimum, can bring about dephasing that broadens the clock resonance by exciting transitions between the Zeeman sublevels.  Gradients inside the trap are an almost unavoidable side effect of miniaturization of the trap and the package surrounding it.

\section{Performance of the test bed clock}

The test bed clock was our basis for developing a clock loop, timing sequence, and software to perform these functions (written by Yuan-Yu Jau).  Once we had a good understanding of our ion trap, we used this system to run the clock for long periods of time.  We compared this clock with our in-lab Cesium beam frequency standard (the Symmetricom 5071A), using the Symmetricom 5125 Phase Noise Test Set to make the comparison by measuring frequency and phase drift and computing the Allan Deviation.  

\begin{figure}
   \centering
  \includegraphics[scale=0.6]{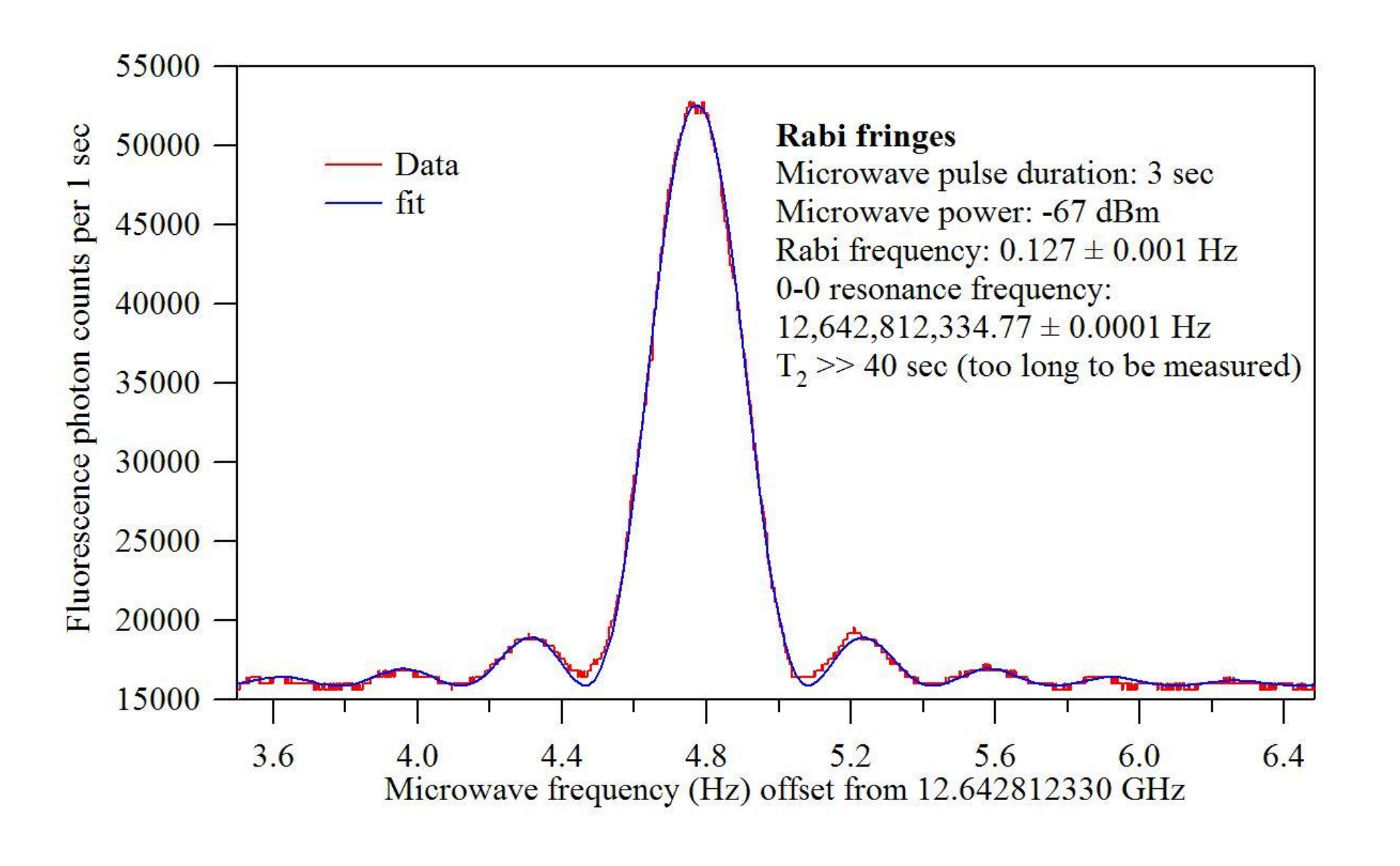}
   \caption{Rabi fringe clock resonance in the test bed trap.}
 \label{fig:OldTrapRabi} 
\end{figure}

An example of the Rabi fringe at the clock resonance for a pulsed microwave is shown in Fig. \ref{fig:OldTrapRabi}.  One can see the excellent signal to noise and signal size possible because of the larger size of this clock compared to later small packages.  This clock resonance is measured by running the microwaves in pulsed mode, as if the clock were running, and sweeping the microwaves across the resonance on a slower timescale, showing the peak of the fluorescence.  For this trap we had no trouble with background so these measurements are made at 369 nm.

\begin{figure}
   \centering
  \includegraphics[scale=0.8]{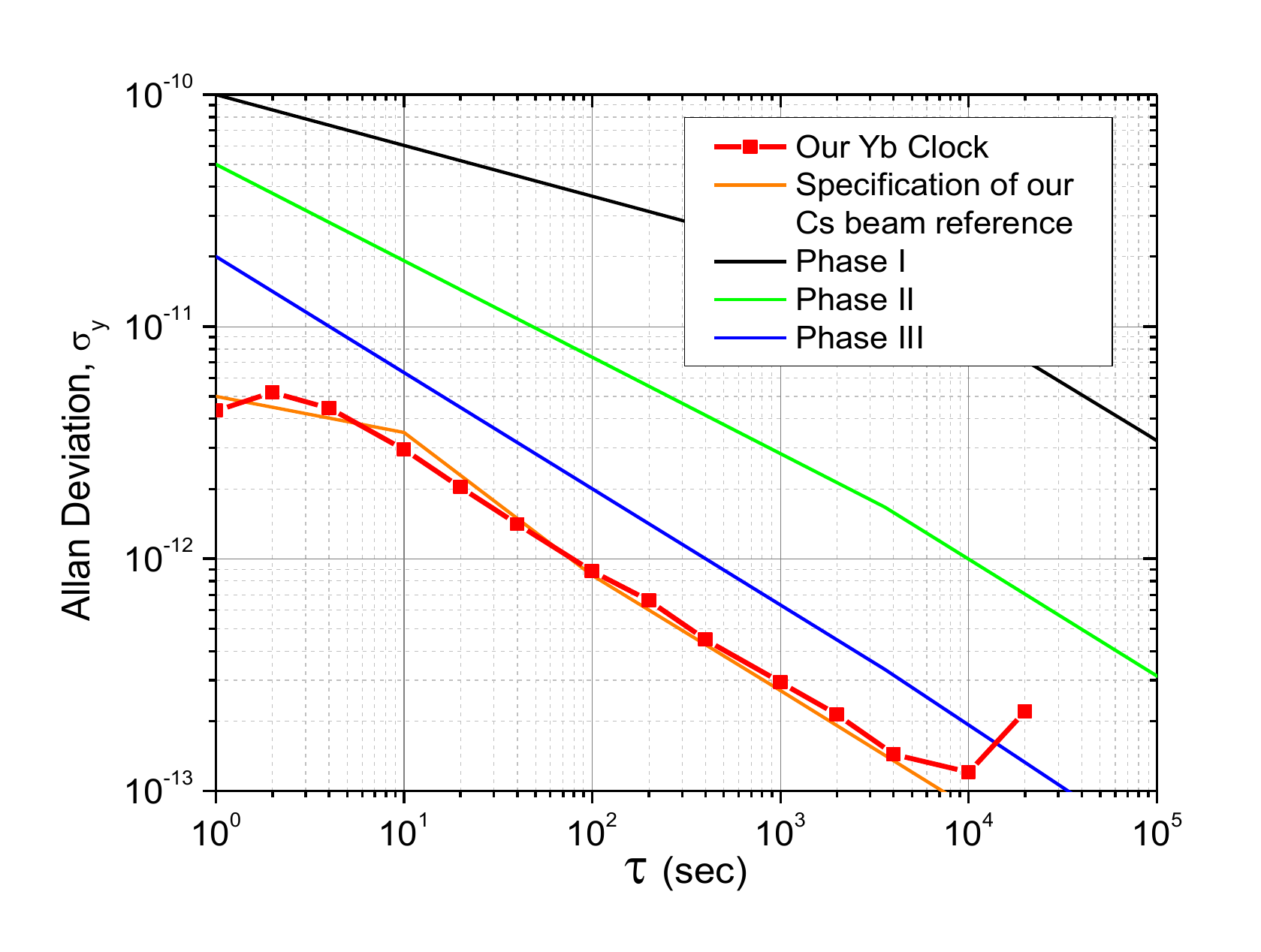}
   \caption{Allan Deviation of our test bed Yb clock.  It meets the Phase III specification of the project out to $10^4$ seconds.}
 \label{fig:OldTrapAD} 
\end{figure}

The Allan deviation for this clock is given in Fig. \ref{fig:OldTrapAD}.  
 Because the Cesium beam clock we use to compare our clock is by nature not very stable at shorter integration times, we are seeing the stability of only the local oscillator at short times.  After about 100 seconds the atoms really begin to have an effect on the lock.  Only when the Cesium beam is better than the LO are we really measuring the stability of our signal locked to the ions.  We see that the Allan deviation stops integrating down after about $10^4$ seconds.  This is most likely completely limited by magnetic field fluctuations, since we were unable to incorporate sufficient, consistent shielding, and the background magnetic field, which was acting as our C-field, can vary significantly on a timescale of hours.  The data shown has the overall drift of the magnetic field, which we measured separately, removed.  In the end, we have shown that we can reach the  Phase III goals of the project (at least the short-term goals, i.e. the time loss for up to one day) with this tabletop clock, which was a great starting point for beginning to integrate and understand the new packages and other components of the miniature clock.

\section{Summary}

We had extremely good results in our test bed clock.  In some ways this is not surprising because the test bed enjoys the luxuries of a large vacuum system, ordinary laboratory lasers and equipment such as the microwave synthesizer.  However, it did prove the validity of an ion trap that is much smaller than the typical traps used, and allowed us to explore our ability to use buffer gas to cool and develop our RF trap driving capabilities, as well as test operating in a sealed environment with only getter pumping for long periods, even though the interior volume was a liter instead of on the order of cm$^3$.  Furthermore, we learned a great deal about magnetic fields and field gradients from this trap, since it was the test system for most of the data shown in Chapter \ref{ch:biasfield}.

%
%
%
%
%
%
%
%

\chapter{Miniaturization and the ``small packages"} \label{ch:packages}


As discussed in Chapter \ref{intro}, the daunting task presented by the IMPACT project is extreme miniaturization of the atomic clock components while simultaneously preserving long-term stability.  This chapter deals with a major component of the miniaturization process: designing and implementing a small, sealed, vacuum package with ion trap inside that serves as the atomic source and is therefore the centerpiece for the miniature clock.  The idea has been to develop, over the three phases of the IMPACT project, a sealed volume that performs optimally as an ion trap (long ion lifetime, low power, easy to load) and allows for stable clock operation.  Since the ultimate volume goal of the IMPACT project is 5 cm$^3$ for the entire device, our goal has been to put the ``physics package" (the vacuum package with ion trap and loading mechanism, lasers and microwave delivery, and detection system) into a 2.5 cm$^3$ (maximum) volume, while along the way considering the characteristics of such a package as it gets smaller.  This led us through several trap technologies and the testing of many different so-called ``small packages", which I will describe in detail in this chapter. 
 \section{Overview}
\begin{figure}
   \centering
  \includegraphics[scale=0.6]{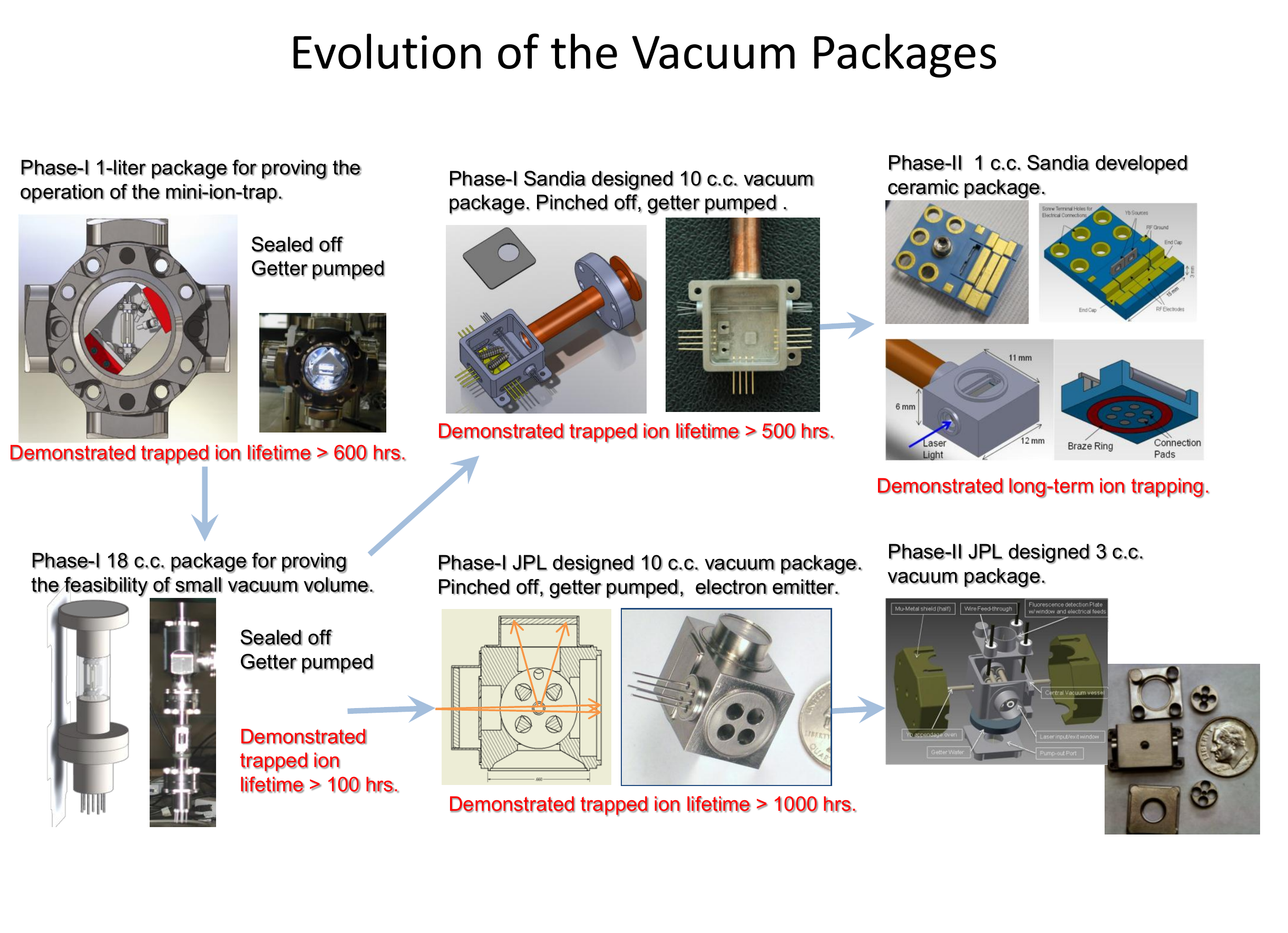}
   \caption{Illustration of our path through various small package designs.}
   \label{fig:packageEvo}
\end{figure}

A graphic showing the progression through the different packages in the first two phases of the project is given in Fig. \ref{fig:packageEvo}.  I will briefly review the progression and then focus on each package separately.  

While we were constructing the test bed clock, we entertained several initial small package designs.  Two of these designs were chosen to move forward and test: one package designed at Sandia and constructed by Pacific Aerospace and Engineering (PA\&E), and the other designed primarily by our collaborators at JPL, who also oversaw its construction.  While these packages were being manufactured, it became clear that we were exploring relatively new territory in constructing UHV-grade metal packages at this scale using, for the most part, conventional machining and assembly techniques, which led to many delays in the process.  During the downtime caused by these difficulties, we built a``commercial-off-the-shelf" (COTS) package, which allowed us to demonstrate trapping in our mini-ion-trap in a small, sealed volume (18 cm$^3$).  With the arrival of the two small Phase I packages, we began testing and evaluating the effectiveness of each, which eventually culminated in a working mobile ion trapping system using one of the pinched-off JPL-designed packages by the end of Phase I.  

With the successful demonstration of the JPL package, we moved on to the planning stages of Phase II.  JPL again designed a metal package for use as the atomic source in the deliverable clock required for the end of Phase II, improving upon the previous design and relying on the robustness of the technology of their Phase I package.  In addition, we designed and tested a new trap geometry made from a very different material (low-temperature co-fired ceramic) in our conventional vacuum chamber.  A plan for a stand-alone package based on this trap was then also also developed.  This ceramic package is our most likely candidate for moving into Phase III and reaching a 1 cm$^3$ package volume. 

In addition to the small packages, we eventually replaced the test bed trap with an all-nonmagnetic version in order to perform measurements as part of the magnetic field effects studies described in Chapter \ref{ch:biasfield}, after we had a far better understanding of the limitations in the original versions due to slightly magnetic parts.

This chapter tells the story of how we progressed through all the different designs, along with their merits, faults and what we learned from each one. Before launching into this story, it is prudent to make a note of what important points of functionality we must consider when designing each version of a small package (many of which have already been discussed in the context of the test bed), to elucidate later discussion of the critical elements.  In order to function as an ion trap and a clock, each miniature ``physics package" must include: 
\begin{enumerate}
  \item An ion trap: typically four electrodes, plus ``endcaps" or effective endcaps, and electrical feedthroughs to outside of the vacuum 
  \item A Yb source: stored raw Yb inside the vacuum and a method for heating to create Yb vapor
  \item A method of ionizing neutral Yb for trapping in the ion trap
  \item Laser and detector access
  \item Microwave access
  \item Connection scheme to conventional vacuum for pumping, baking, buffer-gas filling, with possibility to seal: valve or pinch-off
  \item Getter for passively pumping the sealed volume
\end{enumerate}
  
Other major concerns include the presence of any magnetic materials, whether as bulk parts or as coatings, solder or braze steps, or films, as well as controlling stray and scattered light, and the readiness of the package for imaging, detection, and optical integration in later stages of the project.

\section{``COTS" trap package}


After we had been working with the test bed system for awhile, we were eager to test out ion traps in smaller packages, to see how well a trap could perform with the vacuum walls so close, and to see if we could maintain a good vacuum in a small volume using nonevaporable getters. When it became clear that the ``boutique" manufacturing processes for the two 10 cm$^3$ metal packages was going to cause delays in our investigation,  as a way to move forward while waiting on their production, we began to look for a ``commercial off the shelf" (COTS) test system to use to attempt trapping in a small volume.  

\subsection{Technical elements} 

\begin{figure}
   \centering
  \includegraphics[scale=0.5]{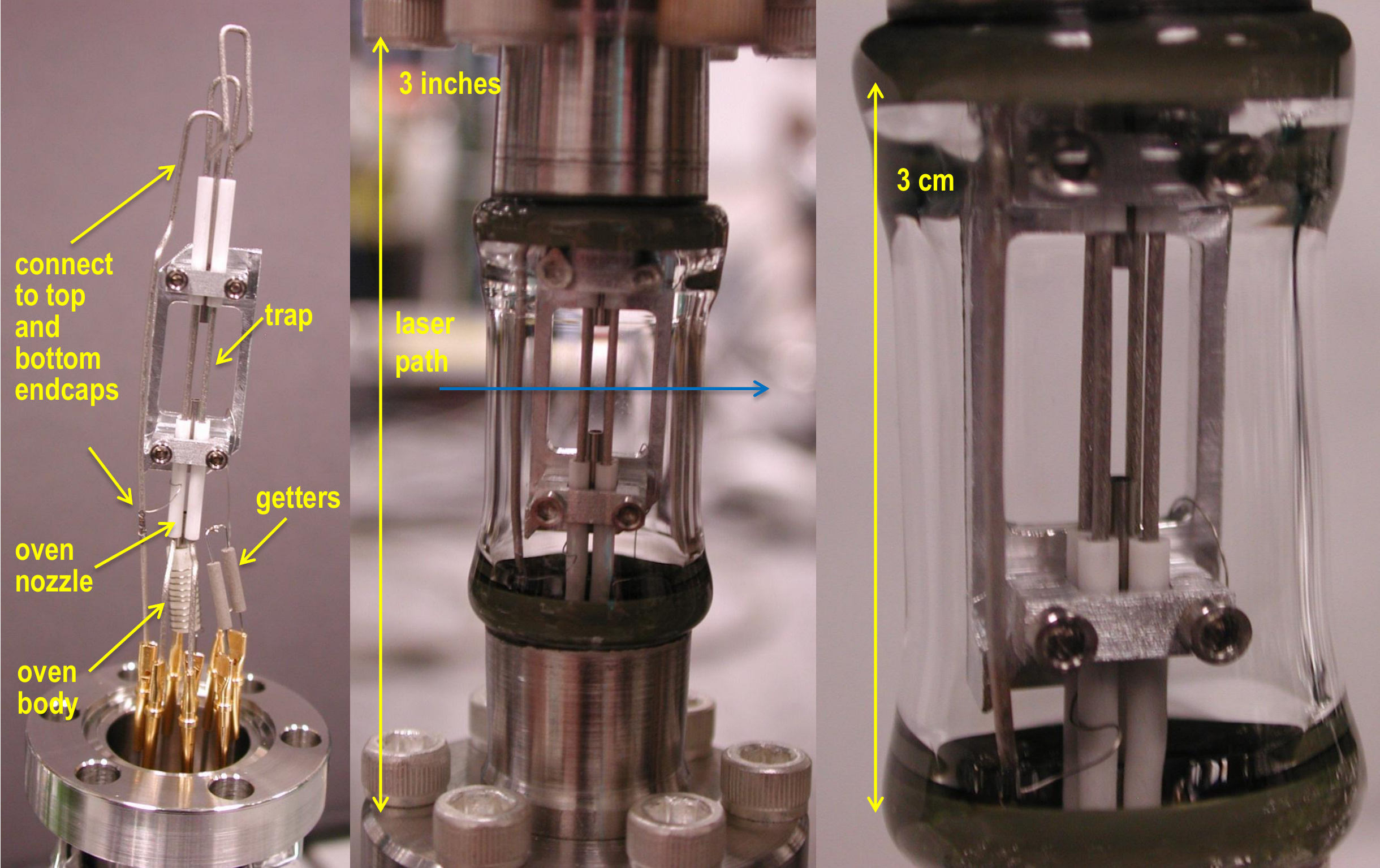}
   \caption{First version of the first small ``package" made from commercial-off-the-shelf parts.}
   \label{fig:COTS1photos}
\end{figure}

Finding easily obtainable parts that could create a very small volume vacuum and have the optical feedthroughs that we needed was a nontrivial task. After much searching, we chose a design that was inspired by a vacuum part consisting of a vacuum nipple with a glass tube divider at its middle.  These are usually used to provide an electrically insulating layer between two sections of a vacuum manifold.  We had to order a custom metal-glass-metal transition in mini flange (1.33") size in order to get a glass tube that was clear enough between the glass-metal transitions be usable as a laser/detection region.  This part was made for us by Precision Glassblowing of Colorado, using a pyrex tube and glass-metal transitions from Larson Electronic Glass.  The part is shown (with trap inside) in Fig. \ref{fig:COTS1photos}A.  It provided a small volume (18 cm$^3$ internal) with 360$^\circ$ optical access for laser light and detection inside of which to place our trap and other internal components and test out a small volume before receiving the more advanced, custom metal versions.  We also progressed later to a second iteration of this trap.

The ion trap used for this package was nearly identical to the test bed trap, constructed by hand using the same basic parts, although the aluminum baseplate that holds the four-rod trap in place had to be reduced in size. Also, it had become apparent in the test bed trap that it was necessary to eliminate all magnetic or nearly magnetic parts (such as stainless steel) close to the trap, so we used as many titanium parts as possible for the COTS version, including replacement of the tungsten wire electrodes with titanium wire (although tungsten is nonmagnetic, it is brittle and difficult to work with), all electrical connection wires and foil for spot welding with titanium instead of steel, titanium endcap tubes, and titanium screws to clamp onto the trap.  The trap, getter, and oven are attached standing up to a miniflange vacuum electrical feedthrough via press-fit crimp pins as pictured in Fig. \ref{fig:COTS1photos}. 
 The oven is integrated under the trap pointing up and the getters are standing up on two of the feedthrough rods.  This somewhat feeble-looking setup is actually very stable when the dual glass-metal transition tube is slipped over the assembly and connected to the feedthrough.  A challenge of this hand-built system was to make the necessary electrical connections and no unwanted ones.  This was accomplished by brute force in the first iteration, and by using alumina stand-offs for isolation, centering, and wire-guiding in the second version.  We connect the miniflange ends of the tube to the electrical feedthrough at the bottom end and to a bakeable valve at the top end, which when closed would seal off the 18 cm$^3$ inner volume.

\begin{figure}
   \centering
  \includegraphics[scale=0.7]{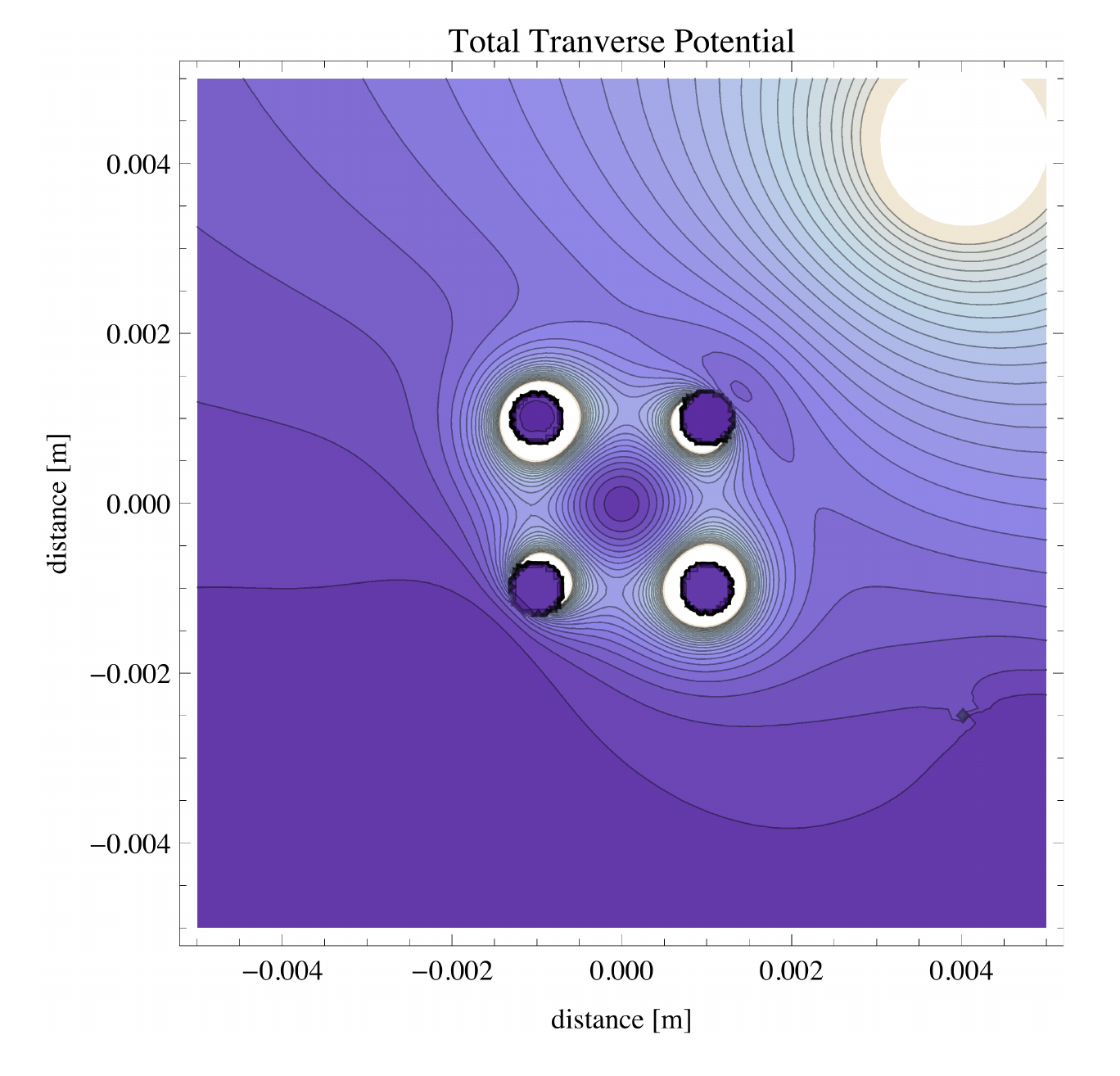}
   \caption{Model of the pseudopotential in the COTS trap. This trap is identical to the test bed trap, but has a wire carrying a DC potential as well as the grounded baseplate that holds the trap (not visible in the lower part of the plot).  Although they distort the field outside significantly, neither of these have much of an effect on the potential inside the trap.}
   \label{fig:COTSmodel}
\end{figure}

A model of the trap is shown in Fig. \ref{fig:COTSmodel}. The electrodes are the same as the test bed trap, but in CPO I have also modeled the presence of an extra, parallel wire carrying DC potential to the upper endcaps from the feedthrough, and the grounded baseplate-clamp that holds the trap.  It is easy to see that these have little or no effect on the middle of the trap potential.  At 3 MHz, with 100 V$_{\rm{RF}}$ and 5 V$_{\rm{EC}}$, this trap has an approximate depth of 1.2 eV and a predicted secular frequency of about 250 kHz.

The Yb source for this trap consisted of the same style of homebuilt oven used in the large test bed.  The oven is positioned below one of the (tube-shaped) trap endcaps so that the Yb beam would ideally travel through the tubular endcap and into the center of the trap, where it could be photoionized using our 399 nm and 369 nm lasers aligned together in the trap center.

For this first small package, in the first iteration we chose to use two of the smaller cylindrical getters, with their heater wires connected in series.  We chose to activate these getters manually by heating the resistive wire inside the getter after the bake was completed.  For the small packages we have also found over time that the bulk Yb in the ovens also tends to absorb a lot of gases during the bake that is released when we heat the oven for the first time (and can also re-saturate the getter), and for this reason we heat the getter and oven simultaneously after the bake to clean out adsorbed gases from each without saturating either.

Since this COTS package, compared with later all-metal packages, contained a lot of glass and less metal near the trap, microwave access was made easily directly using the microwave horn.  Laser access was also straightforward with the glass tube, with the caveat that shining lasers through curved glass can create scattering issues.


\subsection{Trapping Ions}
Since trapping had worked very well in the test bed trap, we expected to have the same ease of use in the package since the trap was near identical in form.  However, we encountered several issues when we tried to use the ion trap.  The first issue was with the oven location/collimation.  The first time we attached the COTS trap tube to the vacuum system, we could not find an ion signal despite all efforts.  One issue was that since for the first time we would only have one oven, we decided to fill it with  isotopically enriched $^{171}$Yb$^+$.  As I have mentioned before, this makes finding the signal extremely difficult, since 3 lasers as well as the microwave frequency have to be on resonance or very close in order to see a signal from the 171 isotope, while the laser alignment is also adjusted\footnote{For a new trap, we have an idea of where the resonant frequencies are, but this can vary somewhat from package to package.}.  In addition, the placement of the oven below the endcap with a gap between the two tubes led to much Yb spraying out between the tubes, meaning that much of the Yb neutral beam was not actually reaching the trap region to interact with the photoionization light.  We know that the Yb was spraying out because it began to coat the walls of the glass tube.  These difficulties, and concerns that something else could be wrong with the vacuum led us to take apart the tube.  

On the second try, we replaced the oven with natural abundance Yb, and inserted a tube to make a continuous closed pathway from the oven opening up to the trap.  After making these modifications, we were still unable to trap any isotopes at first, until we decided to try sending the beam in at a 45 degree angle.  I originally thought that this might help to increase signal by interacting with more of the ion cloud, since this trap, unlike the test bed, could not allow access along the cloud's long axis.  With the 45 degree angle we saw an ion signal almost immediately.  This led us to understand that with photoionization, we needed to consider the angle of the laser beam with respect to the neutral beam.  We can select isotopes by changing the wavelength of the photoionization laser and interacting with a doppler-broadened velocity group that appears on resonance with the light. In the test bed bed, this velocity group was large since the photoionization laser traveled along the long axis of the cloud.  In the COTS tube, the Yb atoms in the vertical beam were not seeing a strong enough parallel component of the photoionization light when we shined the light perpendicular to the neutral beam, so the spread of velocities that could interact with the light was much narrower.  This effect combined with the free-running nature of the 399 nm laser made the neutral fluorescence signal very unstable.  This problem was exacerbated by the long tube that the Yb traveled through (\textgreater 1 cm) which led to a very collimated beam of Yb. Adding some component of the light in the parallel direction allowed us to ionize much more easily.  Due to the constraint provided by the length of glass tube, shining the light at an angle was limited to about 45 degrees, meaning we could not interact with a lot more ions in the trap.  We also tried expanding and focusing the beam into a tall thin cross section in order to talk to more ions, but this had little effect on the signal size.  This may be due to the fact that by expanding the beam, the laser can access a larger extent of ions, but with less intensity.  

Another major issue with this trap stemmed from the round tube and the scattering of light.  The tube was pyrex and uncoated, meaning that it was reflecting about 10\% of light at 369 nm.  At the exit of the tube, this could cause a lot of light to bounce around inside the tube and a similar phenomenon could occur inside the curved surfaces.  Also, because of the geometry it was difficult for us to eliminate room light compared to the test bed trap (and other mini traps), although some attempts at masking the glass with black paper did improve the situation.  All of this difficulty was what first drew us to attempt to observe at 297 nm.  We tested the hypothesis in the test bed trap and found that we could get a signal about 300 times smaller at 297 nm than at 369 nm\footnote{We later improved this ratio by use of better filters and optimizing our detection system for 297.}.  
Looking at 297 nm enabled us to use this trap in the glass tube.  Since this tube was also the first one to use the nonevaporable getters after we established trapped ions in this tube we were able to close the valve connecting it to the pumps and still obtain lifetimes on the order of 100 hours
\footnote{It is worth noting that in a sealed volume we have no gauge of the vacuum besides the ions themselves.  Therefore the only way to confirm that getters are working is by trapping ions and observing their lifetime, etc.}.  This made it possible for us to prove our small trap in a sealed volume for the first time.

\begin{figure}
   \centering
  \includegraphics[scale=0.6]{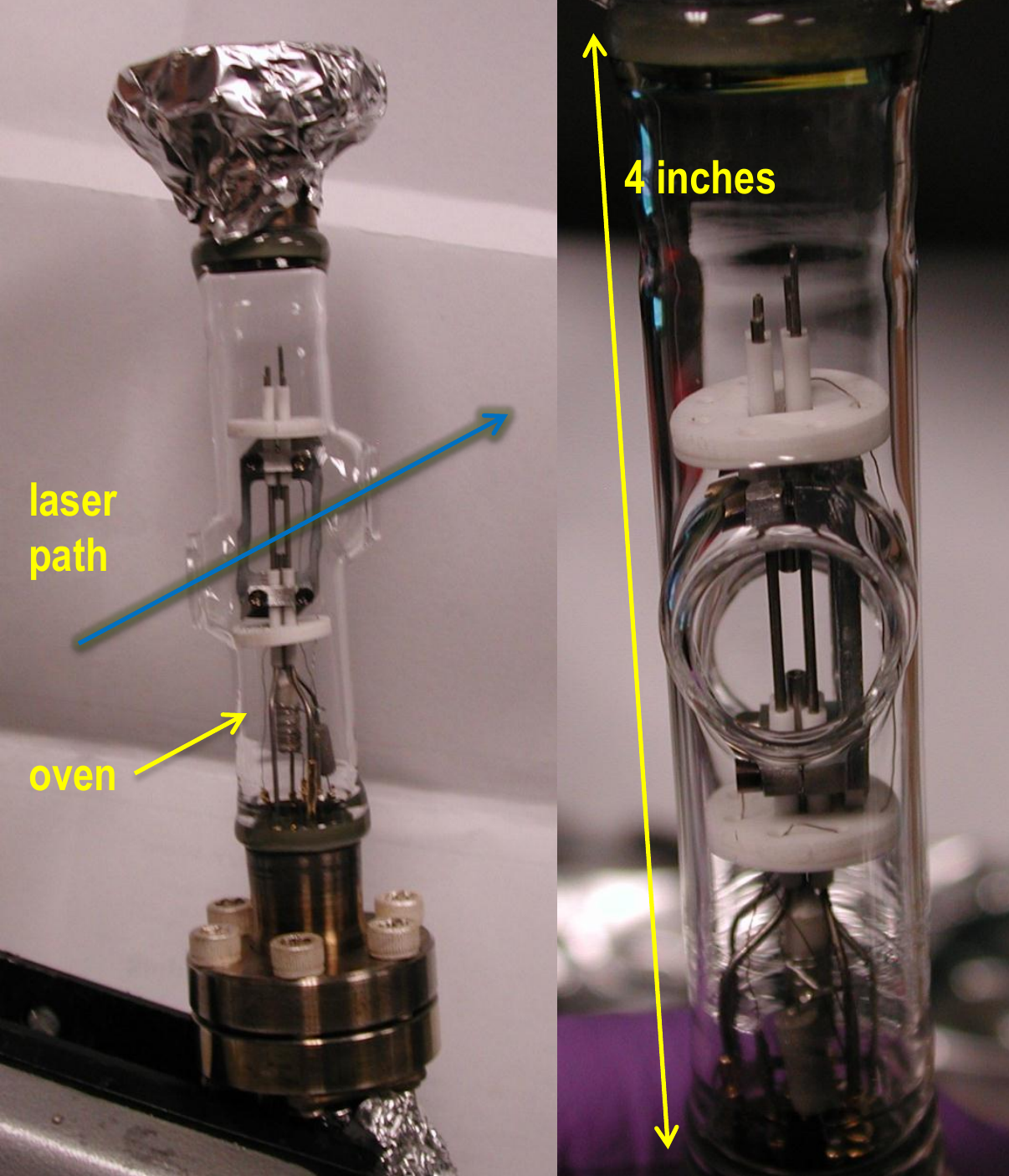}
   \caption{Second iteration of the first small ``package" made from commercial-off-the-shelf parts.}
   \label{fig:COTS2photos}
\end{figure}

The COTS trap used the LC circuit driven by the RF amplifier as a source.  The parameters for its operation after optimization were 3.45 MHz, with 120 V$_{\rm{RF}}$ and 8 V$_{\rm{EC}}$. 
Detecting at 297 nm, we observed trapping of even isotopes as well as the 171 isotope for clocks.  We generated a Rabi fringe signal, but it was too noisy to be used as a clock.

\subsubsection{Later versions}
After realizing that we needed to shine the light at an angle and gaining an understanding of the magnitude of the background light problem, we knew that our 297 nm signal was not strong enough to make a clock, but that 369 nm still had too much background to be useful.  Since we were interested in proving a clock in the small sealed volume, we decided to make a second round of this trap, adding flat windows to the glass tube, staggered so that we could shine the light through at an angle, with the intention of being able to operate with a signal at 369 nm. We also made the interior trap straighter and better isolated by adding alumina supports to hold it in the center of the tube and more easily protect some of the wires from touching each other.   However, we still found that too much light was being reflected at the flat glass surfaces, because of the steep angle used to enter and exit, and so we were still unable to get a good signal at 369 nm. 
Also, light from the exit of the tube was still being reflected within the tube.  We pursued further a third design, where the windows would be oriented for the laser to enter at Brewster's angle\footnote{From classical optics, an incident ray at Brewster's angle transmits perfectly through a medium, without backreflections, if the polarization of the ray is chosen correctly.} in order to eliminate this problem.  By testing a microscope slide oriented at different angles with respect to our 369 nm beam, we estimated that we could reduce the reflection by 30-100 times using this method.  We built a trap with the windows angled properly, but were not able to test this design before our attention turned to the metal packages.  Although it was a valuable test system, we were unable to run a clock with this trap.

\subsection{Lessons from the COTS package} 
To summarize, we learned many new things from this first crude attempt at a small package.  Mostly importantly, we learned we could successfully trap in a \emph{sealed} small volume, and therefore also that our getters were working.  We learned that we need to be careful about how the laser intersects the neutral/ion beam.  Another valuable lesson was that we realized we could trap relatively close (within 1 cm) to dielectric surfaces (the glass tube), which was another concern about bringing vacuum walls very close to the ions.  Last, we learned to detect at 297 nm when the background at 369 nm is too high.

\section{PAE Phase I package}
\begin{figure}
   \centering
  \includegraphics[scale=0.45]{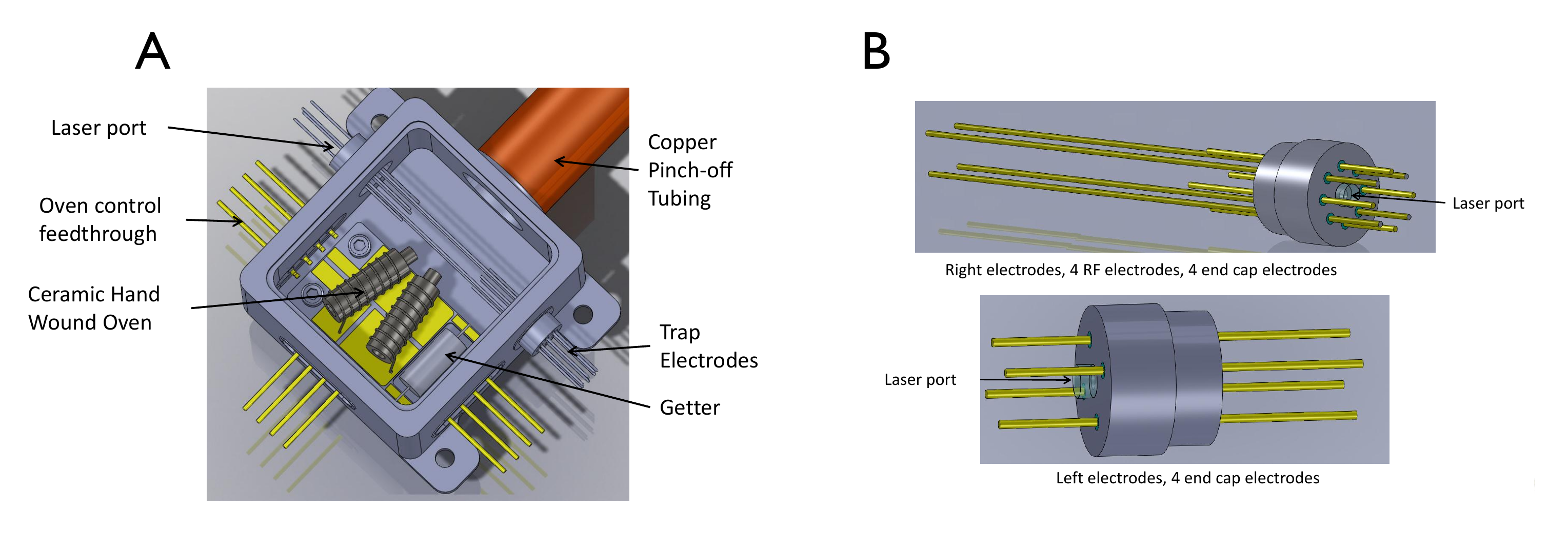}
   \caption{The PA\&E package geometry designed in Solidworks.  A: Physics package containing ion trap, ovens, and getter.  B: Laser window and electrical feedthrough assembly.}
   \label{fig:PAEdesign}
\end{figure}

The first designed Phase I package we received was the so-called PA\&E package.  The package design is shown in Fig. \ref{fig:PAEdesign}.  We chose the company PA\&E to make our package because they believed that they could make the windows and electrical feedthroughs we needed and maintain a relatively high bakeout limit.  Our criteria for the package manufacturer included the ability to machine small parts in titanium and make feedthroughs with a small profile, and to do so with materials that would be absolutely nonmagnetic and allow us to approach the 450$^\circ$C bakeout. (In the end, the limit for this package was 300$^{\circ}$C.)  PA\&E used their ceramic technology dubbed ``kryoflex" to make the glass-to-metal and ceramic-to-metal seals to titanium without taking up extra space with weld collars, etc.  The kryoflex can be identified in the photos as the yellow ring around the windows.

\subsection{Technical elements}
\begin{figure}
   \centering
  \includegraphics[scale=0.55]{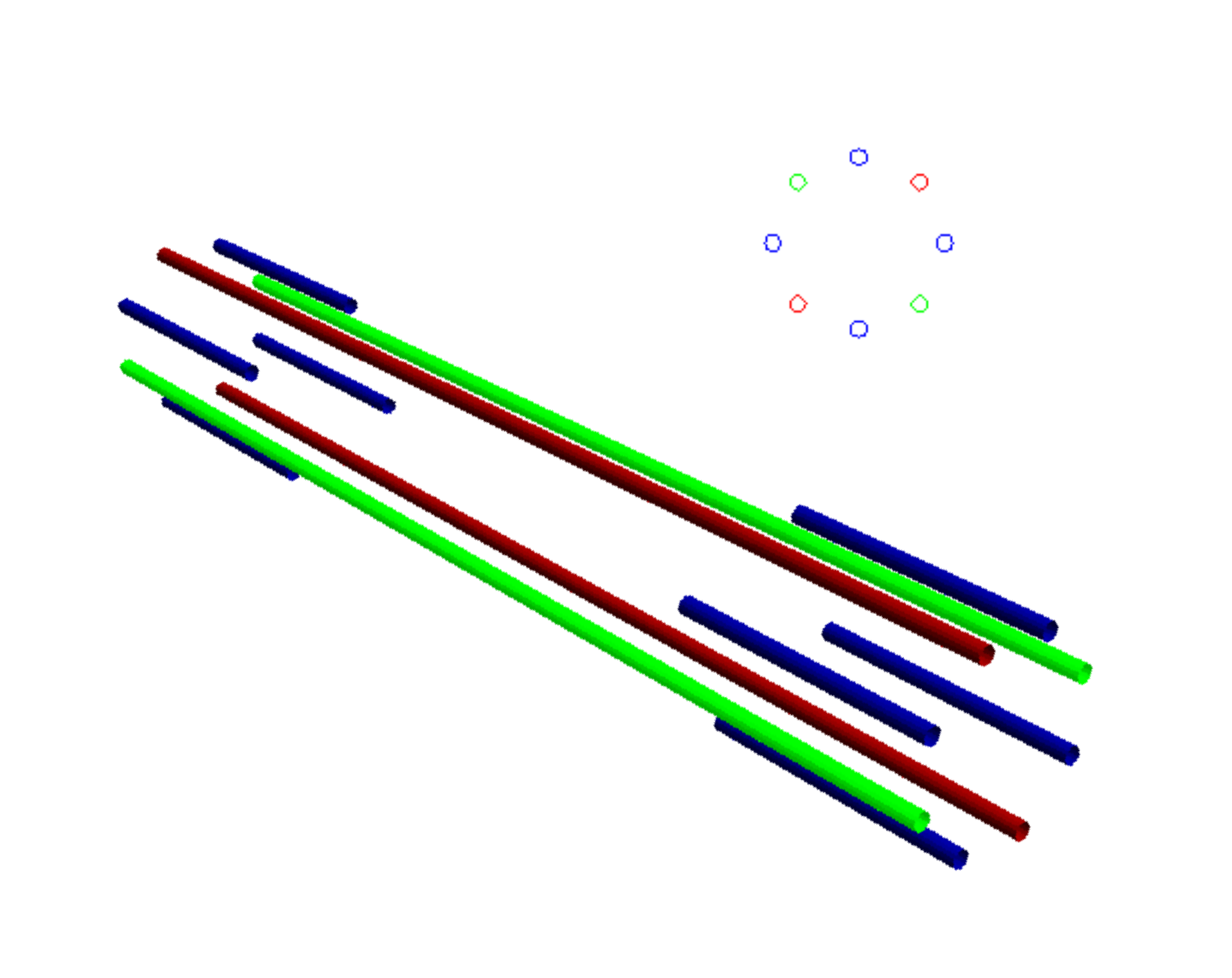}
   \caption{The PAE package geometry modeled in CPO.  The blue endcap electrodes are interlaced between the quadrupole trap electrodes to allow laser access in the center.  The end view of the arrangement of the electrodes in a ring is shown above the perspective view. The spacing of the quadrupole electrodes is 2 mm.}
   \label{fig:PAECPO}
\end{figure}

The package contains a four-rod trap with the same transverse spacing (2mm) as the test bed trap, but with much thinner electrodes (0.010" diameter instead of 0.64").  The geometry is shown in Fig. \ref{fig:PAECPO}.  The endcaps are very different: instead of centered tubular structures, there are four endcap rods at each end, interlaced between the quadrupole rods.  This design was chosen so that in the center of the ring of eight electrical feedthroughs, there could be a window so that we could probe the ion cloud along the long axis.  PA\&E was able to make this dual feedthrough with very close spacing and a very small window (0.5 mm diameter).  In addition, there is a detection window on the front of the package.  Below the trap and window is space for the ovens and getter along with other electrical feedthroughs.  Because this was our first venture into small packages, we were not yet ready to incorporate the micro-ovens we were developing, so we planned an alumina baseplate on which to mount homemade ovens (like the ones in the test bed) and the cylindrical getter (like the one in the COTS package).  This way, PA\&E would send us the package with the lid (with detection window) open, we could populate it with our parts, and return it to them for welding of the lid.  The package included a 3/8" OD copper pinch-off tube, with a miniflange and gasket attached for us to connect it to the vacuum chamber.  As usual we used a miniflange valve to be able to seal off the package.   The exterior physics volume of this package, not including the pinch-off tube, is 10 cm$^3$, so it was a fair step from a tabletop system toward our goal.  

\begin{figure}
   \centering
  \includegraphics[scale=0.55]{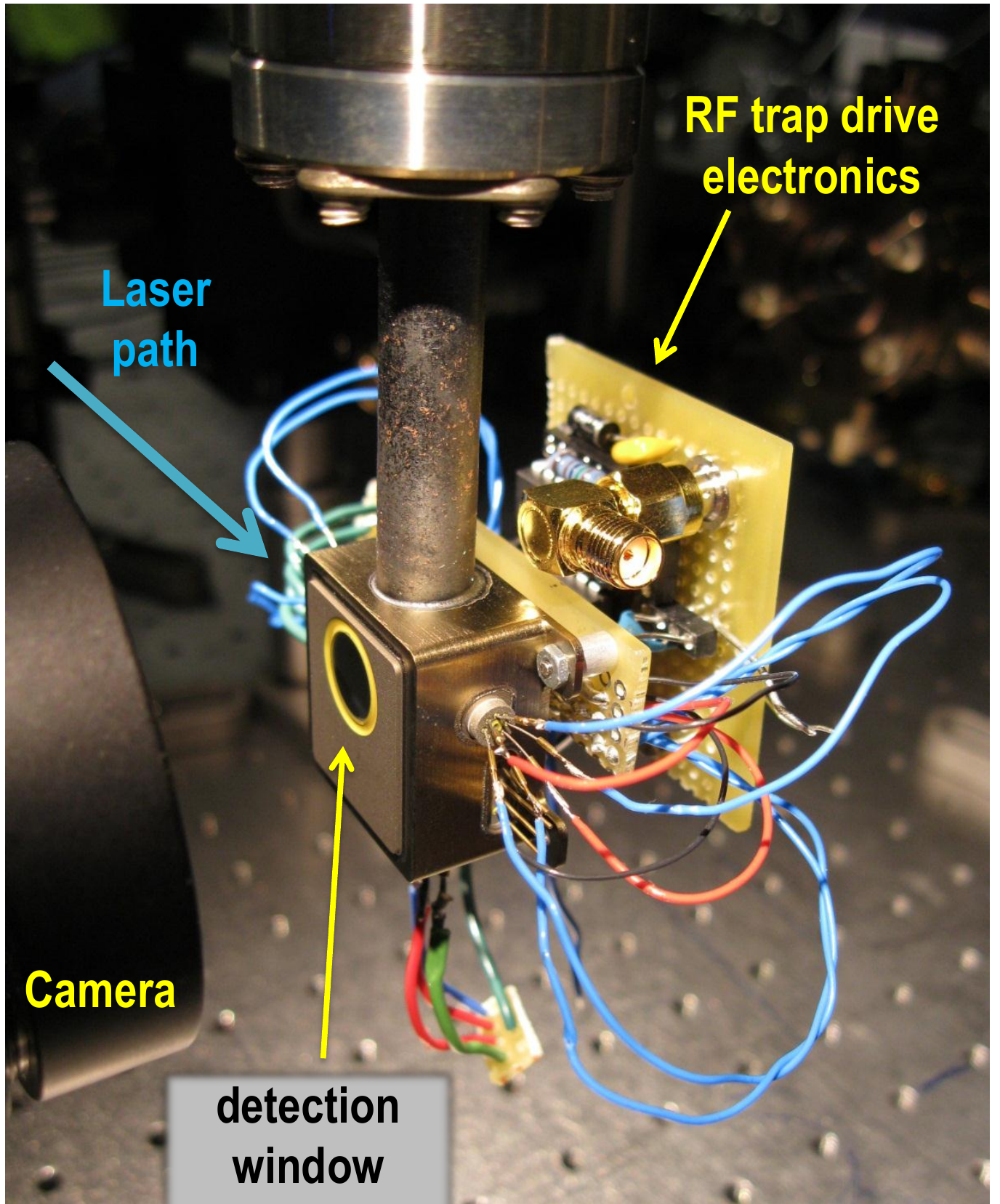}
   \caption{The PA\&E package ``in situ".  The laser passes horizontally through the center of the electrical feedthroughs on both sides of the package.  The trap drive electronics are attached the back of the package. }
   \label{fig:PAEinSitu}
\end{figure}

We connected electrically to the PA\&E package using slip-on connectors on very thin wires that were all attached to a stress-relief board bolted to the package (Fig. \ref{fig:PAEinSitu}).  In the beginning we used the RF amplifier to run the circuit but later we included a prototype of our CMOS circuit which was attached directly to the stress-relief board on the package.

\begin{figure}
   \centering
  \includegraphics[scale=0.45]{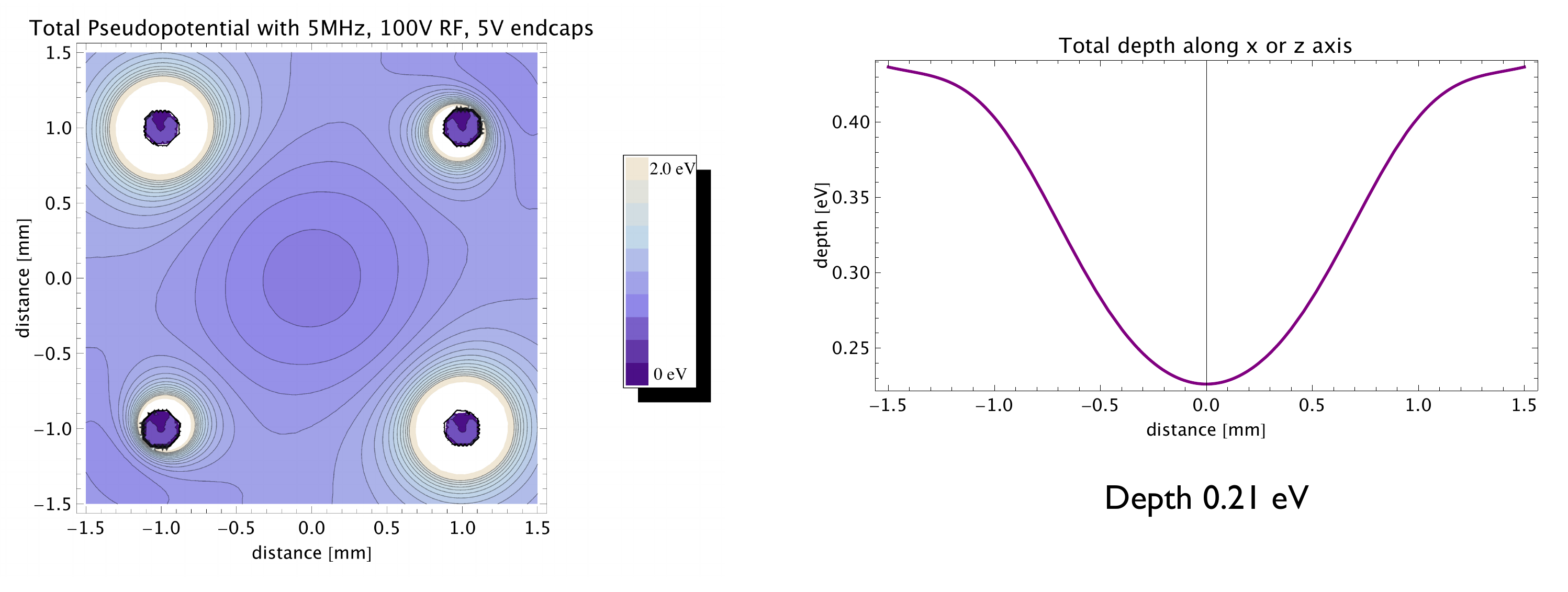}
   \caption{Modeled pseudopotential in the PA\&E package.  This trap is only about 1/5 as deep as the testbed, because of the small-diameter electrodes. }
   \label{fig:PAEmodel}
\end{figure}

A model of the PA\&E trapping potential in the transverse direction is shown in Fig. \ref{fig:PAEmodel}, along with the depth along the $x$ or $z$ axis.  We were concerned at first that the thinner electrodes, which were necessary in order to fit the endcaps and windows all together and preserve the  2mm quadrupole spacing, created a potential about $1/5$ as deep as the test bed trap; this did not prevent us from trapping, but we do believe that a  weaker overall ion signal can be attributed to the shallower well.

In this package we also used a microwave horn to couple the microwaves into the package.  Shining it at an angle toward the detection window was sufficient to saturate the ion signal.  We planned to use photoionization in this package for ion loading.

\subsection{Trapping Ions}

\begin{figure}
   \centering
  \includegraphics[scale=0.55]{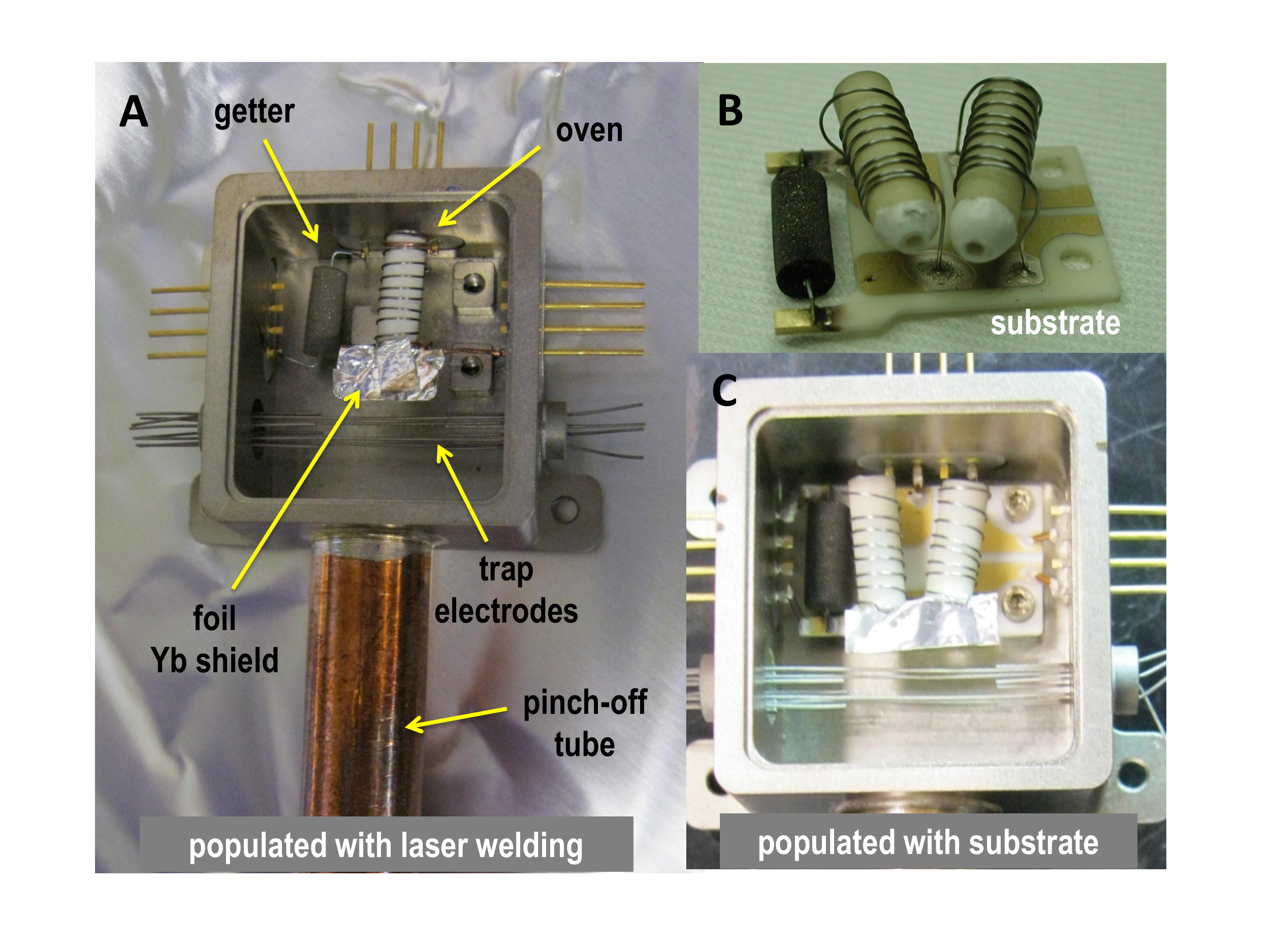}
   \caption{Photos of the PAE package assembled two different ways.  A: A getter and oven are laser welded directly to the electrical feedthroughs. B: The original planned substrate to support two ovens and a getter. C: The substrate installed in the package.  Electrical connections from the substrate to the feedthrough failed after a day of use. }
   \label{fig:PAEphoto}
\end{figure}

In total, three PA\&E packages were made, although some of them were returned and reworked so that overall we received about 6 different versions.  The first one we received sealed and did not populate with oven and getters in order to first test the trap.  Several issues were immediately apparent: the trap electrodes were very thin and not stiff enough to be uniform.  Each trap we received would be warped and need to be adjusted accordingly to try to make the quadrupole close to symmetric.  On the outsides of the feedthroughs, the wires were also very thin and easily broken.  Despite these issues, we attempted to trap in the unpopulated package by inserting a homemade oven into the copper pinch-off tube.  We almost immediately coated the detection window with an opaque film of Yb using the inserted oven.  After trying to remove the coating by rebaking with no success, we removed the package and cleaned the inside with a 2.5\% sulfuric acid solution.  We solved this problem in this package by pulling the oven farther back into the tube, which collimated the beam enough to keep the Yb off the window. From this point, any handmade ovens were also equipped with a foil shield (this can be seen in Fig. \ref{fig:PAEphoto}B) to prevent unwanted spraying of Yb.

For the second iteration of the PAE package we were ready to include our oven and getter so that we could seal off the package with a valve.  The alumina baseplate we planned failed due to difficulties with making electrical connections to the thin- and thick-film pads.  After making one version with this baseplate, the connections quickly failed while we applied current to the ovens.  We also had to insert an oven through the tube into this trap before we found our first ion signal. As an alternative solution in the next package, we used a laser-welder to weld an oven and a getter directly to the feedthrough pins on the inside of the package.  An example of each style is shown in Fig. \ref{fig:PAEphoto}.  After dealing with a series of other setbacks due to leaky windows, welds broken in shipping, incorrectly welded windows, and broken leads on the outside of the package (which we electrically connected to using sharp probes), we were able to successfully trap ions in some of these traps.  The second package (the first populated one) was successful but only using the backup oven inserted into the pinch-off tube.  We were forced to stop using this package after the Yb oven became clogged and we no longer had a working source.

The laser-welded (internal parts) packages were the latest and most robust version of the PA\&E packages.  The first laser-welded package had the first working trap with an integrated, internal oven, but a serious problem with this package was its vacuum performance.  After repairing a significant leak around the window, we found that it was outgassing large amounts of argon from an unknown source.   A second package with laser-welded parts did not exhibit this argon issue, suggesting that the first package simply could not recover from the leak.  We were able to test the second package for some time, and even to seal it off.  However, this package had other problems, for example the Yb was coating the small windows in the feedthrough centers, and causing a lot of reflection from the window, and in some cases preventing sufficient light from reaching the ions.

The (0.5mm) windows unexpectedly caused a lot of reflected light, producing a background as high as the COTS package.  It was difficult to focus the light down to a beam small enough to not touch the edges of the windows on input and/or exit and still have it be very focused in the center of the trap.   We began by focusing into the trap with a 500 mm lens, and later changed to a 250 and then a 150 mm lens which reduced the background by a factor of 4.  In the end, detecting at 297 nm was the solution for this package.  Also, the tiny windows seriously limited our freedom of alignment of the beam, and with the imperfect trap rods we were faced with, it was always possible that the center of the trapping potential was not lined up with the physical center of the feedthroughs, where the windows are located.

The same Yb film coating that was an annoying side effect in terms of window access also provided a surprise advantage when located on the on the back wall of the package and on the electrodes.  We found that by shining a UV broadband light source (such as a UV curing lamp for light-sensitive epoxies) in to the package, we could load the trap without 399 nm photoionization light.  This turns out to be due to the photoelectric effect allowing electrons to be ejected from the Yb surface because of its low work function, as discussed in Sec. \ref{sec:neutralAndIonize}. After using this technique in PA\&E and the JPL traps, this became our ionization method of choice due to its relative simplicity.  UV loading tends to overfill the trap, giving an initial signal that is higher than using photoionization, that quickly decays to be equivalent to the other method.   We also investigated the use of UV light to load the test bed trap for comparison.  At first, it did not work but after using the ovens for awhile we could see some loading from the UV lamp, but with a very weak signal.  This suggests that a fresh coat of Yb (since the last exposure to air) was necessary before UV loading could be used.  It is also possible that the walls must be very close to the trap center in order for the photoelectrons to work efficiently.

This was the first package that we used to test putting the PMT directly at the window of the package without an imaging system.  For this package, we could not get the PMT very close because of a clumsy conventional assembly with the filters, and we believe we actually reduced the solid angle we could observe, so this technique would not be advantageous until later packages where we had a different PMT setup.

We drove this PA\&E trap with an RF amplifier and later a prototype low-power circuit board.  The oscillating frequency can easily be manipulated on the prototype board by changing the capacitor in parallel with the trap capacitance.  We took advantage of this to investigate how the lifetime of the ions was affected by trap frequency.


\subsection{Performance}

In the end, we were unable to seal off several of the PAE packages reliably because of either leaks, faulty ovens causing a need for external electrical connections, or argon outgassing.  However, we explored the trapping parameter space using several different drivers.  We also investigated how the lifetime (and thus the stability) is affected by using different trap frequencies.  

Our initial trapping parameters were: trap frequency $=$ 5.5 MHz, $V_{\rm{RF}}= 350$ V, and  $V_{\rm{EC}}= 6$ V. We had no trouble seeing the $^{171}$Yb$^+$ hyperfine resonance using these parameters.  We tried many different drive styles in this trap, ranging in frequency from  3 to 8 MHz.
Later, using the circuit board, we would typically drive between 3 and 4 MHz and use a proportional amount of voltage amplitude to match the previous trap parameters.  The first populated package worked well with the additional oven inserted through the tube. The first trap that had the welded inside oven  was the best performer, and we were able to trap inside it when it was sealed and removed from the vacuum manifold.  By that time, however, we had much better performance from the JPL traps and were using those as our clock source.

\begin{figure}
   \centering
  \includegraphics[scale=0.55]{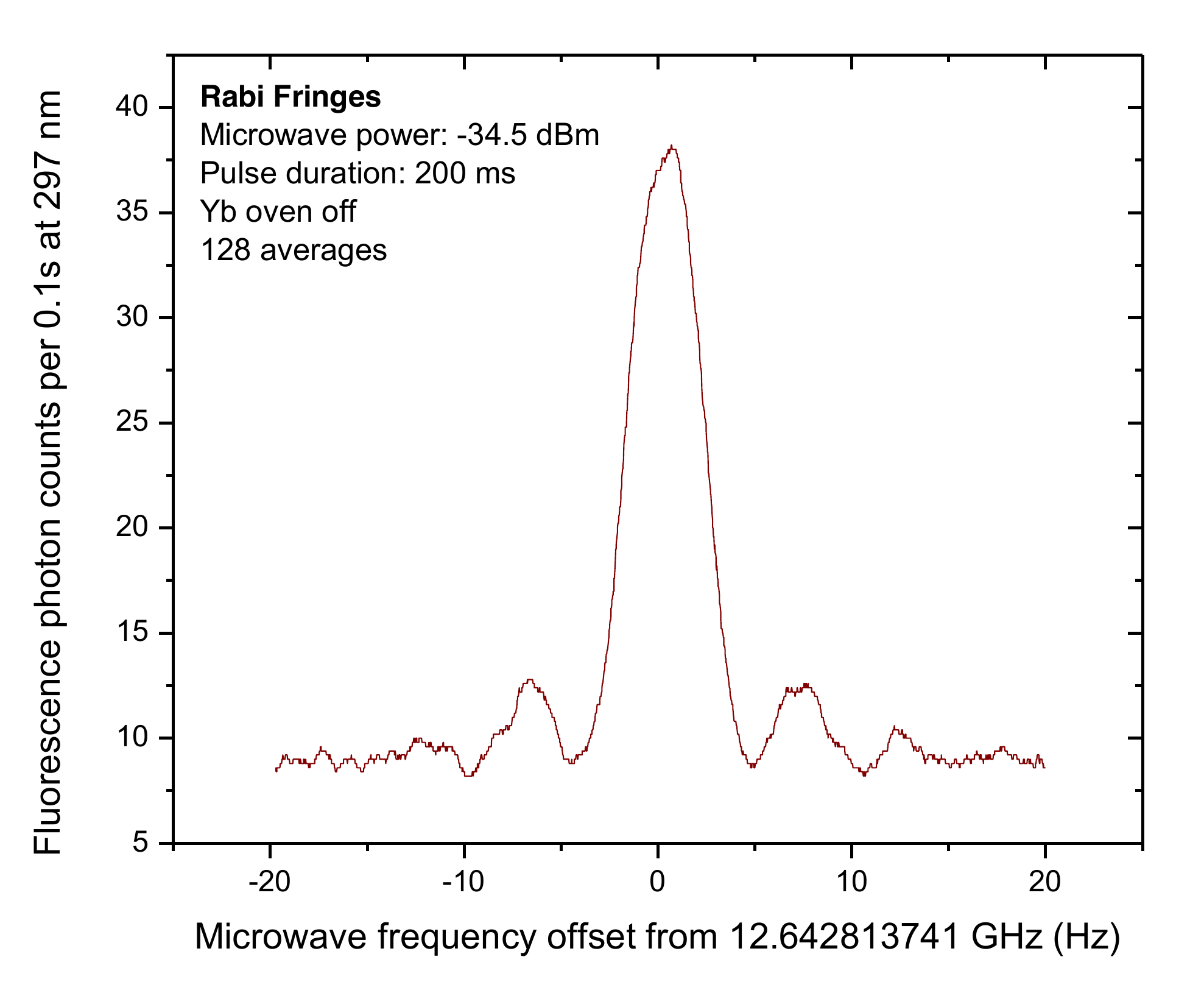}
   \caption{Rabi fringe used as a clock signal from the PA\&E package. }
   \label{fig:PAERabi}
\end{figure}

\begin{figure}
   \centering
  \includegraphics[scale=0.45]{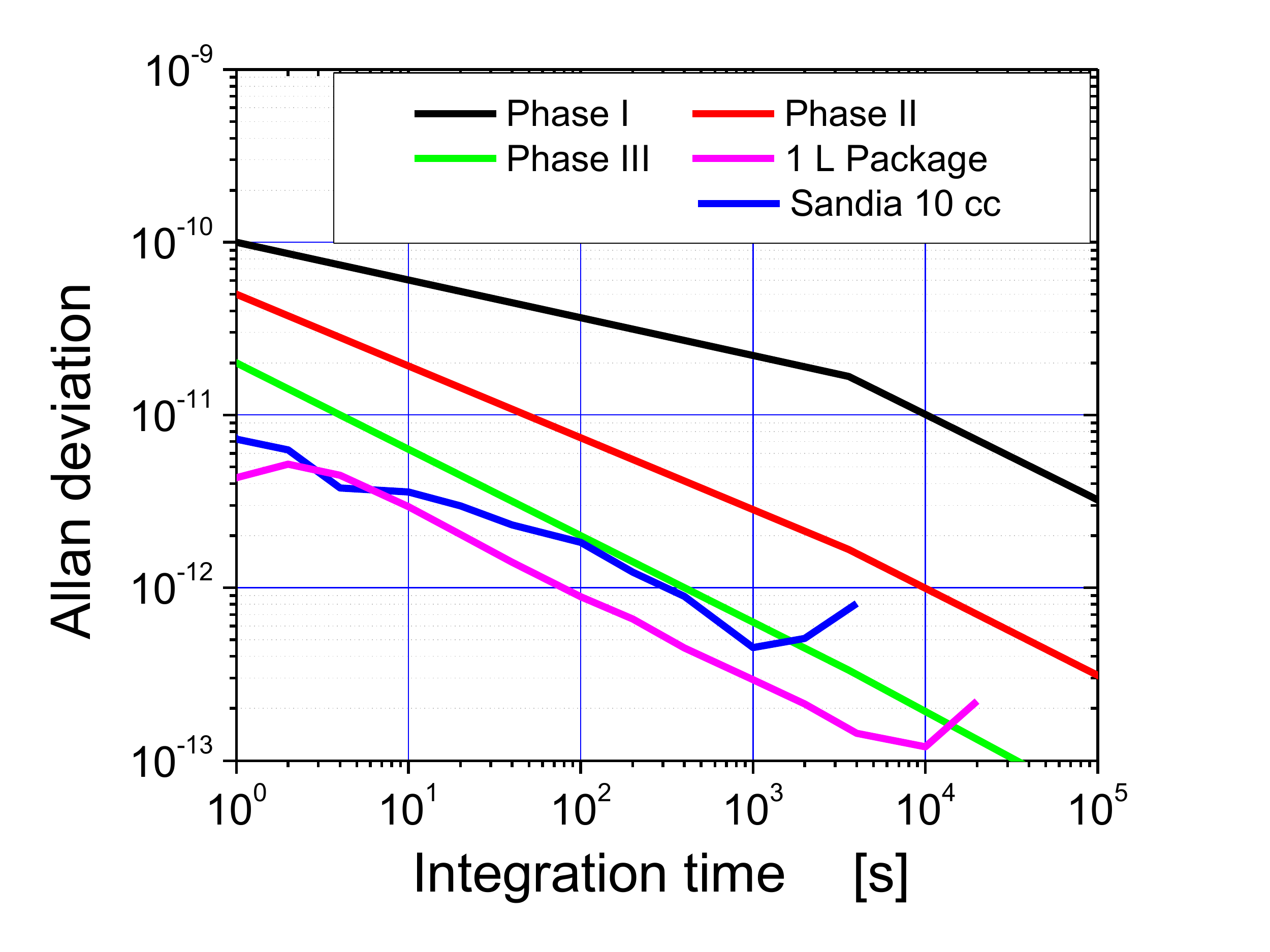}
   \caption{Allan deviation performance of the PA\&E package, compared with the project goals and the test bed clock. }
   \label{fig:PAEALLANDEVIATION}
\end{figure}

We made a clock out of the first PA\&E package to be populated with ovens and getter, even though we used an external oven inserted through the tube due to the broken connections.  We made a series of measurements in that package. Rabi fringe data taken from this package is shown in  Fig. \ref{fig:PAERabi} and the Allan deviation we measured for this trap at this early time is in Fig. \ref{fig:PAEALLANDEVIATION}. After the arrival of the JPL packages, none of the further attempts to improve and make the PA\&E packages function better were successful enough to attempt making a clock, especially while the JPL systems were doing much better. 

\subsection{Lessons from the PAE package}
In this package we learned that very small cylindrical electrodes are not stable and do not create a strong enough trap for our preferences.  We found that we need a different method than kryoflex to attach feedthroughs and windows; this issue is solved by the use of electron-beam welding and brazing to form glass to metal seals in the all-metal packages.  We found that tiny windows create a lot of background light, and do not allow us to explore the trap freely.  We must be especially cautious about the location and direction of the oven so that we do not accidentally coat windows.  We became aware that we can load using photoelectrons ejected from Yb coating with the coated chamber walls and/or electrodes close to the trap.  Last, eliminating almost all magnetic materials in this trap did give an advantage over the test bed (whose endcaps were still made of slightly magnetic stainless steel), although we would see later that the PA\&E package is not as non-magnetic as we had hoped.

\section{JPL Phase I package} \label{sec:JPLphase1}

\begin{figure}
   \centering
  \includegraphics[scale=0.6]{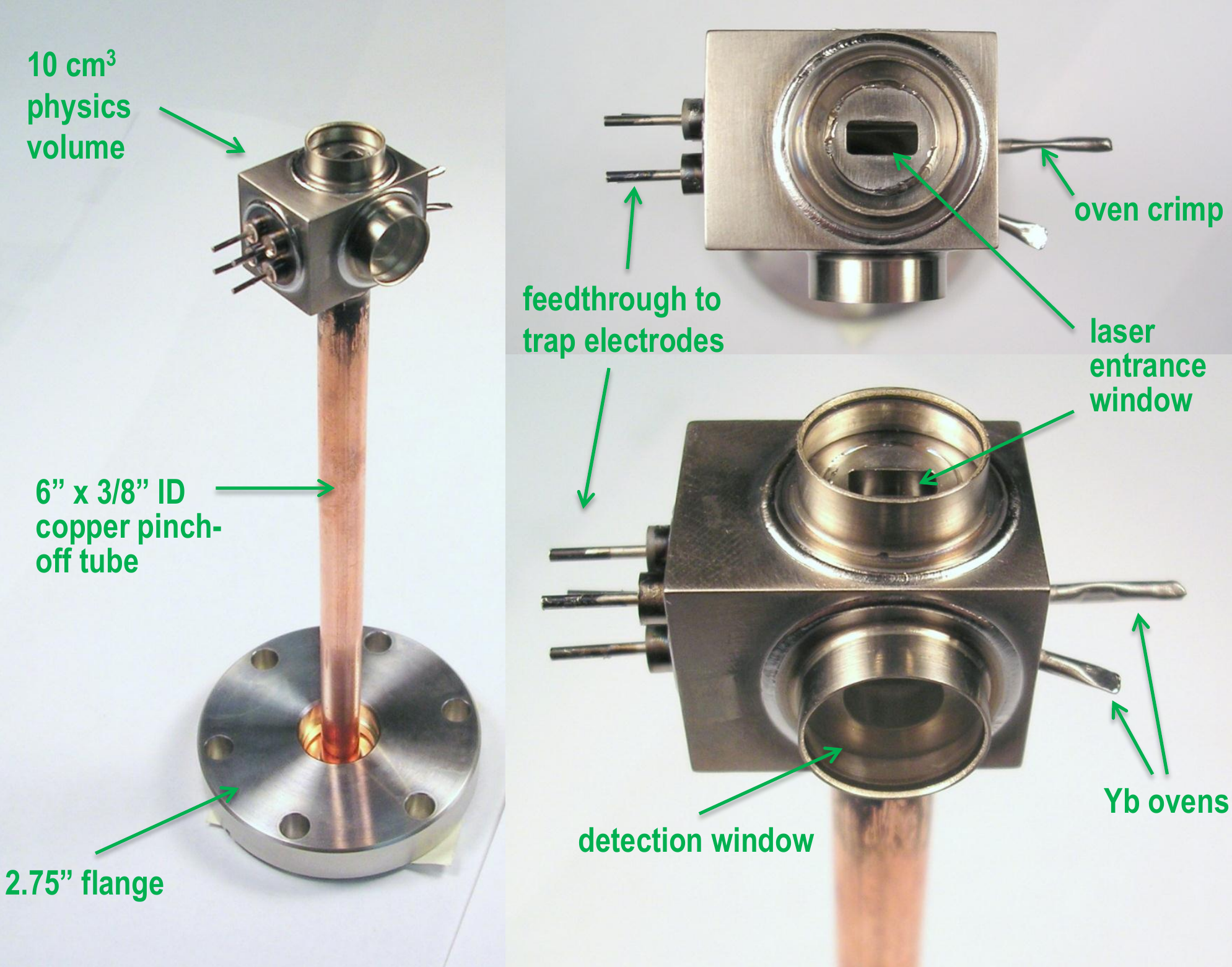}
   \caption{Phase I metal package produced at JPL.}
   \label{fig:JPLphase1photo}
\end{figure}

Our most successful package by far in Phase I was the so-called JPL package.  The completed package is shown in Fig. \ref{fig:JPLphase1photo}. Our collaborators at the Jet Propulsion Laboratory have years of experience in developing Hg ion clocks, including the 1 liter ion trap clock, making them an ideal partner for us in developing a sealed package and a robust, effective ion trap.  As a result, we were not as directly involved in the design and assembly process of this package as we were with the PA\&E package.  However, JPL's expertise shows up abundantly in the packages we received from them.  Having access to JPL's vast knowledge of systems similar to ours had an enormous impact on the success of our work.  The packages designed by JPL are just a part of this impact. 

\subsection{Technical elements}

The main idea of this package was the same as for the PA\&E: to produce an all-metal design, incorporating a strong trap, laser access, and a Yb source.  Four electrical feedthrough wires go directly into the package to four trap rods in the same orientation as the pins.  There are no endcap feedthroughs, since we use a negative bias to create trapping on the long axis.  There is one window for laser access, which faces the pump-out tube so that light entering the package will disappear down the tube.  The electrodes are oriented vertically instead of horizontally when we use this package, and because of the difficulty of putting a window in the center of an electrical feedthrough (this was an advantage of PA\&E), this package is designed for the laser to enter the trap on the transverse axis.  We did expect that we could get a much larger signal by moving back to a design where the laser is along the long axis, which we did in the Phase II package.  A perpendicular (side)  window is used for detection, and on the back wall from the detector is a pyramid-shaped structure designed to deflect unwanted stray light from the camera instead of reflecting it back as a flat wall might do.  This is a structure that JPL had used in previous vacuum packages. 
Both windows have a reduced clear aperture by the design of the walls of the trap.  This helps with light scatter and to prevent window coating.  

The appendages on the bottom of the package are tubes that are used as a Yb source: one is vertical and one at an angle, both attempting to point to the center of the trap.  Yb is placed inside the tubes, so that once under vacuum we can heat the tube externally (we use a cartridge heater and an aluminum adapter for thermal transfer to the appendage).  These two ovens supply natural abundance ytterbium and $^{171}$Yb.

We planned to use photoionization at first but later we were able to use the photoelectron method that we discovered in the PA\&E package.  Microwaves are pumped through a window using a horn, and later on, they are directly coupled to the RF ground electrodes of the trap.  The vacuum package is attached to a pump-out pinch-off tube with a gasket and 2.75" conflat fitting.  The getter in this package is a round getter that is placed out of the way, around the pump-out tube, and is activated passively during bake-out.  This package is all titanium with sapphire windows,  expect for the stainless steel tubes that are used to make the oven appendages. Note, the first package had steel electrodes as well for reasons I will explain in a moment.  The windows and electrodes are brazed together and final assembly is done with electron-beam (e-beam) welding, which is a fast, very localized welding process.  Because of these materials and assembly techniques, the JPL package is bakeable to 450 $^\circ$C.

\begin{figure}
   \centering
  \includegraphics[scale=0.55]{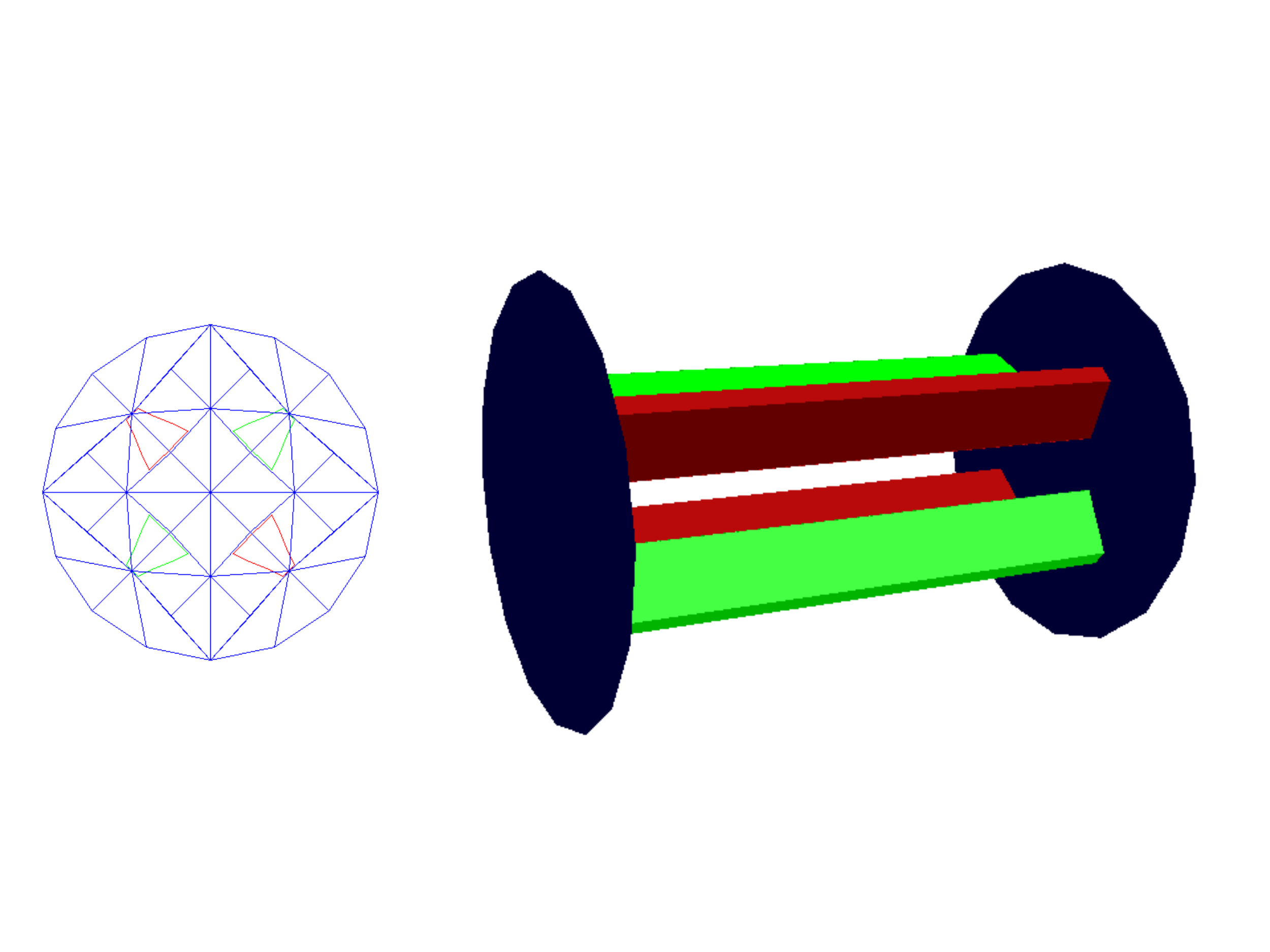}
   \caption{The JPL package geometry modeled in CPO. Left: End view (transverse plane).  Right: Perspective view. The quadrupole electrodes have a trapezoidal shape and the endcaps, represented here by the blue circles, are the trap walls.  This trap is somewhat larger than the other traps.  The spacing between the centers of the faces of the trapezoid facing the center of the trap is 2 mm.}
   \label{fig:JPLP1CPO}
\end{figure}

The trap planned for this package was a quadrupole trap, but was quite different from our other traps in that the electrodes were trapezoidally shaped on the cross section, such that the long side of the trapezoid faces the center of the trap (see Fig. \ref{fig:JPLP1CPO}).  This was a design used by JPL to make a larger, more harmonic well in the trap center.  Although this turns out not to be necessary, it was effective.  The only drawback was that the trapezoid electrodes could have the tendency to limit our visible access to the trap, and thus have some effect on the detection.  

\begin{figure}
   \centering
  \includegraphics[scale=0.45]{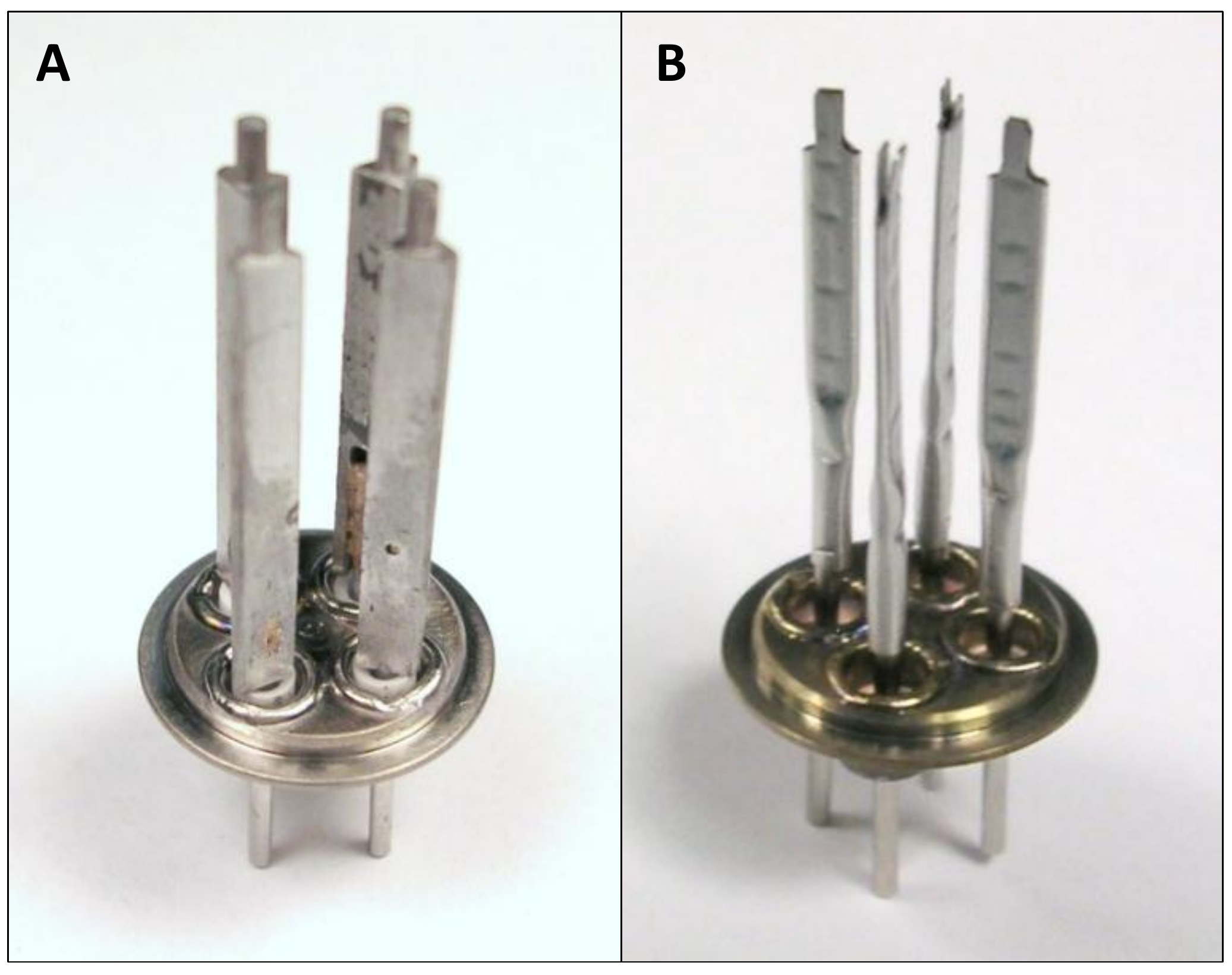}
   \caption{The electrodes in the two JPL Phase I packages.  A: Trapezoidal electrodes that are machined out of titanium.  B: Stainless steel hand-shaped electrodes used to replaced the machined part in the first package after a failure during assembly. }
   \label{fig:JPLelectrodes}
\end{figure}

\begin{figure}
   \centering
  \includegraphics[scale=0.45]{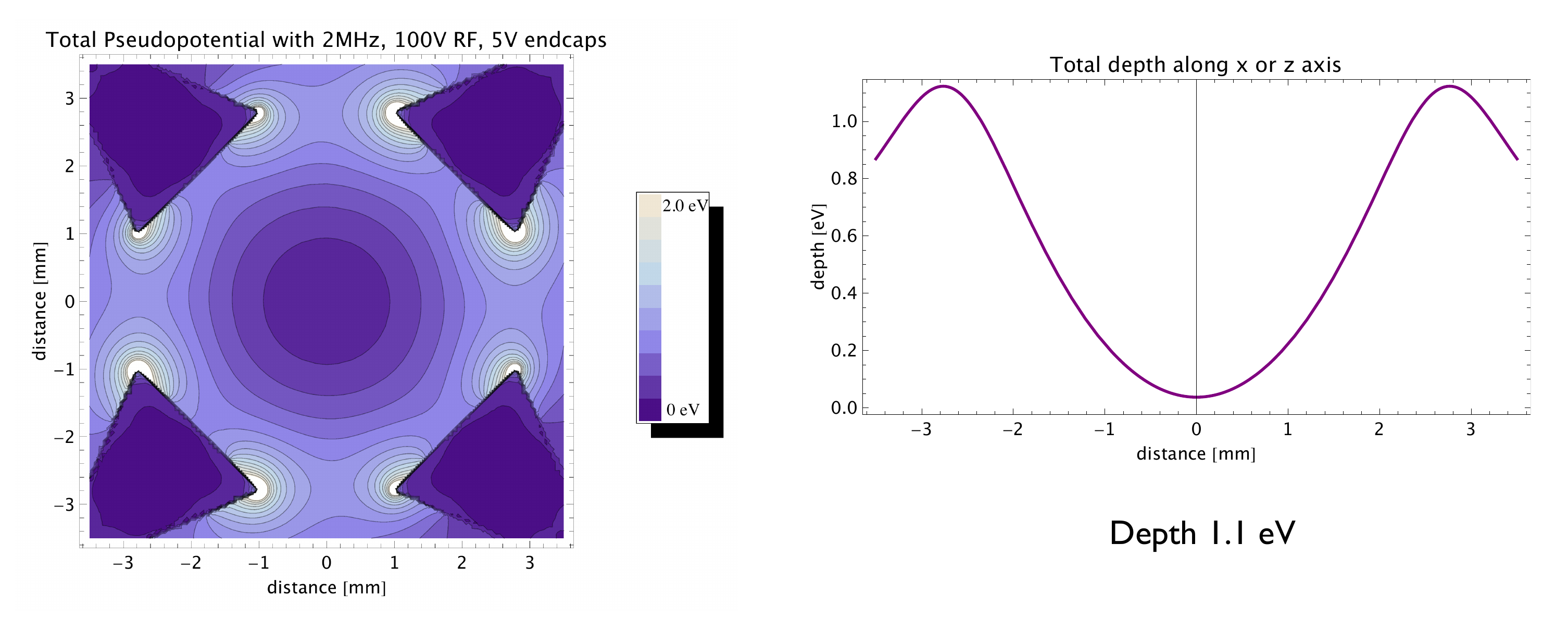}
   \caption{Pseudopotential for the JPL 1.2 package with trapezoidal electrodes.}
   \label{fig:JPLP1model}
\end{figure}

A model of the transverse plane pseudopotential is shown in Fig. \ref{fig:JPLP1model}.  The trap shape does make the center potential very uniform and can be made very deep because of all of the electrode surface area surrounding the trap.  The grounded walls of the chamber at either end of the quadrupole electrodes act as endcaps when the four RF electrodes are negatively biased.

\subsection{Trapping Ions}

There were two major versions of the JPL Phase I package.  The first package we received (``JPL1.1") had some engineering setbacks during construction that  prevented the use of the original trapezoidal electrode assembly, and also damaged the angled oven appendage.  In an attempt to meet a deadline, the package was completed with an alternative electrode assembly, shown in Fig. \ref{fig:JPLelectrodes}B, that consists of flattened stainless steel tubes.  The second oven was cut back to about half the proper length and welded closed where a leak had been.  The main implications were that we had no isotopically purified sample oven, and that we had slightly-magnetic and slightly-nonuniform trap electrodes.  We were eventually able to trap in this first package.  However, we discovered several problems. The first was that the Yb had fallen out of the oven appendage at some point in the package assembly, leaving us with no Yb neutral fluorescence.  This forced us to open the chamber and insert our own Yb bits into the oven through a tiny tube aimed at the oven entrance.  Once we had Yb inside, we found that since the Yb oven was pointing along the axis of the trap to the opposite end of the quadrupole electrodes, the Yb was able to coat and short the electrodes together or to ground. This happened while running the oven,  and also during the bake.  When the electrodes had too low of a resistance between them for the trap to oscillate properly, we ran large amounts of current through them to break open the connection made by coated Yb. When this did not work, we were forced to use drastic measures to clean it; we used sulfuric acid and also an ammonia-peroxide solution to clean the inside of the package.  This worked well and did not damage the package itself, although it likely damaged the getter inside. 

The second package (``JPL1.2") dealt with a few of these problems through design adjustments, and for the rest we adjusted our procedures to avoid them.  JPL placed additional insulating beads on the electrodes to try to shadow them from the Yb coating, and we carefully monitored the electrical connections while running the Yb oven cautiously, to avoid unnecessary coatings.  After experiencing some slight coating on the windows, we decided to heat sink the Yb appendages to a lower temperature (around 250$^{\circ}$C) during bakeouts.  We also had the advantage of having the uniform (titanium) trapezoidal electrodes as well as both types of Yb in the new package.

For a time, we were concerned about the scattered light inside this package.  However, when we decided that detecting on 297 nm was the best way to go forward in miniaturization, we became much less concerned about engineering scattered light controls.  

Since we were illuminating this package in a manner perpendicular to the trap axis, we did make some attempts to communicate with more ions with the laser by using a short-focal-length (75 mm) cylindrical lens to focus the beam into a vertical line at the trap.  This helped somewhat, but because we didn't change the overall intensity we saw a similar amount of flourescence overall.  We also found it convenient to rotate this lens to optimize the signal by matching the focused line to the trap electrode orientation.  After the initial search for ions with unknown trapping parameters, subsequent alignments into the JPL1.2 package were pretty straightforward.  The powerful trap and clean vacuum created in the all-metal surroundings made this package our best performer.

Using this trap, we took the opportunity to examine the problem of ionization, since we knew we had to deliver a portable demo and would not have a miniaturized laser at 399 nm for loading.  We wanted to learn to photoionize with broadband UV light using photoelectrons.  We knew that the trap electrodes were too far away and the Yb beam too collimated for coating the electrodes with Yb in situ; however, Yb coated artificially on the electrodes during assembly would never survive; it could either oxidize in air or, if it survived, be lifted from the electrodes during a 400$^{\circ}$C bakeout.  Therefore, we decided to coat an electrode with calcium.  We chose calcium because it was convenient and already existed in our lab, because it has a low work function, and because it would not migrate away from the electrode during a 400$^{\circ}$C bakeout.  We inserted a homemade oven filled with calcium into the completed JPL package and heated it to eject all the calcium.  Because calcium oxidizes very fast, we had to backfill with argon and pump on the package immediately to avoid it contacting air.  We were able to make a visible spot of calcium on one electrode, which turned out to effectively load the trap when illuminated with UV.  We also eventually performed the same procedure on the JPL1.1 package.

To illuminate with UV, we tried using LEDs at first, but the broad spectra provided even by high-power LEDs did not seem have enough power at the needed wavelengths.  Next we tried mercury lamps sold for use as germicidal lamps.  These have a very bright near-UV spectrum.  We were able to load well by pointing an elongated bulb toward the trap for about ten minutes.  This was the system we used for loading the trap in the portable box. 

This trap's lifetime and behavior improve rapidly whenever we close the valve, when the getter only has the small volumes of the trap and the pumpout tube to clean.  After several bakes related to changes mentioned above, we had a very clean vacuum, and we decided to pinch off a package for the first time.  After pinch-off the trap worked as well as before, and the lifetime continued to improve over time.  We used this package for making a portable demo clock, and were able to use it up to a year later.  The lifetime of ions in the trap is more than a month.  The only issue with the pinched-off trap is that as it gets cleaner and cleaner inside, the effect of the F-state is increased, because there are less gases available to collisionally remove the ions from the F-state.  After about a year, we lost about 50\% of the signal to the F-state.  Again, this could always be solved by the use of a 638 nm laser.

We built a new circuit to run this trap, which we continued to drive with an RF amplifier, that could support the necessity for a negative bias on all four RF electrodes.  We later replaced this with a similar circuit using the CMOS inverters, that could provide a bias and run on a 9 V battery (for about a day).  We started out with the first circuit running at 3.65 MHz, with approximately 1000 V peak-to-peak on the RF electrodes, but we were eventually able to run the JPL package at 1.79 kHZ, with $V_{\rm{RF}} = 500$ V peak-to-peak, and $V_{\rm{EC}} = 4.5 $ V with the CMOS inverters and running on a battery.  These are the approximate parameters we used for our demo and our final Phase I clock.

\begin{figure}
   \centering
  \includegraphics[scale=0.45]{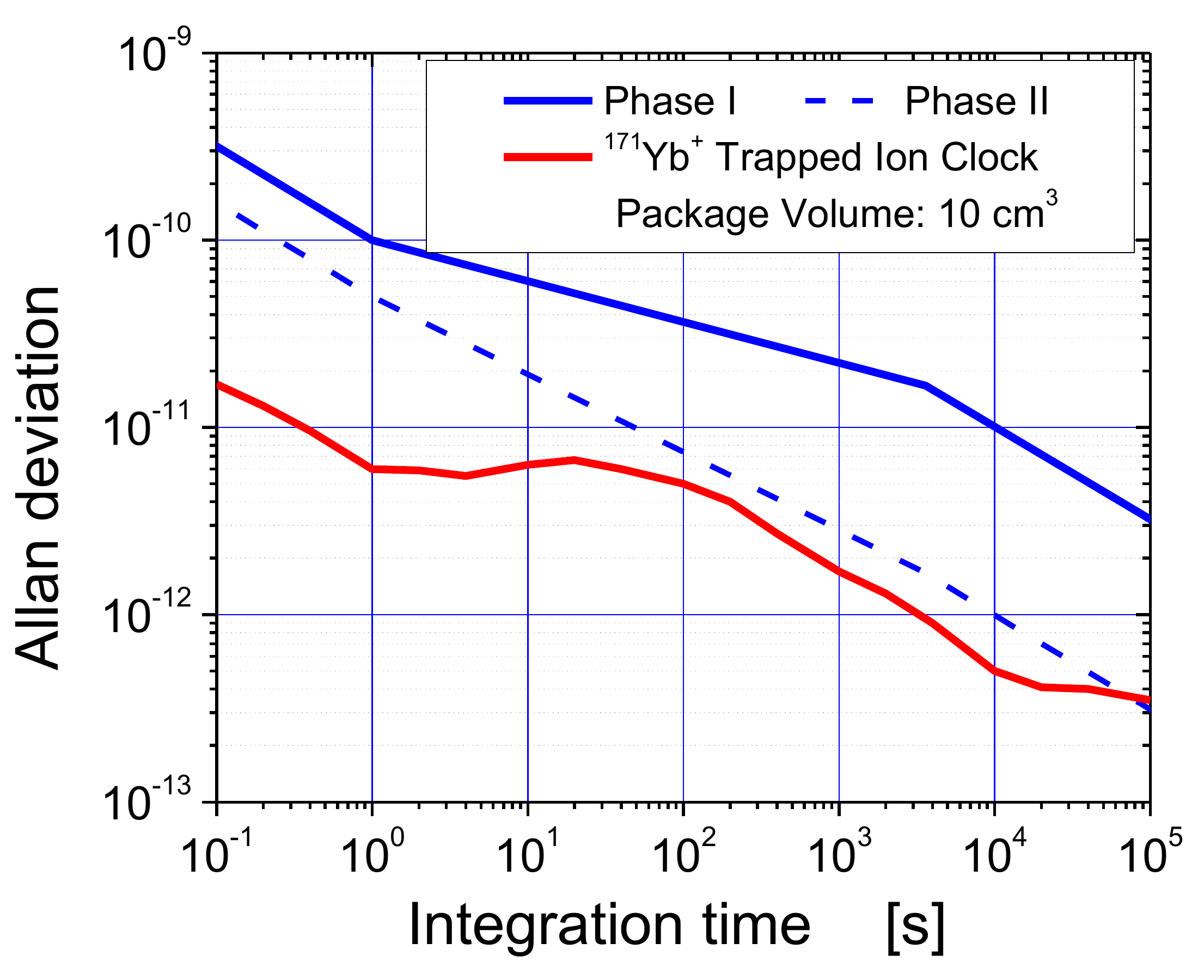} 
   \caption{Allan deviation data from the JPL1.2 ``portable" clock.  This data is taken without field compensation.}
   \label{fig:JPLAD}
\end{figure}

\subsection{Performance}

\begin{figure}
   \centering
  \includegraphics[scale=0.45]{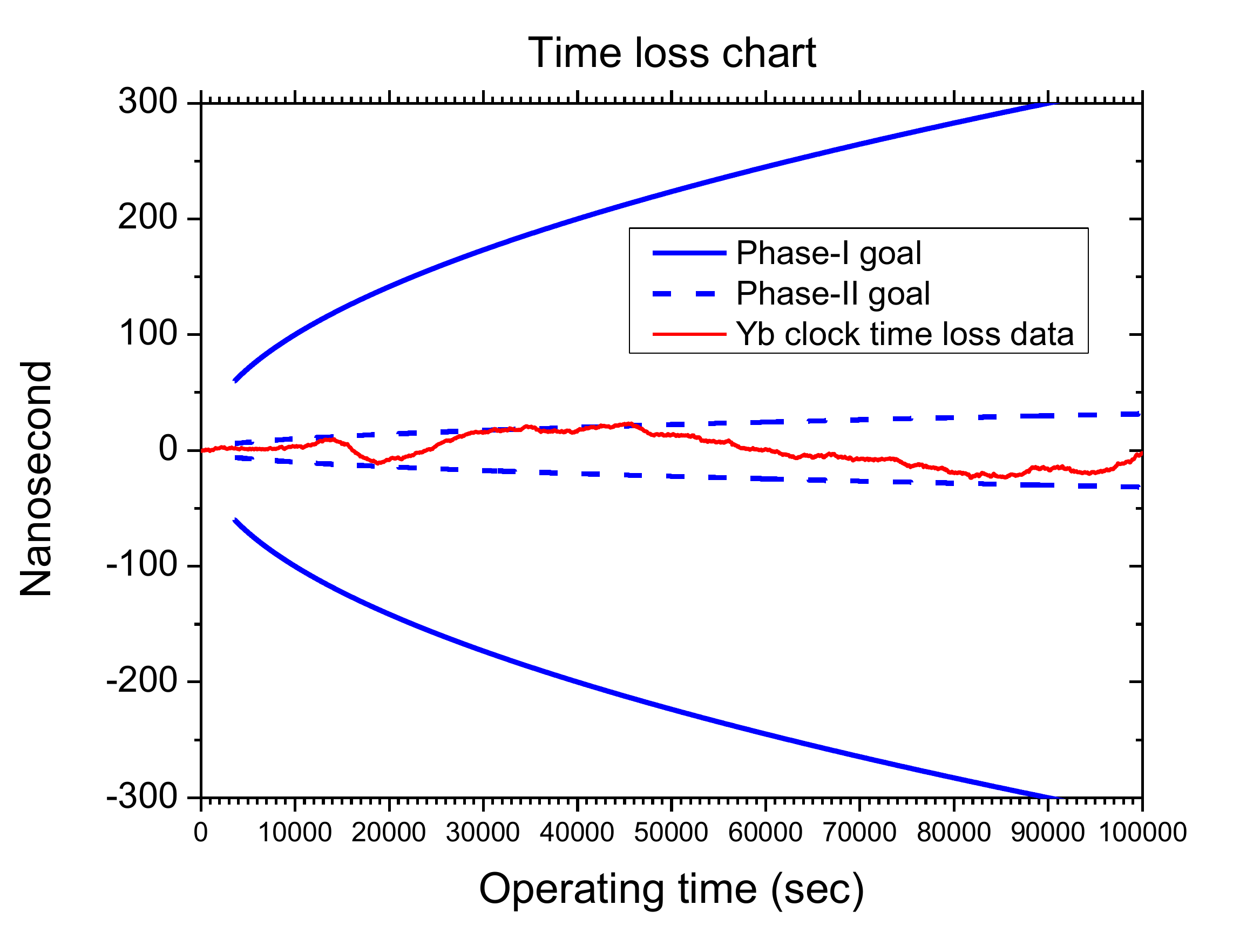} 
   \caption{Time loss data over time plotted against the DARPA metric.  Again we meet the Phase II guideline up to $10^5$ seconds.}
   \label{fig:JPLtimeloss}
\end{figure}

We made several clock measurements with JPL1.2, which was our final  showcase piece for Phase I.  
We were able to show a fractional frequency stability of $4\times10^{-13}$ for $10^4$ s integration time, as shown in Fig. \ref{fig:JPLAD}.  As part of our metric requirements, we also measured the time loss, as shown in Fig. \ref{fig:JPLtimeloss}.  Both metrics are within the Phase II goals using Phase I equipment.  

\subsection{Portability (April demo)}

In April 2011, we were tasked by our funding agency to deliver a demonstration of our progress to a field Principal Investigator meeting.  We took this task as a challenge to show the portability of our device.  We planned to present a working miniature ion trap 2000 miles from our home lab.  This would show that our miniature trap could easily be used for a portable clock system.  We used the pinched off JPL package for this demo.  It proved to be very robust.  Using a mercury lamp for photoelectron ionization and a cartridge heater for heating the oven, we were able to load ions that would stay in the trap for weeks.  We mounted this package in a box with the necessary optics to bring in 935 and 369 nm lasers\footnote{the 935 nm laser was a miniature source; a VCSEL. The 369 nm laser was not yet miniaturized and we carried our laboratory laser system to the presentation to provide this light by fiber into the setup.}.  A detector immediately in front of the window detected ion signal while we swept the 369 nm laser across resonance.  Being able to transport the system and have it continue to work in a completely different environment was an enormous challenge, but it forced us to investigate how to really make everything ready for removal from the lab.  The only element that was not truly portable was the 369 nm source. 

\subsection{Lessons from the JPL package}
From this package we learned the valuable lesson that metal technology with electron-beam welding and brazing produces a excellent vacuum environment for ion trapping. We also learned that ions can have long lifetime in a sealed volume that gets better with time! The package is technologically extremely resilient.  We discovered that we must be careful about where our ovens point with relation to not only windows but also shorting of electrodes together.  We learned to operate with low RF power.  We can load with an electrode coated with a low-work function material such as Yb or Ca (although we could not load directly from the wall coating in this trap as it was too far).  Loading is most efficient with a Hg germicidal or other lamp.  Later, we can load with a high-power LED and eventually a free-running laser diode. Last, we concluded that it is important to shine lasers on the long axis of the trap to make efficient use of the light.

\section{Phase II packages}

In an effort to put all of our knowledge gained into a next generation, we designed two very different packages for Phase II.  They have both been partially tested and I will report on these tests and our outlook for their overall performance. The aim of the all-metal JPL package is to establish a complete sealed package in a deliverable clock, for external testing, at the end of Phase II.  The LTCC trap, on the other hand, is a riskier, more preliminary pursuit which, if successful, will allow us to make a big step forward in Phase III toward a 1 cm$^3$ ion trap ``physics package".
\subsection{JPL Phase II package}
\begin{figure}
   \centering
  \includegraphics[scale=0.6]{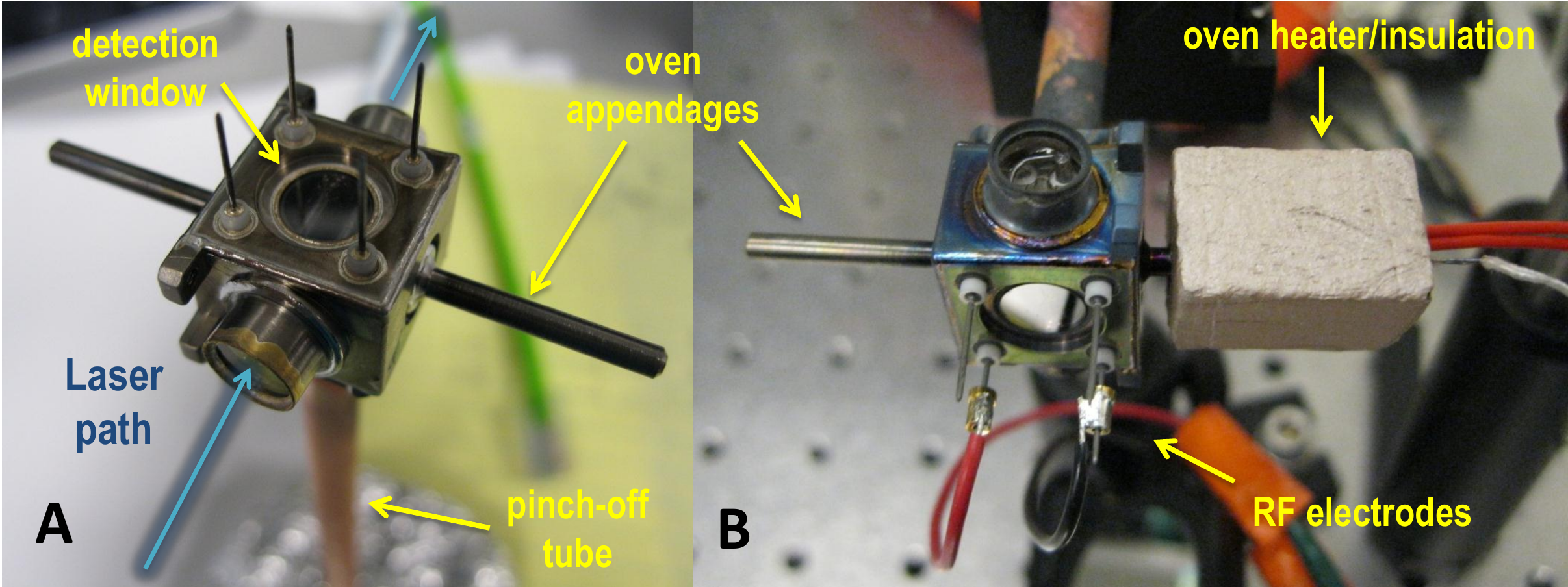} 
   \caption{Photos of the JPL 2.2 package. A: The package before implementation.  The laser enters and exits the package along the long trap axis.  The long tubes on either side are the oven appendages. B: The package in use.  The discoloration is due to the 400$^{\circ}$C bakeout.  Two wires connect to the RF pins surrounding the detection window.  On the right side we are heating one of the appendages with an insulation-covered cartridge heater.  This is the same method we used in Phase I.}
   \label{fig:JPLP2photo}
\end{figure}

\begin{figure}
   \centering
  \includegraphics[scale=0.6]{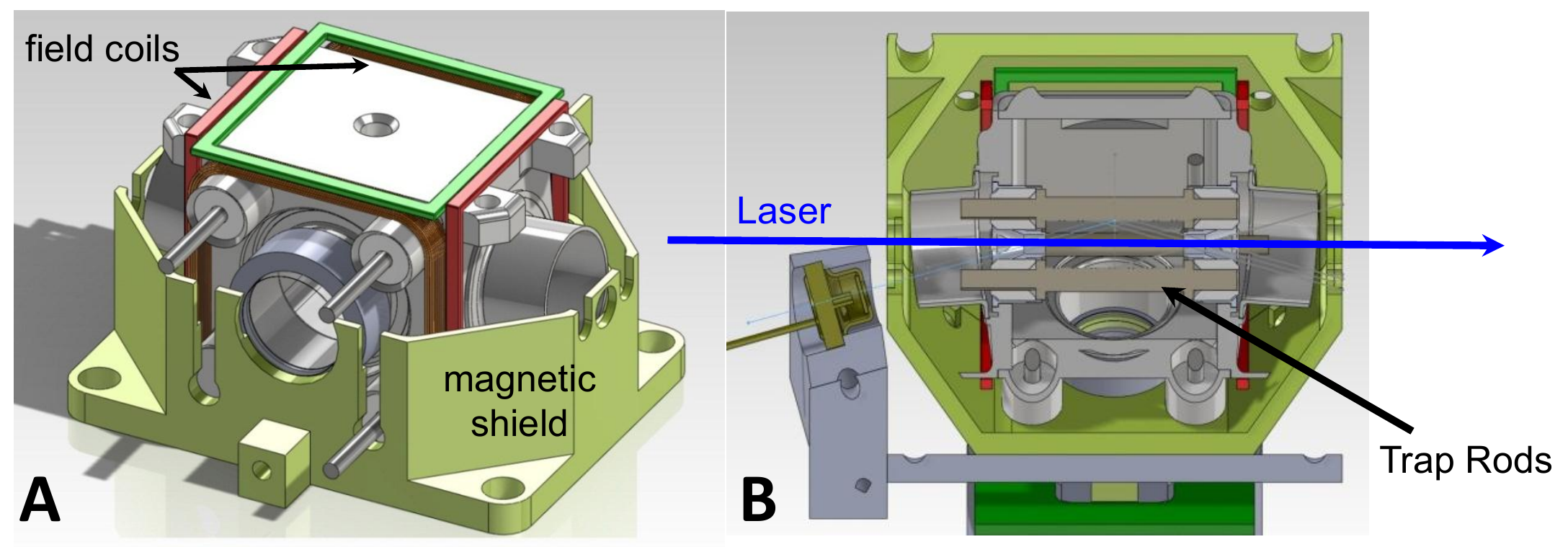} 
   \caption{Solidworks drawing of the JPL Phase II package design.  This version is the JPL 2.1 package which leaked, and varies slightly from the improved 2.2 version pictured in Fig. \ref{fig:JPLP2photo}.  A: Package with one magnetic shield half, showing the coils that will be wrapped and attached to the package inside the shield.  The detection system will be attached directly to the shield. B: Cutaway showing the interior of the package and the laser path through the trap.  The laser diode package on the left is for photoelectron loading.}
   \label{fig:JPLP2SW}
\end{figure}

The Phase II JPL package uses many of the technologies of their Phase I package.  It is all-metal, titanium with sapphire windows, and is assembled with e-beam welding.  Technological improvements include a laser path along the long axis of the trap, tilted entrance and exit windows to prevent backreflections to the miniaturized lasers\footnote{Since we do not have room for actual isolators, we must use polarization optics and tilted surfaces in order to prevent damage and/or instability in the VCSEL/VECSELs due to optical feedback.}, improved oven appendages, and a loop inside the package for the RF servo.   Photos of the JPL Phase II packages are in Fig. \ref{fig:JPLP2photo}.  In conjunction with this package, we are preparing a magnetic shield, attached detection system, and the package is designed for wrapping field coils, so we can deal more directly with the magnetic field issues of our clock.  A Solidworks rendering is in Fig. \ref{fig:JPLP2SW} that shows the shield and coil assemblies and a cutaway showing the trap and laser path.    

The first version of this package had several leaks on the feedthroughs and ovens.  Although we could trap after using vacuum sealant on the leaks, this trap had no chance for a long lifetime or to function when sealed.  Several new parts were sourced and a new version was built (this is JPL 2.2, which is the one shown in Fig.  \ref{fig:JPLP2photo}).  The window and feedthrough assembly was made by IPT-Albrecht GmbH.  The new ovens were made from titanium tubes welded to the titanium package (instead of the steel used in Phase I, or the tantalum used in the failed 2.1 version).  In the previous versions, JPL would insert pieces of raw Yb and crimp the tube so that they would not fall out.  When we received the package we would uncrimp the appendage in order to use the oven.  In this Phase II package there is a much more sophisticated design; the oven appendage, on the inside, has a titanium foil with many laser-drilled holes to hold the Yb inside yet allow a beam to exit the tube when heated.  In addition, the source of Yb this time was not raw pieces of Yb; we used 1.5 mm OD and 1 mm ID glass tubes that had Yb deposited on them.

The external oven for heating up the appendage was also improved over the Phase I JPL setup.  We use a 1/8" diameter cartridge heater and a custom aluminum tube to connect the cartridge heater and the appendage thermally. A sheath of insulating material surrounds this cylinder.  This oven is not only smaller and less clumsy, but also uses a lower voltage (10 V instead of 40 V) than the original, and is much more controllable, with less tendency to overshoot and a fast heatup and cool down compared with previous setups.
  
By design, the JPL Phase II packages include oven and electrodes placed with intention to avoid the problems we have had with unwanted coatings on the wires and windows, while at the same time allowing the Yb to coat as much of the electrode surfaces facing the trap as possible (for photoelectron loading).  We were able to implement photoelectron loading in this package.  The light source we use for this now is the most efficient one we have found: a 405 nm free-running, uncollimated laser diode scavenged from a BluRay disc burner.  We settled on this after testing many LEDs (e.g. 370, 390, 400, 450 nm center wavelengths and others), and using mercury lamps as we did in Phase I.  Compared with an LED that also worked for loading an earlier package, the laser diode is more efficient, smaller, and easier to power.  The wavelength and collimation are not important.  The mount and location of this diode in the Phase II setup can be seen in Fig.  \ref{fig:JPLP2SW}.  Using this diode with 120 mW output, we can fully load the trap in about five minutes.

\begin{figure}
   \centering
  \includegraphics[scale=0.7]{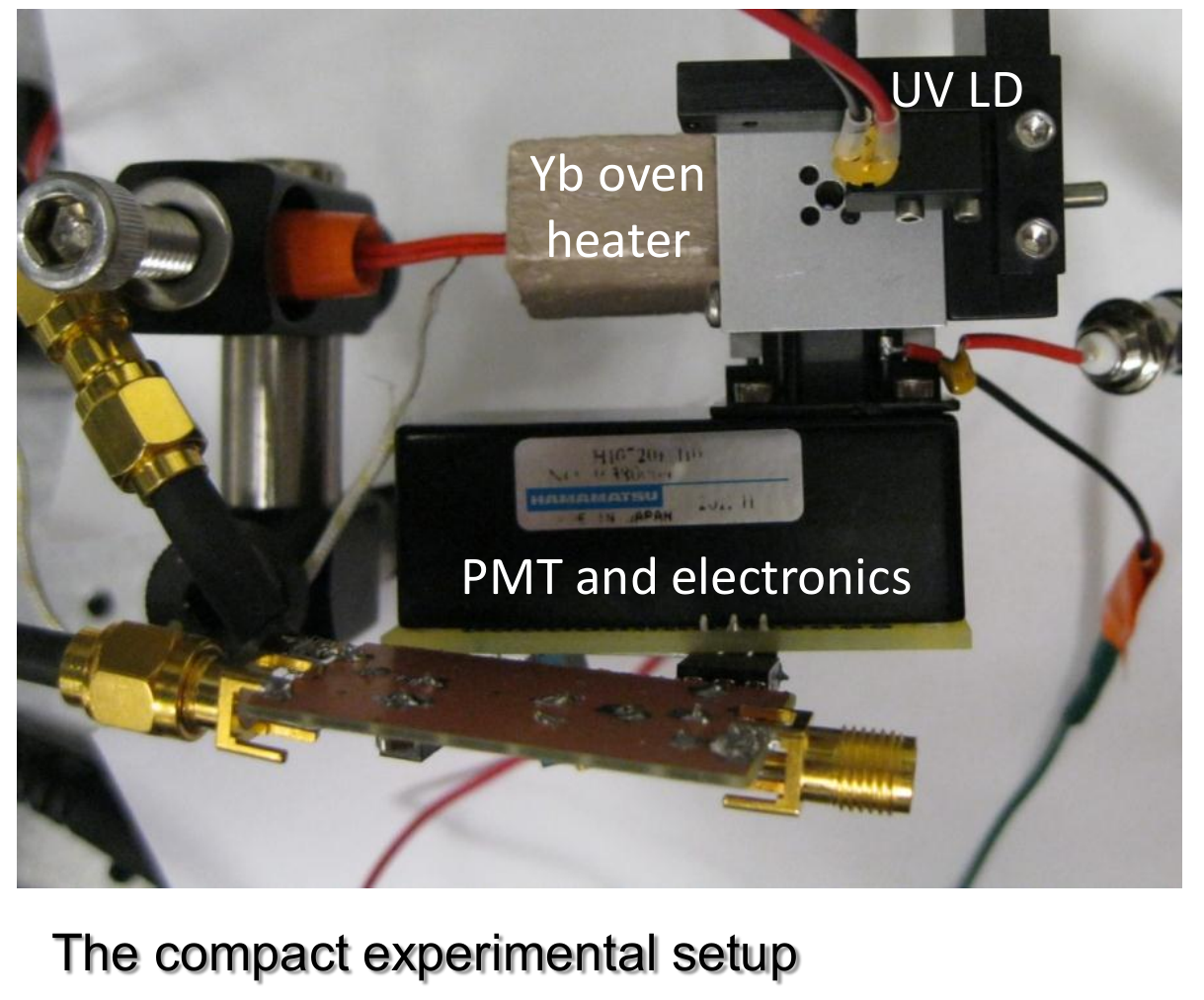} 
   \caption{The physics package component of the compact experimental setup for the deliverable clock with the Phase II JPL 2.2 package.  The package is in the upper right corner of the picture, inside the aluminum casing that acts as a dummy shield until we receive our custom machined magnetic shield.}
   \label{fig:JPL222photo}
\end{figure}

\subsubsection{Performance}
\begin{figure}
   \centering
  \includegraphics[scale=0.5]{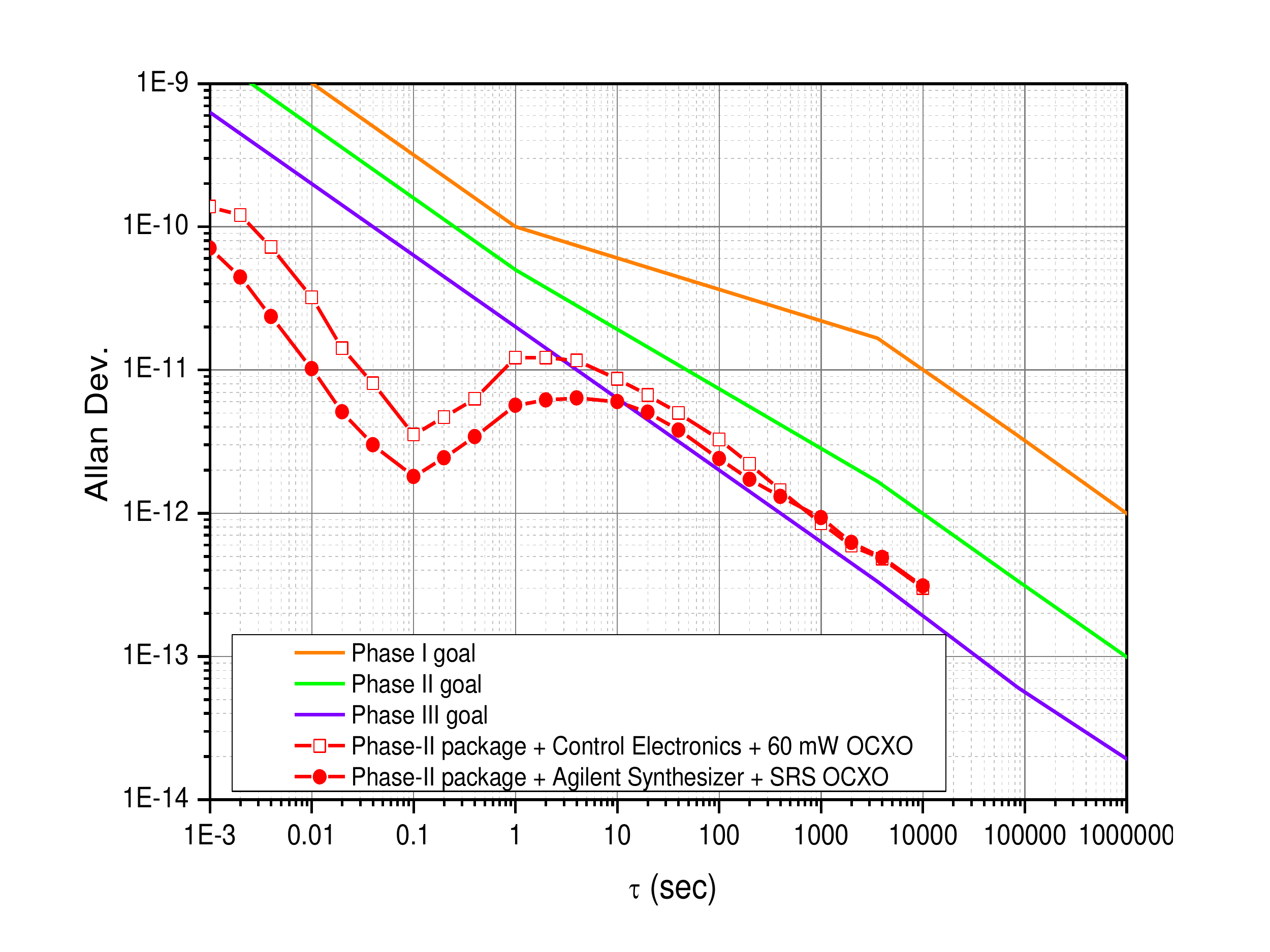} 
   \caption{Initial clock performance data from the JPL 2.2 package, taken with the small, attached detection system.}
   \label{fig:JPL22AD}
\end{figure}

The compact setup including PMT, UV laser diode for loading, and mock magnetic shield is shown in Fig. \ref{fig:JPL222photo}.  We trapped in this package at 1.7 MHz with $V_{\rm{RF}} \leq 180$V peak-to-peak and  $V_{\rm{EC}} = 12$V.  Using the isotopically enriched source, we get a very good Rabi signal for running the clock.  It is at least as good as the JPL 1.2 package in terms of SNR for this signal.  An example of early Allan deviation data obtained by running a clock with this package is shown in Fig. \ref{fig:JPL22AD}.  This data was taken using the PMT (without an imaging system) attached directly to a dummy magnetic shield, making the physics package relatively small and independent. 

\subsection{LTCC/HTCC Packages} \label{sec:LTCC}
%
%
%
%
%
One of the most interesting traps we have worked on because of its promise for a very small ion trap system is a ceramic package for Phase II moving to Phase III.  There are two main technologies for this, Low-Temperature Cofired Ceramic (LTCC) and High-Temperature Cofired Ceramic (HTCC).  Both types of ceramic are layered technologies that can be treated like a ceramic PCB.  In the green state, thin layers of the material are cut and screen printed with metallized inks, including electrical vias that pass through the layers.  When stacked up and fired, the layers can create a nearly arbitrary geometry that hardens into pure glass/ceramic and metal and includes electrically conductive pathways.  This could be an excellent solution to the difficult problem of electrical feedthroughs in the small package.  If the ceramic assembly can both form the ion trap and act as a wall of the package, where it can also bring out the electrical connections, cofired cermanic will be an ideal way to push our package toward the Phase III IMPACT goals.  


A challenge of using the ceramic is finding a trap geometry that is appropriate for the technology.  One vision was to build a stepped structure which would have bonded wires stretched across an open center portion to form a relatively normal quadrupole or octupole trap.  While the geometry is simple, building this is a complicated process because of the individual wire bonding of the electrodes after building the ceramic base, and the openness of the trap would cause a lot of exposure to the dielectric surface of the ceramic.  A better idea was to make trap electrodes that are flat and metallized onto the ceramic surfaces directly, which is a process that could be fully implemented during production of the ceramic part.  

\begin{figure}
   \centering
  \includegraphics[scale=0.5]{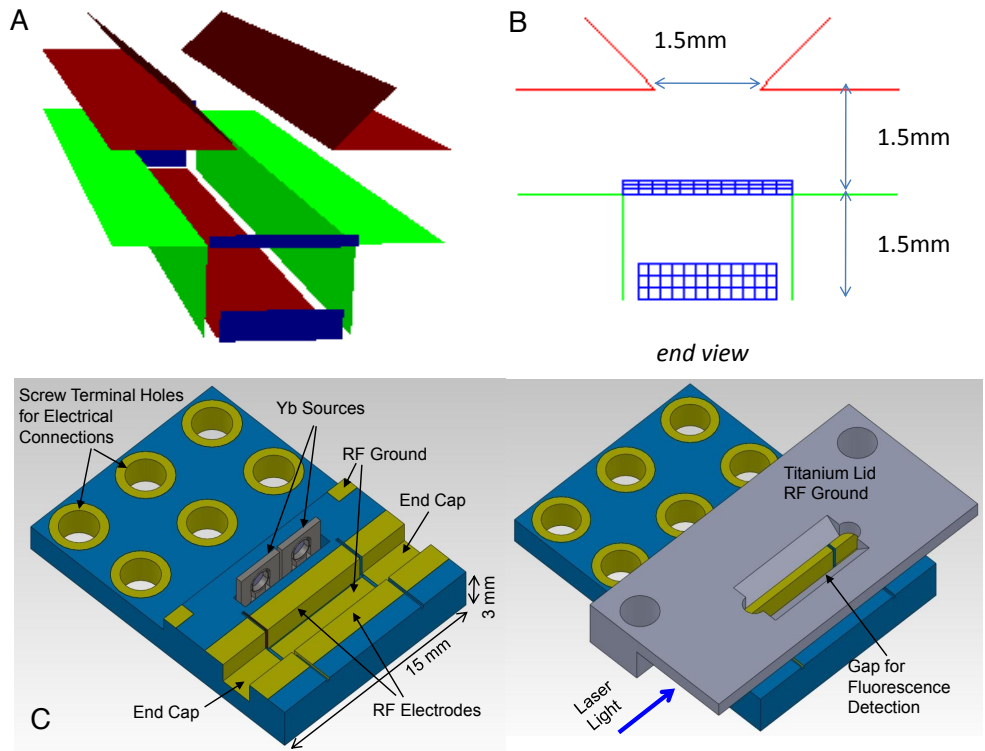}
   \caption{The modeled electrode geometry for the LTCC trap, shown along with its design in Solidworks. A: Perspective view of the LTCC modeled trap geometry.  (The endcaps shown here are a concept from a previous design.) B: Cross-sectional view of the trap. C:  Solidworks model of the trap and lid for testing in a large vacuum.} 
   \label{fig:LTCCCPO}
\end{figure}

\begin{figure}
   \centering
  \includegraphics[scale=0.45]{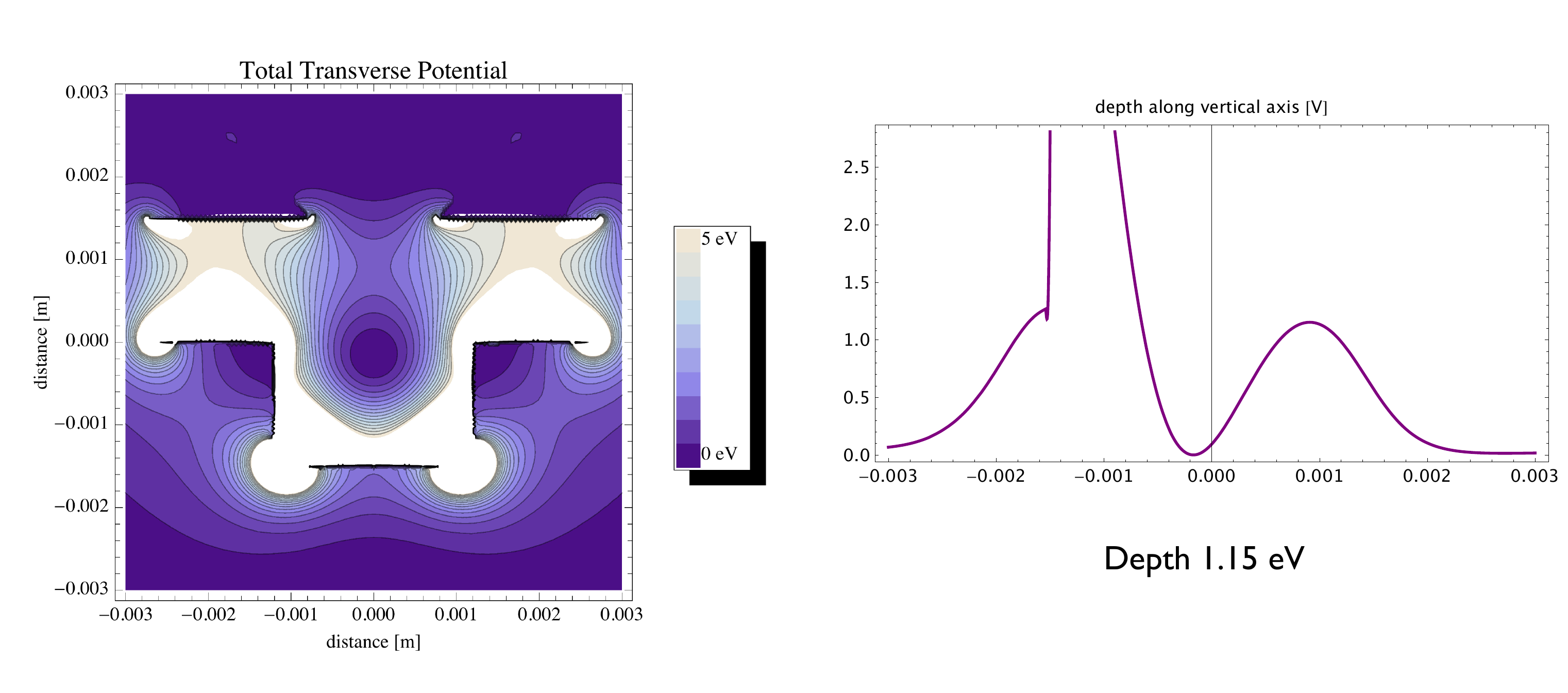} 
   \caption{The pseudopotential in the transverse plane for the LTCC trap geometry.  The depth is plotted along the vertical axis, and the trap depth is taken from the shallow side of the potential well.}
   \label{fig:LTCCmodel}
\end{figure}

For this reason, we designed a completely new trap geometry, whose final result is shown in Fig. (\ref{fig:LTCCCPO}).  The trap is located in a trench in the ceramic, where the bottom of the trench is coated with metallization as a ground electrode, and the ``shoulders" of the trench act as RF electrodes.  Then, there is a metal lid that sits on top of the ceramic, with a slit for fluorescence observation directly over the trap.  When looking at the cross section, we see that the grounded lid provides a split fourth electrode that completes a quasi-quadrupolar geometry for the trap.  There is indeed a trapping minimum between the shoulders of the trench.  A model of the pseudopotential is shown in  Fig. (\ref{fig:LTCCmodel}).  The endcaps of the trap were originally conceived as low metallized surfaces on the ends of the trench; the laser would enter from above, be reflected 90 degrees to pass through the long axis of the trap, and use another 90 degree turn to exit through the slit in the lid.   However, to pass a laser down the axis without the use of micro-sized mirrors, we decided to make the endcaps the same shape as the the trench at either end of the trap.  We know that in practice, the shape of the endcaps for a linear trap is not important as long as they can provide a static potential from one end of the trap to the other.

\begin{figure}
   \centering
  \includegraphics[scale=0.5]{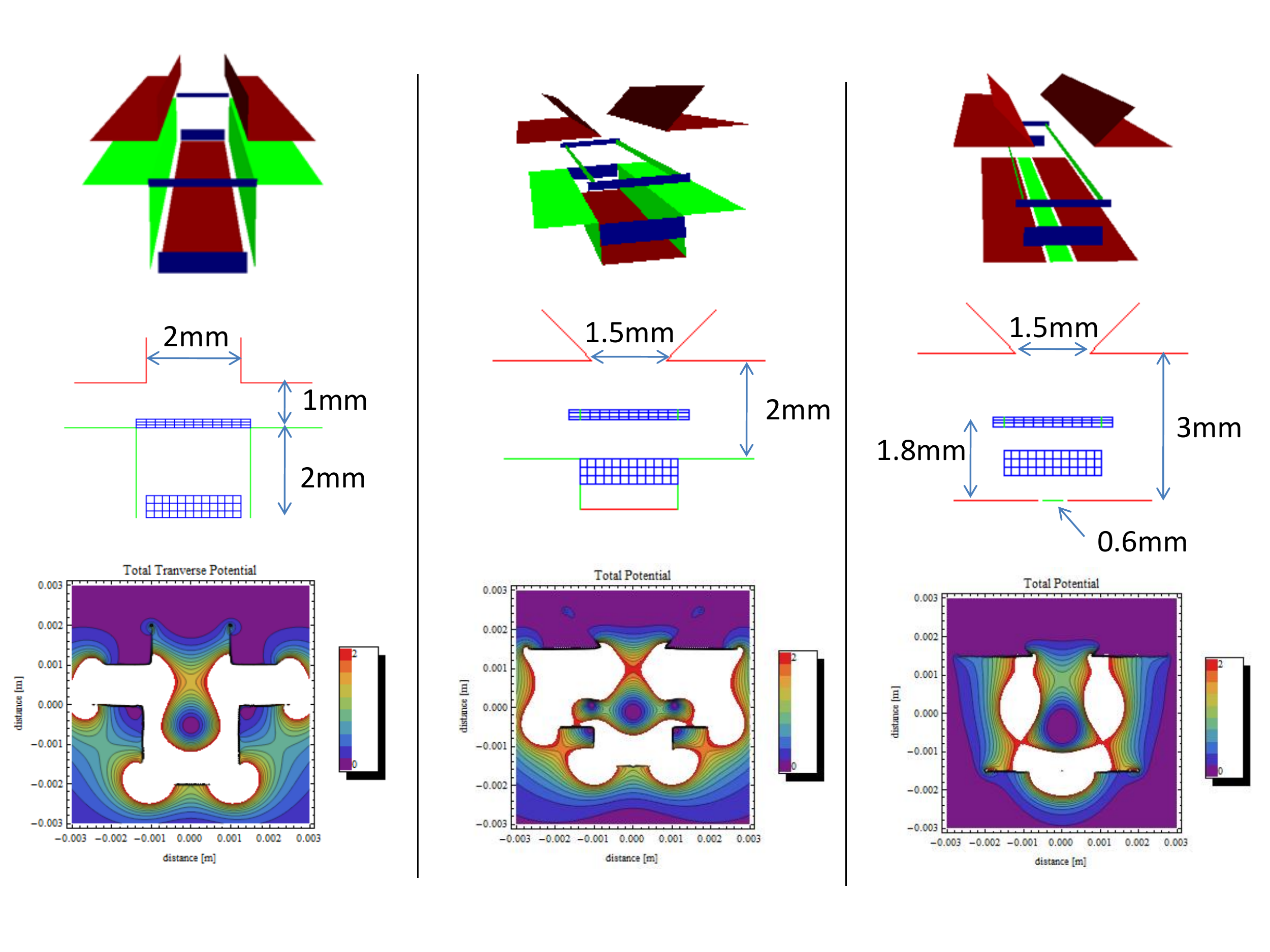}
   \caption{Several designs considered for the LTCC trap geometry, showing the shape, cross-section, and calculated pseudo potential.  The endcaps we chose for our final design are different from what is shown here in our exploratory analysis.  The final design was a modified version of the left-most design, which was the original idea. The final design is slightly shallower and the lid electrodes have 45 degree angles for improving the light collection solid angle.} 
   \label{fig:LTCCothers}
\end{figure}

\begin{figure}
   \centering
  \includegraphics[scale=0.4]{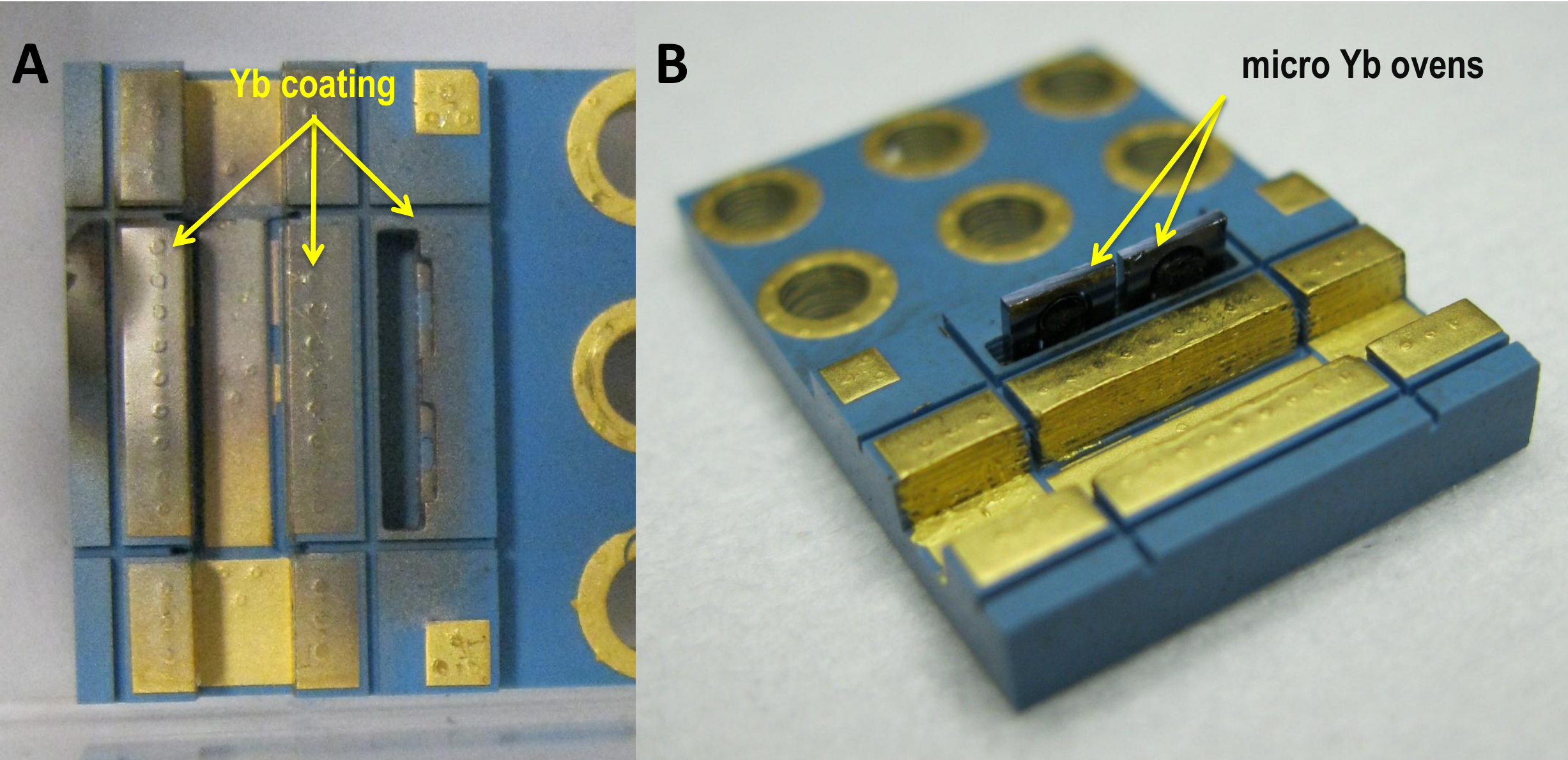} 
   \caption{Photos of our first two traps, tested in the large vacuum.  A: First trap, used with a conventional oven, coated with Yb after removal from the vacuum.  The Yb allowed us to load with UV light but also shorted electrodes together. B: Second trap, with micro-hotplate ovens included.  In the first test the micro-ovens were not successful, either because of breakage or some fault in the electrical connections.}
   \label{fig:LTCCphoto}
\end{figure}

Because of the new geometric nature of the trap, we did make a considerable investigation into its viability in terms of trapping depths and stability.  In an attempt to optimize the geometry to some extent, we modeled many different variations before choosing the combination of the best trapping potential and ease of construction.  Considerations for the design included producing a large trapping minimum, placing the micro-ovens (shown in Fig. \ref{fig:LTCCphoto}, these are being developed at Sandia as a low-power Yb source for our Phase III package)) with access to the trap minimum while avoiding shorting the electrodes together by coating, creating adequate laser access, and stability.  A few of the ideas we tested are shown in Fig. \ref{fig:LTCCothers}, although there were many more. The original trap idea is shown in Fig. \ref{fig:LTCCothers}A.  Our final design looks like this one except the trench is shallower to bring the trapping potential closer to the lid (to provide more access to the Yb source), the lid has 45 degree angles above (to improve our solid angle for detection), and the endcaps are simpler; they are in the same trench shape as the main trap.  Some other designs we considered are shown, such as the trench-with-shoulders shape supplemented by a wire carrying RF above the shoulders (Fig. \ref{fig:LTCCothers}B).  Since our ovens were planned to be located between the lid and the ceramic on one side of the trench, we thought this more open design would make it easier for the Yb to reach the trapping minimum (but it also gives a path for the Yb to short electrodes together).   Also, the flat bottom surface with an RF electrode between two ground electrodes, and an RF wire suspended above, forming more of a hexapole pattern in the transverse direction, is in Fig. \ref{fig:LTCCothers}C.  As is seen in the plot of the pseudopotential, this last idea gives a nice big trapping region and a lot of access for the ovens, but in the end the ease of building this device is limited again by the stringing of the individual wires that must be wire-bonded to some kind of stepped structure.  For this reason we used the straightforward design of painting the electrodes directly on surfaces only.  in addition to being simpler, the structure of the trap shields almost all dielectric surfaces from view of the trap, reducing the risk of surfaces charging up during ionization procedures.

\begin{figure}
   \centering
  \includegraphics[scale=0.6]{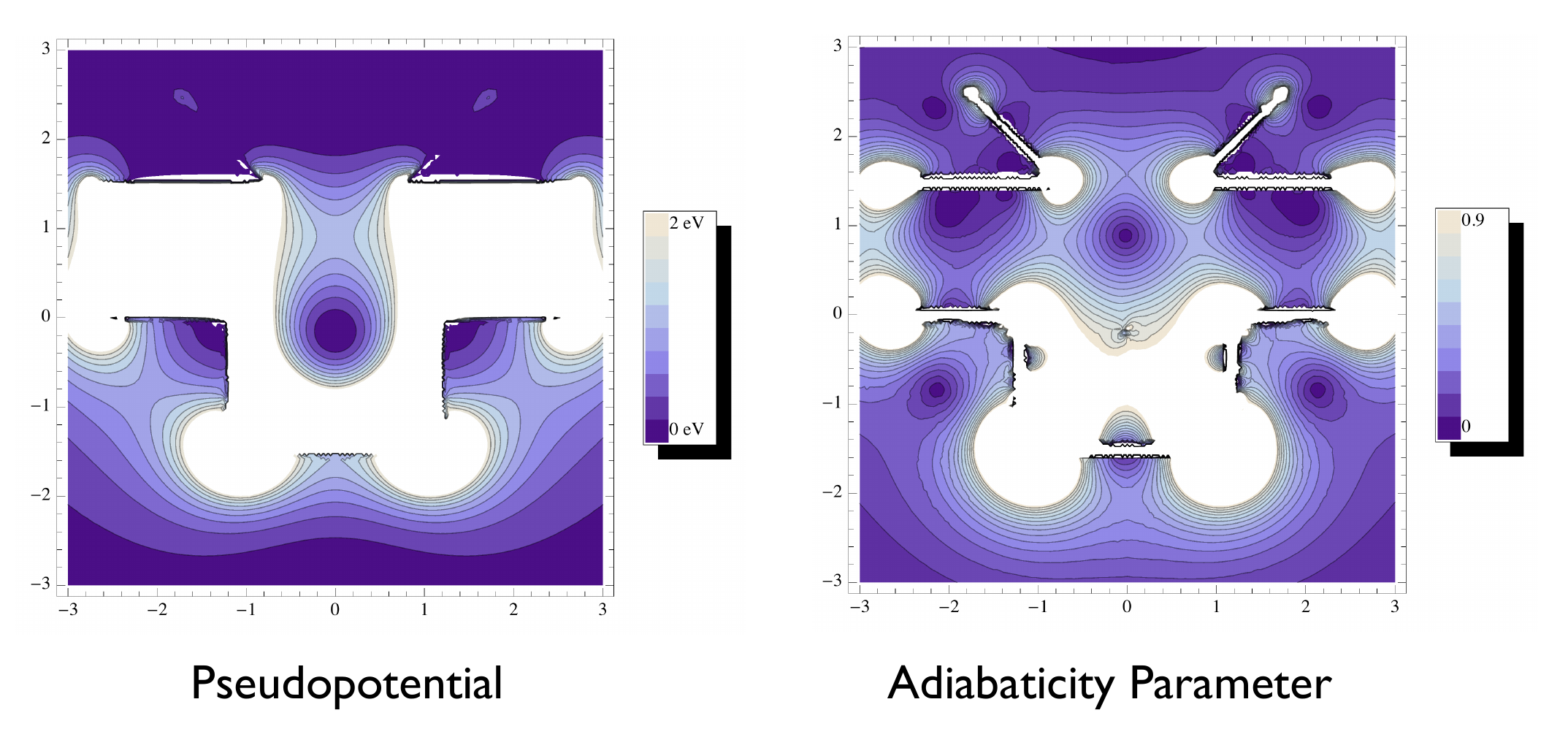} 
   \caption{Model of the adiabaticity parameter $\eta$ discussed in the text.  The parameter is related to the gradient of the pseudopotential at each point.  For a particular set of trapping parameters, the trapping region shown on the left should overlap with stable regions that have $\eta \leq 0.9$ on the right in order for the geometry to create a stable trap.  Axes are in mm.}
   \label{fig:LTCCadiabat}
\end{figure}

Another major concern was the probability for a trap like this one to be stable, due to the irregular shape of the trapping minimum.  It is clear from the pseudopotential model in Fig. \ref{fig:LTCCmodel} that the curvature and depth along the vertical axis is quite nonsymmetric, making the traditional measures of stability (that are usually based on symmetric, hyperbolic or linear traps) less relevant.  We used a more general criterion that is based on the concept that ions will continue to be stable and trapped as long as the ion motion is adiabatic.  This is true if the ion is not experiencing RF heating, i.e. gaining energy from the RF confining field, which is true when the gradient of the pseudopotential field is small.  To see this, we looked at the adiabaticity parameter given by \cite{JPLmultipole2000}
\begin{equation}
\eta = \frac{2q|\Delta E_0|}{m \Omega^2} = k(k-1)\frac{qU_0}{m\Omega^2r_0^2}\hat{r}^{k-2} \ .
\end{equation}
This parameter is a generalized version of the analogous quadrupole parameter.  For stable ion motion in a multipole trap, the trap will be stable for all points where this parameter is $\leq$0.3.  For quadrupoles, this expression reduces to the conventional stability parameter for a quadrupole, in which case the limit for $\eta$ is 0.9. In Fig. \ref{fig:LTCCadiabat} I have plotted this parameter at all points in the transverse plane of the trap.  By comparing with the pseudopotential, we see that there is a likely overlap between the stable regions ($\leq$0.9) and the trapping minimum region.

For the first run, we used an LTCC trap without micro-ovens installed, with the normal handmade oven mounted near the trap, to confirm that trapping in the new geometry would be successful.  We mounted this trap in a large vacuum space (the same chamber used for the test bed trap) to test its validity, and planned to do the same to test the micro-ovens. We were indeed able to trap in the LTCC trap with no problems.  Once we test the ovens, we can move on to putting the trap in its own small package.

\begin{figure}
   \centering
  \includegraphics[scale=0.4]{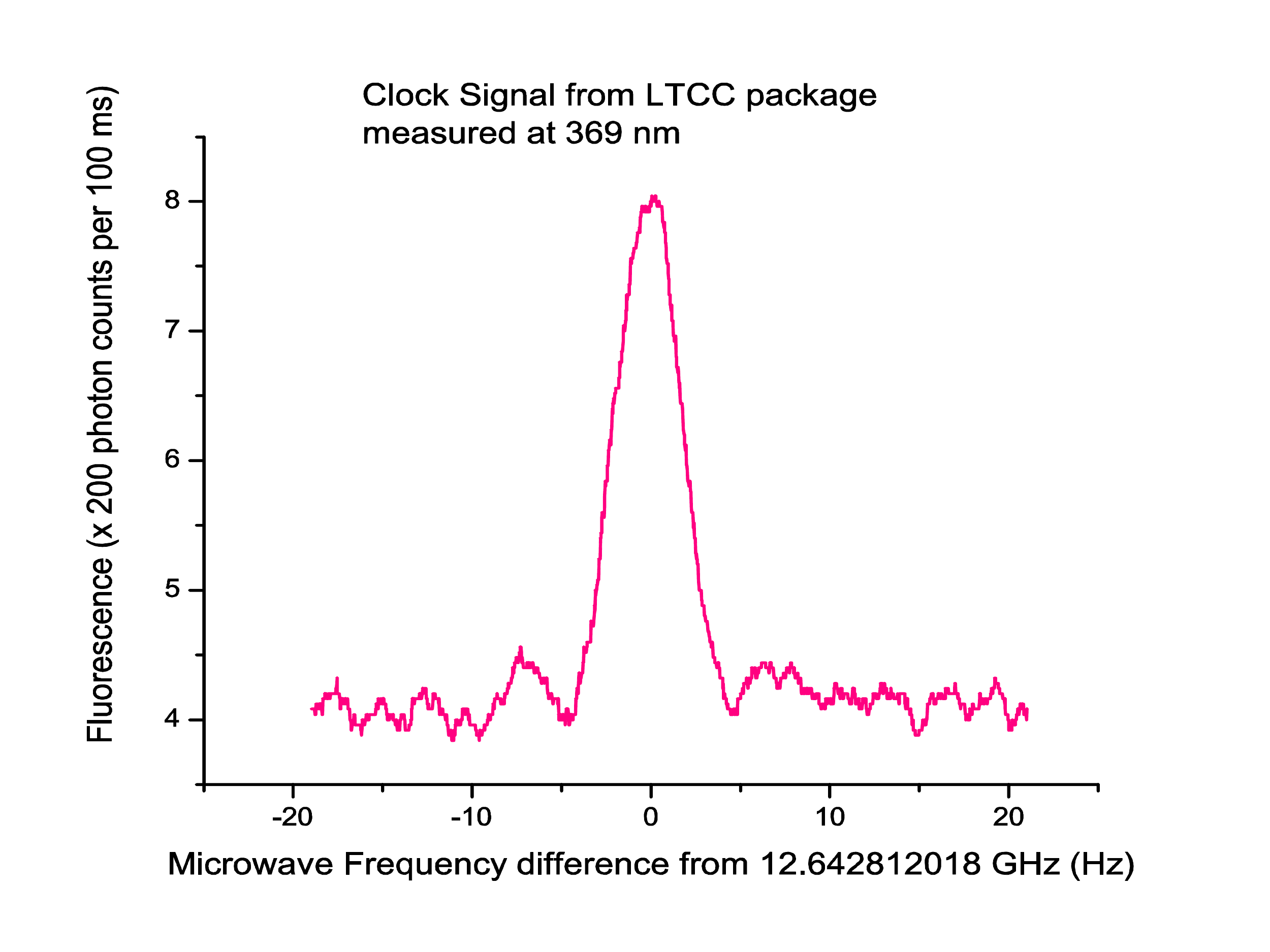} 
   \caption{Rabi fringe from the LTCC trap.  We showed that we could detect $^{171}$Yb$^+$ and run a clock in this trap.}
   \label{fig:LTCCRabi}
\end{figure}

We trapped in the package with the first trap used with a conventional oven.  The trapping parameters were 4.5MHz, $V_{\rm{RF}}=525$V$_{p-p}$, and $V_{\rm{EC}}=20$V.  The initial lifetime we measured was around 45 hours.  We captured a clock signal as shown in Fig. (\ref{fig:LTCCRabi}).  Although we were able to run a clock, we did not make any long-term measurements in this package.  Also, we were able confirm the photoelectron loading in this package with the free-running 405 laser diode.  Our homemade oven was coating the electrodes with plenty of Yb to facilitate this process, and unfortunately also shorted some of the electrodes together with some resistance (RF to RF ground, RF to endcap), although additional RF power enabled the trap drive to oscillate despite the shorts.  A photo of the actual trap after we removed it from the system is shown in Fig. \ref{fig:LTCCphoto}A.  In this picture we can clearly see the Yb from the macroscopic oven coating several surfaces.  We hope that the design  with the micro-ovens will not cause this degree of coating and the associated electrical damage (although we still want them to produce enough coating to do UV broadband loading).   Fig. \ref{fig:LTCCphoto}B shows the second generation LTCC with micro-ovens included in the oven wells facing the trap.  We put this trap into vacuum, but unfortunately the oven connections were faulty and the ovens were broken, so we were unable to see fluorescence or to attempt trapping in the first version with ovens.

\begin{figure}
   \centering
  \includegraphics[scale=0.4]{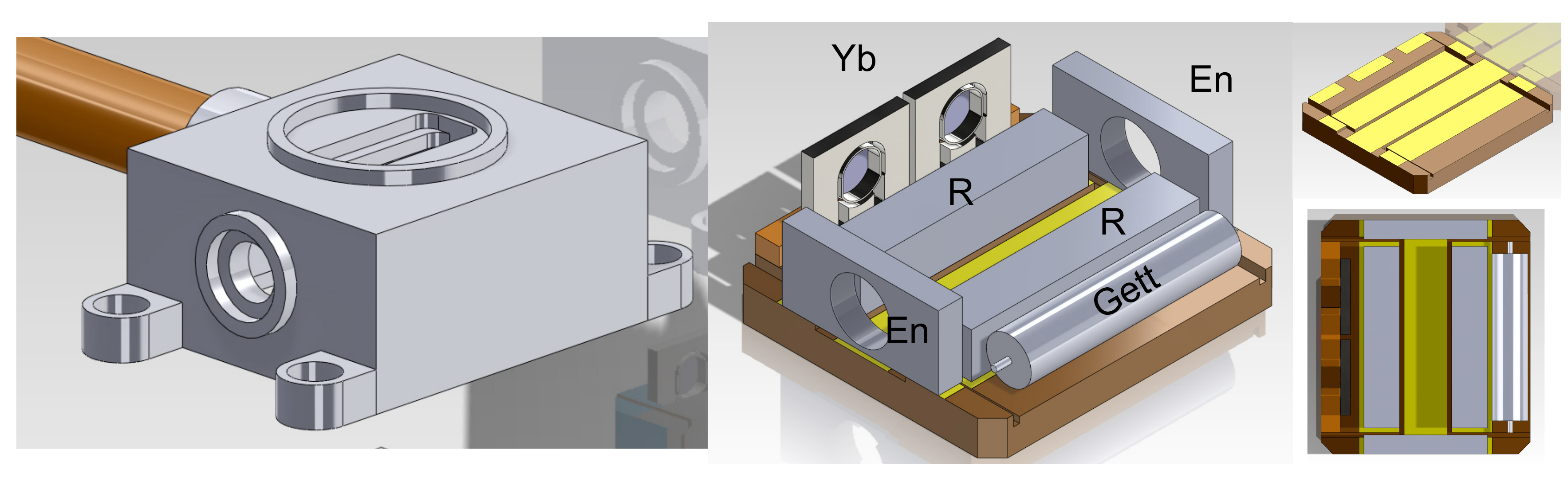}
   \caption{The HTCC trap design that will incorporate the LTCC trap geometry into a standalone titanium package near the end of Phase II. }
   \label{fig:HTCCtrap}
\end{figure}

Last, after testing the LTCC trap in the large chamber, we have designed a version of a very similar trap that goes into its own titanium package with sapphire windows.  For the next phase, the ceramic will likely form part of the package wall.  The final size of the package (excluding the pinch-off tube) is $\leq$ 1 cm$^2$.  Its design is pictured in Fig. \ref{fig:HTCCtrap}.  We elected to use HTCC to build this trap, mainly because of the ability to use high temperature braze processes with the HTCC to attach the titanium base to the ceramic to form a hermetic seal.

\section{Summary} 
\begin{figure}
   \centering
  \includegraphics[scale=0.62]{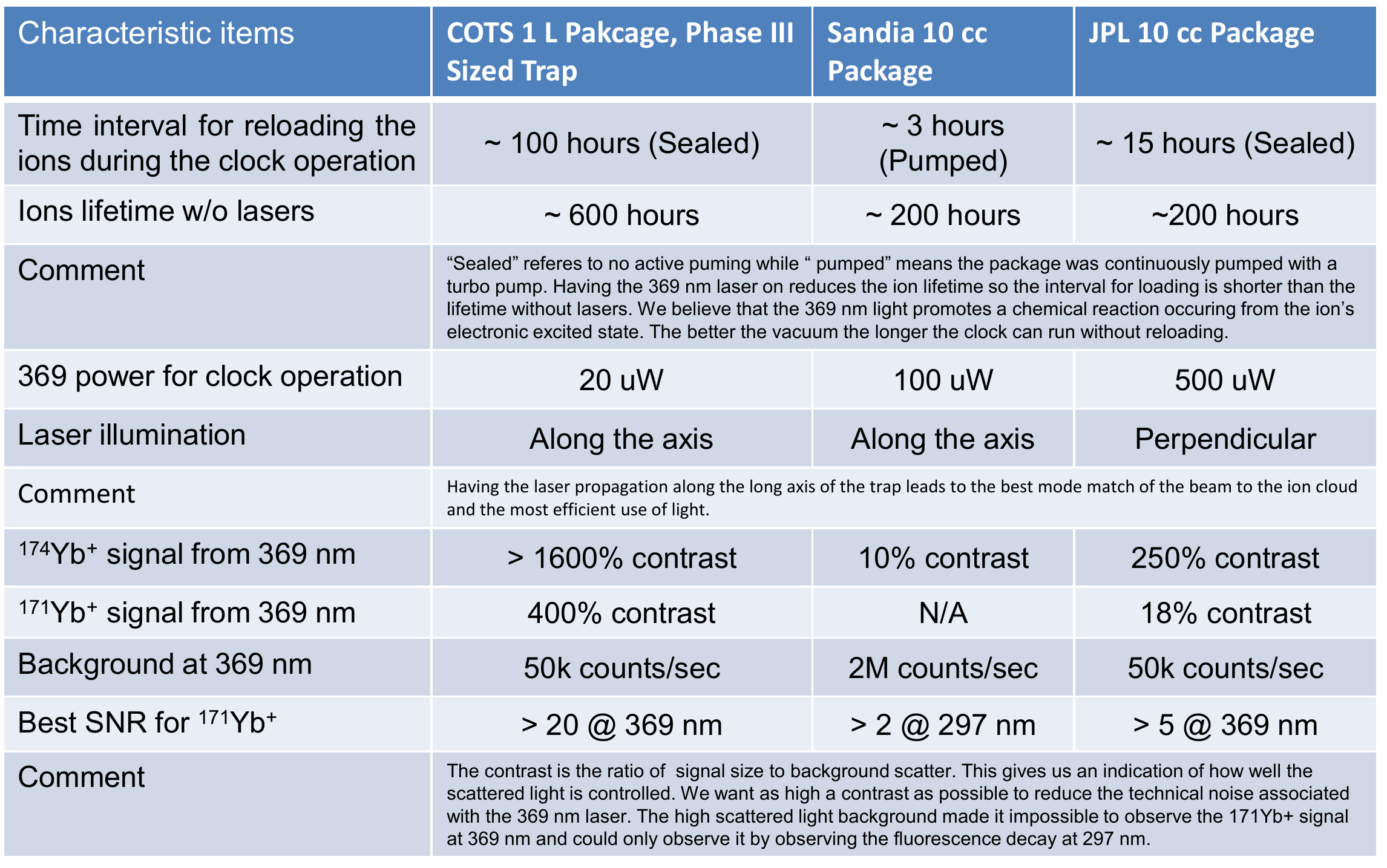}
   \caption{A summary of the comparison between the Phase I traps.}
   \label{fig:TrapCompare}
\end{figure}

Through the progression of these packages, we have gained a lot of wisdom and insight into the proper processes and optimal components to use for making a small package.  Some trap characteristics to be compared in the Phase I packages are summarized in Fig. \ref{fig:TrapCompare}. This work has culminated in the success of the JPL 1.2 package in the portable clock, and the very promising performance of the JPL 2.2 and the LTCC traps.  
The PA\&E package (also identified as the Sandia 10 cc package) and the JPL package both perform better than the Phase II goals, for the integration times we have measured.   From the data it is clear that we are on the right track, but that the long-term stability 
needs to be studied and improved.  These packages are most likely limited by magnetic field instabilities in the long term.  This is a problem we will be working on moving forward to Phase III, and there is much discussion on this topic in Chapter \ref{ch:biasfield}.


\section{Other components} 
Although the main focus of my part of IMPACT was on the vacuum packages, there are many other elements of the clock that must be miniaturized before an integrated solution can be realistic.  The vacuum package with ion trap is the centerpiece of the package, but the clock cannot operate without an LO, lasers, Yb source, electronics, etc., and we had to find a solution for each part of the clock.   Apart from the ``small packages" that are the focus of this chapter, other project time and resources were spent miniaturizing other parts of the clock, which I do not focus on in this work, but will describe briefly here: the MEMS oscillator, the micro-ovens, the miniature light sources, and optics and other common components.

%

A MEMS oscillator to serve as a local oscillator for the clock underwent significant development during the first phase of the project.  The goal was to create a resonator that oscillates directly at 12.6 GHz that was ovenized to a stability surpassing what is currently available.  Although it would be ground-breaking to successfully develop such a device, its development remains incomplete because of both technical and financial obstacles.

A very difficult aspect of using ytterbium as the atomic source is obtaining miniaturized light sources for interrogating the ions.  Ytterbium has a complex level system and requires at least two lasers to utilize the typical clock scheme, one of which is at an inconvenient UV wavelength (369 nm) that is becoming easier to obtain in the macroscopic world but is unheard of in a microscopic version.  To combat this difficulty we had a plan to frequency double an infrared vertical-external-cavity surface-emitting laser (VECSEL) to obtain the 369 nm light.  The development of the VECSEL at 739 nm as well as the performance of the frequency doubling are extremely challenging and the development of this complex miniturized source is still underway for use in Phase III of the project.  The second source, at 935 nm, has been developed as a vertical-cavity surface-emitting laser (VCSEL), of which we have used several prototypes beginning from the end of Phase I.  VCSEL and VECSEL sources are the most convenient for the IMPACT project because they have low lasing threshholds, require less power, and are very controllably tunable using straightforward temperature and current control.  

Development of micro-hotplate Yb-filled ovens will be an extremely important element of a successful final package.  Although these are ultimately an integrated part of the small physics package, we have only used macroscopic ovens thus far to load traps in the small packages, while performing only independent tests on the micro-hotplates ovens themselves.  These micro-ovens consist of a silicon hotplate with a micro-heater patterned on the back.  The cup of the micro-hotplate is filled with ytterbium (small bulk pieces are dissolved onto the surface) and when heated, can produce sufficient vapor of ytterbium with very little electrical power, due to the very small heating load.  We plan to fully implement these hotplates in the Phase III physics packages.

Finding and integrating simple components such as a shutter and detector that are efficient at the right wavelength are significant challenges at the level of miniaturization we need.  A shutter is necessary for turning the interrogation laser on and off during the clock duty cycle.  It is very important that the light be blocked with a high extinction ratio to avoid light shifts in the measured clock resonance.  Some solutions we considered for shuttering included LCD shutters, tunig the laser wavelength away from resonance, misaligning the laser, MEMS mirrors, and a mechanical MEMS shutter that physically blocks and unblocks the light.  We are currently implementing the last option, the mechanical MEMS shutter.  Finding a detector is also difficult, especially considering our preferred detection wavelength at 297 nm.  We continue to use a small photomultiplier tube (PMT) that is approximately the size of the Phase III goal for the entire clock.  We expect a smaller, ``micro-PMT" to come onto the market soon that we hope to test. 
 
 Micro-optics, along with their integration and alignment, are another issue we have to deal with as we begin to integrate our physics package.  We hope to leverage experience with similar systems at Sandia to help us with this task.  Finally, making an electronics package that performs all the functions needed to run the clock and its associated components is another engineering challenge.

Although my focus has been on the vacuum packages, developing an integrated system is always at the forefront and requires thinking about all the other elements.  These elements are a result of many collaborations both within and outside Sandia.

\chapter{Magnetic field gradient effects on the ion-trap frequency standard}
\label{ch:biasfield}

While Chapter \ref{ch:packages} dealt with the problem of miniaturization of the physical clock package, in this chapter we turn our attention to the second primary focus of the IMPACT project goals: the problem of long-term stability.  The quest to preserve long-term clock stability in our miniaturized system necessitates a hard look at the effects of magnetic fields and field gradients in our system, as we have long believed the performance of our Phase I clocks to be limited at long measurement integration times  by magnetic field fluctuations.

%

  \section{Introduction}



A miniaturized ion trap clock system has a particular susceptibility to fluctuations in the static magnetic field and the presence of gradients because of its size.  Bias field fluctuations can occur due to ambient fluctuations that are not sufficiently shielded, or due to variations in the bias field current.  The first can be reduced with improved shielding; however, although smaller shields have more field attenuation, any holes in the shield can allow field lines to enter.  Our shield will be very small and must have numerous holes for laser and detector access, and although the size of those holes would ideally much be much smaller than the dimensions of the shield itself, for a shield at such a small scale this is not possible. The shape of the shield will be irregular as well, making it difficult to know, without actual testing, how well it can perform.  The influence of bias field fluctuations due to instabilities in the bias field coil currents can be reduced by operating at lower field values, or by engineering a more stable current source (this is a nontrivial task when subject to power restrictions such as ours).

Magnetic field gradients are also difficult to avoid.  Because of the very small nature of the package, any magnetic materials used in the manufacture of the trap can cause intrusive gradients in the trapping region.  Even without magnetic materials present, gradients will still come from sources outside the package, such as the laser diode, PMT, and electronics boards.  These and other ambient gradients may not be shielded adequately.  Also, the small coils (on the order of the package size) that will be responsible for setting the bias field will not be able to create a uniform field over all of the ions while staying within the dimensional limitations.

Instabilities in the constant bias field clearly cause a problem because they affect the stability of the clock resonance frequency directly.  Although the clock transition is insensitive to magnetic fields to first order, the frequency can still shift due to a second order effect.  Because of this second-order sensitivity to magnetic fields, the fractional frequency stability is sensitive to magnetic field fluctuations.  This relationship can be quantified by considering the Zeeman effect to second order. 
Then the relationship between frequency stability and field fluctuations is given by
\begin{equation}
\frac{\delta\nu}{\nu} = 4.9\times 10^{-8} [\rm{G}^{-2}] B \delta B
\end{equation}
where $\nu$ is the clock frequency and B the magnitude of the magnetic field.  As an example, with a bias field of 200 mG, with fluctuations on the order of 1 $\mu$G, the fractional frequency stability can change by $1\times10^{-14}$. With higher bias fields, for the same amount of fluctuation, the resulting stability is even worse. The DARPA metric 
 for Phase III of the IMPACT project requires the clock to reach the equivalent of nearly $10^{-14}$ fractional frequency stability at a measurement time of $2\times10^6$ seconds.  This means that if we believe we can maintain fluctuations at or below one $\mu$G, then in order to remove the effect of field fluctuations from limiting the stability of the clock, we must operate at a bias field of 200 mG or below.  If we cannot maintain fluctuations below 1 $\mu$G, we would need to operate at an even lower field to eliminate the influence of these fluctuations on stability.

However, a second, distinct problem arises in the bias field region below 200 mG.  All of the traps we have been studying and using in our miniaturized packages have secular frequencies in the range of 100 to 300 kHz.  The problem arises when these secular frequencies are on the same order as the Zeeman frequencies that separate the $|F=1, m_F=0,\pm1 \rangle$ sublevels. 100 to 300 kHz Zeeman frequencies occur for magnetic fields of 70 to 215 mG, since the  gyromagnetic ratio is approximately 1400 kHz/Gauss.  This means that when we operate with a bias field in this range (in particular, below 200 mG), we must take care to avoid operating at a field that induces Zeeman resonances near the secular frequencies, for reasons I am about to explain.

We can understand the problem in the following way.  Ions in a (symmetric) linear quadrupole trap experience a time-averaged pseudopotential that is similar to a harmonic potential in the center of the trapping region.  The ions undergo thermal motion in this harmonic potential which is known as the secular motion and is characterized in Chapter \ref{ch:trapping}.  If a gradient of the magnetic field is present inside the trap, where the magnitude of the field varies at a distance at least on the same order as the amplitude of the ion motion, then the ion undergoes motion in a spatially varying magnetic field.  From the frame of reference of the ion, the ion experiences a time-varying magnetic field, which is equivalent to an RF field that oscillates at the same frequency as the ion motion.  In other words, the ion feels as if it is being illuminated with an RF field.  

\begin{figure}
   \centering
  \includegraphics[scale=0.7]{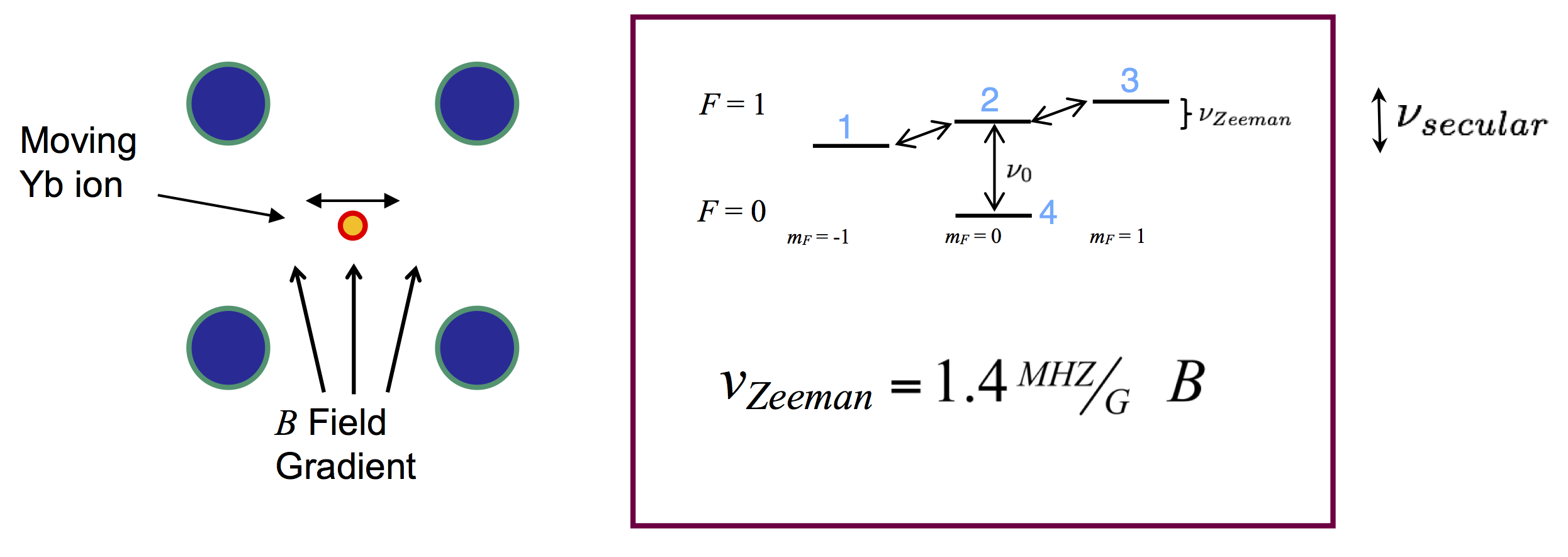} 
   \caption{An illustration of the motion of the ion in the RF quadrupole trap experiencing a gradient and the affected levels.  $\nu_{secular}$ is the RF field driven by the ion motion that causes the Zeeman levels to interact.}
   \label{fig:IonMotionZeeman}
\end{figure}

Now we can see that if the secular motion is on the same order as the Zeeman frequency induced in $|F=1\rangle$ by the bias field, the RF field seen by the ion can induce transitions between the magnetically sensitive sublevels (see Fig. \ref{fig:IonMotionZeeman}).  This additional decay mechanism out of the $|F=1,m_F=0\rangle$ clock state leads to broadening of the clock resonance.  Therefore, we want to avoid the overlap of such frequencies through our design of the clock and its operating parameters in order to prevent broadening.  Increasing the secular frequency enough to move it out of the desired $\leq$200 mG range is not a good option because of the additional power that would be required to do so (and the increased 2$^{\rm{nd}}$ order Doppler effect that comes with it). 

\begin{figure}
   \centering
  \includegraphics[scale=0.63]{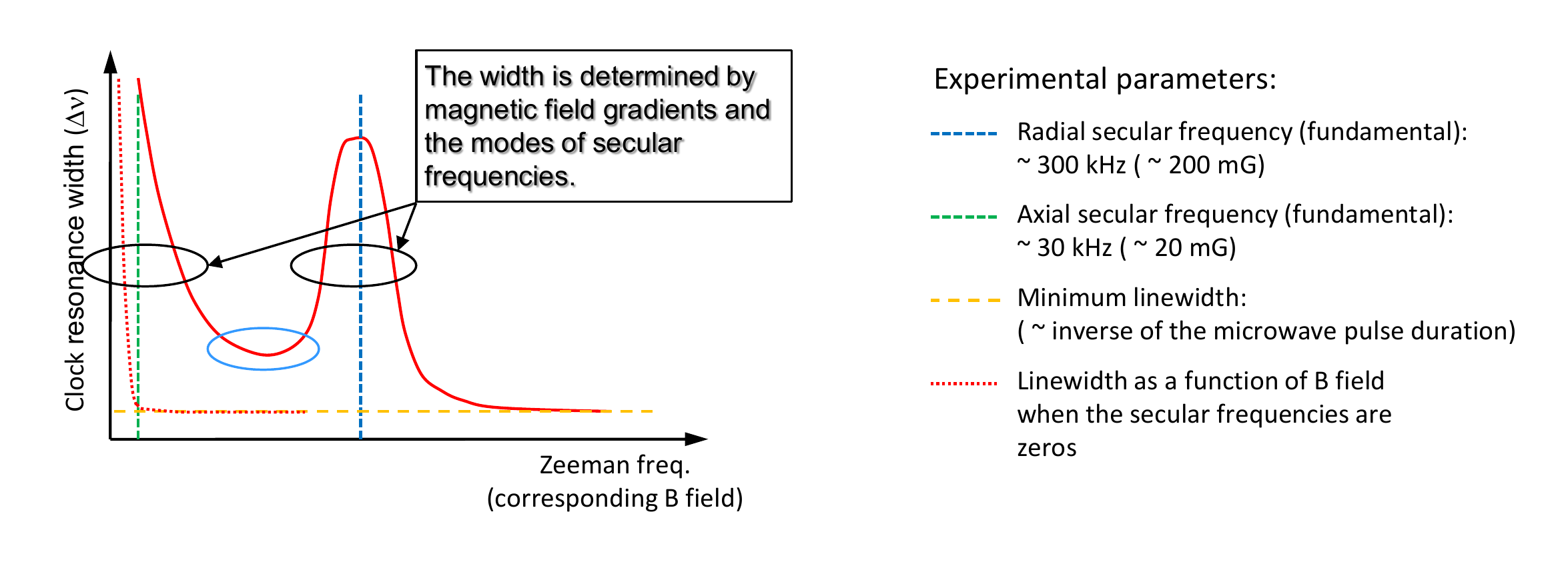} 
   \caption{An illustration of the plot of clock resonance linewidth vs. bias field which is the focus of much of this chapter.  The peak near the left side of the plot is due to the longitudinal motion, and the peak that occurs at a higher bias field is due to the transverse secular motion.  The preferred operating region is circled in blue.}
   \label{fig:BroadeningCartoon}
\end{figure}

Since we want to operate with a bias field of 200 mG or less, we need to understand the broadening caused by the moving ions in the region between a 0 mG and 200 mG bias field.  We have explored the effect of the broadening due to the ion secular motion by making measurements of the broadening mechanism in this region, and we have found that we see at least two distinct peaks in the broadening; one that corresponds to the transverse secular motion, and one that corresponds to the longitudinal motion in the trap (see the cartoon in Fig. \ref{fig:BroadeningCartoon}).  We have concluded that in the region between these two peaks, on the condition that we eliminate all magnetic materials and suppress gradients as much as possible, there exists a ``valley" point corresponding to a bias field strength at which we can operate and still maintain a narrow linewidth.

Many of the ideas about ion motion and fields have been explored in the literature.  Studies of the Doppler effect on ions in a linear quadrupole trap that examine temperature and space charge served as a basis for the space charge and ion number density calculations I did for the trajectory simulations.  In particular, Cutler, et al \cite{springerlink:10.1007/BF00697492} studies a thermalized cloud model in order to use Doppler sidebands to estimate temperature of the ions for the purpose of determining the Doppler shift with good accuracy in ion trap frequency standards.  They develop a model relating ion density, temperature, and space charge in order to calculate the sideband spectra.  Similarly, Meis et al. \cite{springerlink:10.1007/BF00694316} developed a simple analytical model to find the temperature of the ions using other measurable quantities.  Their results agree with earlier numerical and experimental results.  Along the same lines, Prestage et al.  \cite{367391}  performed a Monte Carlo simulation of ions in a linear trap to model the position of Doppler sidebands in order to deduce ion number from temperature information.  These ion motion simulations are very similar to what I have done in this work.  Also relevant to this work are studies of relaxation due to magnetic field gradients.  This effect has been studied in vapor cells by McGregor \cite{PhysRevA.41.2631}, which builds upon the theoretical works of Cates and Happer \cite{PhysRevA.38.5092,PhysRevA.37.2877}, in which a model for relaxation due to magnetic fields, both static and oscillating, and at all pressures, is developed.  


\section{Measurement}

\begin{figure}
   \centering
  \includegraphics[scale=0.1]{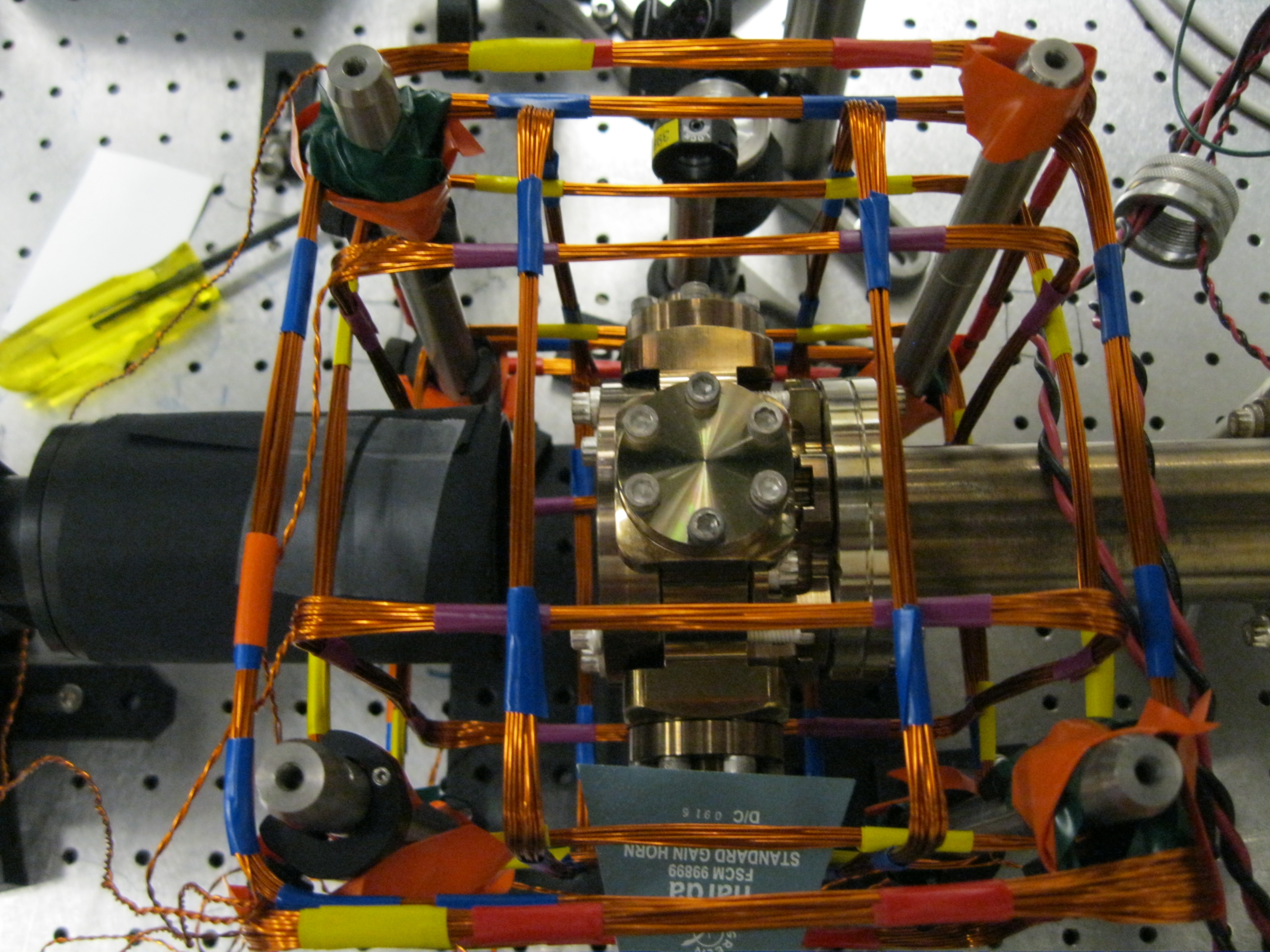} 
   \caption{A photo from above the chamber of the Helmholtz coils used to control and add magnetic fields.  Two additional pairs of coils can add gradients to the trap.  The microwave horn is visible at the bottom of the photo pointing along the trap axis and the camera detects from the left.  }
   \label{fig:MagneticSetup}
\end{figure}

An example of the experimental setup for making measurements related to the ion motion in magnetic fields is shown in Fig. \ref{fig:MagneticSetup}.  Most of our magnetic field measurements were made in the second generation of the test bed trap, in which we replaced the relatively magnetic stainless steel parts with titanium.  We used this trap because of its strong signal to noise ratio for making measurements, and our ability to have the trap at a distance from other magnetic field sources.  We will present some measurements from small packages as well.

We place three pairs of Helmholtz coils on the three major axes around the package.  The coils used for this experiment are square, and about 9 inches on each side.  They were placed at approximately the correct Helmholtz distance, to make the field as uniform as possible.  The coils are not placed with extreme precision, so it is difficult to know definitively if the trapped ions are in the center of the coil pair or not.  In practice, we use measurements on the ions to zero the field, which is possible regardless of whether or not the coil pair is exactly centered (as long as it does not move), but leads to the possibility that the direction of our bias field is not exactly in the direction of our coil axis.  We use two axes to cancel the ambient field, and the third axis to apply the bias and define the quantization axis.   In addition, we add gradient (anti-Helmholtz) coils in two directions for imposing additional gradients on the system.  Typically our bias field is in the vertical direction in the lab (where the long axis of the trap is horizontal).   As I discussed in Chapter \ref{ch:testBed}, although the microwave radiation enters the chamber along the longitudinal axis of the trap, it is difficult to discern what the real microwave field inside the chamber is, in particular at the location of the ion cloud. 



\begin{figure}
   \centering
  \includegraphics[scale=1.0]{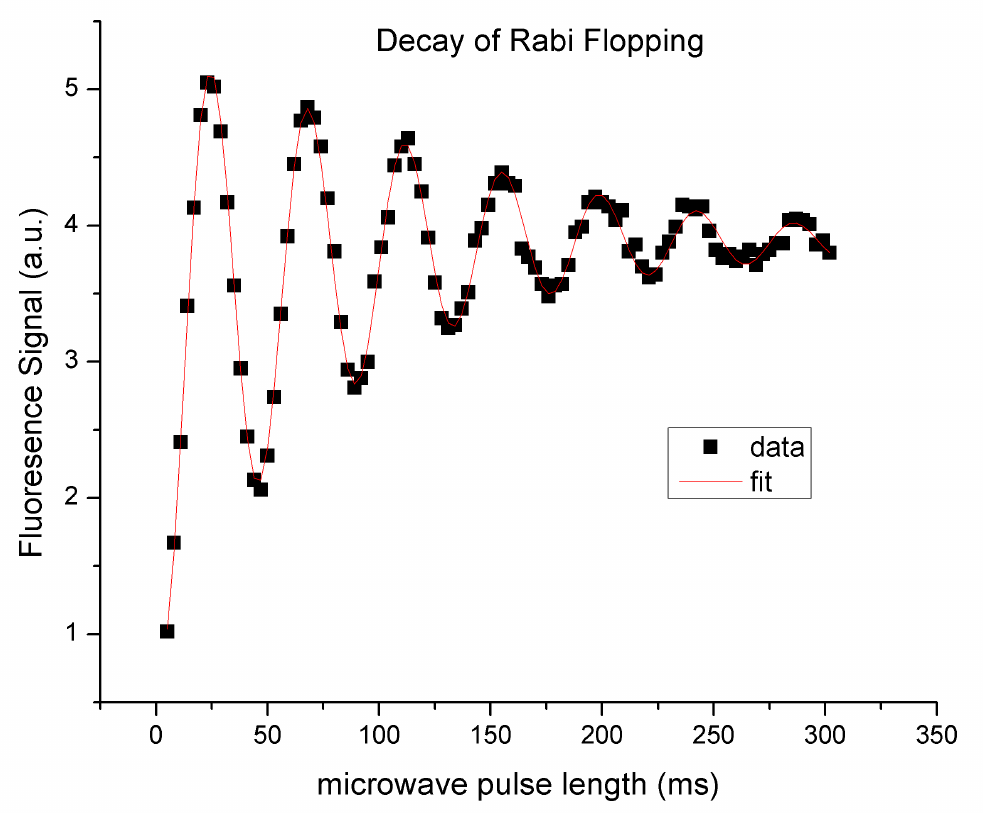} 
     \caption{Measurement of decay of Rabi flopping that we use to estimate the decay rate $\Gamma$ that is dominated by decoherence due to the ion motion.}\label{fig:decayingRabi}
\end{figure}

In order to understand the broadening phenomena, we can observe the linewidth broadening of the clock resonance as a function of bias field.  In the lab, instead of measuring the clock linewidth directly (which is subject to many effects besides the one in question) we look at the decay of Rabi flopping between the $|F=0,m_F=0\rangle$ and $|F=1,m_F=0\rangle$ states and observe the 
 characteristic decay time.  This gives us a measurement that is related to the inherent minimum linewidth due to decoherence and relaxation processes, whereas the actual measured linewidth could be inhomogeneously broadened by light or microwaves, or limited by the pulse length used to make the measurement.   
For a fixed bias field, each Rabi flopping curve is taken by:
\begin{enumerate}
\item{Preparing the ions in the dark state $|F=0,m_F=0\rangle$ by pumping with 369 nm.}
\item{Applying microwave radiation at 12.6 GHz for a fixed pulse length.}
\item{Reading out the fluorescence as we re-prepare the ions in the dark state to observe the amount of ions remaining in the $|F=1,m_F=0\rangle$ state after the microwave pulse and the decay.}
\item{Repeating this sequence for microwave pulse lengths from a few milliseconds up to seconds (the maximum pulse length in a particular measurement depends on the rate of decay and the amount of data needed to make a good fit of the decay time).}
\end{enumerate}
Following these steps results in a decaying Rabi oscillation such as the one shown in Fig. \ref{fig:decayingRabi}.
By fitting a curve (based on phenomenological decay of an oscillation) to this decaying Rabi flopping 
curve, we estimate the decay rate $\Gamma$ out of the $|F=1,m_F=0\rangle$ state, which is directly related to the inherent linewidth of the clock resonance by
\begin{equation}
\frac{1}{\Gamma} =  \frac{1}{2 \pi \Delta_{\rm{HWHM}}} 
\end{equation}
where $\Delta_{\rm{HWHM}}$ is the ``half-width at half-maximum" of the clock resonance.  Therefore, there is a factor of $\pi$ between the decay rate we measure (also referred to as the dephasing rate, or the inverse of the decay time) and the full clock linewidth.
We measure the Rabi flopping for many different bias field strengths (which is equivalent to scanning the Zeeman resonance frequency) and plot the resulting decay rate versus bias field.   
When the clock resonance is broadened by the interaction between the Zeeman levels due to the ion motion, we see an increased decay of the Rabi flopping.   Therefore, we see a peak in the decay rate (corresponding to a peak in the resonance linewidth) that corresponds to the transverse secular frequency of the ion motion, as well as a peak closer to zero that corresponds to the longitudinal motion of the ion, which happens at a much lower frequency. There is a continuous increase in linewidth as we approach zero field due to ambient fluctuations in the field that become intrusive at low field strengths.

%
%
We have tried a variety of measurements of this type to explore this behavior. We have looked at how this curve varies with trap parameters such as RF voltage amplitude, which determines the secular frequency, and endcap voltage.  We have also looked at variations in the curve upon applying additional gradients.  We have observed these effects in several traps, including packages with some magnetic materials.  We have also looked at how the decay rate for a fixed bias field changes with gradient, so we could use that information to predict how much gradient our ion clock can tolerate.  This was helpful in designing the Phase III traps.  In the remainder of this chapter, I will discuss many of these measurements and the simulations we performed in an attempt to reproduce and understand this curve for various circumstances.

\section{Decay rates}

A representative picture of the frequency spectrum of the ion motion along with the spatial distribution of the field strength can give us information about the exact magnetic field that the ions are seeing as a function of time.  We will consider in this section how the decay between the levels is related to the frequencies of the RF fields seen by the ion.  In the next sections we will make a detailed prediction of the spectrum of frequencies of motion undergone by the ions.  Combining this information, we can examine the relationship between the ion's motional frequencies and the decay rate that occurs due to this motion in a gradient.



In the lab, we measure Rabi flopping that we use to estimate the decay rate of the population out of the $|F=1,m_F=0\rangle$ state.  The decay that we measure is a total decay, which includes ordinary relaxation mechanisms as well as decoherence caused by the motionally induced RF fields.  As put forward in the last section, this decay rate is directly related to the linewidth of the clock transition between $|F=0,m_F=0\rangle$ and $|F=1,m_F=0\rangle$. 
We want to relate the theoretical evolution of the system to the decay rate $\Gamma$ estimated from the measurements, taking into account this decoherence, which has its strongest effect when the secular frequency is similar to the Zeeman frequency.  

To make a reasonably complete analysis, we must consider the four-level ground state system that was shown in Fig. \ref{fig:IonMotionZeeman}.  We use lasers with the excited state to read out the population in the lab, but this does not contribute to the behavior we want to examine.  In the four level system, we take into account the microwave interaction with the clock state as well as transitions among the Zeeman levels by the RF frequencies of the ion motion.  After calculating the spectrum of frequencies of that motion, for a known gradient in all space, we could calculate the exact evolution of the ion.  However, we do not know the gradient at all points, and this treatment is very complex.  We will instead make the following simplifying assumptions: 
\begin{enumerate}
\item{We assume that at late times, the total decay rate due to relaxation and motionally induced RF fields can be estimated from the Rabi flopping measurement in the same way as the decay rate $T_2$  due to relaxation only, even though the dynamics are actually more complex.  That is, when we fit the Rabi flopping data, we do not consider the nonlinear part of the evolution at short times that occurs due to the fact that decoherence caused by ion motion can only occur for ions in the upper state.  Since we start with the ions in the lower state, the decoherence effect we are interested in is evident only after population begins to be transferred to the upper level and the populations in the $m_F=0$ levels begin to mix. This assumption amounts to ignoring that nonlinear behavior and considering the late-time decay only.}
\item{We assume that at late times in the measurement, the evolution of the populations in the $|F=1\rangle$ Zeeman levels can be characterized by a simple decay rate between the levels.  Although the driving between these levels due to the motional RF fields may start as a coherent process, the coherence is rapidly lost at the ensemble level because different ions experience varied gradients at different times relative to each other.  The coherent part of the process is a transient at the time scales we consider for the Rabi flopping.  Thus, we can characterize these dynamics by a decay out of the $|F=1, m_F=0\rangle$ level.}
\end{enumerate}

By considering the late-time decay behavior of the system, we eliminate the nonlinear behavior and simplify our fitting model. 
In the picture of the Rabi flopping measurement, this means that the exponential decay envelope in curves like Fig. \ref{fig:IonMotionZeeman} is the same, after some time, as it would be for a decoherence process that behaves equally in both the upper and lower levels.



\subsubsection{Density matrix formalism}

We will consider the evolution of our system in the density matrix formalism.  The Schr\"{o}dinger equation for one atom is
\begin{equation}
i \hbar \frac{d}{dt} |\psi \rangle = H |\psi \rangle
\end{equation}
where $|\psi\rangle$ is the state of the individual atom and $H$ is the total Hamiltonian for the atom.  The Hamiltonian in this case is $H = H_0 + H_1$, where $H_0$ is the ``free" Hamiltonian and $H_1$ is the Hamiltonian for the atom's interaction with RF (and/or microwave) fields. 
Since we have a large ensemble of ions, we can represent the state of the ensemble by a density matrix $\rho_{ij}$,
and the equation of motion for $\rho$ is given by the Von Neumann equation:
\begin{equation}  \label{eq:vonneumann}
\dot{\rho} = \frac{1}{i\hbar} [H,\rho]
\end{equation}
where we also note that $\rho$ is positive semi-definite and Hermitian, and $\rm{Tr}(\rho) = 1$.
The probabilities of populations to occupy each level are given by the diagonal elements $\rho_{ii}$, and the off-diagonal elements $\rho_{ij} = \rho_{ji}^*$ represent the coherences between levels.  Since we have assumed that the relationship between the Zeeman levels rapidly becomes incoherent and is then represented by a decay, we only need to consider the coherences between the hyperfine levels $2$ and $4$ due to the microwave radiation (see Fig. \ref{fig:IonMotionZeeman} ).  
The evolution of this system can thus be represented by a 4$\times$4 density matrix with elements
\begin{equation}
\rho = 
 \left( 
\begin{array}{cccc} \label{eq:densitymatrix}
\rho_{11} & 0 & 0 & 0\\
0 & \rho_{22} & 0  & \rho_{24} \\
0 & 0  & \rho_{33} & 0 \\
0 & \rho_{42}  & 0 & \rho_{44}
\end{array} 
\right)
\end{equation}
where the basis labels are indicated in Fig. \ref{fig:IonMotionZeeman} and we are only considering interactions between the $|F=1,m_F=0\rangle$ level and each of the other levels.

\subsubsection{Rate equations}

We can consider this system logistically to write down basic equations for the above populations and coherences. To begin, let there be an overall decay rate $\Gamma$ between the $F=1$ Zeeman sublevels ($|F=1,m_F=0,\pm1\rangle$).  This overall decay rate will include both relaxation due to collisions and other mechanisms  as well as any changes due to driving of the system.  Since a measurement in the lab does not distinguish between these sources of decay, this $\Gamma$ represents the combined effect.  With this in mind, we can write down the rate equations for the populations going into and out of these levels as follows (considering only the $|F=1\rangle$ states):
\begin{align}
\dot{\rho}_{11} &= -\Gamma \rho_{11} + \Gamma\rho_{22}       \label{eq:rho1pop1}  \\
\dot{\rho}_{22} &= -2\Gamma \rho_{22} + \Gamma \rho_{11} + \Gamma \rho_{33}       \label{eq:rho2pop1}  \\
\dot{\rho}_{33} &= -\Gamma \rho_{33} + \Gamma\rho_{22}    \label{eq:rho3pop1} \ .
\end{align}

These equations indicate the population decay in and out of the states, for example in Eq. (\ref{eq:rho1pop1}) we see that the change in population in state $|1\rangle$ is equal to the loss from that state at rate $\Gamma$ plus the population leaving state $|2\rangle$ and going to $|1\rangle$  at the same rate.  Similarly, in Eq. (\ref{eq:rho2pop1}) there is a factor of 2 in front of the decay out of that level, since it can decay to either $|1\rangle$ or $|3\rangle$.

We will also consider 
a driving interaction between the hyperfine levels.   If we add a microwave driving field that is near resonance with the clock transition ($|F=0, m_F=0 \rangle$ to $|F=1,m_F=0 \rangle$) frequency, it can drive transitions (Rabi flopping) between states $|2\rangle$ and  $|4\rangle$, which act as a two-level system since we consider $|1\rangle$ and $|3\rangle$ to behave as a loss channel.
In the presence of a near-monochromatic oscillating field (perpendicular to the quantization axis and with a frequency near the resonance of the transition frequency) the driving interaction is governed by the equations 
\begin{align}
\dot{\rho}_{44} &= -\frac{i \Omega_{\rm{mw}}}{2} \left(   \rho_{42} - \rho_{24}  \right)  \label{eq:bloch1}  \\
\dot{\rho}_{22} &= -\frac{i \Omega_{\rm{mw}}}{2} \left(   \rho_{24} - \rho_{42}  \right) \label{eq:bloch2}\\
\dot{\rho}_{24} &=  i \Delta_{\rm{mw}} \rho_{24}-\frac{i \Omega_{\rm{mw}}}{2} \left(   \rho_{22} - \rho_{44}  \right) = \dot{\rho}_{24}^*  \label{eq:bloch3}
\end{align}
where $\Omega_{\rm{mw}}$ is the Rabi frequency due to the microwave field 
for the hyperfine levels $|2\rangle$ and $|4\rangle$. The detuning $\Delta_{\rm{mw}}$ of the oscillating field frequency from resonance with the hyperfine transition is defined by $\Delta_{\rm{mw}} = (\omega_{\rm{mw}} - \omega_0)$ where $\omega_{\rm{mw}}$ is the microwave frequency and $\omega_0$ is the resonant frequency of the transition.   In order to write these equations we have transformed to a rotating frame and performed the rotating wave approximation.

Putting these effects together, we can describe the system by
\begin{align}
\dot{\rho}_{11} &= -\Gamma \rho_{11} + \Gamma\rho_{22}  \\   \label{b1}
\dot{\rho}_{22} &= -\frac{i \Omega_{\rm{mw}}}{2} \left(   \rho_{24} - \rho_{42}  \right)   - 2\Gamma \rho_{22} + \Gamma \rho_{11} + \Gamma \rho_{33}        \\ \label{b2}
\dot{\rho}_{33} &= -\Gamma \rho_{33} + \Gamma\rho_{22}  \\  \label{b3}
\dot{\rho}_{44} &= -\frac{i \Omega_{\rm{mw}}}{2} \left(   \rho_{42} - \rho_{24}  \right)  \\  
\dot{\rho}_{42} &=  i \Delta_{\rm{mw}} \rho_{42}-\frac{i \Omega_{\rm{mw}}}{2} \left(   \rho_{44} - \rho_{22}  \right) = \dot{\rho}_{24}^*  \ . 
\end{align}

These equations approximate the behavior of the decaying Rabi fringes that we measure such as the one shown in Fig. \ref{fig:decayingRabi}.  With the optical readout in the lab, we measure the populations of states $1$, $2$, and $3$ simultaneously, so we see a sum of the populations given by Eqs. (\ref{b1}-\ref{b3}).  Although we are interested in the decay out of the $|F=1,m_F=0\rangle$ state, the decay rate at late times in the measurement is the same for both cases.  
We want to use the physics of the system to find the form of the decay rate $\Gamma$, which will directly relate the time-dependent magnetic field seen by the ion to the decay out of the $|F=1,m_F=0\rangle$ state.  Then we will be equipped to compare our ion motion simulation with the decay rate measurement.

\subsubsection{The decay rate}
%
%
We must create a link between the ion motion simulation and the measured decay rate.  One way to make an estimate of the decay rate as a function of the spectrum of motional frequencies of the ion is to start by considering the coherent effect of a single-frequency field inducing transitions between the Zeeman levels. 
If we consider this system in steady state, we can compare the steady-state population dynamics to the phenomenological decay discussed in the last section to find an expression for the decay rate at late times in the evolution.  Then we can generalize the single frequency case to a spectrum of many frequency components by doing a weighted average over all frequencies in that spectrum.    This is a straightforward approximation that allows us to see the decay behavior.  A more rigid approach is to consider the magnetic field seen by the ion as an arbitrary function of time and consider the equations of motion in the frequency domain in order to find the decay rate in frequency space.  This approach is the subject of future work.

Although we are treating the gradient-ion-motion effect as a resonant RF interaction at first, in reality, the ions are moving around at room temperature and experiencing different gradients throughout the trap.  That is, individual ions are not experiencing the same gradients.  This is why, on average over the ions in the trap, we expect to see a decoherence effect that leads to an overall decay in the observed population.


Now we consider only the three upper levels of the four-level system, and look at the phenomenon of driving transitions between the Zeeman levels.  Let�s consider adding a radio frequency (RF) driving field that is near resonance with the Zeeman frequency, that is, it can drive Larmor precession between states $|1\rangle$ and $|2\rangle$ and states $|2\rangle$ and $|3\rangle$. In the presence of a single-frequency oscillating field, the driving interaction is governed by
\begin{align}
\dot{\rho}_{11} &= -\frac{i \Omega_{\rm{rf}}}{2} \left(   \rho_{21} - \rho_{12}  \right)  \label{eq:bloch4}  \\
\dot{\rho}_{22} &= -\frac{i \Omega_{\rm{rf}}}{2} \left(   \rho_{12} - \rho_{21}  \right) - \frac{i \Omega_{\rm{rf}}}{2} \left(   \rho_{32} - \rho_{23}  \right)   \label{eq:bloch5}\\
\dot{\rho}_{33} &= -\frac{i \Omega_{\rm{rf}}}{2} \left(   \rho_{23} - \rho_{32}  \right)  \label{eq:bloch6}\\
\dot{\rho}_{12} &=  i \Delta_{\rm{rf}} \rho_{12}-\frac{i \Omega_{\rm{rf}}}{2} \left(   \rho_{11} - \rho_{22}  \right) = \dot{\rho}_{21}^*  \label{eq:bloch7}  \\
\dot{\rho}_{32} &=  i \Delta_{\rm{rf}} \rho_{32}-\frac{i \Omega_{\rm{rf}}}{2} \left(   \rho_{33} - \rho_{22}  \right) = \dot{\rho}_{32}^*  \label{eq:bloch8}  
\end{align} 
where $\Omega_{\rm{rf}}$ is the Larmor frequency and the detuning $\Delta_{\rm{rf}}$ is the difference of the oscillating field frequency from resonance with the Zeeman transition $\Delta_{\rm{rf}} = (\omega_{\rm{rf}} - \omega_0)$.


We also want to include a relaxation term to represent the intrinsic decoherence induced by collisions and other effects. 
We take note of the two typical relaxation types: decay of the populations and decay of the coherences.  These are traditionally associated with the characteristic relaxation times $\rm{T_1}$ (sometimes known as ``longitudinal" relaxation in NMR terminology) and $\rm{T_2}$ (``transverse" relaxation).  If we call the population relaxation $\gamma_{\rm{pop}} = 1/ \rm{T_1}$, then we can include this in the equations for the change in population by adding a term $-\gamma_{\rm{pop}} \rho_{ii}$.  Similarly, the decay in coherence can be given by  $\gamma_{\rm{coh}} = 1/ \rm{T_2}$ and described by adding an additional decay term  $-\gamma_{\rm{coh}} \rho_{ij}$. 
The decay of the population for these states is very small because there is no spontaneous decay between the states.  Thus, we only include the decay of the coherences, and we will simplify the notation by calling this coherence decay $\gamma_0$ ($=1/\rm{T_2}$). 
Now our coherence evolution equations are
\begin{align}
\dot{\rho}_{12} &= ( i \Delta_{\rm{rf}}-\gamma_0) \rho_{12}-\frac{i \Omega_{\rm{rf}}}{2} \left(   \rho_{11} - \rho_{22}  \right) = \dot{\rho}_{21}^*  \label{eq:bloch13}  \\
\dot{\rho}_{32} &= ( i \Delta_{\rm{rf}}-\gamma_0)  \rho_{32}-\frac{i \Omega_{\rm{rf}}}{2} \left(   \rho_{33} - \rho_{22}  \right) = \dot{\rho}_{23}^*  \label{eq:bloch14}  
\end{align}
which, along with the population equations (\ref{eq:bloch4}-\ref{eq:bloch6}), can be used to deduce the character of the decay. 

We can solve for the steady state solution of this system  by assuming that at later times the coherences stop changing and reach some equilibrium value; then we can find the change in population for these equilibrium values.  For example, setting $\dot{\rho}_{12}=0$ in Eq. (\ref{eq:bloch13}) results in 
\begin{equation}
\rho_{12} = \frac{ i \Omega_{\rm{rf}}/2 }{\Delta_{mw}-\gamma_0} \left( \rho_{11}-\rho_{22} \right) =\rho_{21}^* \ .
\end{equation}
Solving  $\dot{\rho}_{32}=0$ gives a similar expression, and when these are put into Eq. (\ref{eq:bloch5}) for $\dot{\rho}_{22}$, which is the population whose decay we are interested in, we obtain
\begin{equation}
\dot{\rho}_{22} = 
  \frac{(\Omega_{\rm{rf}}^2/2) \gamma_0}{\Delta_{\rm{rf}}^2+\gamma_0^2} (2\rho_{22}-\rho_{33}-\rho_{11}) \ .
\end{equation}
This is the decay term we are looking for.  We can compare this with the phenomenological form found earlier in Eq. (\ref{eq:rho2pop1}):
\begin{equation}
\dot{\rho}_{22} = \Gamma \left( \rho_{11} + \rho_{33} -2\rho_{22} \right) 
\end{equation}
and we can now write down the form of the ``overall" decay constant $\Gamma$ between the Zeeman sublevels:
\begin{equation}
\Gamma = - \frac{(\Omega_{\rm{rf}}^2/2) \gamma_0}{\Delta_{\rm{rf}}^2+\gamma_0^2} \label{eq:gammaplain}
\end{equation}
where we recall that $\Delta_{\rm{rf}} = (\omega_{\rm{rf}} - \omega_0)$ and $\omega_0$ is the frequency of the Zeeman splitting.  In summary, this is the decay rate between levels $|1\rangle$, $|2\rangle$, and $|3\rangle$, considering RF fields and relaxation, in a system subjected to a static field strength $B_0$, which determines $\omega_0$, and an RF field given by $B(t)$, which determines $\Omega_{\rm{rf}}$ as well as $\omega_{\rm{rf}}$.

Now we consider that we are illuminating the ions with RF fields of widely varying frequencies and amplitudes.  This identifies our analysis with the spectrum of the fields seen by the ion due to its motion in a gradient.  If we know the form of this spectrum, we can do a weighted average over the contribution of the different field frequencies by integrating over the spectrum of RF fields in frequency space.  We start by writing $\Omega_{\rm{rf}}$, instead of as a monochromatic Rabi frequency due to a monochromatic RF field, as a spectrum of Rabi frequencies due to a spectrum of RF fields:
\begin{equation}
\Omega_{\rm{rf}}(\omega) = \gamma_{\rm{gyro}} B_{\rm{rf}}(\omega) \label{eq:proportional}
\end{equation}
where  $\gamma_{\rm{gyro}}$ is the gyromagnetic ratio which here is $ \gamma_{\rm{gyro}} = \mu_B g_e / \hbar$, with $\mu_B$ being the Bohr magneton and $g_e$ being the electron g-factor. Using this spectrum we can employ Eqs. (\ref{eq:gammaplain}) and (\ref{eq:proportional}) to calculate the amplitude of the decay rate for a particular static field $B_0$ (that is, for a particular Zeeman splitting value) using
\begin{equation}
\Gamma_{\rm{rf}}(\omega_0) = \int_{\text{all}\  \omega}  - \frac{(\gamma_{gyro}^2 B_{\rm{rf}}^2(\omega)/2) \gamma_0}{ (\omega - \omega_0)^2+\gamma_0^2} \ d\omega \ . \label{eq:gammaintegral}
\end{equation}
Since a measurement of the ions gives us a decay value that is an averaged effect of all of the RF frequencies that the ions are seeing, this integral is a representation of our measurement of the decay rate at one bias field value $B_0$.  We looked at the decay phenomena in the lab for a range of bias field values, and we can use this expression to approximately reproduce the decay rate for a range of bias fields (i.e. a range of $\omega_0$ values) if we know the RF spectrum.  The spectrum of the RF fields is determined by the spectrum of the frequencies of the motion of the ions in the RF trap, and on the strength and shape of the gradient in which their motion takes place.  One way to determine this theoretically is through Monte Carlo simulations of the ion motions in the trap.  The spectrum is dominated by the secular frequencies of the trap in each direction, since the trap is harmonic in the center and the ions perform an oscillatory motion in each direction.  However, since our ions are in a large cloud, are not laser cooled and occupy a large extent in the trap, there is a spread of frequencies experienced by the ions that can only be predicted through simulation.  This simulation is the subject of the next section. 

 \section{Simulations}

Simulations of the ion motion were carried out using Matlab, to examine the ion motion in our 3-D numerical potential, considering to some extent the space charge and actual ion density distribution.  From this simulation we have gleaned the spectrum of motional frequencies that the ion experiences, which can then be translated into a spectrum of magnetic field frequencies when the ion is moving in a gradient.  This spectrum can tell us how much linewidth broadening we will see due to transitions through the Zeeman sublevels.

\subsection{Ion number density distribution}
 \label{sec:density}
Before we can  do a representative calculation of ion trajectories, we must have an understanding of the distribution of the ions in the trap.  The ion distribution carries information about the regions in which the ions are moving within the trap and also about how the space charge from the cloud itself will affect the trap potential.  Each ion in the vicinity of the trap is influenced by two primary potentials: one due to the trapping potential of the RF and endcap electrodes $\phi_{trap}$, and the other due to the space charge of the ions in the cloud surrounding it $\phi_{sc}$.  We know  $\phi_{trap}$ numerically, but we do not know the value of the space charge contribution.  For our first order calculation we calculate the density based on a cylindrically symmetric potential, with no angular dependence and no variation on the long axis of the trap.  This approximation is reasonable for a linear trap where the ratio of the length to the radial distance between the electrodes is large and the trap is harmonic.  For us, this means that the ions are very near the center of the trap in the transverse direction, since the center of the trap is very harmonic.  As our intention with the simulation is to understand the influences of the asymmetries as the ions get farther from the center of the trap, this approximation may seem out of line with our goal.  However, we consider that the space charge effect is small (no more than a few percent change in the trap depth due to the space charge effect for the ion numbers we expect) and utilize this cylindrical symmetry only for determination of the number density and the consequent space charge.  This space charge will then be  overlaid onto our calculated numerical potential.  Therefore, the potential shape will be dominated by the ``real" numerical pseudopotential, and influenced by the approximate space charge of a cylindrically symmetric ion cloud.  This means that the ends of the trap, where the real cloud deviates the most from the cylindrical model, will have the most distortion from the approximated space charge.  

The cylindrical assumption in the calculation of the density will also come into play when we choose initial conditions of the ion trajectories, which will be drawn from a distribution that represents the calculated density.  The implications of this are less severe, since in the course of the simulation, ions that start at a position that is not realistic in the numerical potential will simply leave the trap.  Any ions that leave the trap in the course of the ion trajectory simulation are not used in the final calculation of the spectrum.

The structure of this analysis follows the work of \cite{367391}  and appears in some form in several other works I have already mentioned \cite{springerlink:10.1007/BF00694316,springerlink:10.1007/BF00697492}.  Our contribution is to include, after evaluating the space charge, a real 3D numerical potential in the trajectories instead of assuming spherical or cylindrical symmetry and therefore perfect harmonic motion.

We begin by assuming that the ions are in thermal equilibrium, and therefore the number density is Boltzmann-distributed (Gaussian) according to
\begin{equation}
n(r) = n(0) \exp{\left(\frac{-\Phi_{total}(r)}{k_B T}\right)}
 \label{eq:nDistribution}
\end{equation}
where 
\begin{equation}
\Phi_{total}(r) = q \phi_{trap}(r) + q \phi_{sc}(r) 
  \label{eq:PhiTot}
\end{equation}
 is the total potential energy felt by the ion due to both the trap electrodes and the space charge from the ion cloud, and $q$ is the charge of the ion. 
 The constant $n(0)$ is a some ``initial" density value, corresponding to the central density along the node of the linear trap, that acts as a normalization factor for the distribution.  The space charge potential $\phi_{sc}$ is also governed by the Poisson equation 
 \begin{equation}
 \nabla^2 \phi_{sc}(r) = -\frac{\rho_{ions}}{\epsilon_0} 
  \label{eq:Poisson}
 \end{equation}
 where the charge distribution $\rho_{ions}(r)$ is given by $q n(r)$.  Since these quantities depend only on $r$, this equation becomes
 \begin{equation}
 \phi_{sc}''(r) + \frac{1}{r}  \phi_{sc}'(r) = -\frac{q}{\epsilon_0} n(r) \ .
  \label{eq:PoissonSC}
 \end{equation}
 Using this equation and the expression for $n(r)$ one can construct a nonlinear differential equation for the density $n(r)$.  Combining Eqs. (\ref{eq:nDistribution}) and (\ref{eq:PhiTot}), taking derivatives, and solving for $\phi_{sc}$ yields
 \begin{equation}
 \phi_{sc}'(r) = -\phi_{trap}'(r) - \frac{k_B T}{q} \left( \frac{n'(r)}{n(r)} \right)
 \end{equation}
 and
 \begin{equation}
 \phi_{sc}''(r) = - \frac{k_B T}{q}  \left( \frac{n''(r)}{n(r)} \right) + \frac{k_B T}{q} \left( \frac{n'(r)}{n(r)} \right)^2 -  \phi_{trap}''(r) \ .
 \end{equation}
 Substituting these derivatives into the Poisson equation (\ref{eq:PoissonSC}) one obtains the differential equation for $n(r)$:
 \begin{equation}
 n'' - \frac{(n')^2}{n}  + \frac{n'}{r} + \frac{q}{k_B T} \left(  \phi_{trap}'' + \frac{1}{r} \phi_{trap}'  \right) n - \left(\frac{q}{\epsilon_0}\right) \frac{q}{k_B T} n^2 = 0
  \label{eq:densityDE1D}
 \end{equation}
 where the dependence of $n$ and $\phi_{trap}$ on $r$ is understood and the derivatives are also with respect to $r$.  If we have numerical or analytical expression for $\phi_{trap}$ as a function of \emph{only} $r$, we can solve this equation numerically for the number density distribution $n(r)$.  Once this distribution is found we can calculate the space charge by inverting the distribution in Eq. (\ref{eq:nDistribution}).

\begin{figure}
   \centering
  \includegraphics[scale=0.5]{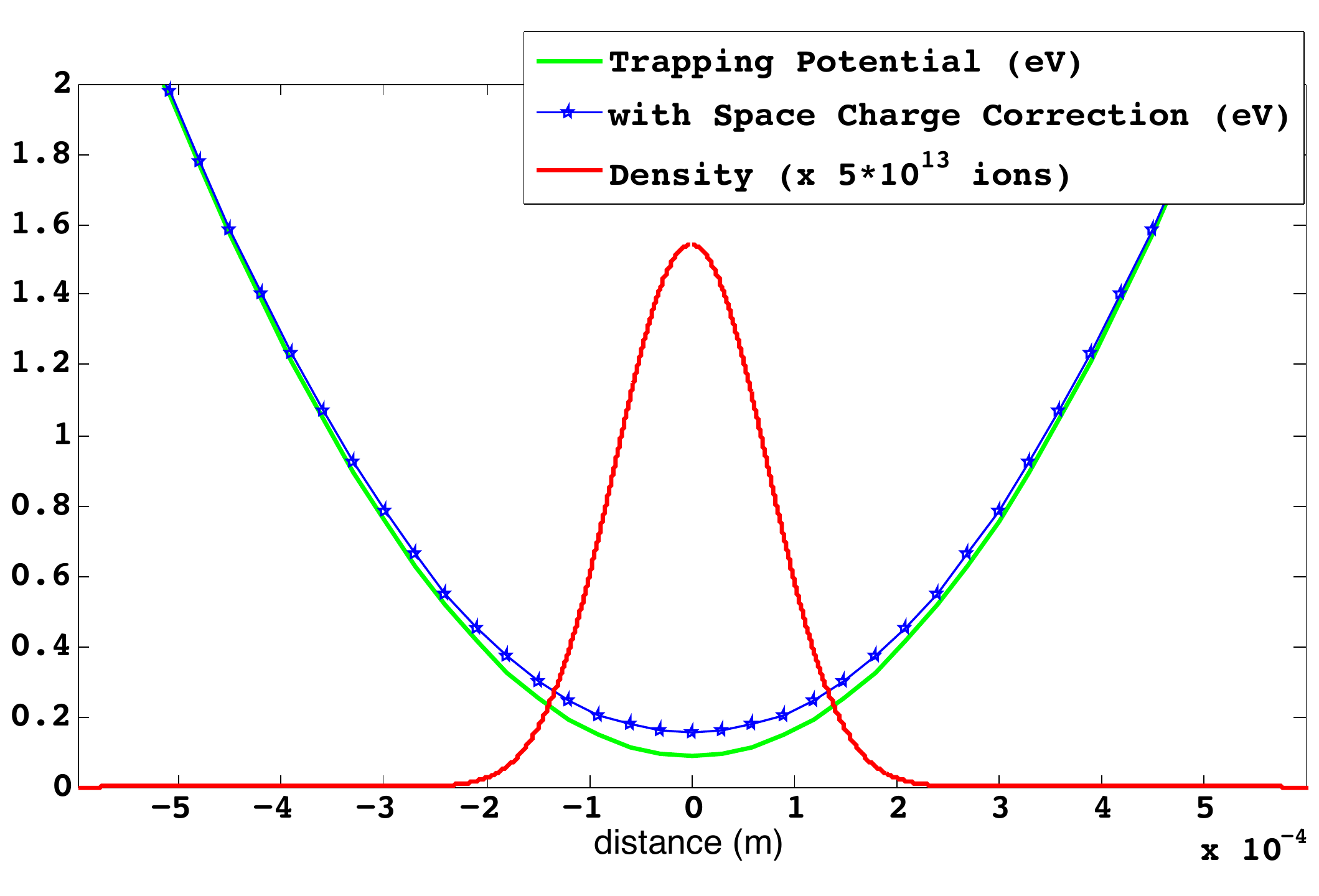} 
   \caption{An example of the ion density calculated from the trap well and the corresponding space charge correction.  Note the difference in scales between the density and the potentials.}
   \label{fig:SpaceChargeExample}
\end{figure}

Now, when we assume for the purpose of deriving the ion density distribution and the space charge that the trap and thus the density are cylindrically symmetric, we can replace the trap potential with a harmonic oscillator potential, so that $q \phi_{trap}(r)$ in Eq. (\ref{eq:PhiTot}) becomes $ \frac{1}{2}m \omega^2 r^2$.  Replacing the derivatives of $\phi_{trap}$ in Eq. (\ref{eq:densityDE1D}) with the harmonic potential results in
\begin{equation}
n'' - \frac{(n')^2}{n}  + \frac{n'}{r} + \frac{2 m \omega^2}{k_B T} n - \left(\frac{q}{\epsilon_0}\right) \frac{q}{k_B T} n^2 = 0
\end{equation}
which can then be solved numerically and used to calculate the space charge as described above.  Fig. \ref{fig:SpaceChargeExample} gives an example of the calculated potential corrected for space charge. 

%

Of course, one would prefer to consider a three-dimensional model using our numerically calculated potential to find the real density of ions.  We could find this by using Poisson's equation in three dimensions, that is, by solving a 3-dimensional version of Eq. (\ref{eq:densityDE1D}) for the density.  To do this, we would start from Eq. (\ref{eq:Poisson}) and instead of assuming exclusive radial dependence we assume a dependence on three dimensional variables.  Practically speaking, we know our potential distribution from CPO in Cartesian coordinates $x,y,z$ instead of spherical coordinates, so the space charge version of Poisson's equation is straightforward: Eq. (\ref{eq:PoissonSC}) becomes
\begin{equation}
\frac{\partial^2}{\partial x^2} \phi_{sc}(x,y,z) + \frac{\partial^2}{\partial y^2} \phi_{sc}(x,y,z) +\frac{\partial^2}{\partial z^2} \phi_{sc}(x,y,z) = -\frac{q}{\epsilon_0} n(x,y,z)
\end{equation}
which through the same process leads to an equation for arbitrary $n(x,y,z)$ 
\begin{equation}
\sum_{i = x,y,z} \left[  \frac{k_B T}{q}  \left(\left( \frac{\partial_i n}{n} \right)^2 - \left(  \frac{\partial_{ii} n}{n}  \right) \right) - \partial_{ii} \phi_{trap} \right] = -\frac{q}{\epsilon_0} n
\end{equation}
where $\frac{\partial f}{\partial x}$ is denoted by $\partial_x f$ and so on.  We could now include our arbitrary numerical potential $\phi_{trap}(x,y,z)$ in the calculation since there is no cylindrical limitation.  This would of course give us the best estimate of the actual density and cloud distribution in our real trap.  However, this is a nonlinear, partial differential equation in 3 dimensions that is very difficult to solve.   Based on ease of evaluation and the reasonable approximation of the cylindrical symmetry for the density calculation, we assume that solving the equation in one dimension is sufficient as a first approximation for our purposes. 

\subsection{Trap potential}

The trap potential is calculated with the program CPO as discussed in Chapter \ref{ch:trapping}.  This program and its processing in Mathematica provides us with a 3-dimensional grid of potential values in a region that includes the trapping region, the trap electrodes, and some additional space in all directions.  Treatments in the literature I have mentioned examine the analytical or numerically calculated motion in a cylindrically or even spherically symmetric potential, but this allows us to numerically examine motion in an arbitrary 3-dimensional geometry for any set of electrodes we can input into CPO.

Of course, as discussed in Chapter \ref{ch:trapping}, the real ion motion consists of two main components, the so-called ``secular motion" component and a ``micromotion" component.  By nature of how we calculate and process the potentials using CPO and Mathematica, we ignore the micromotion component, which occurs at a frequency (the RF trap driving frequency) much higher than we are interested in for the phenomena we want to explore in this simulation.  Therefore, we assume that the micromotion is averaged out and we only care about the secular motion due to the harmonic-oscillator-like potential provided by the ponderomotive RF pseudopotential in the transverse dimension and the DC endcap potential in the third dimension.  Others have considered micromotion by using a smaller step size in the ion trajectories and then averaging over 10 or more periods and found that the secular motion is unaffected \cite{367391}.

We take the arbitrary trap potential calculated by CPO and combine it, mesh element by mesh element, with the space charge potential calculated using the results discussed in Sec. \ref{sec:density}. The space charge correction in Figs. \ref{fig:SpaceChargeExample}  
is shown using the real CPO potential calculated for the test bed trap. This gives a realistic whole potential as it looks from the point of view of the ion and should help us to see the true motion of the ion, on average (neglecting micromotion).

\subsection{Trajectory calculation}
\begin{figure}
   \centering
  \includegraphics[scale=0.5]{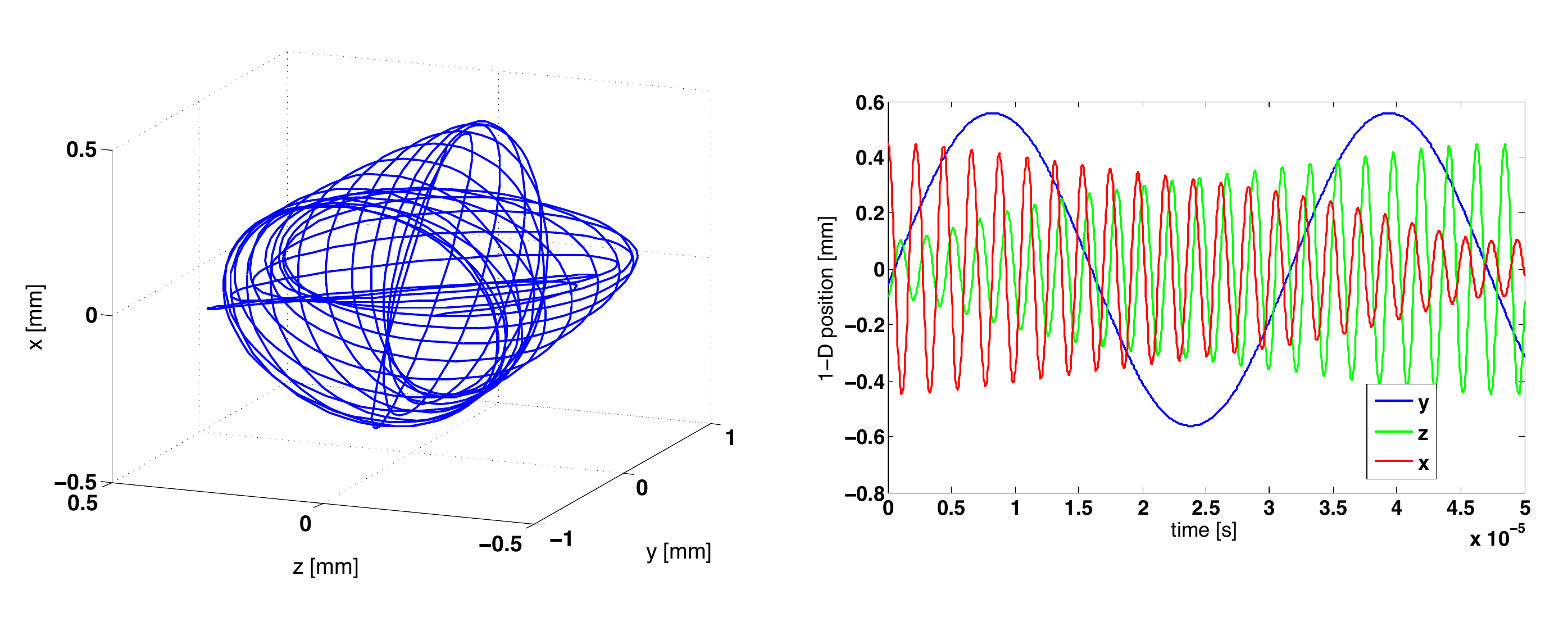} 
   \caption{Example trajectory of the ion for 0.05 ms in 3D space (left) and in a projection along each axis (right).  The $y$-axis is the longitudinal axis of the trap.}
   \label{fig:TrajectoryExample}
\end{figure}

The trajectory calculation is carried out by placing an ion in the trap with an initial velocity and allowing it to evolve in the whole potential (which includes both trap and space charge effects). An example trajectory for a short time is shown in Fig. \ref{fig:TrajectoryExample}. The ions that leave the trap in this process are not used in the final calculation of frequencies.  Many ions selected from a representative set of initial conditions (see below) are run for a trajectory of 1 to 5 milliseconds.  In the absence of collisions (which we assume), averaging these individual trajectories in the presence of the space charge gives a representative look at the collective motion.

\subsubsection{Initial Conditions}
The initial conditions are chosen as follows.  Our potential is based in a Cartesian coordinate system.  Therefore, we choose $x$,$y$, and $z$ position and velocity components for each ion.  The positions in the transverse trap direction ($x$ and $z$) are chosen using the radial distribution calculated using the methods described in  Sec. \ref{sec:density}, which is meant to provide a representative selection of actual ion locations within the cloud.  In particular, a radial coordinate is chosen randomly from the probability distribution derived from the density number distribution $n(r)$, an angle $\theta$ is chosen randomly from $[0, 2\pi]$ and these coordinates are transformed to $x$ and $y$.  Because we do not consider the longitudinal dependence of the cloud density in our density calculation, the $z$ coordinate is chosen from a normal distribution with a width that estimates the cloud size in that dimension.  

The ions are assumed to be in thermal equilibrium at a temperature $T$ that is determined separately (see Sec. \ref{sec:temp}), and the velocity components are randomly chosen from the Maxwell-Boltzmann distribution parametrized by that temperature, i.e. a normal distribution with mean $\mu_v = 0$ and standard deviation $\sigma_v = \sqrt{\frac{k_B T}{m}}$ for each dimension.

An alternative to sampling the positions and velocities in this way would be to calculate the trajectories for a uniformly sampled set of initial positions and velocities selected from some reasonable range, and then weighting the resulting averaged ion motion spectra by a Boltzmann factor using the initial energy indicated by the selected initial conditions.  Performing some examples of this type of calculation confirmed the validity of our sampling method, however in the interest of computational resources, the original method was used for the remaining calculations.


\subsubsection{Energy calculations}
\begin{figure}
   \centering
  \includegraphics[scale=0.3]{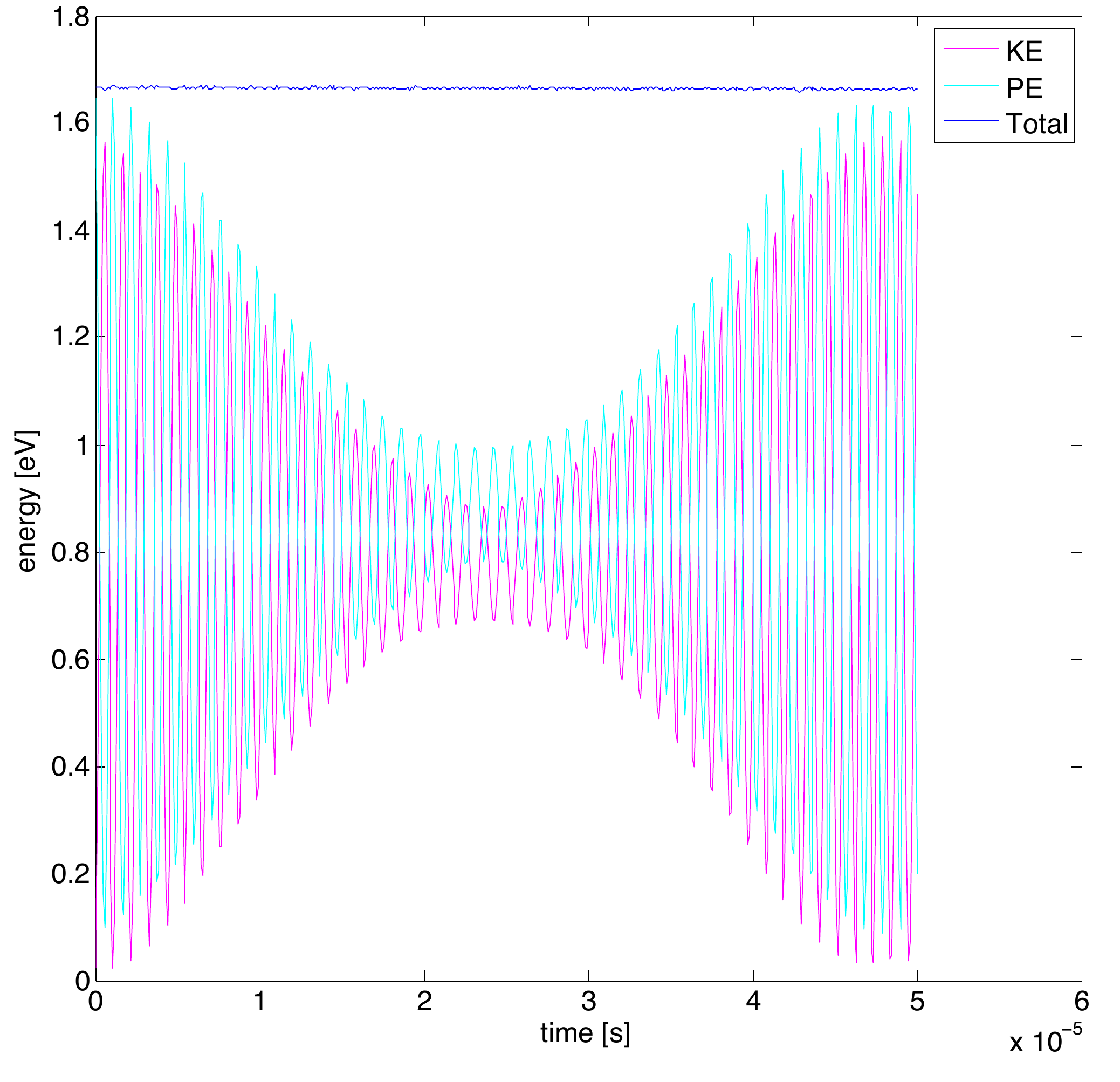} 
   \caption{The kinetic, potential, and total energy calculated at each step of the same 0.05 ms trajectory shown in Fig. \ref{fig:TrajectoryExample}.} 
   \label{fig:EnergyExample}
\end{figure}

In order to make a judgment about the validity of the trajectory calculations we calculated the total energy of the ion during the entire trajectory.  The kinetic energy and potential energy of the ion was computed at each step of the trajectory.  A sample of the energy as a function of time is given in Fig. \ref{fig:EnergyExample} (this example matches the example trajectory in Fig. \ref{fig:TrajectoryExample}).  As the ion moves through its orbits inside the trap, the kinetic and potential energies oscillate, but the total energy remains constant as expected.  We can also use this information to calculate statistics on the energies of the ions in the trap.

\subsection{Frequency spectrum} 

\begin{figure}
   \centering
  \includegraphics[scale=0.6]{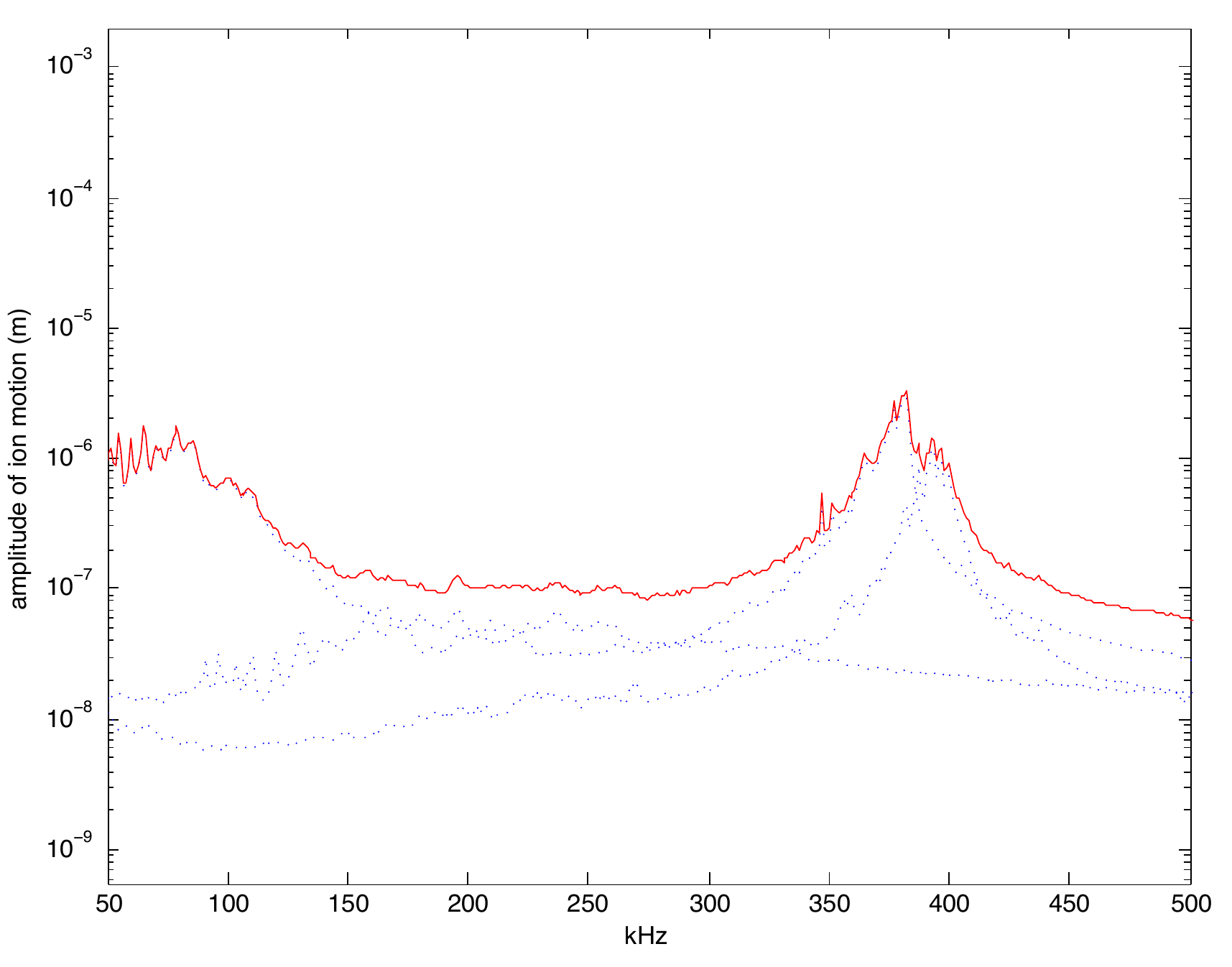} 
   \caption{An example of the ion motion spectrum calculated by taking the fast Fourier transform of the ion motion along each direction (in the LTCC trap).  The dotted lines show the spectra on each axis, and the red line is the total spectrum.}
   \label{fig:ExampleSpectrumLTCC}
\end{figure}

After the calculation of the ion trajectory for a length of time $t$, we have a time series for the ion's motion in each principal direction. 
Performing an FFT on this time series results in a spectrum of the frequencies of motion that the ion in question has undergone.  When performing the FFT it can be necessary to use a window function to make our amplitudes more accurate, but usually we use a flat-top window (i.e., no window),   which gives the highest resolution of peaks in the frequency domain.  Also, it may be necessary in some traps (for example, the test bed trap) to rotate the axes so that the frequencies along the principal directions are separated before performing the Fourier transform.  In this way we can apply magnetic fields along the principle directions and sum the contributions.  An example spectrum for the ion motion frequencies is shown in Fig. \ref{fig:ExampleSpectrumLTCC}.

\subsection{Decay rate}

As discussed in the previous section, we will estimate the decay rate for each bias field strength using Eq. (\ref{eq:gammaintegral}), repeated here for clarity:
\begin{equation}
\Gamma_{\rm{rf}}(\omega_0) = \int_{\text{all}\  \omega}  - \frac{(\gamma_{gyro}^2 B_{\rm{rf}}^2(\omega)/2) \gamma_0}{ (\omega - \omega_0)^2+\gamma_0^2} \ d\omega \label{eq:gammaintegral}
\end{equation}
In this equation, $ B(\omega)$ is given by the ion motion overlaid onto a magnetic field grid.  In one dimension, we assume a linear gradient and multiply amplitude of the motion by the gradient slope.  It can also be calculated exactly for the trajectory if the magnetic field amplitude is known for all of the space in question.  
The dimension for which we will see decay due to the motion is along an axis perpendicular to the quantization axis.  In order to see a contribution from motion along all three axes, our quantization axis may be fixed along a direction such that the ion motion has a component perpendicular to the quantization axis in all three directions. 
 Then we calculate these components using our simulation.

%
%

\section{Discussion}
We have confirmed the presence of this effect through simulation, since agreement of the experiment and the simulations shows that the peaks we see in the linewidth broadening are in fact due to the the ion motion.  In this section I will review our experimental measurements and related conclusions and compare them with simulation.

\subsection{Experiment}

\subsubsection{Nonmagnetic test-bed trap}

\begin{figure}
   \centering
  \includegraphics[scale=0.4]{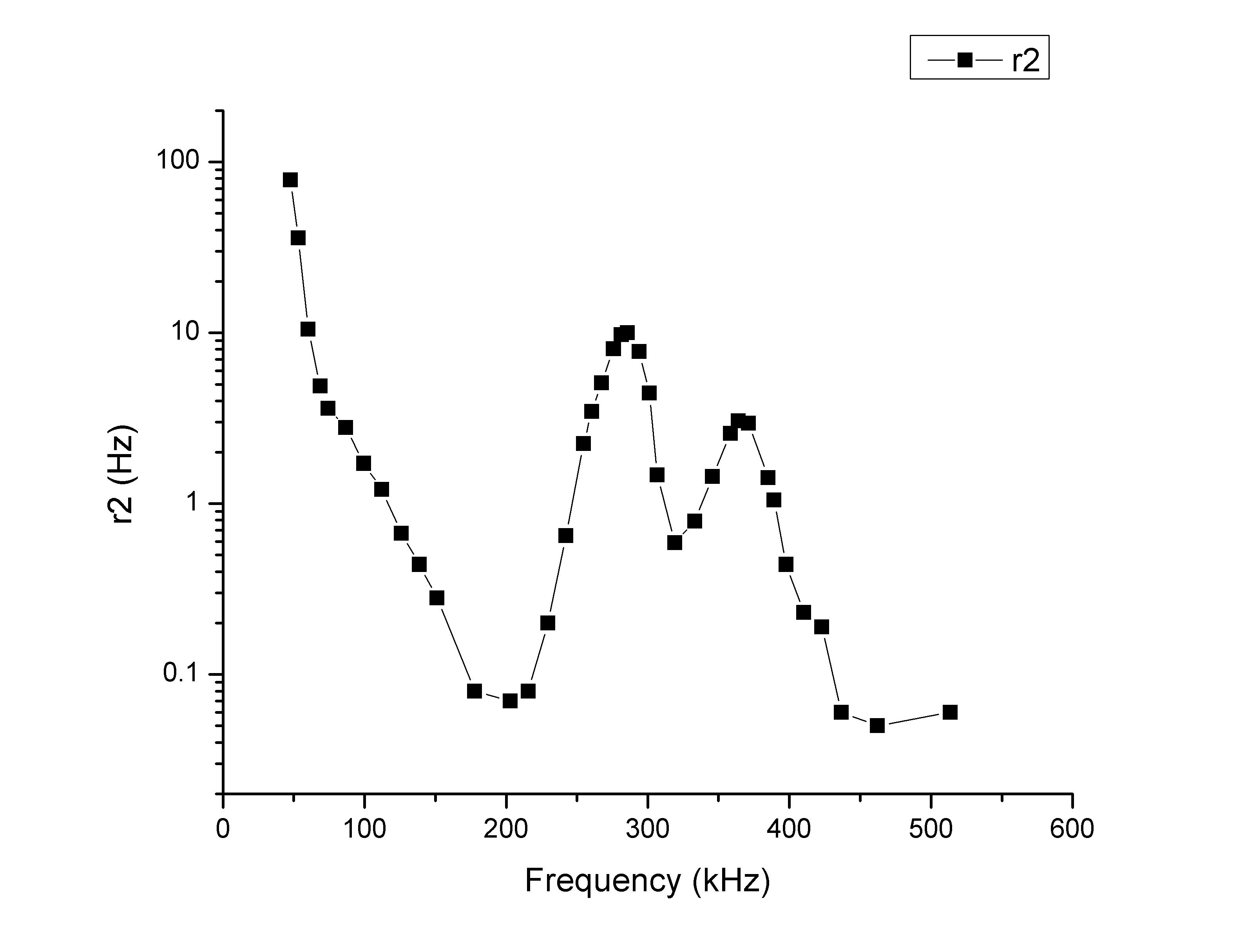} 
   \caption{An example measurement showing the double-peak feature that is due to the presence of two different transverse secular frequencies in the test bed trap.  This feature was caused by a stray DC voltage and appears in many of our data sets.}
   \label{fig:DoublePeak}
\end{figure}

\begin{figure}
   \centering
  \includegraphics[scale=0.8]{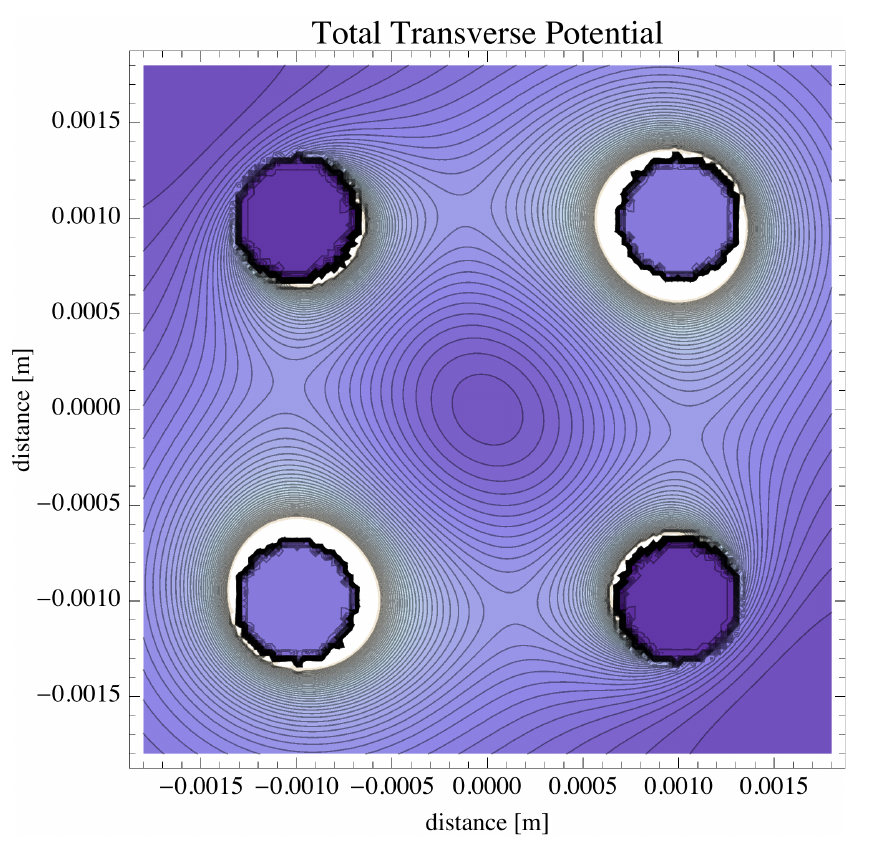} 
   \caption{The pseudopotential in the transverse plane for the test bed trap with a 2.4 V DC bias on two diagonally opposing RF trap rods.  This ``squeezed" potential is what causes the dual traverse secular frequencies evident in Fig. \ref{fig:DoublePeak}.}
   \label{fig:BiasSqueeze}
\end{figure}

We performed the majority of our magnetic field measurements in the test bed trap after replacing all of the magnetic parts with nonmagnetic ones.  Unfortunately, there was an unusual feature in this trap that appears in all the measurements: a double-peaked structure whose source is trivial, but that was not understood until a long time after the measurements were completed.   By that time, our experimental system had changed and we were unable to repeat the measurements.  An example of a measurement made in this trap of decay rate vs. bias field is given in Fig. \ref{fig:DoublePeak}.  The double-peak structure, if our hypothesis were correct, would indicate that the trap has more than one major secular frequency in the transverse direction, due to, for example, some asymmetry in the trap.  We would expect to see two peaks (in some cases) for an asymmetric trap, but since the test bed trap is extremely symmetric, and it was difficult to identify another possible source for the discrepancy (e.g., a stray electric field due to some element near the trap), it took some time to identify the source of this unexpected structure.  In the end, the source of the double-peak feature was found in the course of the simulations, which we used to model many asymmetries before discovering the problem.  The driving circuit we used for this trap had a small unexpected DC offset (about 2.6 V) which added a DC voltage to the two ungrounded RF driving electrodes.  This resulted in a diagonal ``squeezing" of the central potential which led to the asymmetry causing the degenerate secular frequencies and thus the double-peak feature (see Fig. \ref{fig:BiasSqueeze}).  For this reason, many of the plots seen in this chapter will have this two-peak structure.  The resulting analysis is the same, since the secular frequencies apparent from the two peaks were a real phenomenon.  

\begin{figure}
   \centering
  \includegraphics[scale=1.2]{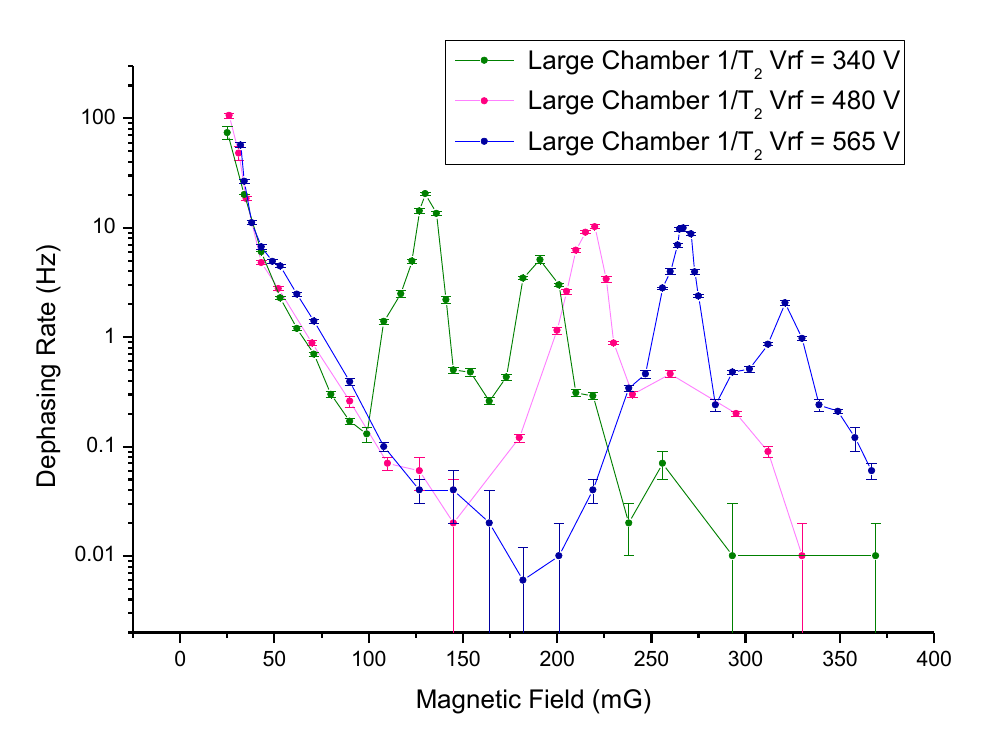} 
   \caption{Measurements in the test bed trap for several different RF voltages, which affect the location of the secular frequency peaks.}
   \label{fig:TiVaryVoltage}
\end{figure}

We know that for a given driving frequency, we can increase or decrease the transverse secular frequency of the trap by increasing and decreasing the overall RF voltage amplitude, which has a direct effect on the steepness of the harmonic potential walls.  This means that, in the broadening-vs-bias-field picture, we can move the transverse secular frequency peak left and right.  We have observed this experimentally as shown in Fig. \ref{fig:TiVaryVoltage}. As the voltage and therefore the secular frequency is lowered, the transverse secular frequency peak approaches the longitudinal one.  This can affect the valley point as seen in the figure.  
 Since we want to operate at a low voltage in order to conserve power, we must consider this effect if we want to operate in between the peaks.  If the peak is moved significantly to the left, then we may be able to operate at a bias field above the peak, but being able to operate below the peak is preferable because the overall bias is lower and therefore less sensitive to field fluctuations.  

\begin{figure}
   \centering
  \includegraphics[scale=0.5]{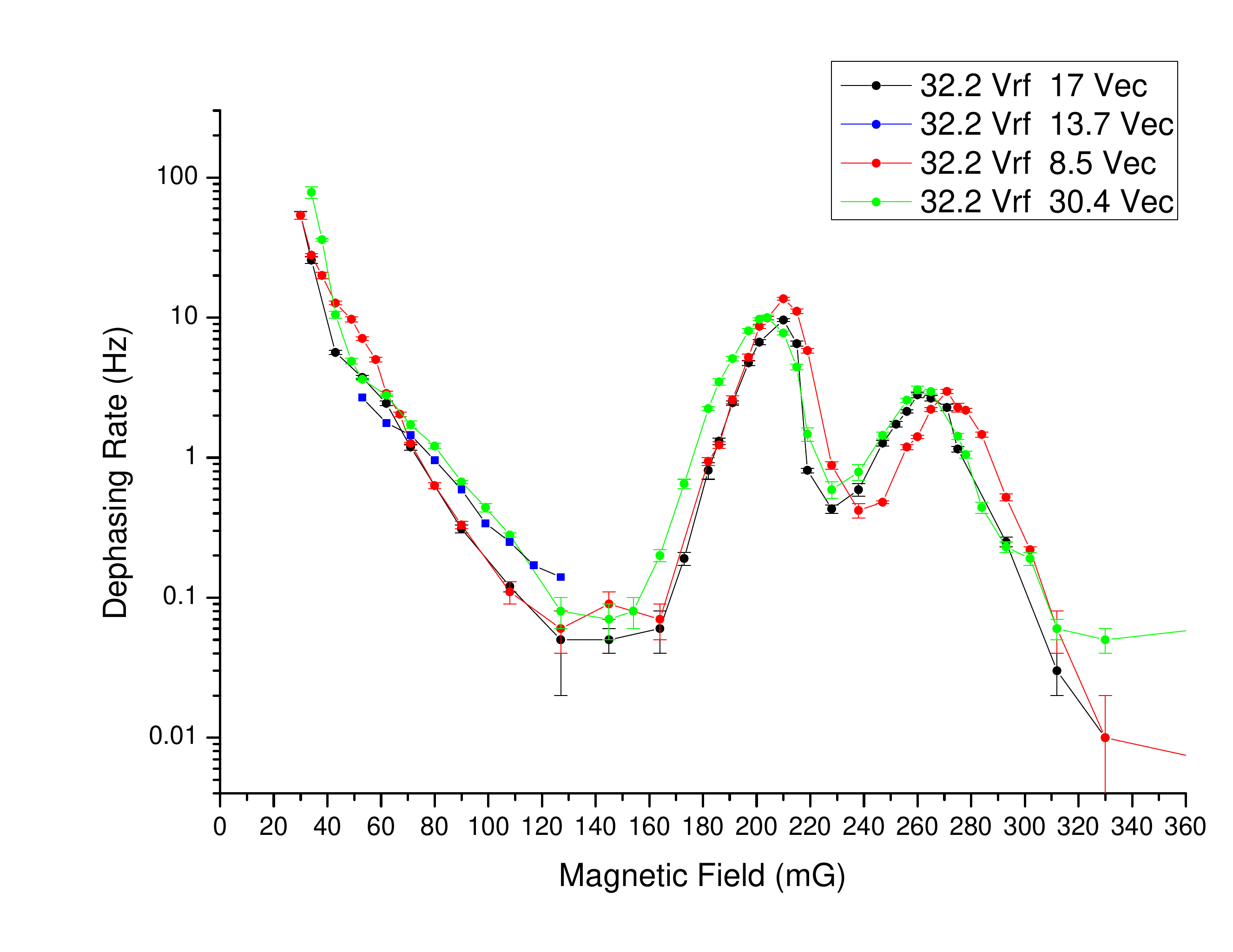} 
   \caption{Varying end cap voltage for the same measurement.  This had very little effect, as discussed in the text.}
   \label{fig:VaryEndcaps}
\end{figure}

We also tested adjustment of the endcap voltages to see if we could influence the peak that is allegedly due to the longitudinal secular motion.  We found that this did not have an effect.  This can be due to the fact that we need much larger changes in the endcap potential to see the effect (which is both impractical and unnecessary for us), or it can be that the peak near zero is completely dominated by ambient field uncertainties that we cannot influence in this way.  A figure showing the measurements at various endcap voltages is shown in Fig. \ref{fig:VaryEndcaps}.

To understand how the linewidth is affected by severe gradients, we placed an additional coil in two directions that is operated in an anti-Helmholtz configuration, in order to \emph{add} additional gradients to the ions.  The effect of additional gradients is to lift the whole curve upward, meaning that the linewidth is broadened at all bias fields while the peaks of the secular frequencies remain.  We have seen similar behavior when comparing the broadening curve for packages that have magnetic materials to packages where almost all magnetic materials have been eliminated. 
The conclusion from this is that in order to feasibly operate between the ion motion peaks, and therefore in the optimum bias field range, we must take extreme steps to eliminate magnetic materials in the construction of the clock.  The presence of any (even slightly) magnetic materials in the direct vicinity of the trap could prevent operation of the clock at all, except when using very high bias fields.  This would create an additional difficulty to the advancement of the IMPACT clock technology.

\begin{figure}
   \centering
  \includegraphics[scale=0.6]{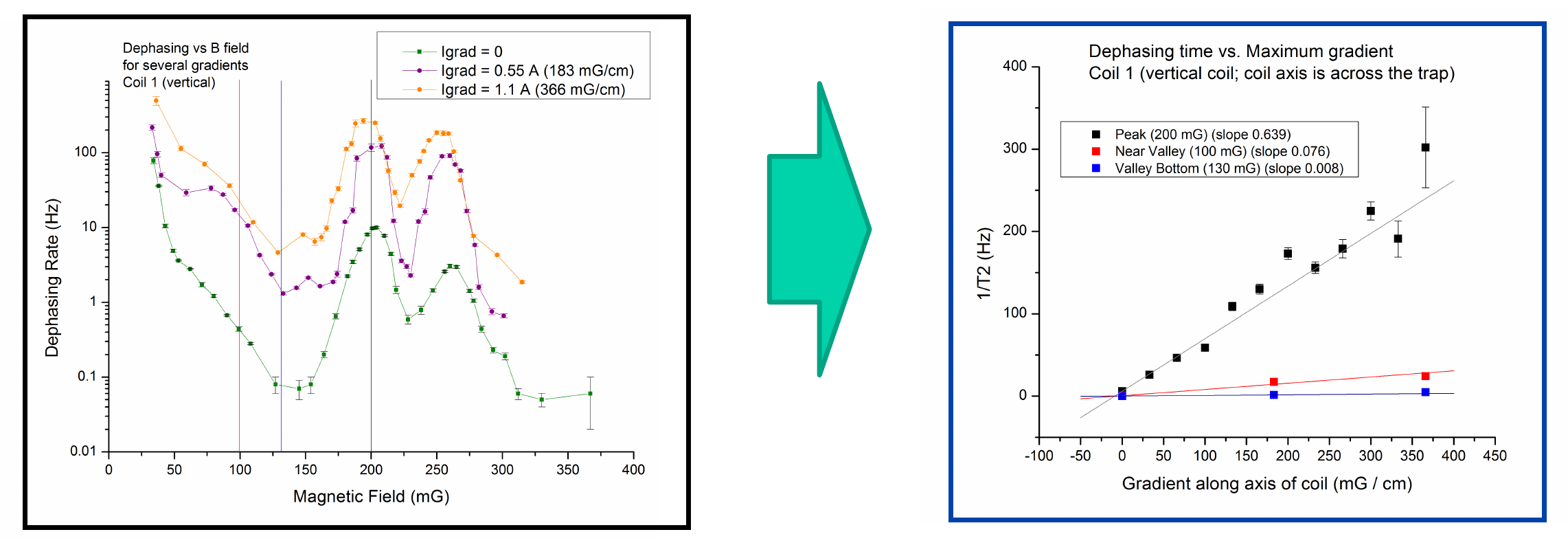} 
   \caption{Adding magnetic field gradients to the ions to determine the tolerable gradient for future traps.}
   \label{fig:VaryGradient}
\end{figure}

We have also used the method of adding gradients to make a practical estimate of how much of a gradient can be tolerated by ions inside of a similar trap.  This information helps us in designing next generation traps, especially when we are designing coils for the system that, because of their size, may not be able to create a uniform field across the entire cloud despite being in a Helmholtz configuration.  By estimating the slope of the linewidth versus gradient curve, we can estimate how much gradient can be tolerated by the system while still being able to produce desirable linewidths.  A demonstration of this estimate is shown in Fig. \ref{fig:VaryGradient}.  We assume that the gradient produced by the coil pair has its theoretical value in the location of the ions, which may not be true if the field is distorted by nearby objects or fields.  Nonetheless, we feel these numbers can be used as a rough guide.  For example, in the nonmagnetic version of the test bed trap, we find that in order to maintain a 1 Hz intrinsic linewidth, 
for operation at 130 mG we must engineer and maintain a (first order) gradient at or below 100 mG/cm.  

\subsubsection{PA\&E trap}

\begin{figure}
   \centering
  \includegraphics[scale=1.2]{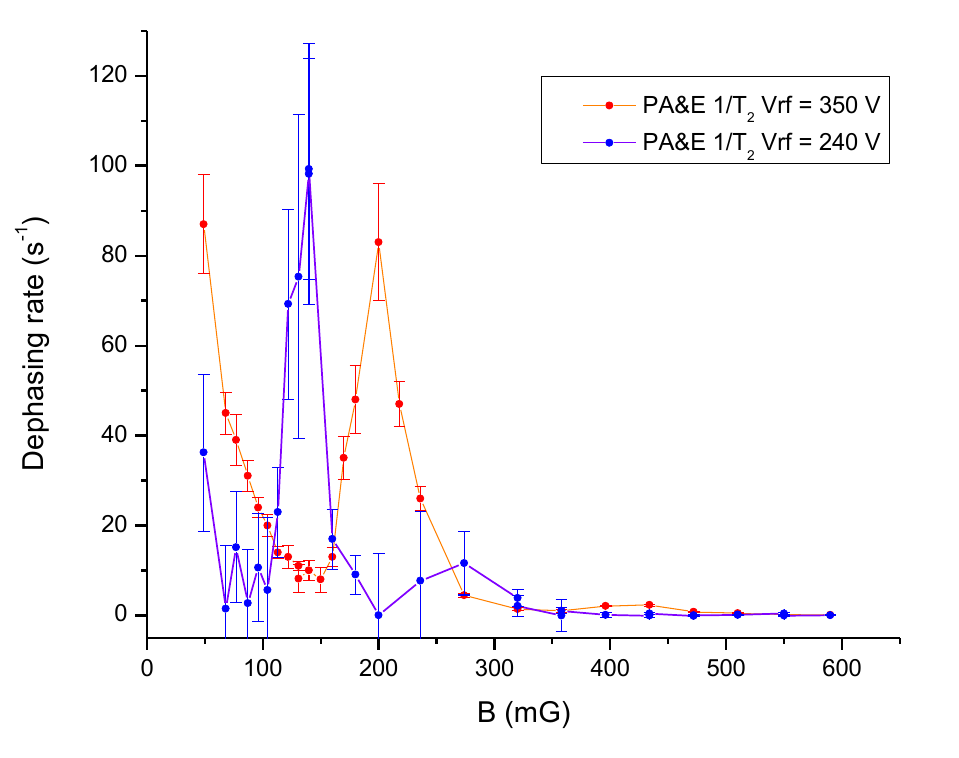} 
   \caption{Varying RF voltages in the PA\&E package.  The linewidths were larger overall in this trap compared to the others (after the test bed was replaced with all nonmagnetic materials).  This plot is on a linear scale to more clearly demonstrate the features.}
   \label{fig:PAEvaryVoltage}
\end{figure}

\begin{figure}
   \centering
  \includegraphics[scale=0.4]{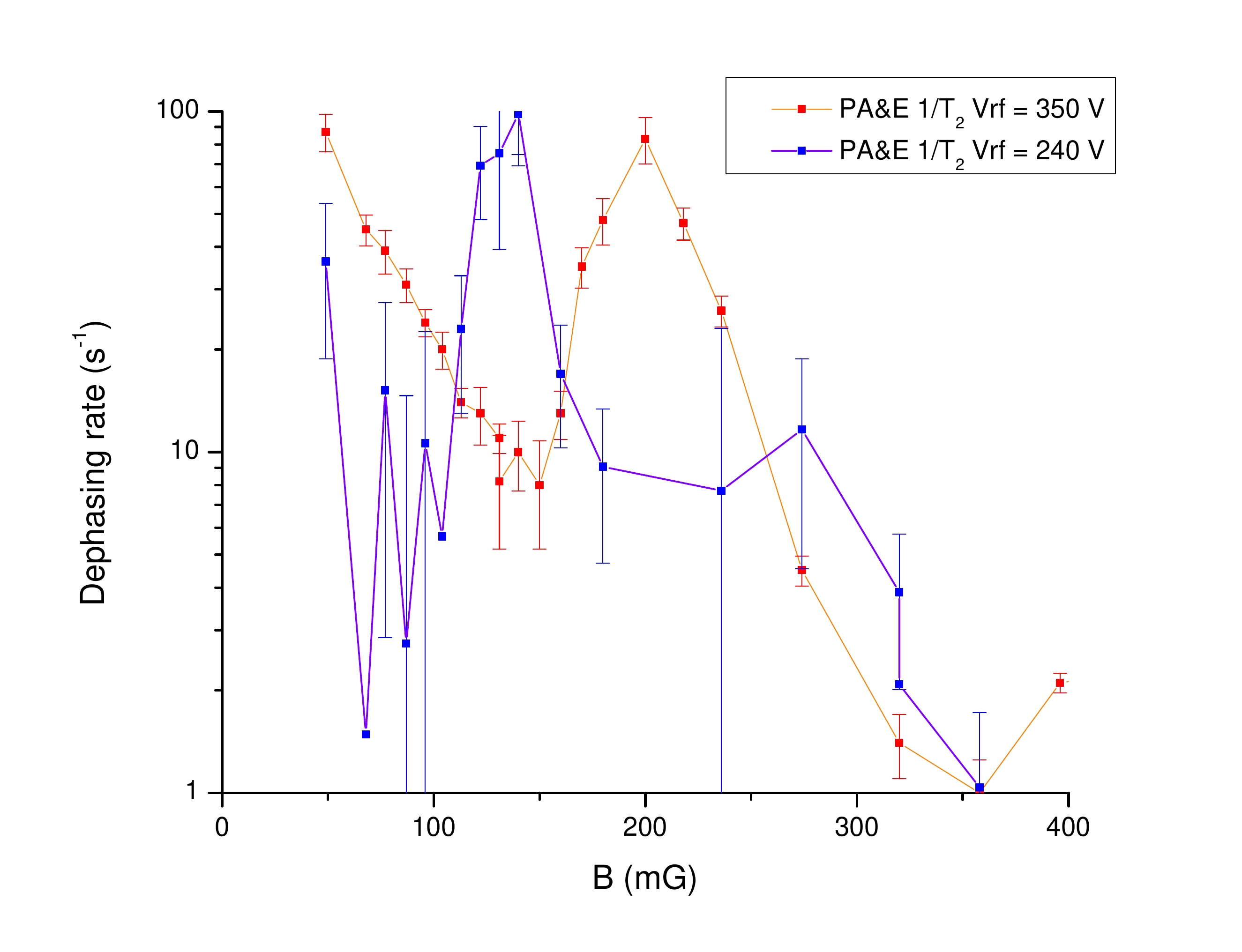} 
   \caption{Logarithmic plot zooming in on the measurement in Fig. \ref{fig:PAEvaryVoltage}.  This was the first package we measured using this method and the data is noisier than for the other packages.  Some error bars have been removed for clarity.}
   \label{fig:PAElog}
\end{figure}

\begin{figure}
   \centering
  \includegraphics[scale=0.7]{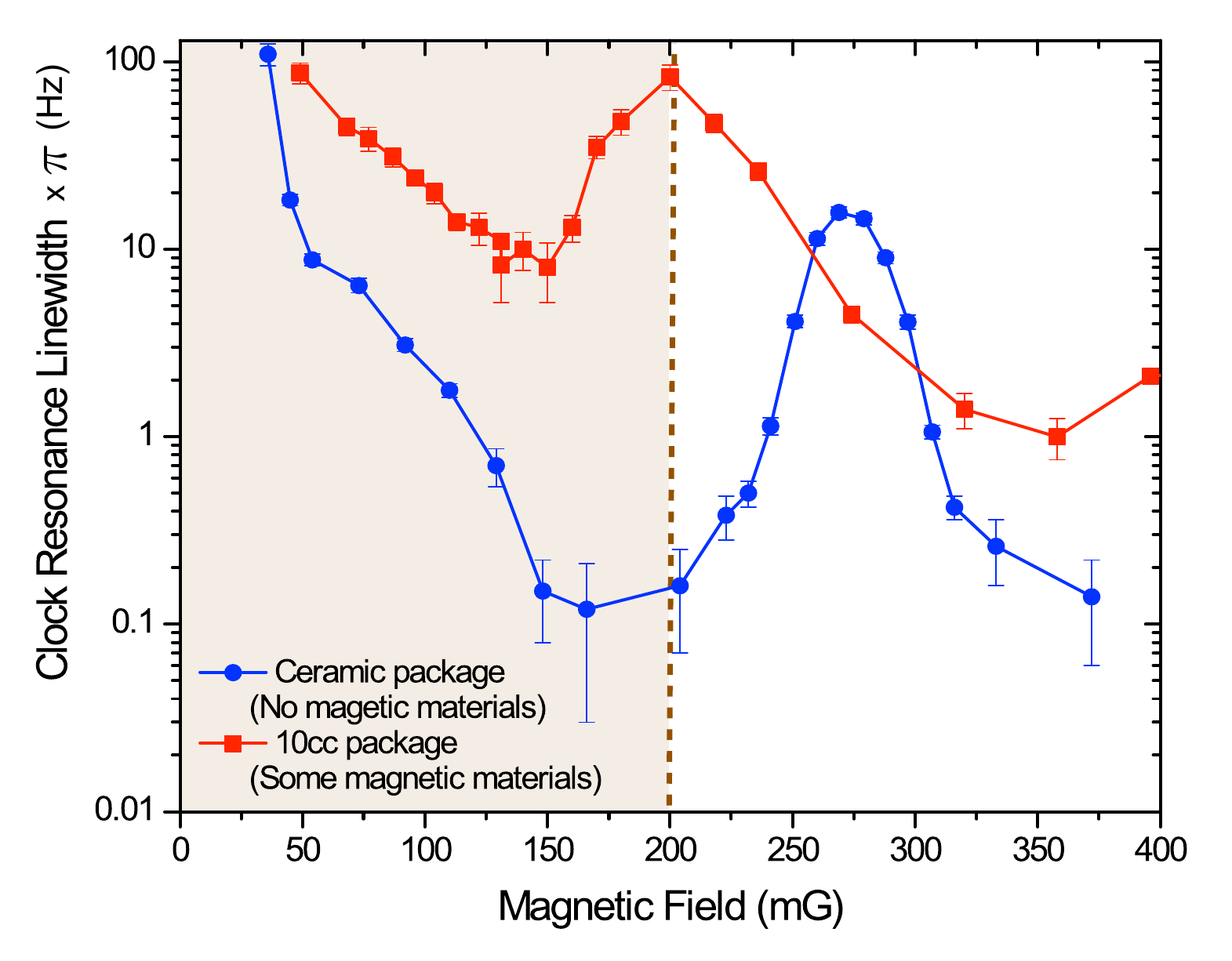} 
   \caption{Comparison of linewidth vs. bias field curves for the LTCC (``ceramic") package with the PA\&E (``10 cc") package demonstrating the difference due to their magnetic properties.  The preferred region for bias operation ($\leq 200$V) is shaded. }
   \label{fig:ComparePAEtestBed}
\end{figure}

In the PA\&E trap we also observed the phenomenon of the moving secular frequency peak as we vary trap voltage.  This measurement is shown in Fig. \ref{fig:PAEvaryVoltage}.  We notice that in this trap, the overall curve has an increased value (see the zoomed-in logarithmic version in Fig. \ref{fig:PAElog}).  This is due to the presence of magnetic materials as discussed earlier.  Although a great effort was made to avoid magnetic materials in the design of this package, the gold coating on the electrical feedthroughs requires a thin film of nickel (which is considerably magnetic), and the feedthrough itself uses a material called Nitronic 50, which is supposed to be a very nonmagnetic alloy form of stainless steel (much better than 304 stainless for example).  It is hard to know which causes the increased broadening.  However, comparing the possible PA\&E linewidths (for varying bias fields) with the possible linewidths of the more nonmagnetic systems (the improved test bed and the LTCC) makes it clear that magnetic materials must be eliminated in order to optimize the performance of the clock.  A plot of the PA\&E data against the test bed data elucidates this in Fig. \ref{fig:ComparePAEtestBed}.

\subsubsection{LTCC/HTCC traps}

\begin{figure}
   \centering
  \includegraphics[scale=0.4]{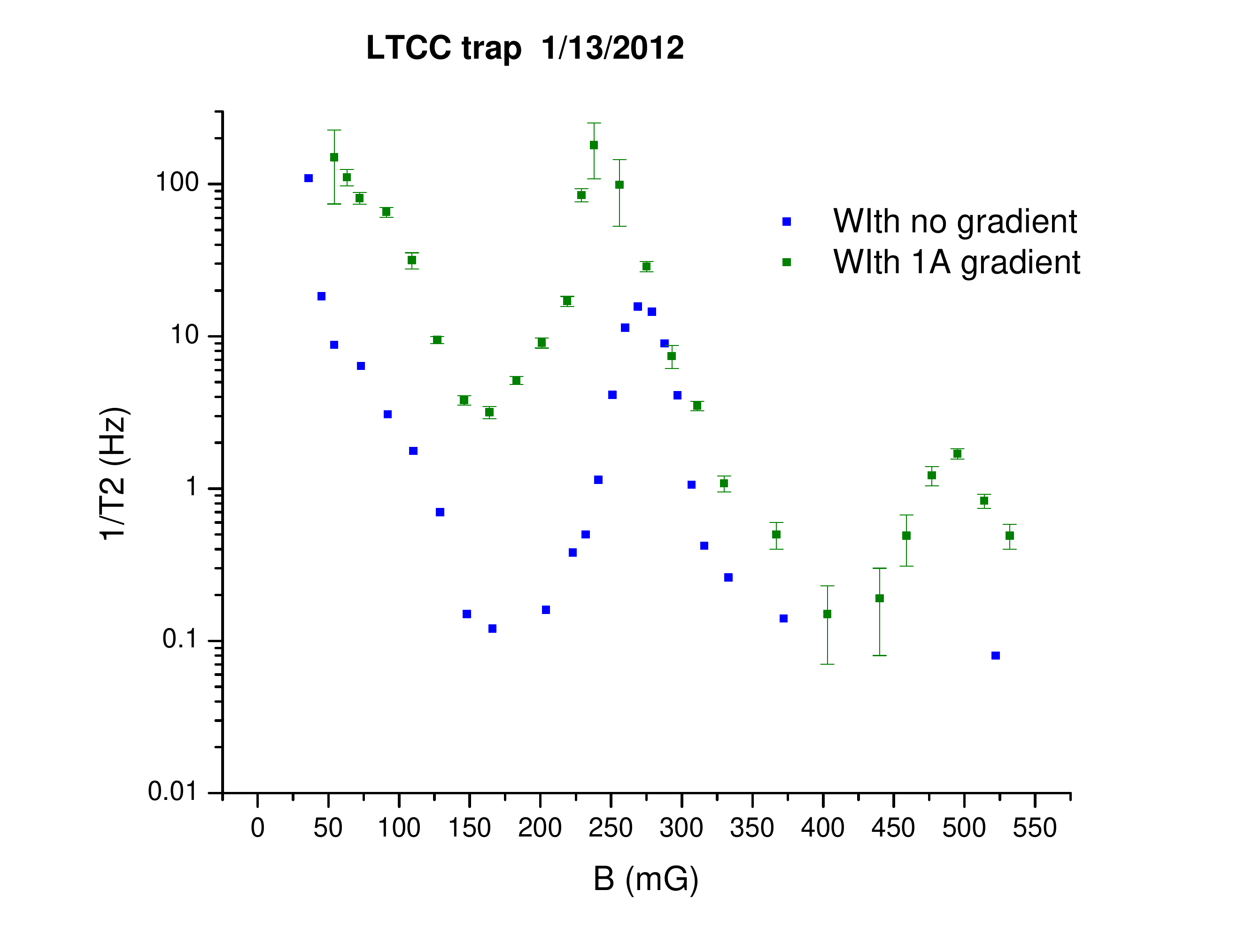} 
   \caption{Measurement in the LTCC package, normally and with an added gradient of 350 mG/cm.}
   \label{fig:LTCC_gradient}
\end{figure}

We were able make a few measurements in the recent  LTCC trap, which has no magnetic materials (Fig. \ref{fig:LTCC_gradient}).  We see that this curve is very well behaved in the nonmagnetic trap, despite the asymmetry in the two transverse axes of the trap geometry.  We do not see two distinct secular frequencies because the frequency values are very close together and are blurred by the decay mechanism.

I have also plotted the curve under the influence of an additional gradient; by adding an artificial gradient to the LTCC trap, we can see how the linewidth in the ``valley" point is quickly degraded.  This is another confirmation of the necessity to avoid magnetic materials.  

%

\subsection{Simulation}

Our simulation has been able to approximate the measurements made in two traps.  As a summary of the process discussed above, a simulation of the potential with the addition of space charge is used to calculate trajectories and obtain the spectrum of motional frequencies in the cloud.  This information is then translated, through information about the gradient, to a decay rate that is a direct indication of linewidth broadening. 

\begin{figure}
  \centering
  \includegraphics[scale=0.6]{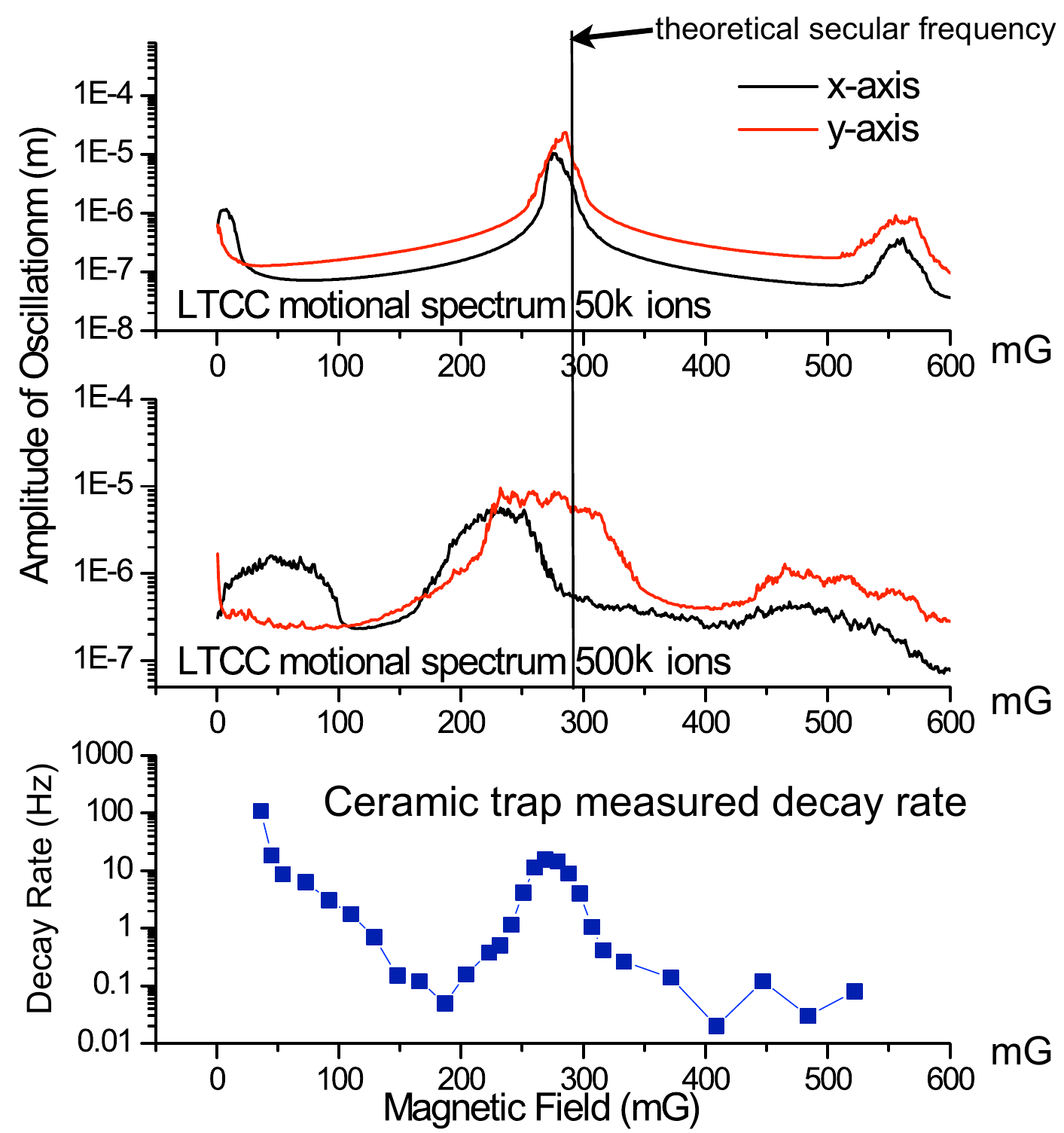} 
   \caption{A comparison of the spectrum calculated in LTCC for different numbers of ions as discussed in the text.  The experimental data is plotted as well for comparison.}
   \label{fig:ExampleSimu}
\end{figure}

 Ion number can be estimated from these curves, based on the reduction of the secular frequency from the theoretical ``empty trap" value, which is due to space charge causing a flattening of the central potential.  An example of the reduction and broadening of the secular frequency peak due to space charge can be seen in Fig. \ref{fig:ExampleSimu}.   When we match the simulation to the experimental data, the field gradient in each direction and the transverse relaxation  decay rate ($\gamma_0$) are unknown.   The width of the motional peaks reflects the amount of relaxation (which broadens the secular frequency resonance) and the height is related to the magnitude of the gradient in the trap. 

\begin{figure}
   \centering
  \includegraphics[scale=0.5]{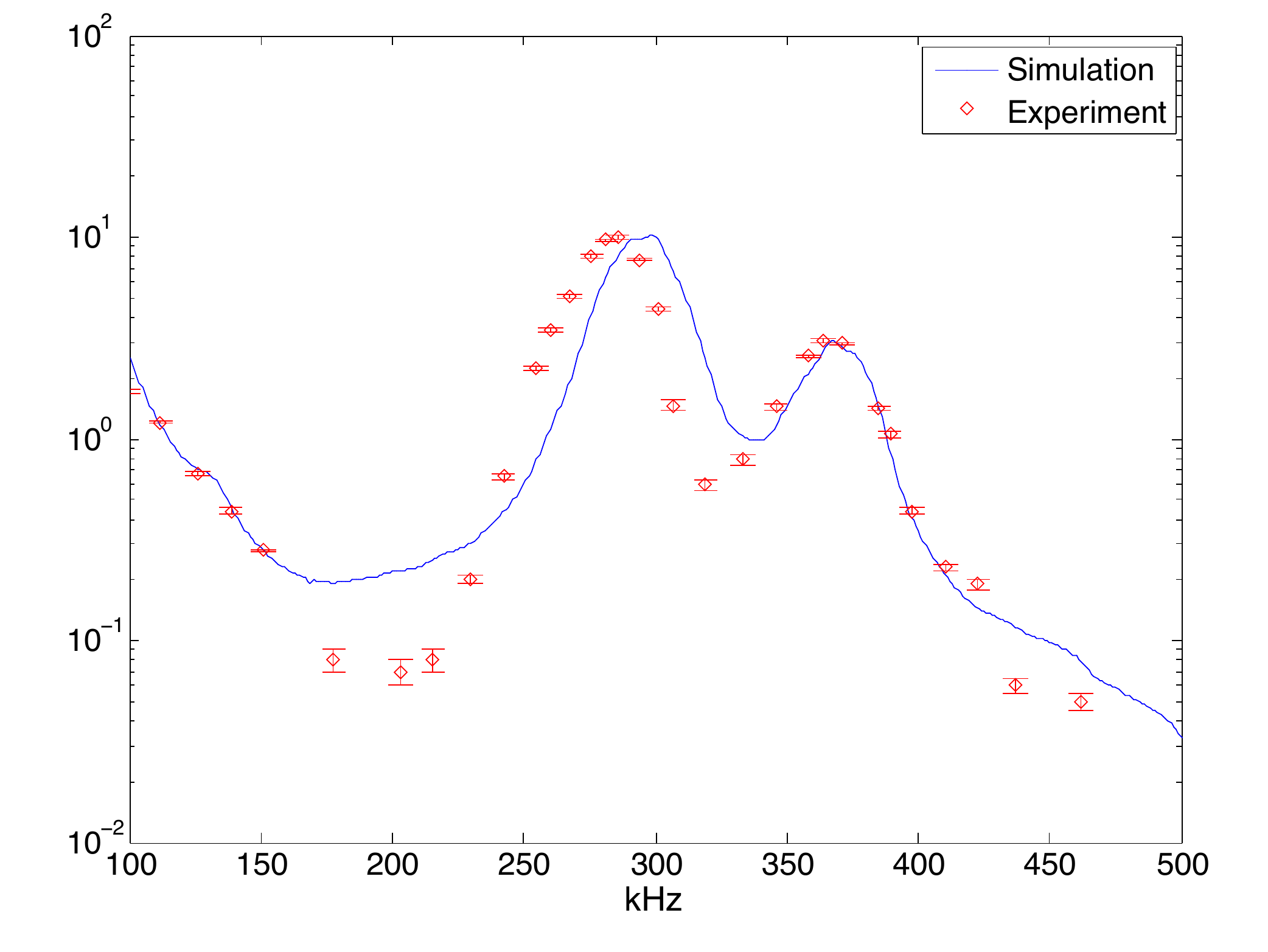}
   \caption{Simulation data for the decay rate we measure compared with experimental data for the test bed trap. }
   \label{fig:TiSimu}
\end{figure}

\begin{figure}
   \centering
  \includegraphics[scale=0.5]{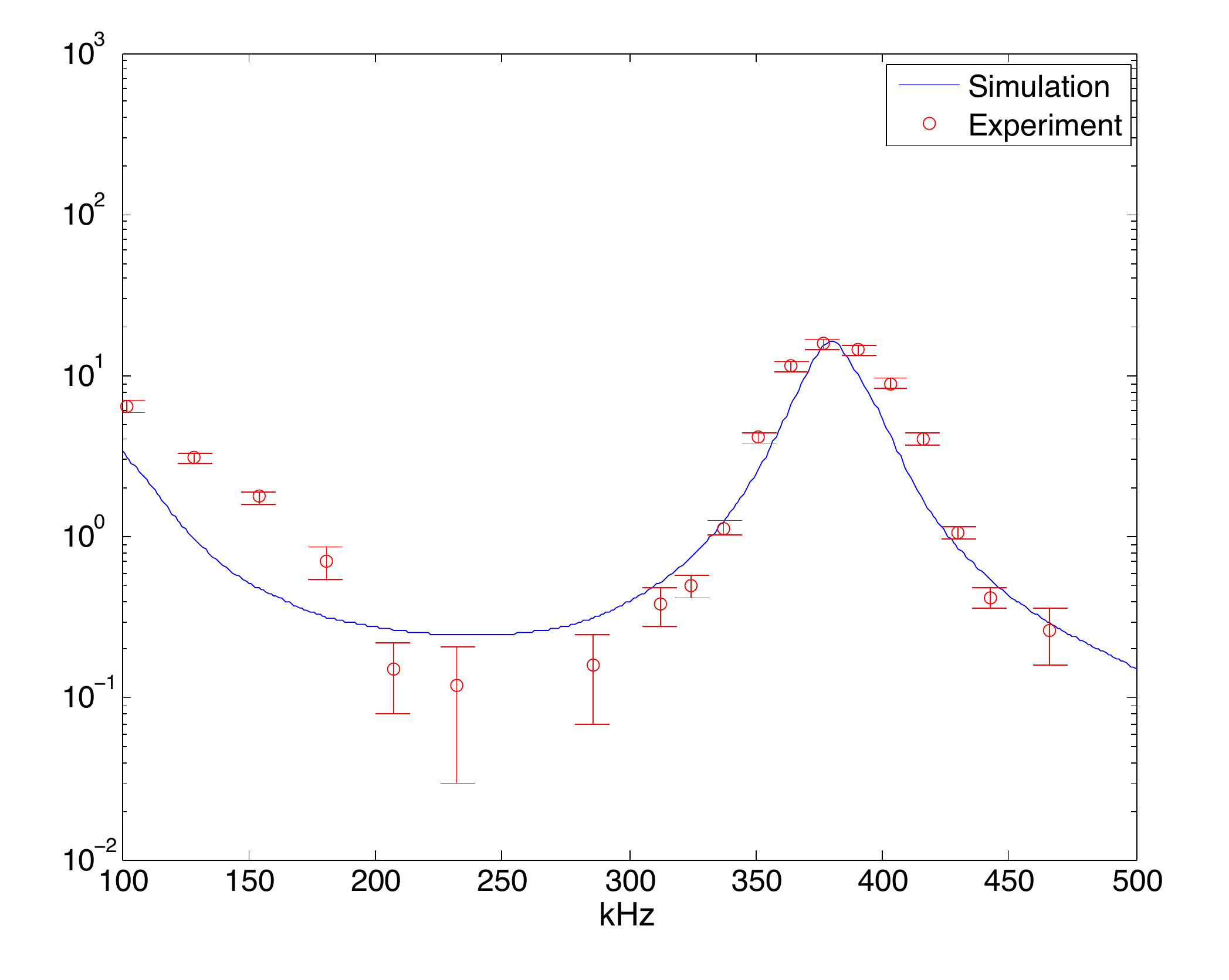}
   \caption{ Simulation data for the decay rate we measure compared with experimental data for the LTCC trap.}
   \label{fig:LTCCSimu}
\end{figure}

Fig. \ref{fig:TiSimu} shows the simulated decay rate calculated from the 3-D trajectory model versus bias field plotted against the experimental data for the test bed trap.  The trap parameters are those used in the experiment, and the number of ions in the trap is estimated to be on the order of 100,000, based on the nominal value of the motional secular frequency peaks. 
Comparison of the model with the experimental data suggests that we have gradients at the location of the ions on the order of 1.4 mG/cm.  The fact that the left-hand peak associated with the transverse secular frequencies is not as well aligned with the experimental data suggests that we had a bias of more than the estimated 2.6 V on the trap, which would cause the peaks to be somewhat more separated.  It can be difficult to measure these experimental values exactly as they are at the location of the trap. 
A similar calculation for the LTCC trap is shown in Fig. \ref{fig:LTCCSimu}.  Again, the trap potential is calculated using the real parameters used in the experiment and the number of ions in this trap is found to be approximately 50,000. 
The gradients in this trap are determined to be on the order of 0.7 mG/cm 
based on the height of the peaks.  In both cases, the qualitative agreement shown here of the simulation with the experiment under the discussed approximations strongly suggests that our understanding of the motionally-induced linewidth broadening phenomena is correct.  A more detailed calculation that aims for better agreement is also underway.

\subsection{Conclusions}

Having a model of the ion motion that can approximately reproduce the linewidth broadening (or decay rate) data not only confirms our hypothesis about the source of the phenomenon, but allows us to explore the parameter space for new traps in order to make good design choices and find optimal operating parameters.  We have shown that operating at a bias field at the ``valley" point between the secular frequency peaks in the linewidth-versus-bias-field picture is the optimal choice for a miniaturized ion clock.  This is true because we are able to reduce sensitivity to field fluctuations by operating at a low bias field, while still being able to operate at a lower power, despite the fact that the secular frequency corresponds to Zeeman frequencies produced by bias fields in the $\leq$ 200 mG bias field range.  Further, we have shown that the narrowest linewidths in the ``valley" region are obtained when magnetic materials are completely eradicated in the manufacturing and processing of the trap and package. 

\chapter{Conclusions} \label{ch:conclusions}

To conclude, I have reported many accomplishments in this thesis.  We have assembled and tested a tabletop ``test bed" clock as a proof of concept for the ytterbium clock based on a buffer-gas-cooled cloud of $^{171}$Yb$^+$ ions in a miniature ion trap.  We have designed and tested many small ion trap vacuum packages both for their ion trapping qualities as well their potential for clock stability.  This progression led to many successes:
\begin{enumerate}
\item{A sealed ion trap package, 10 cm$^3$ in size, with ion lifetime of over 30 days with no active pumps}
\item{A sealed  ion trap package, 3 cm$^3$ in size, which is on track for similar or better performance}
\item{Several small packages that have demonstrated clock stabilities exceeding \\
$10^{-10}\  \tau^{-1/2}$ without magnetic field stabilization or shielding}  
\item{Development of the photoelectron loading technique using Yb coating directly from the trap ovens}
\item{Development and study of a novel trap geometry that is constructed from low- or high-temperature cofired ceramic, which is a nonmagnetic material that solves the vacuum electrical feedthrough problem and is built using a process that is friendly toward mass production}
\item{Development of a process by which the important parameters of the system (magnetic field, laser, and microwave frequency) are stabilized actively during clock interrogation using only the ion signal}
\item{Development of a low-power RF trap driver using CMOS inverters}
\end{enumerate}

Although our clock stability is still limited by magnetic fields, we are mitigating this problem in the current and next phase of the project, which will allow us to significantly improve the potential for long-term stability in these clocks.  

Furthermore, an in-depth simulation of the ion motion in the different traps led to a deeper understanding of the subtleties of using a cloud of ions in an ion trap as a frequency standard.   Understanding these effects is critical to maintaining a narrow linewidth while operating in a low-power environment and avoiding stability drift due to field fluctuations.  This model can be used to interpret and plan for various aspects of the future traps, and has provided insight into the necessary precautions that must be taken in order to preserve long-term stability in the final phases of the project.  

The success of the IMPACT project could lead to the commercial availability of a small, long-term-stable atomic frequency reference that rivals the performance of a cesium beam clock.  Since the cesium beam standard has been the industry workhorse for decades, 
the implications of such a small and low-power technology, if cost-effective, could be game-changing in the timing and frequency community, not to mention its potential influence on telecommunications and GPS.  Although these potentially dramatic influences are still a long way off, this dissertation is my contribution toward that goal.

\appendix



\chapter{Mathematica for processing CPO output} \label{app:mathematica}

The CPO data is imported into Mathematica as an array of 2 or 3 dimensional coordinates, electric field values (for the RF electrodes), and potential values (for the end cap electrodes) for unit voltage.  In Mathematica, the pseudopotential is calculated from the electric field of the RF electrodes and added to the potential of the endcaps to obtain the potential in 3 dimensions for the trap parameters.  Each dimensional or geometrical change requires a new simulation in CPO, but varying RF and end cap voltages can be done in Mathematica.  Then, the potential is plotted in several planes, the depth is calculated along the shallow axis in the transverse plane, and the secular frequency is calculated in the transverse plane.  The stability parameter and the longitudinal secular frequency can also be calculated but these do not provide a good characterization for all geometries.  

\chapter{Matlab for Monte Carlo ion motion simulation} \label{app:matlab}
%
%
%
\lstset{basicstyle=\scriptsize,tabsize=2}
\begin{lstlisting}[language=Matlab]
%%%%%%%%%%%%%%%%%%%%%%%%%%%%%%%%%%%%%%%%%%%%%%%
% densityAndIonNumber.m
% %
% finds the density for a range of specific intitial conditions,
% calculate the number of ions for that initial condition and make a data
% set that maps initial conditions to ion number which we can interpolate
% and use to predict initial condition based on ion number.
%
% uses 
%%%%%%%%%%%%%%%%%%%%%%%%%%%%%%%%%%%%%%%%%%%%%%%%%%

% set number of densities to calculate
indx0 = 1;
indx = 5;
%%
% distance to integrate and plot out to
maxDist = 1e-3;
Temperat = 730;
trapFreq = 2*pi*360e3;
%%
% allocate vectors to initial densitya and corresponding ion number
initDensity = zeros(1,indx-indx0+1);
Number = zeros(1,indx-indx0+1);

%figure(1), hold off

for i = indx0:indx
    
% set initial density (density at cloud center)
initDens = 1e14+i*1e14

% set initial conditions for ODE solver (init. density above and zero slope)
init = [initDens 0];


% calculate the density

setTolerances = odeset('Reltol',1e-14,'Abstol',1e-14);%,'NonNegative',[1]);


[t,x] = ode45(@(t,r)NumDensity3(t,r,Temperat,trapFreq),[1e-10 maxDist], init, setTolerances);


% plot all of them together
% convert aaxes to ions per mm^3 vs mm
figure (2)
plot(t/1e-3,x(:,1)/1e9), axis([0 maxDist/1e-3 0 initDens/1e9])
xlabel('radial distance from center [mm]') 
ylabel('density [ions/mm^3]')
hold on
%%

% Integrate over the newly found density profile to see the number of ions

q = x(:,1);

tic
[Q,fn] = quad(@(ti)intIonNumberFun(ti,t,q),1e-10,maxDist,1e-5);
toc

Length = 5e-3;

% save number and initial density in corresponding vectors
Number(i-indx0+1) = 2*pi*Q*Length;
initDensity(i-indx0+1) = initDens;

end

combo = [Number/1e4;initDensity/1e13]'

% % plot initial condition vs. number of ions
%  figure(2), plot(Number, initDensity)
\end{lstlisting}

\lstset{basicstyle=\scriptsize,tabsize=2}
\begin{lstlisting}[language=Matlab]
%%%%%%%%%%%%%%%%%%%%%%%%%%%%%%%%%%
% function for the ODE that is solved to find density
%%%%%%%%%%%%%%%%%%%%%%%%%%%%%%%%
function dr = NumDensity3(t,r,Temp,omega)


kb = 1.38e-23; %[J/K]
eps0 = 8.85e-12; %[C/V]
charge = 1.602e-19; %[C]
mass = 0.171 / (6.02*10^23); % 199Hg+
% Temp = 500; % [K]
% omega = 2*pi*62e3; % [rad/s]

n0 =  2*eps0*mass*omega^2 / (charge^2); % 8.4224e+14 for Yb at 330 kHz
lambdaDsq = kb*Temp / (2*mass*omega^2);


dr(1) = r(2);
dr(2) = r(2)^2/r(1) - r(2)/t - r(1)*(n0-r(1))/(n0*lambdaDsq);


dr = [dr(1);dr(2)];

\end{lstlisting}

\lstset{basicstyle=\scriptsize,tabsize=2}
\begin{lstlisting}[language=Matlab]

 %%%%%%%%%%%%%%%%%%
 % Interpolates to
 % find the central density for a particular "real" ion number and then
 % recomputes the density function, makes it into a distribution (and reflects
 % about the axis) and normalizes it.  The outcome is a real probability
 % distribution for ion number in the x and z directions.  
%% 
maxDistSpec = 1e-3;
RealIonNumber = 5e4; 
Temperat = 730;
trapFreq = 2*pi*405e3;
Length = 5e-3;

setTolerances = odeset('Reltol',1e-20,'Abstol',1e-20);

InitialDensity  = interp1(Number, initDensity, RealIonNumber)
[t0,x0] = ode45(@(t,r)NumDensity3(t,r,Temperat,trapFreq),[1e-10 maxDistSpec], [InitialDensity 0], setTolerances);

figure (3)
semilogy(t0/1e-3,x0/1e9), axis([0 maxDistSpec/1e-3 0 InitialDensity/1e9])
xlabel('radial distance from center [mm]') 
ylabel('density [ions/mm^3]')

distribution = x0(:,1); %[[flipud(x0(:,1))];[x0(:,1)]];
radial = t0; %[[-1.*flipud(t0)];[t0]];
%figure(3),plot(radial,distribution)

% check if integrating this chosen distribution gives the total ion number
% we asked for
Q = quad(@(ti)intIonNumberFun(ti,t0,x0(:,1)),1e-10,maxDistSpec,1e-5);
Numbercheck = 2*pi*Q*Length

figure(12),plot(t0,x0), axis([0 maxDistSpec 0 InitialDensity])
hold on

% integrate over the distribution and normalize it.
% normConst = quad(@(ri)intDistFun(ri,radial,distribution),1e-10,maxDist,1e-6);
% ionDistribution = distribution/normConst;
% one = quad(@(ri)intDistFun(ri,radial,ionDistribution),1e-10,maxDist,1e-6);


\end{lstlisting}

\lstset{basicstyle=\scriptsize,tabsize=2}
\begin{lstlisting}[language=Matlab]
%%%%%%%%%%%%%%%%%%%%%%%%
% Create cumulative distribution function from the density for drawing initial conditions
%%%%%%%%%%%%%%%%%%%%%%%%
% generate CDF, create fn from 0,1 to CDF to PDF
%%%%%%%%%%%%%%%%%%%%%%%

maxDist = 1e-3;

totalRadius = maxDist;
radius = 1e-7:1e-7:totalRadius;  % create a fixed step of radius

% interpolate and resample the density in uniform step size
ioninterp = interp1(radial, distribution, radius); 
% Make the unnormalized probability distribution by multiplying by the new-stepped radius
ionDistinterp = radius.*ioninterp;

% % plot the probability distribution


 
% find the area under the density curve to normalize to a distribution
normConst = quad(@(ri)intDistFun3(ri,radius,ionDistinterp),1e-7,maxDist-1e-7,1e-6);

ionDistnormalized = ionDistinterp/normConst;

 figure(1),hold off
 plot(radius, ionDistinterp/1e7, 'r')

% double check that the area under the new curve is one
one = quad(@(ri)intDistFun3(ri,radius,ionDistnormalized),1e-7,maxDist,1e-6)

CDF = cumtrapz(ionDistnormalized)*1e-7;
figure(2),hold off, plot(radius, CDF,'g')

%%
% for i = 1:1e5;
% % test the CDF
% y = sum(rand < CDF);%/length(CDF);
% yy = radius(y+1)-1e-7;
% %hist(y)
% yyy(i) = yy;
% end
% 
% figure(3), hold off
% hist(yyy,100)


%%
% use this to find what element of the CDF stops having unique values (by changing lim below), then
% scale rand by that value, and cut both the radius and the CDF off at
% that element
lim = 6000;
CDF1 = CDF(1:lim);%3053
radius1 = radius(1:lim);
randRadius = interp1(CDF1, radius1, CDF(lim)*rand(1,1e6));
figure(3), hold off, hist(randRadius,100),xlim([0 1e-3]), hold on
plot(radius, ionDistinterp/5.4e5, 'r')

\end{lstlisting}

\lstset{basicstyle=\scriptsize,tabsize=2}
\begin{lstlisting}[language=Matlab]
%%%%%%%%%%%%%%%%%%%%%%%%%%
% Calculates the space charge using the density and adds it to the potential.  The offset is changed manually to make the corrected potential line up with a harmonic oscillator after reaching a certain radius, to prevent numerical errors at the edges.  The potential is plotted against the space charge correction, and the center point of the space charge can be adjusted for potentials that are not symmetric.  The whole corrected potential is saved.
%%%%%%%%%%%%%%%%%%%%%%%%%%
% NOTE: if there is some wigglies between the space charge peak and the
% zero values you probably need to reduce maxDistSpec in the previous file
% (selectDensity.m)

mass = 2.84e-25;
trapFreq = 2*pi*405e3;
kb = 1.38e-23;
Temp = 730;
charge = 1.602e-19;

%Potfine = SHOpotential3D;
% have to find offset manually by running with non-offset extranr and
% PotSHO and find the difference in y for a point far from center
offset = 0.113e-19;%0.87e-19;%0.9660e-19;%0.94e-19;%0.104e-19;%0.009e-19;;%0.111e-19;
offsetx = 0e-4;
offsetz = 0e-4;

% create SHO-based density for values of r outside our boundary from
% "selectdensity.m"
extraRadius = maxDistSpec:0.0001e-3:3.0e-3;
%extranr = InitialDensity*exp(-(1/2*mass*trapFreq.^2.* extraRadius.^2)/(kb*Temp));
extranr = InitialDensity*exp(-(1/2*mass*trapFreq.^2.* extraRadius.^2-offset)/(kb*Temp));

% concatenate the real density values with the SHO density
radialRad = [0; radial; extraRadius(2:length(extraRadius))'];
radialPot = [InitialDensity; distribution; (extranr(2:length(extranr)))'];

% plot the combined version
figure(1)
plot(radialRad, radialPot), xlim([maxDistSpec - 0.01e-3,maxDistSpec + 0.01e-3])

%%

% now interpolate in r and pull values for all the xyz points
densitynr = getnri(Xfine,Zfine,radialRad,radialPot);

% plot this on top of the radial version
hold on
plot(Zfine(:,26,61), densitynr(:,26,61),'g:')

% find total potential from the density
PotTOT = -(kb*Temp) * log(densitynr/InitialDensity);
figure(2)
plot(Zfine(:,26,61), PotTOT(:,26,61))

% find an SHO potential with the named frequency
% PotSHO = (getpotSimple(Xfine,Zfine,mass,trapFreq,0));
PotSHO = (getpotSimple(Xfine,Zfine,mass,trapFreq,0)-offset);

% plot this on top of the total potential
hold on
plot(Zfine(:,26,61), PotSHO(:,26,61),'r:')

% shift the SHO down and subtract to get the space charge potential
PotSC = PotTOT - PotSHO;

% Convert the space charge to eV for plotting and adding to the CPO values
PotSCeV = PotSC/charge;

%% this cell helps you center the estimated space charge on the real potential
% modify the PotSCeV2D potential to line up its center with the real
% potential center, which may not be at (0,0) because of an asymmetry
% Do this by permuting the matrix, since we know all around the sides is
% zero anyway and we don't have to shift a long way. We are limited by the
% resolution of the grid, which could in theory always be interpolated and
% resampled for more resolution.  But this gets us pretty close (much
% better than without any correction). In 121 grids each step is 3e-5 m.
shiftCorrection = [-6 0 0];
PotSCeVmod = circshift(PotSCeV,shiftCorrection);

% plot this on its own for better scaling
figure(3)
%hold off
% plot(Zfine(:,26,61), PotSC(:,26,61),'g')

Contos1 = [0.2 0.4 0.6 0.8 1 2 3 4 5 6];
Contos2 = [0.02 0.04 0.06 0.08 0.1 0.12 0.14 0.16 0.18 0.2 0.25 0.3];
contour(squeeze(Xfine(:,26,:)),...
squeeze(Zfine(:,26,:)),squeeze(PotSCeVmod(:,26,:)),Contos2,'r')
hold on
% figure(6)
contour(squeeze(Xfine(:,26,:)),...
squeeze(Zfine(:,26,:)),squeeze(Potfine(:,26,:)),Contos1,'b')

% plot the surface
% surf(squeeze(Xfine(:,26,:)),squeeze(Zfine(:,26,:)),squeeze(PotSCeV(:,26,:)))

% add to the CPO potential
WholePotential = PotSCeVmod + Potfine;

% plot the CPO potential and the whole potential on the same graph, in 1D
figure(4)
plot(Zfine(:,26,61), Potfine(:,26,61),'g')
hold on
plot(Zfine(:,26,61), WholePotential(:,26,61),'p:')


\end{lstlisting}

\lstset{basicstyle=\scriptsize,tabsize=2}
\begin{lstlisting}[language=Matlab]

%%%%%%%%%%%%%%%%%%%%%%%%%%%%%%%%%%%%%%%%%%%%%%%%%%%%%%%%%%%%%%%%%
% TrajectoryPowerSpectrumLoopV3.m  
% 2 January 2012
% 
% This script will run the TrajectoryPowerSpectrum.m in a loop, getting
% different initial conditions each time and then averaging the power
% spectra at the end.
% 
% This script calls the function POT3Dmod.m
% also, POT3Dmod.m requires the modified interpolation function interp3HP.m
%
%Initial Conditions can be chosen uniformly and scaled or drawn from distributions.    
%%%%%%%%%%%%%%%%%%%%%%%%%%%%%%%%%%%%%%%%%%%%%%%%%%%%%%%%%%%%%%%%%%

% Potfine = SHOpotential3D;

%%%%%%%%%%%%%%%%%%%%%%%
% DATA - Open the 3-D potential generated using CPO and Mathematica and the
% CDF for selecting from the density

Potential = WholePotential;

%%%%%%%%%%%%%%%%%%%%%%%
% Constants

kb = 1.38e-23; % Boltzmann's constant [J/K]
mass = 2.84e-25; % 171Yb+ [Kg]
charge = 1.602e-19; % excess electron [C]

%%%%%%%%%%%%%%%%%%%%%%%%
% Other values

Temp = 730; % temperature of ions [K]

numSpectra = 15; % number of spectra to average (# of passes through loop)

sampleTime = 1e-7; % uniform sampling for the fft
tfinal = 1e-3; % time length for each trajectory

t0 = 0;  % start from zero
timeListLength = round(tfinal/sampleTime) + 1; % number of values in each trajectory
time = (0:sampleTime:tfinal)';  


% Allocate arrays for storing initial conditions, position/velocity data, energy, and power
% spectrum data
InitialCondslist = zeros(7, numSpectra);
Trajlist = zeros(timeListLength, 6, numSpectra);
Energylist = zeros(timeListLength, 3, numSpectra);
Spectrumlist = zeros(timeListLength, 3, numSpectra);


tic % time the whole loop for all i's

parfor i = 1:numSpectra

    
    % spit out the loop number
    % i
    
   
    
%%%%%%%%%%%%%%%%%%%%%%%%
% Get initial values from distributions (intialConditions.m)

%%%%%
% Velocity distribution

% distribution parameters
meanVelocity = 0;
sigmaVelocity = sqrt(kb*Temp/mass);

% get 1 value for each velocity component, put in a vector, readout
velocities = meanVelocity + sigmaVelocity.*randn(1,3);

% calculate the speed (magnitude)
vMag = sqrt(velocities(1).^2+velocities(2).^2+velocities(3).^2);

% plot histograms of velocity component and speed distributions
% figure, hist(velocities(1),40)
% figure, hist(vMag,40)

%%%%%
% Position distribution
% For now, just draw from a gaussian distribution, leaving these separate
% since the dist is diff't in diff't directions until I get the real thing.
% distribution parameters - different for y vs (x,z)
% meanPositionxz = 0;
% sigmaPositionxz = 0.3e-3;

meanPositiony = 0;
sigmaPositiony = 3.0e-3;

randRadius = interp1(CDF1, radius1, CDF1(lim)*rand(1));
randTheta = 2*pi*rand(1);
randZ = meanPositiony + sigmaPositiony.*randn(1);


% get 1 value for each position component
px = randRadius*cos(randTheta);
py = randZ;
pz = randRadius*sin(randTheta);

% define position vector in m and readout the positions in mm
positions = [py pz px];
% positionsinMM = [px py pz]*1e3;

 % plot histogram of position distribution
 % figure, hist(p2,40)


%%%%%%%%%%%%%%%%%%%%%%%%%%
% Get the 3D ion trajectory (TrajectoryFine.m)

%%%%%
% Calculate an ion trajectory and plot in 3D and 1D




% inital coordinates (m) and velocities (m/s)
% x0 = px; y0 = py; z0 = pz;
% vx0 = vx; vy0 = vy; vz0 = vz;
initConds = [positions velocities];

% derivative interval size (m)
int = 3e-5;

% solve the DE
tolerance = odeset('Reltol',1e-4,'Abstol',1e-6);
[t,x] = ode45(@(t,r)POT3Dmod(t,r,Yfine,Zfine,Xfine,Potential,int),...
    [t0 tfinal],initConds,tolerance);



% check if the trajectory terminates before the full time, by looking to
% see if the there are NaNs in the x vectors of the trajectory (by checking
% whether "isnan" returns any ones).  If the ion goes out of the potential
% given, "Ion left trap" is printed and this trajectory is ignored (not
% recorded)

if isequal(isnan(x),zeros(size(x))) == 0, disp('Ion left trap')
    continue
end 



% plot the trajectory in 3D
figure(1), plot3(1e3*x(:,1),1e3*x(:,2),1e3*x(:,3))
% view([270,0]) % show the xz (transverse) trap plane
xlabel('y [mm]')
ylabel('z [mm]')
zlabel('x [mm]')
% axis equal

% plot the components of the trajectory in x,y,z vs. time
figure(2)
hold off
plot(t,1e3*x(:,1),'b')
hold on
plot(t,1e3*x(:,2),'g')
plot(t,1e3*x(:,3),'r')
legend('y','z','x')
xlabel('time [s]')
ylabel('1-D position [mm]')







%%%%%%%%%%%%%%%%%%%%%%%%%%%%
% Interpolate and sample from the trajectory in order to get a uniform step
% size for the fft (and for storage, etc.)




% choose the sampling time by determining the fixed step size of time for the interpolation
% and make it a column
% time = (0:sampleTime:tfinal)';  
% interpolate and pick out the values that match the fixed-step time series
x1interp = interp1(t, x(:,1),time); 
x2interp = interp1(t, x(:,2),time);
x3interp = interp1(t, x(:,3),time);
xallinterp = [x1interp x2interp x3interp];



% do the same thing for velocities
v1interp = interp1(t, x(:,4),time); 
v2interp = interp1(t, x(:,5),time);
v3interp = interp1(t, x(:,6),time);
vallinterp = [v1interp v2interp v3interp];

% % Plot the integrated solution and the interpolated sampling to compare
% figure(8), plot(t, x(:,4)) % plot the original version with varying time stepsize
% figure(9), plot(time, v1interp) % plot the interpolated sampled version 


% Use interpolated sampled data for calculating energy
% calculate the total energy, in eV (this is why the divide by charge)
KE = (1/2)*mass*(v1interp.^2+v2interp.^2+v3interp.^2)/charge;
PE = charge*interp3HP(Yfine,Zfine,Xfine,Potential,x1interp,x2interp,x3interp)/charge;
totalEnergy = KE + PE;

% % plot the KE, PE, and total energy on the same plot
% figure(3)
% hold off
% plot(time,KE,'m')
% hold on
% plot(time,PE,'c')
% plot(time, totalEnergy,'b')
% legend('KE','PE','Total')
% xlabel('time [s]')
% ylabel('energy [eV]')


%%%%%%%%%%%%%%%%%%%%%%%%%%%%
% Get the powerspectrum and plot it (powerspectrumscript.m)

M = xallinterp;

freq = (1/sampleTime)* 1/(length(M)-1) * (0:length(M)-1);
FM = fft(M)/length(M);


% % mheight=max(abs(FM));
% figure(4), plot(freq/10^3,2*abs(FM)), axis([0 500 0 2e-3]),xlabel('kHz')
% figure(5), semilogy(freq/10^3,2*abs(FM)), axis([0 500 0 2e-3]),xlabel('kHz')
% % this is for absolute value

%%%%%%%%%%%%%%%%%%%
% Put relevent data from this loop into arrays

% notice that "time" is the interpolated timestep that is appropriate for
% plotting these against (not "t").  "time" will be saved after the loop;
% it is the same for all iterations in each run of this program becasue it
% is based on sampletime and tfinal.

% Store the initial conditions (3 pos, 3 vel, plus velocity magnitude)
InitialCondslist(:,i) = [initConds'; vMag];
% Store the raw value of the FFT matrices in an array, these values are
% divided by length of M but still complex; I take the abs and mult. by 2
% after in case I want the complex values sometime.
Spectrumlist(:,:,i) = FM;
% Store the interpolated sampled trajectory information which includes 
% 3 positions and 3 velocities.
Trajlist(:,:,i) = [xallinterp vallinterp];
% Store the energy values; include time, PE, KE, total
Energylist(:,:,i) = [KE PE totalEnergy];

    % readout initial conditions to keep tabs on it
    %  InitialCondslist' 

end

toc % time required for all i loops


%%%%%%%%%%%%%%%%%%%%%%%%%%%%%%%%%
% Now average the power spectra that were stored

ionsSurvived = length(find(InitialCondslist(1,:)))
ionsLeftTrap = numSpectra - ionsSurvived

FMavgGen = sum(2*abs(Spectrumlist),3) ./ numSpectra; % convert to 2*abs and average

freq1 = (1/sampleTime)* 1/(timeListLength-1) * (0:timeListLength-1);
% plot the average of the spectra, linear and log scale
% figure(6), plot(freq1/10^3,FMavgGen), axis([0 500 0 2e-3]),xlabel('kHz')
figure(7), semilogy(freq1/10^3,FMavgGen), axis([0 1000 0 2e-3]),xlabel('kHz')


% save the initial conditions, the trajectories, the power spectrum, the
% energies and the averages

% This will save the 3-d arrays that contain the data from loop i in
% (dim 3) column "i" (except initial conditions, which are a matrix where
% the i-th column contains the conditions from loop i) and the time
save test_2_Ti_360_120kions_730deg_3D_Vbias2.3V InitialCondslist time Spectrumlist Trajlist Energylist




\end{lstlisting}

\lstset{basicstyle=\scriptsize,tabsize=2}
\begin{lstlisting}[language=Matlab]

% This function is a modification of my original Pot3D ODE function which gets
% the potential during a trajectory by interpolating.  This file calls interp3HP which is my
% modified interp3D which causes the trajectory to just stop when it
% reaches the edge of the given potential (the interpolation region)

function dr = POT3Dmod(t,r,x,y,z,v,int)

charge = 1.602e-19;
mass = 2.84e-25;

dr(1) = r(4);
dr(2) = r(5);
dr(3) = r(6);
dr(4) = -(charge/mass)*(interp3HP(x,y,z,v,(r(1)+int/2),r(2),r(3))...
-interp3HP(x,y,z,v,(r(1)-int/2),r(2),r(3)))/int;
dr(5) = -(charge/mass)*(interp3HP(x,y,z,v,r(1),(r(2)+int/2),r(3))...
-interp3HP(x,y,z,v,r(1),(r(2)-int/2),r(3)))/int;
dr(6) = -(charge/mass)*(interp3HP(x,y,z,v,r(1),r(2),(r(3)+int/2))...
-interp3HP(x,y,z,v,r(1),r(2),(r(3)-int/2)))/int;

dr = [dr(1); dr(2); dr(3); dr(4); dr(5); dr(6)];



\end{lstlisting}

\lstset{basicstyle=\scriptsize,tabsize=2}
\begin{lstlisting}[language=Matlab]
%%%%%%%%%%%%%%%%%%%%%%%
% Rotates the spectrum around an angle to separate principle axes for applying individual gradients on different dimensions.
%%%%%%%%%%%%%%%%%%%%%%%

% Rotate axes for FFT along some angle A

Adeg = -45;
A = pi/180 * Adeg;

RotationA = [cos(A) 0 sin(A); 0 1 0; -sin(A) 0 cos(A)];

% for(i=1:size(InitialCondslist,2))
% tt(:,i) = time;
% end
xx = squeeze(Trajlist(:,3,:));
yy = squeeze(Trajlist(:,1,:));
zz = squeeze(Trajlist(:,2,:));
TrajCat = cat(3,xx,yy,zz); 
% for i = 1:size(InitialCondslist,2)
%     TrajCat(:,:,i) = [Trajlist(:,1,i) Trajlist(:,2,i) time];
% end
        
   

ShiftedTrajCat = shiftdim(TrajCat,2);
ShiftedTrajRot = zeros(size(ShiftedTrajCat));

for i = 1:size(InitialCondslist,2)

ShiftedTrajRot(:,:,i) = RotationA*ShiftedTrajCat(:,:,i);

end


TrajRot = permute(ShiftedTrajRot,[2 1 3]); 
time;

%%

%%%%%%%%%%%%%%%%%%%%%%%%%%%%%%
% This part does the FFT and average

sampleTime = 1e-7; % uniform sampling for the fft
tfinal = 1e-3; % time length for each trajectory
timeListLength = round(tfinal/sampleTime) + 1;

FMlist = zeros(size(TrajRot));

for j = 1:size(InitialCondslist,2)
    M = TrajRot(:,:,j);
    FMlist(:,:,j) = fft(M)/length(M); % now FMlist is just like Spectrumlist
end

ionsSurvived = length(find(InitialCondslist(1,:)))
FMavgGen = (sum(2*abs(FMlist),3) ./ ionsSurvived); % convert to 2*abs and average
freq1 = ((1/sampleTime)* 1/(timeListLength-1) * (0:timeListLength-1))';

% plot the average of the spectra, linear and log scale
% figure(6), plot(freq1/10^3,FMavgGen), axis([0 500 0 2e-3]),xlabel('kHz')
hold off
figure(7), semilogy(freq1/10^3,FMavgGen,':'), axis([0 500 0 2e-3]),xlabel('kHz')
hold on

FMSum = FMavgGen(:,1) + FMavgGen(:,2) + FMavgGen(:,3); 
semilogy(freq1/10^3,FMSum)




\end{lstlisting}

\lstset{basicstyle=\scriptsize,tabsize=2}
\begin{lstlisting}[language=Matlab]
%%%%%%%%%%%%%%%%%%%%
% Calculates the decay rate from the spectrum for a certain gamma_0 and specified gradients in each direction and plots the experiment vs the simulation
%%%%%%%%%%%%%%%%%%%%%%%%%%%%
% load a dataset, then get the spectrum as a function of frequency.
% multiply by some gradient to get magnetic field spectrum as a fn of
% frequency.
% Assemble the expression for the decay rate in a separate function. 
% integrate over the function for some B0
% make a for loop to integrate for many B0's
% plot decay rate vs B0

gamma0 = 8.5e3;

linearGradientX = 0.065; %gauss/meter. this = 100 mG/cm, a somewhat realistic number
linearGradientY = 0.065;
linearGradientZ = 0.015;

% load('100_TiAssVDC_RFBias3V_450kHz_100Kions_730degrees_CDF.mat')


sampleTime = 1e-7; % uniform sampling for the fft
tfinal = 1e-3; % time length for each trajectory
timeListLength = round(tfinal/sampleTime) + 1; % number of values in each trajectory
% time = (0:sampleTime:tfinal)';  
ionsSurvived = length(find(InitialCondslist(1,:)));

%FMavgGen = sum(2*abs(Spectrumlist),3) ./ ionsSurvived;

FMavgGenX = FMavgGen(:,3);  %sum(2*abs(Spectrumlist),3) ./ ionsSurvived; % convert to 2*abs and average
FMavgGenY = FMavgGen(:,1);
FMavgGenZ = FMavgGen(:,2);
%freq1 = (1/sampleTime)* 1/(timeListLength-1) * (0:timeListLength-1);

% % plot the average of the spectra, linear and log scale
% figure(4)
% hold off
% semilogy(freq1/10^3,FMavgGenX), axis([200 600 0 2e-3]),xlabel('kHz')
% hold on
% semilogy(freq1/10^3,FMavgGenY)
% % semilogy(freq1/10^3,FMavgGenZ)

RFspectrumX = linearGradientX*FMavgGenX;
RFspectrumY = linearGradientY*FMavgGenY;
RFspectrumZ = linearGradientZ*FMavgGenZ;

figure(4)
hold off
semilogy(freq1/10^3,RFspectrumX,':'), axis([0 500 0 2e-3]), xlabel('kHz')
hold on 
semilogy(freq1/10^3,RFspectrumY,':')
semilogy(freq1/10^3,RFspectrumZ,':')

RFspectrumTOT = RFspectrumX + RFspectrumY + RFspectrumZ;
semilogy(freq1/10^3,RFspectrumTOT,'r')
%%

% Now integrate over the B(omega) for some omega0 and gamma0
% number of points in the gamma evaluation
%points = 100;
% initialize decay as a function of frequency and list for storing chosen frequencies
decay = zeros(400,1);%decay = zeros(points,1);
freqlist = zeros(400,1);%freqlist = zeros(points,1);


tic
for k = 1:500
    omega1 = 0e3+1*k*1e3;
    gamma1 = quad(@(omegai)...
IntDecayFun(omegai,freq1,RFspectrumTOT(:,1),omega1,gamma0),0e3,500e3);
    
    freqlist(k,1) = omega1;
    decay(k,1) = gamma1;
    

end
toc


% plot the decay vs frequency (freqlist)

figure(2)
hold off
semilogy(freqlist/10^3,decay,'b'), xlim([100 500]), xlabel('kHz')
hold on

semilogy(BLTCC/1000*1400,decayLTCC,'r')



\end{lstlisting}


%
%

\bibliographystyle{ams}  
\bibliography{myBib}    

\end{document}